\documentclass{aa} % for the abstract without structuration

\usepackage{graphicx,txfonts}
\usepackage{natbib}

\usepackage{xcolor} % for colours

\newcommand{\simgreat} {\mathbin{\lower 3pt\hbox{$\rlap{\raise 5pt\hbox{$\char'076$}}\mathchar"7218$}}}
\newcommand{\simless}{\mathbin{\lower 3pt\hbox {$\rlap{\raise 5pt\hbox{$\char'074$}}\mathchar"7218$}}}

\begin{document}

\title{ Large Interstellar Polarisation Survey.}

\subtitle{III. Observational constraints on the structure of grains}

\authorrunning{R.~Siebenmorgen et al.}
\titlerunning {LIPS~III: Observational constraints on the structure of grains}

  \author {Ralf~Siebenmorgen\inst{1}, Stefano~Bagnulo\inst{2},
    Lapo~Fanciullo\inst{3,4}, Thomas Vannieuwenhuyse\inst{1}, Vincent
    Guillet\inst{5, 6} }

\institute{{European Southern Observatory, Karl-Schwarzschild-Str. 2,
85748 Garching, Germany {\tt email: Ralf.Siebenmorgen@eso.org}}
\and {Armagh Observatory and Planetarium, College Hill, Armagh BT61 9DG, UK}
\and {National Chung Hsing University, 145 Xingda Rd., South Dist., Taichung City 402, Taiwan}
\and {Tamkang University, 151 Yingzhuan Rd., Tamsui Dist., New Taipei City 251301, Taiwan}
\and {Institut d'Astrophysique Spatiale, CNRS, Univ. Paris-Sud, Universit\'{e} Paris-Saclay, B\^{a}t. 121, 91405 Orsay cedex, France}
\and {Laboratoire Univers et Particules de Montpellier, Universit{\'e}
  de Montpellier, CNRS/IN2P2, CC 72, Place Eug{\`e}ne Bataillon, 34095
  Montpellier Cedex 5, France }
}

\date{Received: June 6, 2025/ Accepted: March 30, 2026 }

\abstract{Dust in the diffuse interstellar medium remains incompletely
  understood with regard to the structure, composition, size
  distribution, and alignment properties of the grains. Joint
  observations of reddening, starlight polarisation spectra, and
  polarised dust emission for individual sightlines provide essential
  constraints on such properties. We study a far-UV selected sample of
  96 reddening curves, for which optical linear polarisation spectra
  were obtained with FORS at the VLT as part of the Large Interstellar
  Polarisation Survey (LIPS). Starlight polarisation spectra for 60
  stars are presented in this work. These data are combined with Gaia
  distance estimates and Planck thermal dust emission. A
  three-component dust model is made publicly available. It consists
  of nanoparticles, amorphous grains, and micrometre-sized dust
  agglomerates, varying axial ratios, porosities, sizes,
  element abundances, and alignment efficiencies to match the
  observations. The diversity of reddening and polarisation spectra
  is well reproduced by prolate grains with typical axial ratios of
  two, a porosity of 10\,\%, and high alignment efficiencies. Such
  efficiencies can be achieved with radiative torque alignment theory
  (RAT), but not with imperfect Davis–Greenstein (IDG) alignment,
  except when adjusting the magnetic-field orientation to maximise the
  polarisation. Micrometre-sized dust contributes
  wavelength-independent grey extinction in the optical, accounts for
  about one-third of the visual extinction, and carries one-third of the
  dust mass. A follow-up submillimetre survey with high-resolution
  polarimetry will further constrain grain shapes and alignment
  physics.}


\maketitle
\nolinenumbers

\section{Introduction}

Dust is ubiquitous in the interstellar medium (ISM) and plays a
significant role in many astrophysical processes. Clues to the
composition of interstellar dust come from various sources, such as
the observed elemental depletions in the gas phase and spectroscopic
signatures of dust \citep{HD21}. The observed extinction curves
display several spectral features, the most prominent being the
2175\,\AA\ bump, commonly attributed to graphite or polycyclic
aromatic hydrocarbons (PAHs). Another important constraint is provided
by dust emission, with conspicuous mid-infrared (IR) emission bands
often assigned to PAHs \citep{Allamandola1989, Puget89}. In addition,
strong dust bands at 9.7\,$\mu$m and 18\,$\mu$m arise from the Si--O
stretching and O--Si--O bending modes of silicate minerals
\citep{Dorschner95}, respectively. The extinction curve gives the dust
extinction as a function of wavelength and provides strong constraints
on dust models, in particular on the size distribution of the grains
\citep{Mathis77}.

The polarisation of starlight by dust extinction \citep{Hall_49,
  Hilt_49} and of polarised thermal dust emission \citep{Hild_88} show
that interstellar dust grains are non-spherical and aligned with
respect to the interstellar magnetic field. This is the result of a
rotating grain's axis of maximum inertia $\mathbf{a}$ aligning with
its angular momentum $\mathbf{J}$ (\textit{internal} alignment), while
$\mathbf{J}$ aligns with the magnetic field-line direction
$\mathbf{B}$ (\textit{external} alignment). Internal alignment is
generally agreed to be driven by the Barnett effect
\citep{Purc_79}. Theories suggested to explain external alignment
include paramagnetic relaxation \citep[Davis--Greenstein or DG
  effect;][]{DG_51} and the torque exerted by an anisotropic radiation
field on a helical grain due to the differential scattering/absorption
of left- and right-handed circular polarisation \citep[Radiative
  Torque or RAT theory;][]{DM_76, DW_96, Lazarian97}.

%%%%%%%%%%%%%%%%%%%%%%%%%%%%%%%%%%%%%%%%

\begin{table*}[!htb]
\begin{center} 
{\caption {Stars with derived Planck, reddening, FORS and Serkowski
    parameters. }}  \label{Tab1org.tab}
\scriptsize   
 \begin{tabular}{l  r | c c c c |l c r  r  | c  c  c  c | c  c  c }
%                 1  2   3 4 5 6  7 8 9 10    11 12 13 14  15 16 17 \\
    \hline\hline
    1 & 2& 3& 4& 5& 6& 7& 8& 9& 10& 11& 12& 13& 14& 15& 16 & 17 \\
    \hline
    \multicolumn{2}{c|}{Star}  &   \multicolumn{4}{c|}{PLANCK}  & \multicolumn{4}{c|}{Reddening}  &   \multicolumn{4}{c|}{FORS}   & \multicolumn{3}{c}{Serkowski} \\
\hline    
Name     & $||b||$  & $I_{850}$ & $p_{850}$  & $\theta_{850}$ & $A^{850}_{V}$ & $A_{V}$ & $A_{V}^{\rm ref}$& Ref & SM & Date & $p_{V}$ & $\theta_{V}$ & ${\rm{d}\theta}/{\rm{d}\lambda}$ &
   $p_{\rm {max}}$ & $\lambda_{\rm {max}}$ &  $k_{\rm {pol}}$ \\
         &  &MJy/sr  &\% & $^{\circ}$ &mag&mag&mag & & & &\% &$^{\circ}$ & $^{\circ}/\mu$m&\%&$\mu$m& \\
\hline
HD~024263  & 35 &   1.00 &    6.0$\pm$   2.4 &    77$\pm$   18 &    0.8 &   $-$ &    0.7 &   V & S &  2019-02-24 &    1.1  $\pm$  0.1 &   149  $\pm$  0.5 &      2 $\pm$  0.7 &    1.08 &    0.58 &    1.04 \\
$\cdots$  & &    &     &  &    &  &   &   &  &  & &  & &  &  &   \\
Walker~67  &  1 &   58.9 &    1.1$\pm$   2.3 &    91$\pm$   16 &    49 &   $-$ &    0.7 &   F & S &         B17 &    4.1  $\pm$  0.3 &    17  $\pm$  0.6 &     -6 $\pm$  1.2 &    5.17 &    0.81 &    1.47 \\
\hline
 \end{tabular}
\end{center}
%  \scriptsize
{\textbf{Notes.} The entries for the 96 stars are provided in the
  appendix Table~\ref{Tab1.tab}. The columns are explained in
  Sect.~\ref{sample.sec}, B17 referes to \citet{B17}, and S14 to
  \citet{S14}.}
\end{table*}
%%%%%%%%%%%%%%%%%%%%%%%%%%%%%%%%%%%%%%%%

In the imperfect Davis--Greenstein (IDG) alignment \citep{HG80, Vosh12}
the grain wobbles and rotates around its axis of greatest momentum
while also precessing around the magnetic field vector. A major
criticism pointed out by \cite{JS67} and \cite{RL99} is that the DG
model ignores internal alignment and therefore is a physically
incorrect simplification.

Following the development of an analytical model for RATs
\citep{Lazarian07}, and due to its good qualitative agreement with
polarimetric observations \citep[e.g.,][]{Andersson15}, interest in
RATs has increased significantly in the last two decades, and RAT
theory itself has been significantly expanded. The high polarisation
ceiling observed in dust thermal emission
\citep[$\sim20$\%;][]{PlanckXXI} requires that dust grains be more
efficiently aligned than either the DG effect or the RAT alignment in
its original form can account for. A solution found by \cite{HL_16} is
to consider grains with iron inclusions, resulting in increased grain
magnetic susceptibility, and therefore alignment: this is the
Magnetically Enhanced RAT, or MRAT, theory. The MRAT theory
successfully accounts for grain alignment from the diffuse ISM to
dense star-forming regions \citep{Giang_2025}. Note that, while DG
torques are also enhanced by iron inclusions, they are typically
significantly weaker than MRAT torques.

In this article observational constraints are correlated with the
physical properties of interstellar dust. A dust model is tested
against representative element depletions, stellar distance estimates,
and the characteristics of reddening and polarisation in absorption
and, where available, in emission, along individual sightlines through
the diffuse ISM. Particular attention is given to micrometre-sized
dust agglomerates that absorb a fraction of the interstellar radiation
field (ISRF, \cite{Mathis83, Bianchi24}). Because they are large, they
remain cold and emit at long wavelengths. Initially, very cold (10\,K)
dust emission was detected in our Galaxy toward high-density regions
\citep{30} and in non-active galaxies \citep{Chini95}. This cold dust
was later confirmed by ISO \citep{32,33}. More recently, excess
emission at 0.5\,mm observed by Herschel could not be explained by a
single modified blackbody temperature component \citep{34, 35,36},
with similar results confirmed using ALMA \citep{37} and LABOCA
\citep{38} at even longer wavelengths. Furthermore, micrometre-sized
particles originating from the local interstellar cloud surrounding
our solar system ISM were directly measured in situ by the Ulysses,
Galileo, and Stardust space probes \citep{26, 27, 28}. They appear in
sightlines associated with the cold ISM \citep{S20}.

In the ISM, a grey component of micrometre-sized grains was introduced
by \cite{Mathis77} and by \cite{WL15a, WL15b} to account for the
observed IR extinction. Such grains have also been incorporated into
other dust models \citep{Vosh04, KS94, K08, Ormel11,
  Ysard24}. Recently, the impact of grey extinction on Type Ia
supernova distance measurements was analysed by the Dark Energy Survey
Collaboration \citep{Popovic24}. The submillimetre excess continuum
emission in the Milky Way detected by Planck \citep{PlanckXII} can be
matched by adjusting the grain emissivity at these wavelengths
\citep{HD21}. However, such models fail to resolve the discrepancy
between trigonometric distance estimates provided by the \cite{DR3}
and the overprediction of the luminosity distance of the same stars
\citep{S25}. Unification between luminosity and trigonometric distance
estimates could be established by considering a population of
micrometre-sized dust, which provides the necessary additional dimming
of starlight. These grains are large enough to produce consistent
reddening and grey extinction at wavelengths shorter than 1\,$\mu$m.

For this purpose the Large Interstellar Polarisation Survey (LIPS) was
performed to measure the starlight polarisation spectra of 161 stars
using the FORS instrument \citep{Appenzeller98} on the ESO Very Large
Telescope. The observations covered a wavelength range of $0.38 -
0.95\,\mu$m at a spectral resolving power of $\sim880$. In LIPS~I
\citep{B17}, a catalogue of 127 linear polarisation spectra
corresponding to 101 sightlines was published.

\section{The sample and data \label{sample.sec}}

Observing sightlines that intersect different components of the ISM
introduces complexities in relating extinction and polarisation data
to physical dust parameters. To address this issue in LIPS~II
\citep{S18} stars were observed with the high-resolution spectrograph
UVES, which offers a resolving power of $\lambda /\Delta \lambda \sim
75,000$ \citep{UVES1, UVES2}. These spectra were used to confirm the
spectral type and luminosity class of the stars used for the reddening
curve determination and to examine the profiles of interstellar
absorption lines, particularly K{\sc i}. The concept of "single-cloud
sightlines" was introduced, referring to cases where a dominant
Doppler component accounts for more than half of the observed column
density. A total of 65 such rare single-cloud sightlines were
identified. It was found that interstellar polarisation is lower for
multiple-cloud sightlines compared to single-cloud sightlines,
indicating that the presence of additional clouds depolarises the
transmitted radiation. Furthermore, significant variations in dust
properties between different clouds were inferred from dust modelling.

In this work (LIPS~III), the sample is expanded with additional
starlight polarisation spectra for 60 stars. The same observing
strategy, data reduction, and calibration procedures as detailed in
LIPS~I are applied \citep{B17}.  Combining reddening and polarisation
continuum observations is necessary to constrain the nature of
interstellar dust, including its chemical composition and size
distribution. For instance, the reddening rise in the far-ultraviolet
(UV) is indicative of very small nanoparticles, and so is the
2175\,$\AA$ bump \citep{SD65, Blasberger17} which is tied to
carbonaceous nanoparticles. The shape of the polarisation curve in the
optical \citep{Serkowski} can be used to constrain the size
distribution of aligned dust grains \citep{KM95, Vaillancourt20}.

The reddening curves have been derived in the near-IR ($J H K$) using
the Two Micron All Sky Survey (2MASS) \citep{Cutri}, in the optical
($U B V$) from ground-based facilities \citep{V04}, and in the far-UV
below $0.3\,\mu$m down to the Lyman limit from space-based
observations. At these short wavelengths, the International
Ultraviolet Explorer (IUE) and the Far Ultraviolet Spectroscopic
Explorer (FUSE) observed spectra for 417 stars \citep{V04}, 328 stars
\citep{FM07}, and 75 stars with FUSE \citep{G09}. Furthermore,
distances derived from Gaia parallaxes were used to estimate the
reddening at infinite wavelength, providing an estimate of the visual
extinction $A_{V}$ (\cite{S25}, Eq.~\ref{Av.eq}).

The LIPS sample is further complemented by observations of polarised
dust emission obtained from the Planck observatory at $850\,\mu{\rm
  m}$ \citep{PlanckXII}. The Planck total intensity and polarisation
data is derived following the procedure outlined by \cite{Guillet18}
and colour corrected \citep{PlanckXVII}.

The available $0.09$–$2.3\,\mu$m reddening curves, complemented by
UVES spectroscopy, Planck 850\,$\mu$m (353\,GHz) polarimetry, and
$0.38$–$0.92\,\mu$m FORS spectropolarimetry, constitute the sample
under investigation. It includes 96 stars, comprising 36 FORS
polarisation spectra previously published in LIPS~I and LIPS~II, and
60 FORS polarisation spectra presented here. The characteristics of
the sample are summarised in Table~\ref{Tab1org.tab}, which lists the
following 17 columns: For each star (col.~1), we specify the absolute
Galactic latitude $|b|$ (col.~2). The Planck results are presented in
four columns: surface brightness ($I_{850}$ in MJy/sr) in col.~3,
fractional polarisation ($p_{850}$ in \%) in col.~4, the polarisation
angle in equatorial coordinates ($\theta_{850}$ in $^{\circ}$) in
col.~5, and an estimate of the visual extinction $A_V^{850}$ (col.~6),
which is based on the Planck map of the dust optical depth at
850\,$\mu$m \citep{Planck15}.

The visual extinction $A_{\rm V}$ as determined from the GAIA parallax
$\pi$ (col.~7), and the reference extinction $A_{\rm V}^{\rm ref}$
(col.~8) are provided. The latter is estimated by extrapolating
optical/near-IR reddening to infinite wavelength, as given in the
reddening curve catalogues by \cite{V04} (labelled V), \cite{FM07}
(labelled F), and \cite{G09} (labelled G) in col.~9. We classify 55
single-cloud sightlines as 'S' and 41 multi-cloud sightlines as 'M'
(col.~10).

The results of the FORS spectropolarimetry are summarized across seven
columns. Observing dates are listed in col.~11. For stars observed
multiple times, the final polarisation spectra are derived by
averaging the Stokes parameters from individual observations. The
fractional polarisation ($p_{\rm V}$) and polarisation angle in
equatorial coordinates ($\theta_{\rm V}$) at $0.55\mu$m are provided
in cols.~12~-~13. The gradient in the polarisation angle along the
spectrum, ${\rm{d}\theta}/{\rm{d}\lambda}$ ($^{\circ}/\mu{\rm m}$) is
given in col.~14. In the optical, the observed interstellar
polarisation spectra can be well approximated by a matematical
expression known as Serkowski (1973) formulae:
\begin{equation}
p(\lambda) = p_{\max}\,\exp \left[ -k_{\rm {pol}} \ \ln^2
    \left( \frac{\lambda_{\max}}{\lambda} \right) \right]\,,
\label{serk.eq}
\end{equation}
The Serkowski parameters ($p_{\rm {max}}$, $\lambda_{\rm {max}}$, and
$k_{\rm {pol}}$) derived from spectral fits to the FORS polarisation
spectra are provided in cols.~15–17. The Serkowski fits for 43 stars,
for which the available data do not permit detailed dust modelling, are
shown in Fig.~\ref{Fig1.pdf}. The spectral variation of the FORS
polarisation angle, corrected for the optical reference value ($\theta
- \theta_V$) is shown in Fig.~\ref{FigTheta.pdf}.

In the LIPS sample, 27 stars are included in the stellar polarisation
catalogue by \citet{Heiles}, who detect 19 of these stars with high
confidence at $p \gtrsim 0.6\,\%$. For these 19 stars, the linear
polarisation agree in both catalogues, at $|p_{\rm {FORS}} - p_{\rm
  {Heiles}}| = 0.14 \pm 0.08\,\%$. The polarisation angles are
consistent within $3^{\circ}$, except for HD~092044, where the
polarisation angles differ by $18^{\circ}$.

%%%%%%%%%%%%%%%%%%%%%%%%%%%%%%%%%%%%%%%%%%%%%%%%%%%%%%%

\section{Dust model}

We apply the dust model by \citet{S23}, which is consistent with
current observational constraints on dust in the diffuse ISM
\citep{HD21}. The model adopts representative solid-phase elemental
abundances and successfully reproduces the observed
wavelength-dependent reddening, emission, and polarisation from
interstellar dust, spanning from the UV to microwave
wavelengths. Furthermore, the model includes grey extinction by
micronmetre grains, which reduces the luminosity distance, enabling
consistency with trigonometric distances derived from Gaia parallaxes.

\subsection{Grain composition and size distributions}

The number densities of grains follow a power-law size distribution,
$n(r) \propto r^{-q}$, where $r$ is the grain radius, and $q$ the
power-law exponent assumed to be the same for each of the three dust
components: {\it {i) Nanoparticles}} ($r \simless 6$\,nm), including
very small silicate (vSi), graphite (vgr), and polycyclic aromatic
hydrocarbon (PAH). {\it {ii) Amorphous grains}} of silicate (aSi) and
carbon (aC) ($6\,\rm{nm} \simless \ r \simless 250$\,nm). These grains
are considered to have prolate shapes, rather than oblate, as the
former provide a better fit to observed linear polarisation spectra
\citep{S14}. The mean radius of the amorphous grains, averaged over
the size distribution, is typically $\bar{r}_{\rm{aC, aSi}} \sim
30$\,nm.  {\it {iii) Micrometre-sized dust agglomerates}}
($250\,\rm{nm} \simless \ r < 3\,\mu$m) are treated as porous
composites of amorphous silicate and carbon grains The mean radius of
these micrometre-sized prolate shaped grains remains below
$1\,\mu$m. We denote $r_i^-$ and $r_i^+$ the low and upper limit of
grain sizes for population $i$, respectively.  The radius of spheroids
is defined as that of a sphere of same volume ($r^3 = ab^2$), where
$a$ is the grain major axis and $b$ its minor axis.

Optical constants are adopted from \citet{Zubko96} for amorphous
carbon, \citet{Draine03,DH21} for graphite and astro-silicate, and
\citet{Demyk22} for amorphous silicates, assuming a 97:3 mix in mass
of MgO$-$0.5 SiO$_2$ and Mg$_{0.8}$Fe$^{2+}_{0.2}$ SiO$_3$. The
optical constant of porous and composite grains with vacuum inclusions
are computed using the Bruggemann mixing rule. The molecular weights
are $\mu_{\rm C} = 12$ for carbon materials, $\mu_{\rm{Si}} = 135$ for
astro-silicate, and $\mu_{\rm {aSi}} = 100$ for amorphous
silicates. The bulk densities (g/cm$^{3}$) are for nanoparticles
$\rho_{\rm {vgr}} = 2.2$, $\rho_{\rm {vSi}} = 3.5$, carbon particles
$\rho_{\rm {aC}} = 1.6$, amorphous silicates $\rho_{\rm {aSi}} = 2.7$,
and in micrometre-sized grains $\rho_{\mu \rm{Si}} = 3.4$.

%%%%%%%%%%%%%%%%%

\subsection{Grain alignment}

\cite{Draine09} introduced an alignment function where $f_{\rm
  {align}} = 0$ represents random orientation and ${f_{\rm {align}}} =
1$ corresponds to perfect spinning alignment. The alignment function
${\tilde{f}_{\rm {align}}}(r)$ is strongly size-dependent: ${f_{\rm
    {align}}} = 0$ for $r \simless 50$\,nm, then increases for larger
grains, reaching its maximal value for $r_{\rm {pol}}^- \sim 100$\,nm,
remaining constant for larger grains. This modelling is consistent
with the size-dependence of alignment efficiency observed for the
individual sightlines analysed in LIPS~II \citep{S18}: the starlight
polarisation spectra are well reproduced when only large particles, at
$r \simgreat r_{\rm {pol}}^-$, are aligned, while smaller particles
remain unaligned.

Radiative torques can align the angular momentum vector $\bf J$ of the
grains with the magnetic field $\bf B$ in two distinct regimes
characterised along the high-$\bf J$ and low-$\bf J$ attractor
points. The high-$\bf J$ attractor corresponds to perfect alignment and
applies to materials with ferromagnetic inclusions. The low-$\bf J$
point is less well constrained, and so is the fraction of high-$\bf J$
to low-$\bf J$ attractors, which is of interest when applied to
paramagnetic materials. We apply a simplified RAT alignment model
similar to \cite{Reissl20}, with alignment efficiency

\begin{equation} \label{eta.eq}
 {\tilde{f}_{\rm {align}}}(r) = \begin{cases} {f_{\rm {align}}} &: r^{-}_{\rm {pol}} \simless r \simless
  r_{\rm {des}} \,, \\ 0 &: \rm{otherwise.} \end{cases}
\end{equation}

Grains below a minimum alignment radius $r^{-}_{\rm {pol}}$ are
randomly oriented. Suprathermally rotating particles with radius above
$r_{\rm {des}} \sim 1\,\mu$m disrupt because of the centrifugal
stress \citep{Hoang21}. The destruction radius depends on material
properties of the grains, such as tensile strength, composition, and
structure, as well as on environmental parameters, including
intensity, hardness, and anisotropy of the ISRF. Such large grains do
not contribute to the optical polarisation. For simplicity, we set
$r_{\rm {des}}$ to the maximum grain radius. Only grains with radii
$r^{-}_{\rm {pol}} \simless r \simless r_{\rm {des}}$ are
aligned at constant efficiency ${f_{\rm {align}}}$, where we neglect a
mild size dependence or a fraction of low-$J$ attractors
\citep{HoangTruong24}.

\begin{figure}
\center{
\includegraphics[width=9cm]{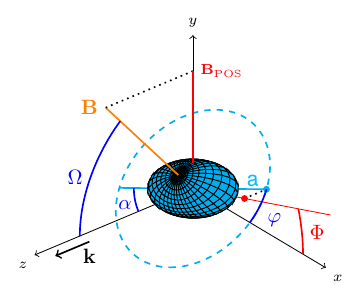}}
\caption{Geometry of a perfectly aligned spinning prolate particle
  with long-side symmetry axis $\bf a$. The sky is in the $xy$ plane.
  The angle $\Phi$ (red) is between the $x$ axis and the projection of
  $\bf a$ onto the plane of the sky. The sightline is the $z$ axis,
  with electromagnetic wavevector $\vec k$ approaching towards us. The
  magnetic field $\bf B$ lies in the $yz$-plane with magnetic field
  angle $\Omega$ measured from $z$ towards $\bf B$. The grain
  spinning plane is perpendicular to $\bf B$ (blue dashed circle) and
  includes $\bf a$. The grain spinning angle $\varphi$ (blue) is
  measured to the $x$ axis. The angle of incidence $\alpha$ is given
  between $\bf a$ and $\bf k$. Note the relation between $\Omega$,
  $\alpha$, and $\varphi$ as given in
  Eq.~\ref{Eq.angle}. \label{fig:spinning}}
\end{figure}

\subsection{Computing grain cross-sections over grain dynamics}

Figure \ref{fig:spinning} presents the geometry of the problem for a
prolate grain with a spinning axis perfectly aligned with the magnetic
field. The plane of the sky is the $xy$-plane. The sightline is along
the $z$ axis and light coming towards us ($\bf k$). The magnetic field
$\bf B$ lies in the ($yz$) plane. The magnetic field angle $\Omega$ is
measured from the sightline $z$ towards $\bf B$. Maximum polarisation
is observed for $\Omega=90^{\circ}$ and no polarisation for
$\Omega=0^{\circ}$. The spinning plane of the grain is perpendicular
to $\bf B$, with $\varphi$ the spin angle of the grain symmetry axis
$\bf a$ with respect to the $x$ axis. To compute the grain extinction
and polarisation cross-sections, we consider a linearly polarised
electromagnetic wavevector $\vec k$. The plane of incidence is the
plane containing $\bf k$ and $\bf a$. We denote $\alpha$ the angle
between $\bf a$ and $\bf k$, whose sector is contained in the plane of
incidence. We also define $\Phi$ as the angle between the $x$ axis and
the projection of $\bf a$ onto the plane of the sky. The angles
$\alpha$ and $\varphi$ are related to $\Omega$ and $\Phi$:

\begin{align} \label{Eq.angle}
\cos\alpha & = \sin\Omega \cos\varphi\,,\\
%\cos2\Phi & = 1- \frac{2\cos^2\varphi\cos^2\Omega}{1-\cos^2\varphi\sin^2\Omega} \,.
\tan\Phi & = \cos\Omega\tan\varphi \,.
\end{align}

For symmetry reasons, for the calculations one only needs to consider
values for $\Omega$ and $\varphi$ between 0 and $\pi/2$, resulting in
values for $\alpha$ and $\Phi$ also between 0 and $\pi/2$. The
polarisation cross-section, integrated over the spinning angle
$\varphi$, expresses (focusing here on the dependence on angles and
radius only):
\begin{equation}\label{Cpol.eq}
C_{\rm {pol}}(\Omega , r)=2 \, {f_{\rm {align}}}\,r^{2}\int_0^{\pi/2}
\bigl(Q_{\rm {ext}}^{\rm TM}(\alpha ,r)-Q_{\rm {ext}}^{\rm TE}(\alpha ,r)\bigr)
\cos 2\Phi\,\mathrm{d}\varphi\,.
\end{equation}

where $Q_{\rm {ext}}^{\rm TE}$ (resp. $Q_{\rm {ext}}^{\rm TM}$)
is the Transverse Electric (resp. Transverse Magnetic) grain
extinction cross-section computed for an electromagnetic wave of
wavevector $\vec k$ with an oscillating electric field vector $\vec E$
(resp. oscillating magnetic field vector $\vec H$) perpendicular to
the plane of incidence as defined by \citet{BH83}. The efficiency
factors $Q(\lambda, r, a/b, m, \alpha)$ depend on the wavelength
$\lambda$, the grain radius $r$, the axial ratio $a/b$, the optical
constants $m$, and the angle of incidence $\alpha$, which depends in
the given alignment model on the magnetic field angle $\Omega$
(Eq.~\ref{Eq.angle}). Computation of the efficiency factors for large
spheroids becomes difficult. They are computed using \cite{V93} and
custom software provided by \cite{Vosh04}. The code converges for size
parameters $x = 2\pi a/\lambda$ up to $|m - 1| \, x \sim 22$. At such
or even larger values of $x$, we replace the extinction cross-section
of the spheroids with that of spheres using Mie theory and set the
polarisation $Q_{\rm {pol}}(\lambda) = 0$. At $x > 22$, such
large grains do not contribute significantly to the observed starlight
polarisation in the optical. The absorption and scattering
cross-sections for aligned particles is
{\small{
\begin{equation}\label{Cextalign.eq}
C^{\rm {align}}_{\rm {abs, sca}}(\Omega , r)=4 \, {f_{\rm {align}}}\,r^{2}\int_0^{\pi/2}
\bigl(Q_{\rm {abs, sca}}^{\rm TM}(\alpha ,r)+Q_{\rm {abs, sca}}^{\rm TE}(\alpha ,r)\bigr)
\cos 2\Phi\,\mathrm{d}\varphi\,.
\end{equation}
}}

For randomly aligned grains, $C^{\rm {rand}}_{\rm {abs,sca}}$, we
apply Eq.~3.34 by \cite{Vosh12}. The grain absorption and scattering
cross-sections are
\begin{equation}\label{Cext.eq}
C_{\rm {abs, sca}}(\Omega, r) = {\tilde{f}_{\rm {align}}}(r) \ C^{\rm {align}}_{\rm {abs,sca}}(\Omega , r) \ +
\ \left(1 - {\tilde{f}_{\rm {align}}}(r)\right) \, C^{\rm {rand}}_{\rm {abs,sca}}(r) \,,
\end{equation}
and $C_{\rm {ext}} = C_{\rm {abs}} + C_{\rm {sca}}$ the grain extinction
cross-section (cm$^2$).

%%%%%%%%%%%%%%%%%%%%%%%%%%%%%%
\subsection{Cross-sections per unit mass of dust}

The total absorption, scattering, or polarisation cross-section
$K_{i}(\lambda)$ per unit mass (cm$^2$/g-dust) for dust component $i$
is
\begin{equation} \label{Eq.6}
K_{i}(\lambda) = \frac{3}{4 \pi} \ \frac{m_i}{\rho_i}\ \frac {\int_{r^-_{i}}^{r^+_{i}} C_{i}(r, \lambda) \ r^{-q}
    \ \mathrm{d}r} {\int_{r^-_{i}}^{r^+_{i}} r^{3-q} \ \mathrm{d}r}\ ,
\end{equation}

where $C_i(r,\lambda)$ is the corresponding absorption, scattering, or
polarisation cross-section for a grain of radius $r$ at wavelength
$\lambda$ (Eqs.~\ref{Cpol.eq}–\ref{Cext.eq}), $\rho_i$ is the bulk
density for population $i$, and $m_i$ is the mass of component $i$ per
unit mass of dust:
\begin{equation} \label{Eq.mi}
  m_{\rm {i}} = \mu_{\rm i} \ {\frac{ [\rm{X_{\rm {i}}}]}{[\rm{H}]}}
  \bigg{/} \sum_{i}{\mu_{\rm i} \ {\frac{ [\rm{X_{\rm {i}}}]}{[\rm{H}]}}} \,.
\end{equation}

The elemental abundance in the dust relative to hydrogen in the gas
phase $[\rm{X_{\rm {i}}}]/[\rm{H}]$ is constrained to respect the
depletion limits \citep{HD21,S23} so that
\begin{equation}
  \frac{[\rm{C}]}{[\rm{Si}]} < 5.2 \ .
  \label{abu.eq}
\end{equation}

The total dust cross-section, $K(\lambda)$, is the sum of $K_{i}$ from
all components\footnote{A suffix or index `{abs}' indicates
absorption, `{ext}' extinction, `{pol}' polarisation, `{sca}'
scattering, `{des}' destruction, `{tot}' total, and `$\mu$A'
micrometre-sized agglomerates.}.

\subsection{Dust observables in extinction}

The optical depth $\tau_V = A_{V}/1.086$ is
\begin{equation}
\tau_V = N^{\rm na} \ K^{\rm na}_V \ + \ N^{\mu{\rm A}} \ K^{\mu{\rm A}}_V \ ,
\end{equation}

where $N^{\rm na}$ represents the sum of the dust column density of
nanoparticles and amorphous particles, while $N^{\mu{\rm A}}$ is the
dust column density of micrometre-sized grains. The corresponding mass
extinction cross-sections are denoted by $K^{\rm na}_V$ and
$K^{\mu{\rm A}}_V$, respectively. At infinite wavelengths, $K(\infty)
= 0$, so that $A_{V}=-E(\infty - V) > - E(H -V)$ in the $H$-band. The
reddening $E(B-V) = 1.086 \ (\tau_B - \tau_V)$ expresses with our
notations
\begin{equation}
E(B-V) = N^{\rm na} \ (K^{\rm na}_B - K^{\rm na}_V) \ + \ N^{\mu{\rm A}} \ (K^{\mu{\rm A}}_B - K^{\mu{\rm A}}_V) \ , 
\end{equation}

providing a second constraint for estimating the relative mass
fraction of the micrometre-sized grains, $m_{\mu{\rm A}} = N^{\mu{\rm
    A}} / (N^{n} + N^{\mu{\rm A}})$. From this we derive the absolute
reddening of the model
\begin{equation} \label{absredd.eq}
  E(\lambda - V) = \frac{2.5}{\ln 10} \ \left(\tau_{\lambda} - \tau_V \right) 
\end{equation}

and the starlight polarisation spectrum 
\begin{equation} \label{polabs.eq}
  p(\lambda) = N \ K_{\rm {pol}}(\lambda) \ ,
\end{equation}

where $N = N^{\rm na} + N^{\mu{\rm A}}$\,(g-dust/cm$^2$) is the total
dust column density and $K_{\rm {pol}}$\,(cm$^2$/g-dust) the total
linear polarisation cross-section (Eq.~\ref{Eq.6}).

The dust model accounts for representative solid-phase element
abundances of the main absorbing dust components of the assumed grain
stoichiometry and explains phenomena such as wavelength-dependent
reddening, starlight polarisation, and the emission of unpolarised and
polarised light. It also provides the necessary grey extinction for
reconciling the luminosity distances with the Gaia parallaxes
\citep{S25}. We will confront it to the LIPS sample.

\subsection{Infrared emission per gram of dust}

The emissivity $\epsilon_i(r)$ per gram of dust for a grain of
population $i$ and particle radius $r$ is determined from the energy
balance between emission and absorption of photons from the mean
intensity $J^{\rm{ISRF}}(\lambda)$ of the ISRF \citep{Mathis83}.

\begin{equation} 
  {\int} {K_{{abs}, i}} (\lambda, r) \, J^{\rm{ISRF}}(\lambda) \,
    \mathrm{d}\lambda = {\int} {K_{{abs}, i}} (\lambda, r) \, P(T) \,
    B_{\lambda}(T) \, \mathrm{d}T \, \mathrm{d}\lambda \,,
\label{emis.eq}
\end{equation}

where $\epsilon_{i}(r)$ is given by the right-hand side, $T$ refers to
the temperature of material $i$ and particle radius $r$,
$B_\lambda(T)$ is the Planck function, and $P(T)$ is the temperature
distribution function \citep{GD89, K08}. This function is evaluated
using an iterative scheme by \cite{S92}. Quantum heating of the dust,
and thus $P(T)$, needs to be evaluated only for nanoparticles, as
$P(T)$ approaches a $\delta$-function for the larger amorphous and
micrometre-sized grains. The total emissivity, $\epsilon$, is the sum
of the emissivity from all dust components.

The total polarised emission $\epsilon_{\rm {i,pol}}$ of population
$i$ is computed by summing the contributions from the minimum
alignment radius, $r_{\rm {i,pol}}^-$, to the maximum radius
$r_{i}^+$:

\begin{equation}
  \epsilon_{\rm {i,pol}} (\lambda, r) = {f_{\rm {align}}} {\int}_{r^-_{\rm {i,pol}}}^{r^+_{\rm i}} 
    K_{\rm {i,pol}} (\lambda , r) \, B_{\lambda}(T) \, \mathrm{d}r \ , 
\label{emispol.eq}
\end{equation}

where $K_{\rm {i,pol}}$ is given by Eq.~\ref{Eq.6}, and $T$ refers
to the temperature of material $i$ and particle radius $r$. The total
polarised dust emission, $\epsilon_{\rm {pol}}$, is the sum of the
polarised emission from all components contributing to the
polarisation, which include the amorphous and the micrometre-sized
grains. The corresponding fractional polarisation from dust emission
is

\begin{equation} \label{polemis.eq}
  p = \frac{\epsilon_{\rm {pol}}}{\epsilon}  \ .
\end{equation}

\subsection{Infrared emission per H atoms}

More material along a given sightline will increase both $E(B - V)$
and $N_{\rm H}$. Observationally, it is assumed that the reddening
scales approximately linearly with the dust column density and, if
well mixed, also with the hydrogen column density such that $N_{\rm H}
/ E(B - V)$ remains roughly constant. \cite{Bohlin78} derived $N_{\rm
  H} / E(B - V) = 5.8$ (with this value and subsequent values given in
$10^{21} \text{H-atoms cm}^{-2}\, \text{mag}^{-1}$), which is close
to the value of 5.9 found for translucent clouds
\citep{Rachford09}. However, significantly different values have been
reported: 4.9 \citep{Diplas94}, 7.5 \citep{Ensor17}, 9.4
\citep{Nguyen18}, $N_{\rm HI} / E(B - V) = 8.3$ \citep{Liszt14} and 8.8
by \cite{Lenz17}. These differences are consistent with systematic
variations in the gas-to-dust mass ratio, with lower values in the
Galactic plane and higher values at high Galactic latitudes.

The total dust mass, $M_{\rm {dust}}$, is estimated by summing all
atoms depleted from the gas phase and scaling by the molecular weights
corresponding to the assumed grain stoichiometry. The gas mass,
$M_{\rm {gas}} \sim 1.4 \, M_{\rm {H}}$, is calculated by summing the
contributions of helium and hydrogen, assuming a He:H ratio of
1:10. At high Galactic latitudes, the derived gas-to-dust mass ratio
is $M_{\rm {gas}} / M_{\rm {dust}} \sim 125$ \citep{HD21,S23}. This
ratio may vary by up to $\sim 50\%$ while still being consistent with
elemental depletion. However, because we use relative dust abundances
in our dust model, such variations, if applied to all components
(Eq.~\ref{Eq.mi}), do not affect the fit to the reddening curves.

The gas-to-dust mass ratio and the hydrogen column density are used to
scale the dust emission in the model $\epsilon$ (Eq.~\ref{emis.eq}) in
erg s$^{-1}$ Hz$^{-1}$ sr$^{-1}$ per g-dust to the Planck surface
brightness $I$ (erg s$^{-1}$ Hz$^{-1}$ sr$^{-1}$ cm$^{-2}$ per
H-atom) at 353\,GHz, with atomic mass unit $m_{\rm u}$:

\begin{equation}
N_{\rm H} = \frac{1}{m_u} \ \frac{M_{\text{gas}}}{M_{\text{dust}}} \ \frac{I}{\epsilon} \ . 
\label{NH.eq}
\end{equation}

For our nominal dust composition (Sect.~\ref{sightlines.sec}), we find
that the $N_{\rm H} / E(B - V)$ ratio is 6.3 for HD~027778, 7.6 for
HD~108927, and 7.3 for HD~287150, in units of
$10^{21}\,\text{H-atoms cm}^{-2}\,\text{mag}^{-1}$. These values
are consistent with the reference values and fall within the
uncertainty range of the gas-to-dust mass ratio
$M_{\text{gas}} / M_{\text{dust}}$.

%%%%%%%%%%%%%%%%%%%%%%%%%%%%%%%%%%%%%%%%%%%%%%%%%%%%%%%%%%%%%

\subsection{Dust Model Fitting Procedure \label{pro.sec}}

The dust model is applied to sightlines with available high-quality
far-UV selected reddening curves, visual extinction values
derived from Gaia distance estimates ($A_V$, Eq.~\ref{Av.eq}), near-IR
reddening using 2MASS, starlight polarisation spectra obtained with
FORS, and the colour-corrected \citep{PlanckXVII} total and polarised
emission observed by Planck. Best-fit dust parameters are derived
using a three-step iterative procedure, under the assumption that
along a given sightline where dust extinction dominates, the magnetic-field
direction does not vary significantly. For sightlines where the
optical polarisation spectrum does not follow the Serkowski curve
(Sect.~\ref{indi.sec}), the inferred magnetic-field direction should
be regarded with caution.

Initially, the reddening curve is fitted using the publicly available
$\chi^2$ minimisation tool \texttt{absredgaia} \citep{S25code}. This
tool returns the $\chi_r^2$ of the best fit to the reddening curve and
the seven model parameters: the exponent of the size distribution
($q$) and the relative mass fractions of the different dust components
$m_{\text{vgr}}$, $m_{\text{vSi}}$, $m_{\text{PAH}}$, $m_{\text{aC}}$,
$m_{\text{aSi}}$, $m_{\mu{\rm A}}$. These relative dust masses are
linked to the element abundances (Eq.~\ref{Eq.mi}). The tool adheres
to the depletion limits set by Eq.~\ref{abu.eq}. Dust parameters of
the general field of the ISM \citep{S23} are chosen as starting
parameters, with an upper radius $r^+_{\mu{\rm A}} = 3\,\mu$m.

In the second step, the starlight polarisation spectrum is fitted by
varying the minimum alignment radii of the amorphous carbon and
silicate grains, $r^-_{\rm {pol},\,aC}$ and $r^-_{\rm {pol},\,aSi}$,
together with the magnetic-field orientation, using $\Omega =
45^{\circ}$ as the initial value. The polarisation spectra of the dust
model $p(\Omega, r^-_{\rm {pol},\,aC}, r^-_{\rm {pol},\,aSi},
\lambda)$ are compared with the FORS data, and the corresponding
$\chi_{\rm {pol}}^2$ values are computed. The grain radii in the
model grid increase by 5\% from one bin to the next. The dust model is
evaluated for $40 \times 40$ pairs of $r^-_{\rm {{pol}, i}}$ with $i
\in \{{\rm {aC, aSi}}\}$ in the range 50 - 250\,nm. The values of
$r^-_{\rm {pol},\,aC}$ and $r^-_{\rm {pol},\,aSi}$ that best match the
shape of the observed polarisation spectra are identified. Next, the
magnetic-field orientation $\Omega$ is varied and $C_{\rm
  {pol}}(\Omega, r)$ (Eq.~\ref{Cpol.eq}) computed. The model that best
reproduces the Serkowski maximum $p_V$ (col.~15 of
Table~\ref{Tab1org.tab}) is selected. For this value of $\Omega$, a
new set of $r^-_{\rm {pol},\,aC}$ and $r^-_{\rm {pol},\,aSi}$ is
derived. The best-fitting parameters ($r^-_{\rm {pol},\,aC}$,
$r^-_{\rm {pol},\,aSi}$, $\Omega$) are extracted from the model that
minimises $\chi_{\rm {pol}}^2$. Because the extinction cross-section
depends on the magnetic-field orientation (Eq.~\ref{Cext.eq}), both
steps—fitting the reddening and fitting the polarisation
spectra—are iterated.

\cite{PlanckXII} introduced two criteria for constraining dust models
by examining the ratio of submillimetre-to-optical polarisation. These
are: the ratio of the fractional polarisation at 850\,$\mu$m to the
optical polarisation efficiency and the ratio of the polarised
emission intensity $P_{850} = p_{850} \ I_{850}$\,(MJy/sr) to the
optical polarisation,
\begin{align} 
R_{\rm{S/V}}   & = p_{850} / (p_V / \tau_V) \,\label{Eq.RSv} ,\\  
R_{\rm{P/p}} & = P_{850} / p_V \ . \label{Eq.Rpp}
\end{align}

In the third step, the best-fit parameters from step~2 are retained,
and the upper grain radius is varied within the range $0.25 <
r^+_{\mu{\rm A}} \lesssim 3\,\mu$m. This results in 50 models for
which the Planck-to-FORS polarisation ratios $R_{\rm{S/V}}$ and
$R_{\rm{P/p}}$, along with their corresponding goodness-of-fit
parameters $\chi_r^2$, $\chi_{\rm {pol}}^2$, $\chi_{R_{\rm{S/V}}}^2$,
and $\chi_{R_{\rm{P/p}}}^2$, are computed. These $\chi^2$ values are
normalised to their respective median. A total goodness-of-fit
parameter is then derived by assigning equal weight to the reddening
curve, the optical polarisation spectrum, and the Planck ratios:
$\chi_{\rm {{tot}}}^2$. The upper radius of the micrometre-sized dust
$r^+_{\mu{\rm A}}$ corresponds to the minimum of
$\chi_{\rm {{tot}}}^2(r^+_{\mu{\rm A}})$. Steps~1--3 are iterated to
determine the final set of ten model parameters (columns 4--13 in
Table~\ref{Para.tab}) for a given axial ratio $a/b$ and porosity
$V_{\rm Vac}$ of the grains. The upper grain radius is typically
$r^+_{\mu{\rm A}} = 1\,\mu$m, and we identify it with the destruction
radius $r_{\rm {des}} = r^+_{\mu{\rm A}}$ in Eq.~\ref{eta.eq}. Even
larger, non-aligned grains may exist, but they do not, or only
marginally, contribute to the observations.

%%%%%%%%%%%%%%%%%%%%%%%%%%%%%%%%%%%%%%%%%%

\begin{figure*} [!htb]
\begin{center}
  \includegraphics[width=6.0cm, clip=true,trim=3.cm 2.8cm 2.cm 1.9cm]{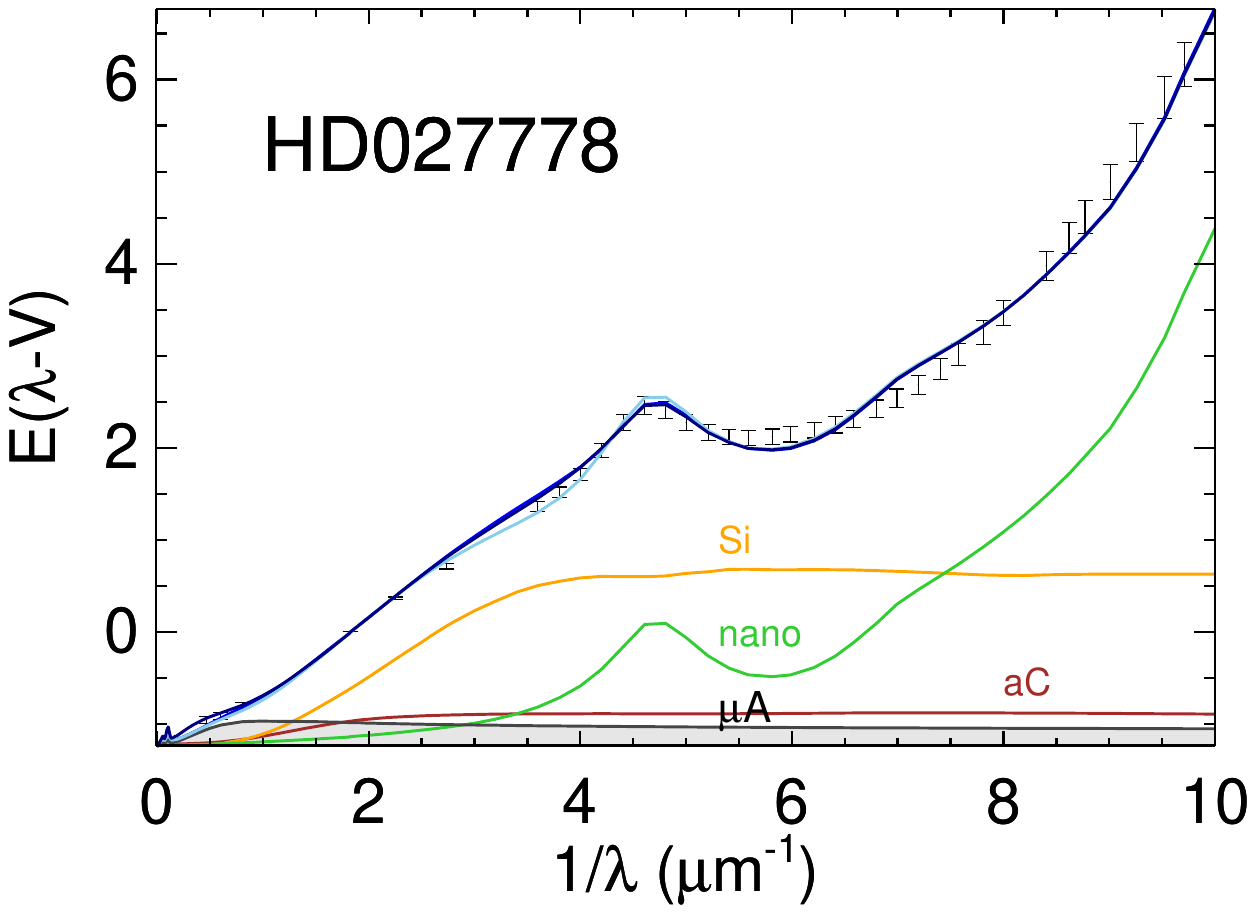}
  \includegraphics[width=6.0cm, clip=true,trim=3.cm 2.8cm 2.cm 1.9cm]{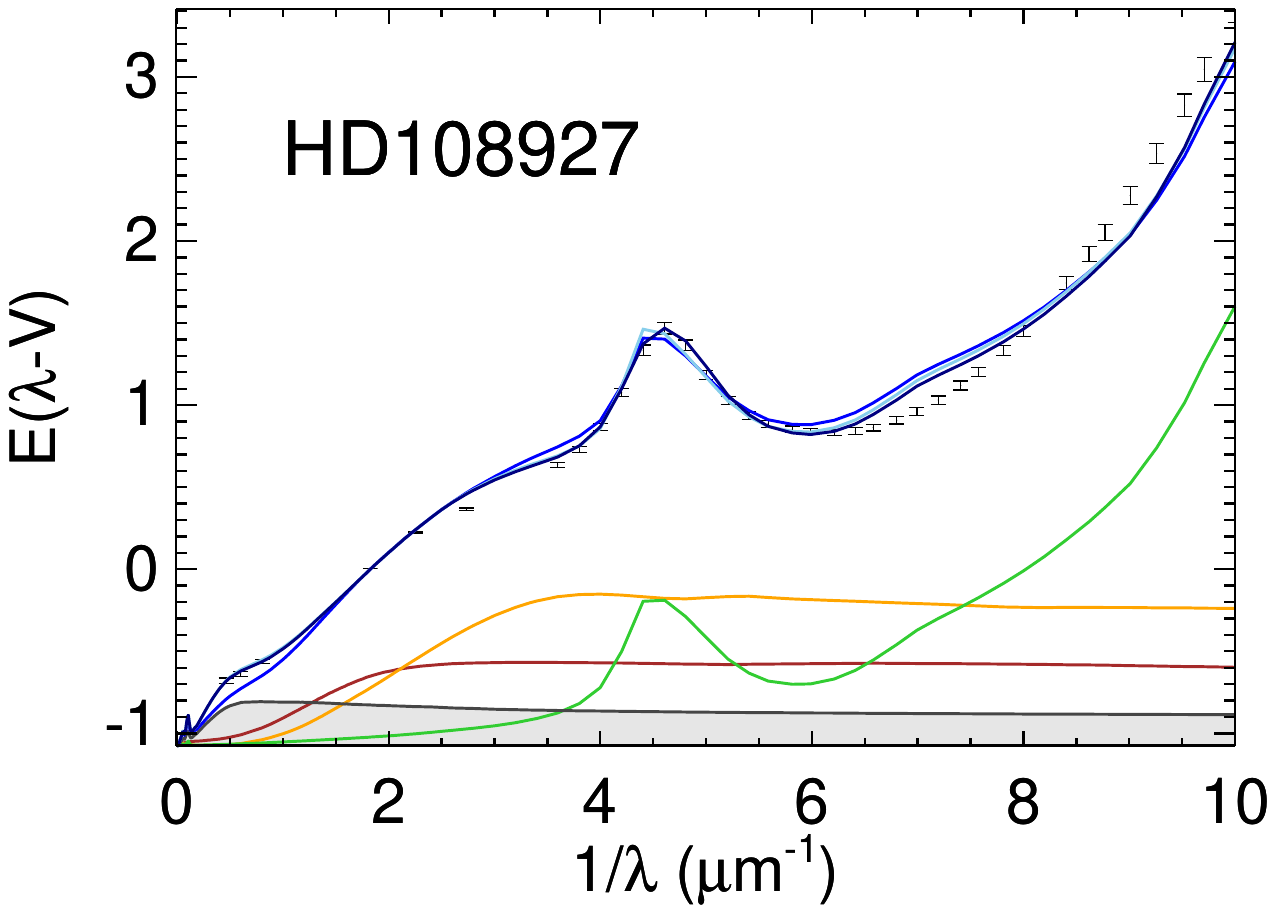}
  \includegraphics[width=6.0cm, clip=true,trim=3.cm 2.8cm 2.cm 1.9cm]{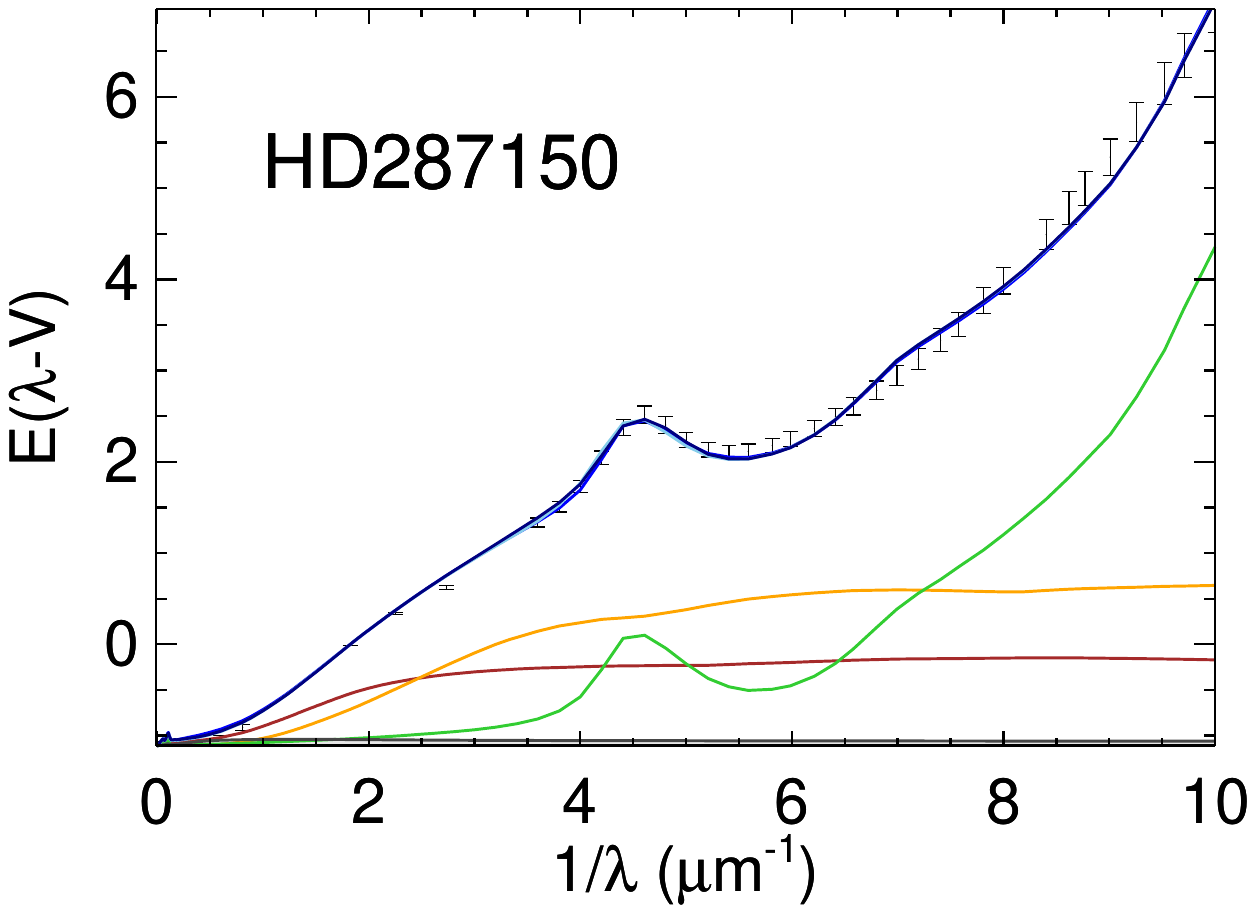}
  
  \includegraphics[width=6.0cm, clip=true,trim=3.cm 2.8cm 2.cm 1.9cm]{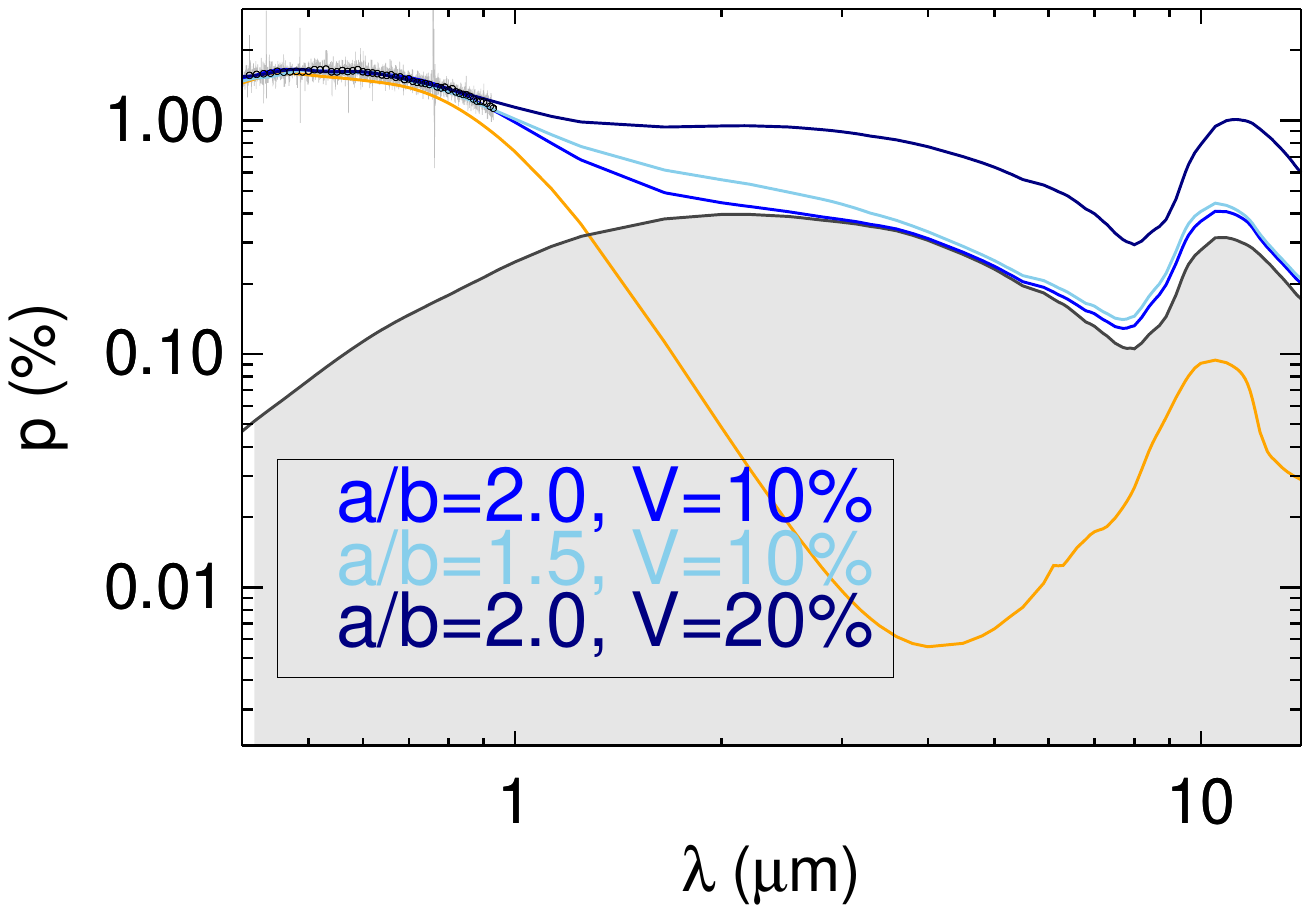}
  \includegraphics[width=6.0cm, clip=true,trim=3.cm 2.8cm 2.cm 1.9cm]{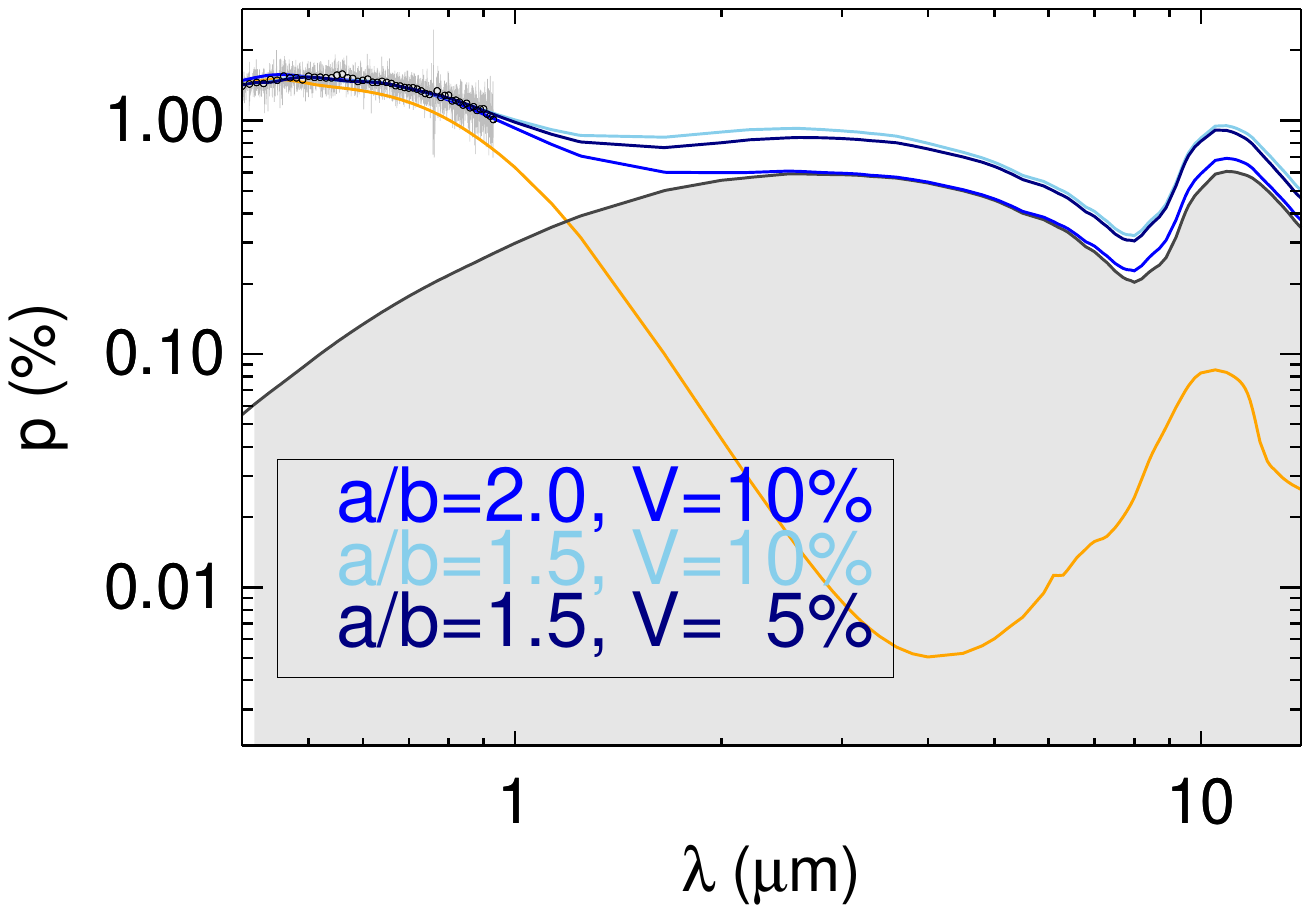}
  \includegraphics[width=6.0cm, clip=true,trim=3.cm 2.8cm 2.cm 1.9cm]{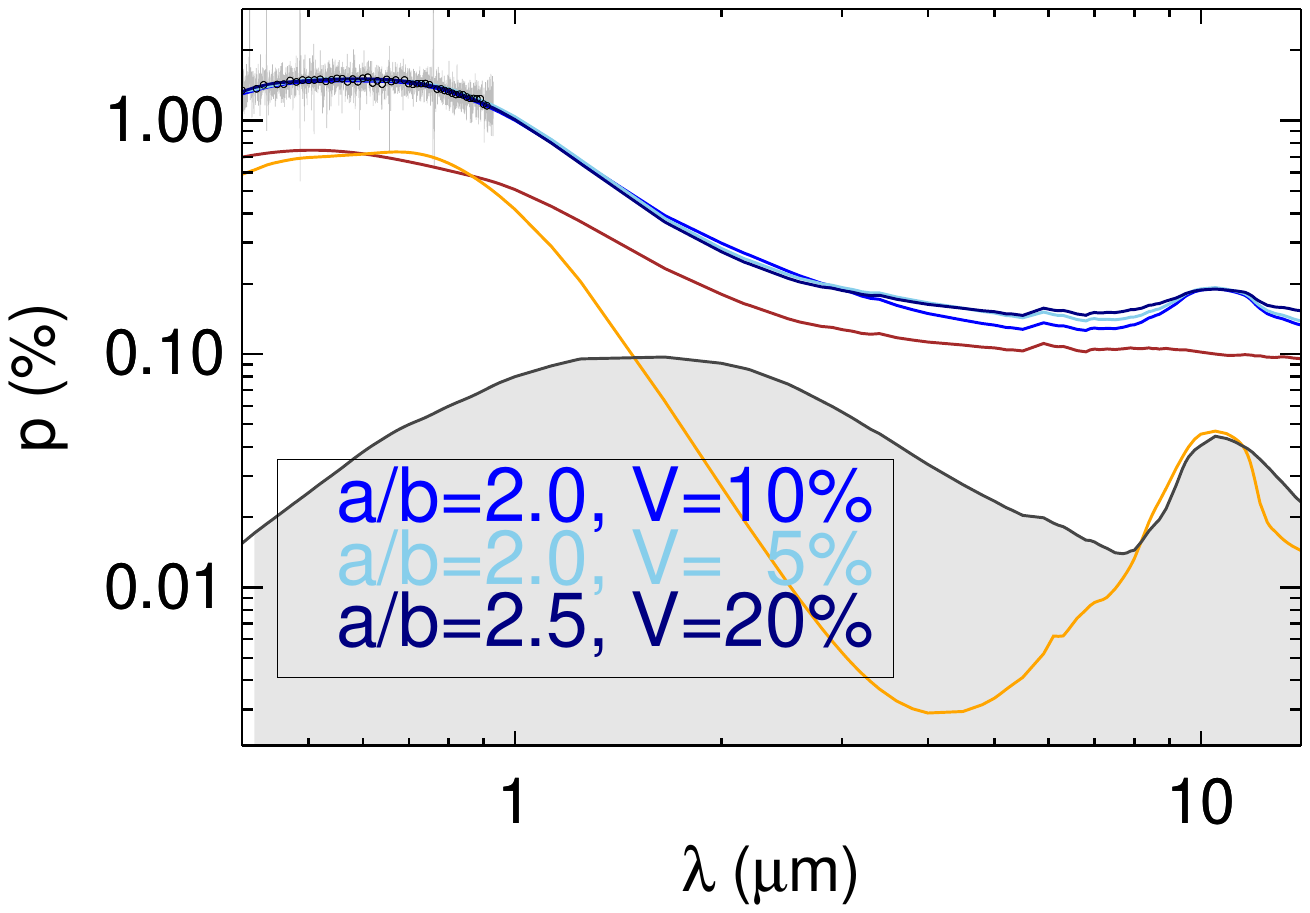}
  
  \includegraphics[width=6.0cm, clip=true,trim=3.cm 2.8cm 2.cm 1.9cm]{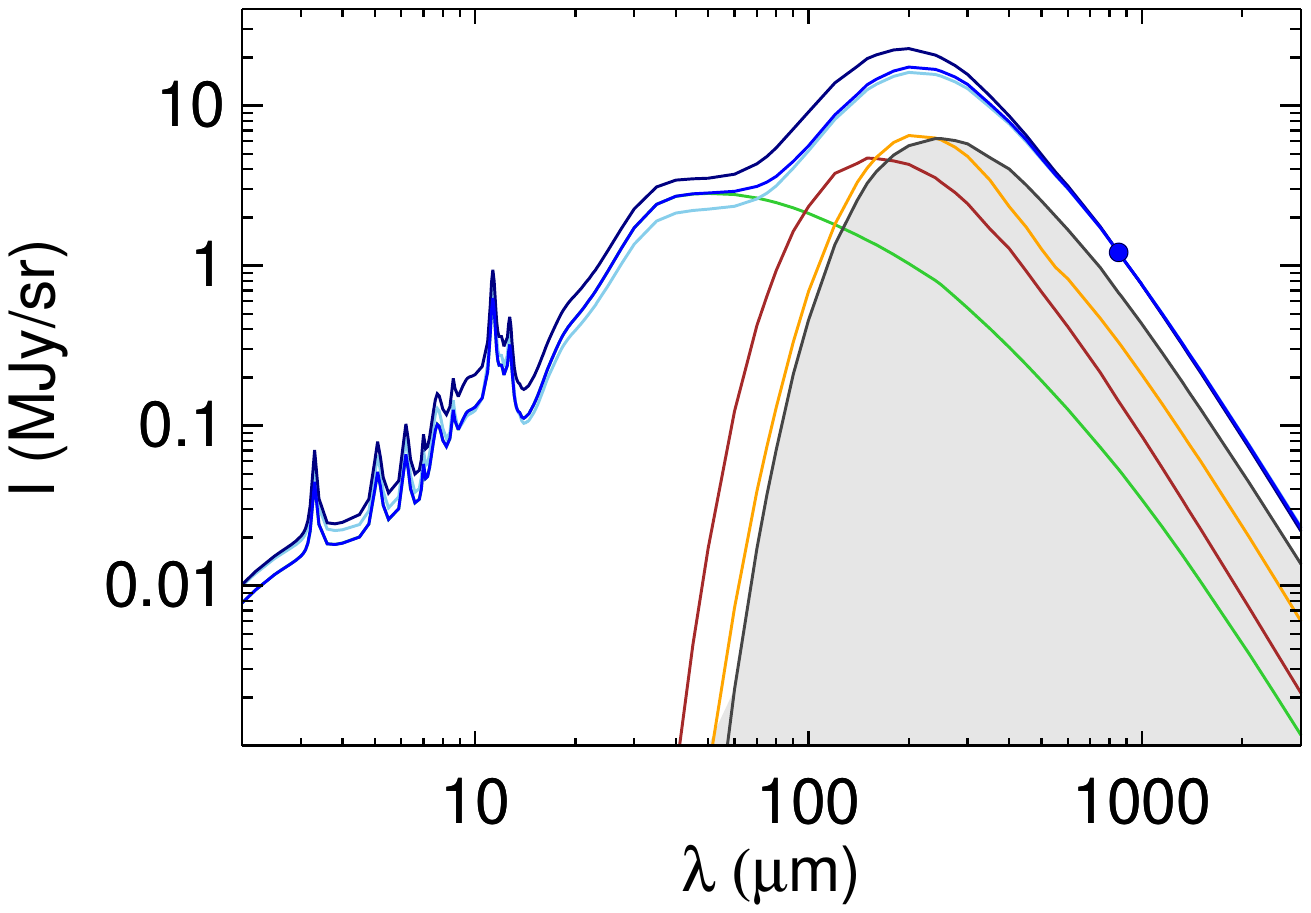}
  \includegraphics[width=6.0cm, clip=true,trim=3.cm 2.8cm 2.cm 1.9cm]{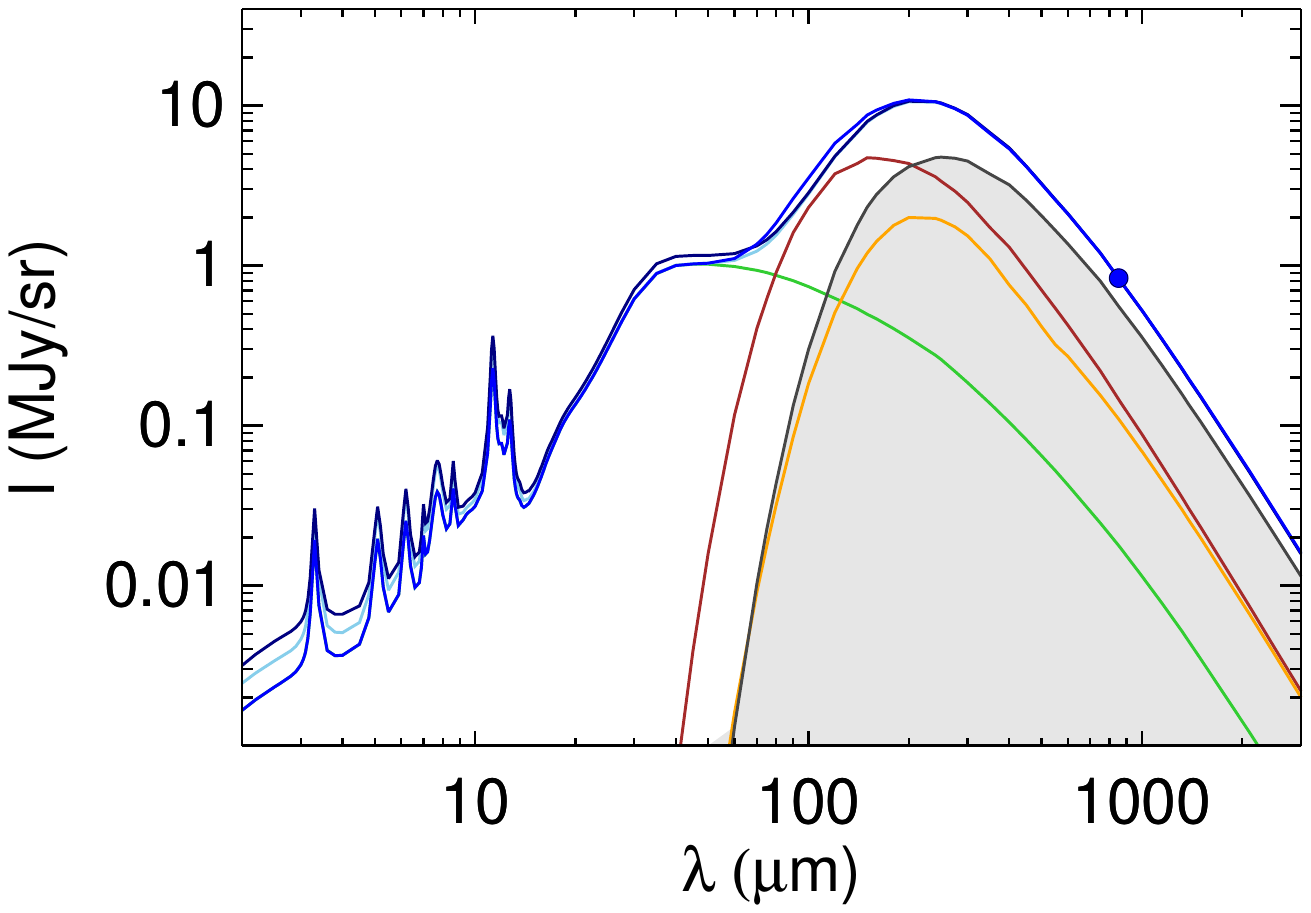}
  \includegraphics[width=6.0cm, clip=true,trim=3.cm 2.8cm 2.cm 1.9cm]{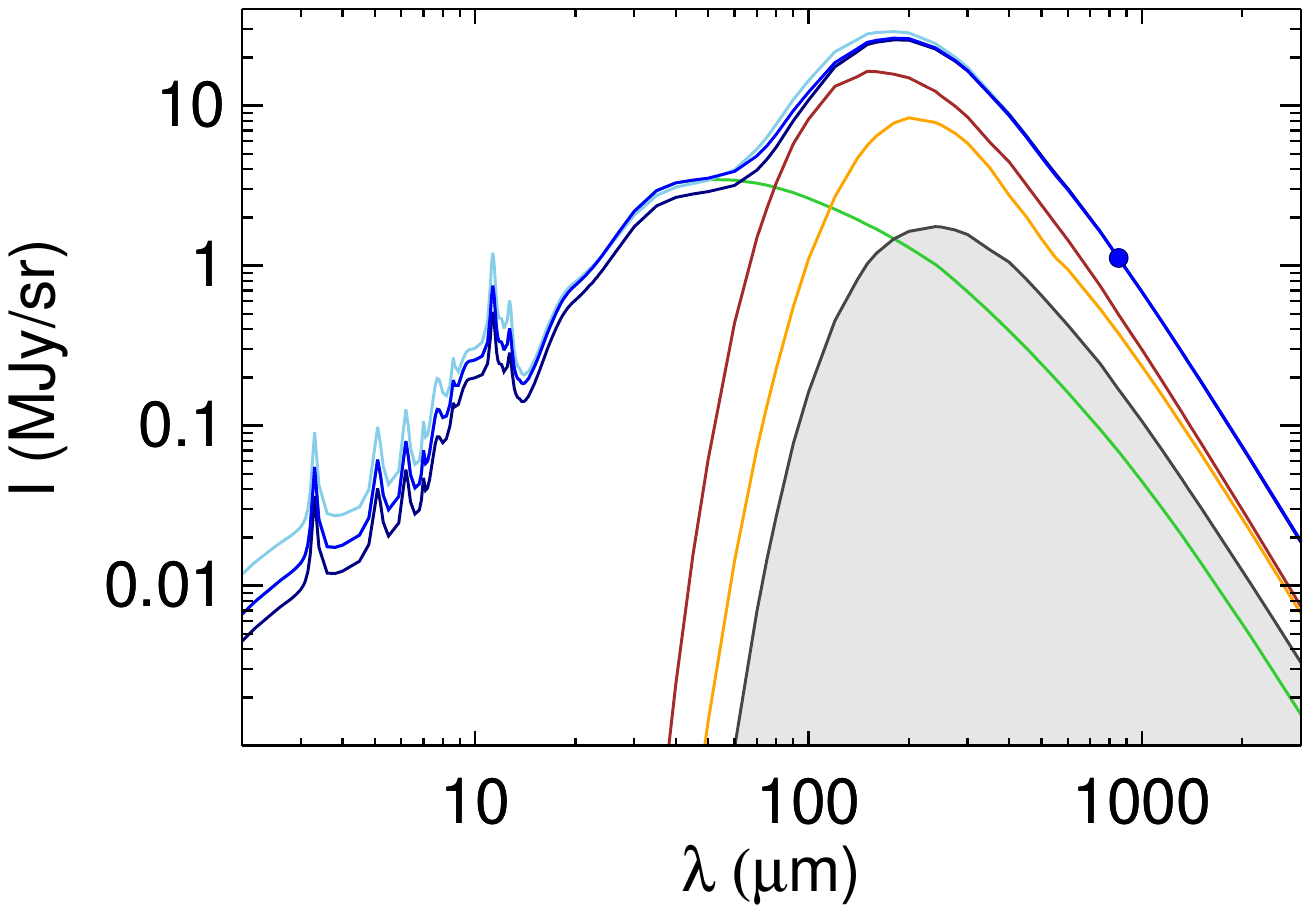}
  
  \includegraphics[width=6.0cm, clip=true,trim=3.cm 2.8cm 2.cm 1.9cm]{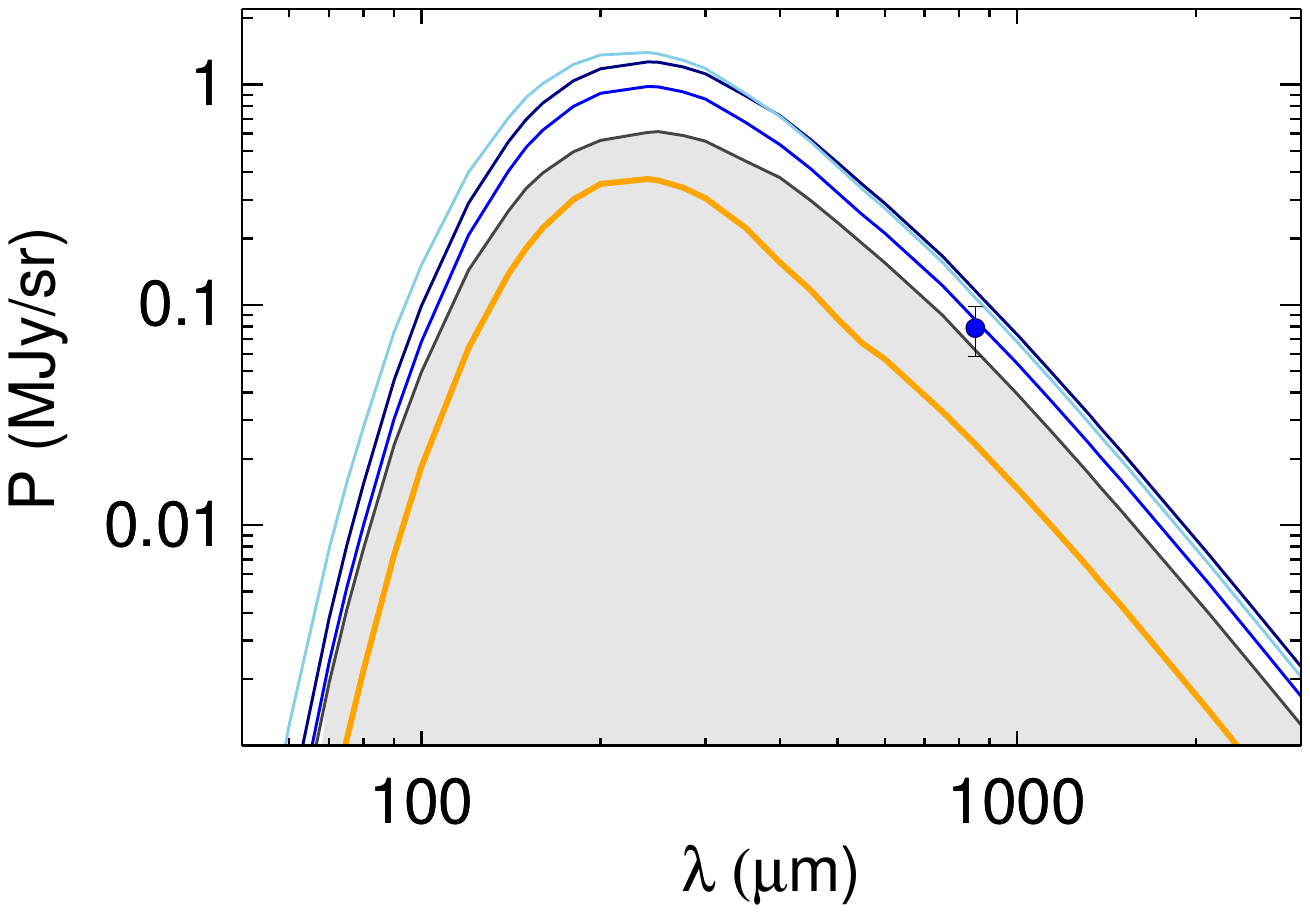}
  \includegraphics[width=6.0cm, clip=true,trim=3.cm 2.8cm 2.cm 1.9cm]{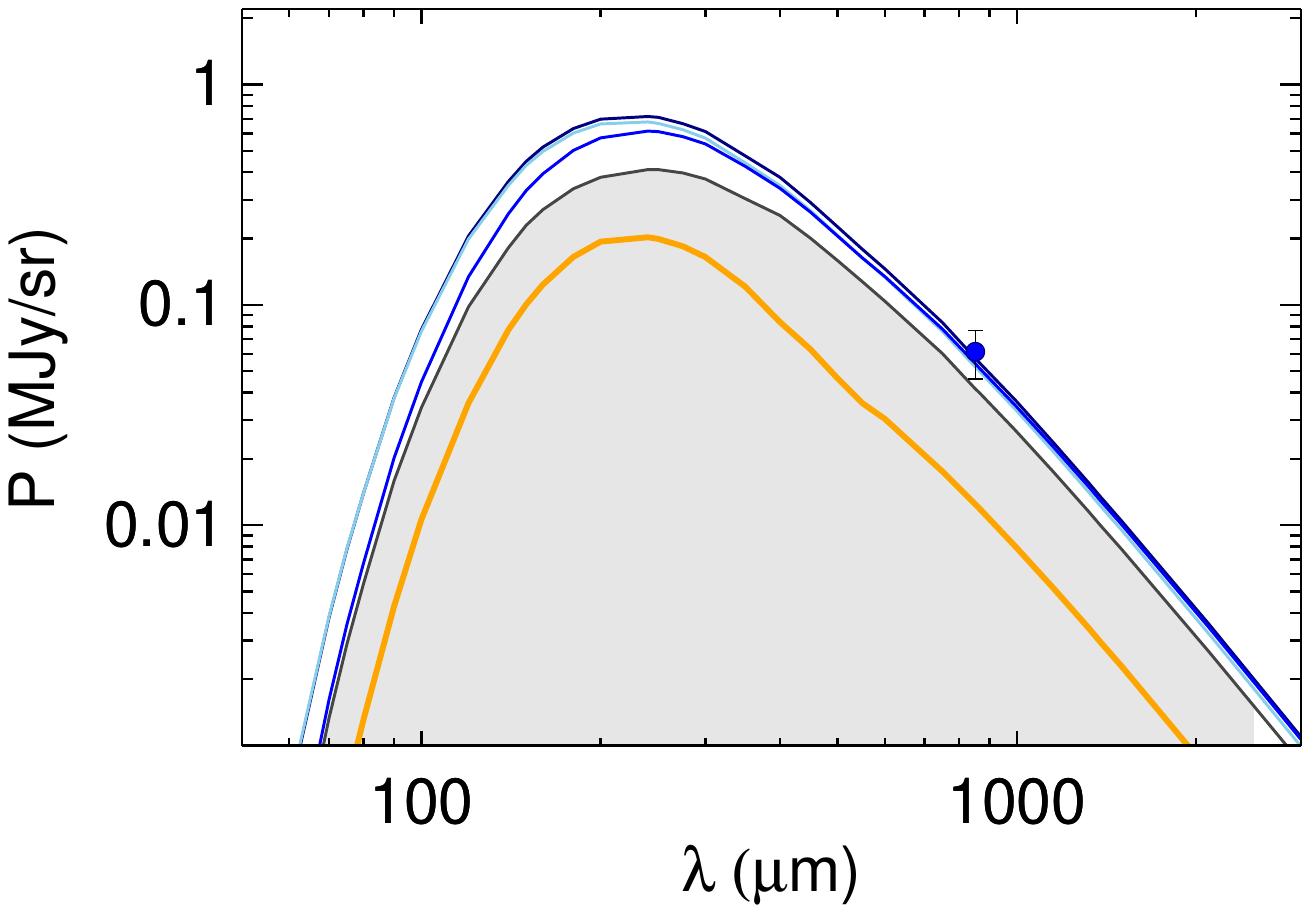}
  \includegraphics[width=6.0cm, clip=true,trim=3.cm 2.8cm 2.cm 1.9cm]{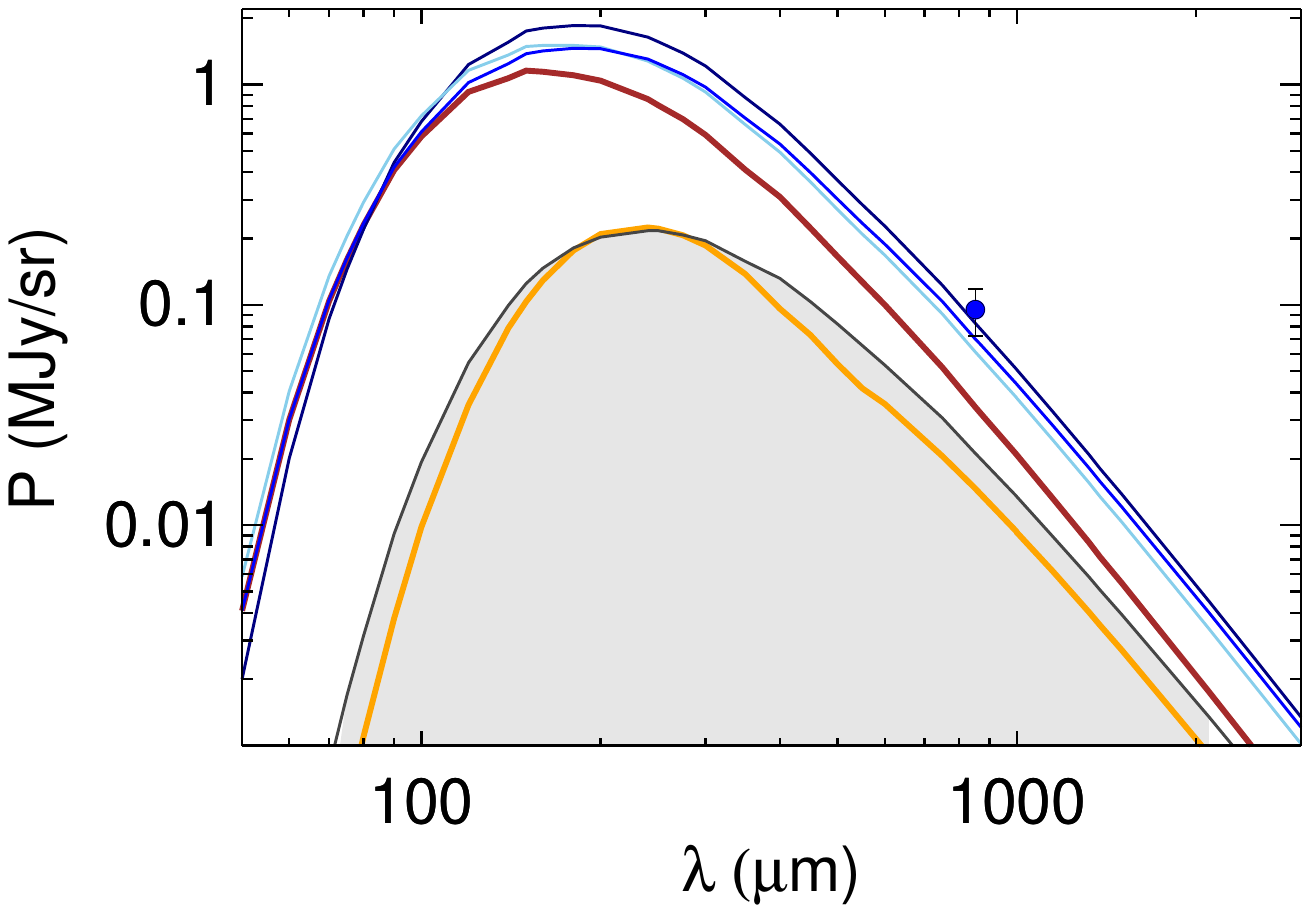}
  
\end{center}
\caption{Dust models for HD\,027778 (left), HD\,108927 (middle), and
  HD\,287150 (right). For each sightline, from top to bottom, the
  reddening curve, the optical polarisation spectrum, and the total
  and polarised ($I$ and $P$) dust-emission spectrum are shown. The
  reddening-curve data with error bars are described in the text. The
  observed starlight polarisation is shown for the unbinned spectrum
  (grey lines) and for data binned to a spectral resolution of
  $\lambda/\Delta\lambda \sim 50$ (black circles). The dust emission
  is consistent with Planck 353\,GHz measurements (filled
  circles). Model curves with the adopted $a/b$ ratio and porosity $V$
  are indicated in the legend. The individual contributions to the
  observables are shown for nanoparticles (green), amorphous carbon
  grains (brown), silicates (orange), and micrometre-sized
  agglomerates (grey shading) for the best-fit model with parameters
  listed in Table~\ref{Para.tab}.}\label{best.pdf}
\end{figure*} 

%%%%%%%%%%%%%%%%%%%%%%%%%%%%%%%%%%%%%%%%%%

The polarised emission spectrum is shown for three sightlines in
Fig.~\ref{best.pdf}. At 850\,$\mu$m, the polarisation is dominated by
the micrometre-sized dust particles. By increasing the radius
$r^+_{\mu{\rm A}}$, these grains become cooler, and their polarisation
spectrum shifts to longer wavelengths, so that the polarised intensity
$P_{850}$ increases as long as one remains in the Rayleigh part of the
spectrum of the micrometre-sized dust. On the other hand, by
decreasing $r^+_{\mu{\rm A}}$, the micrometre-sized agglomerates
approach temperatures of the amorphous components, causing the
polarisation spectra of both components to merge.

In the optical range, the micrometre-sized dust provides a grey
(constant) reddening and has a marginal impact on the best-fitting
model parameters, whereas in the near-IR, the reddening of sightlines
with a significant amount of micrometre-sized grains is strongly
impacted by $r^+_{\mu{\rm A}}$. Therefore, varying $r^+_{\mu{\rm A}}$
will change the fitting parameters of the reddening curve,
necessitating repetition of the procedure. Fortunately, in the three
steps, the dependencies on the free model parameters are weak and
converge after one to two iterations. The FORS and Planck data are
fitted without considering polarisation by nanoparticles ($r <
r^{-}_{\rm {pol}}$).

%%%%%%%%%%%%%%%%%%%%%%%%%%%%%%%%%%%%%%

\section{Observational constraints on grain structure and alignment efficiency \label{GrainStruc.sec}}

Observations of dust polarisation in both absorption and emission
provide complementary, in principle orthogonal, perspectives on dust
grains, allowing constraints to be placed on their shapes, porosity,
and alignment efficiencies.

\subsection{Pristine sightlines \label{sightlines.sec}} 

A high-quality sample of far-UV selected reddening curves was derived
by \cite{S23red}. Stars with multiple bright objects in the IUE
\citep{V04,FM07} and FUSE \citep{G09} apertures were excluded. Only
stars for which the spectral type and luminosity class, as derived
from UVES high-resolution spectroscopy, confirm those used in the
reddening estimation were retained. Furthermore, the photometric
variability of stars in the high-quality sample, both in the $V$ and
$G$ bands, and in the $B-V$ colour, was restricted to $\lesssim 0.03$
mag.

The visual extinction $A_{\rm V}$ (Table~\ref{Tab1.tab}) was derived
following \cite{S25} by inserting the absolute magnitude $M_{V}$ and
Gaia distance estimates $D_{{\rm Gaia}}$ into the photometric
equation:
\begin{equation} \label{Av.eq}
  A_{V} = V - M_{V} - 5 \log D_{{\rm Gaia}} + 5 \ . 		
\end{equation}

The absolute magnitude $M_{V}$ was extracted from the catalogues of
\cite{Bowen08} and \cite{Wegner06} for the spectral type and
luminosity class provided by \cite{S23red}. The distances were
estimated using Data Release~3 (DR3) by \cite{DR3}. To ensure a
reliable astrometric solution, only stars with a renormalised unit
weight error (RUWE) below 1.2 were included \citep{Luri18}. In
addition, the $G$-magnitude-dependent parallax error $\sigma(\pi,G)$
was computed following \cite{MA22}, and only stars with a parallax
precision of $\pi / \sigma(\pi,G) > 10$ were considered. The simple
inverse of the DR3 catalogue parallax typically agrees with
$D_{\rm Gaia}$ within $2\%$. Since parallactic distances inherently
depend on priors, we verified that our distance estimate
$D_{\rm Gaia}$ aligns with other probabilistic distance estimates
within 1--2\% \citep{BailerJones21}. Three stars exhibiting $H$-band
extinction greater than their visual extinction $A_V$ were removed
from the subsample, as their luminosity distance is smaller than the
trigonometric distance. These selection criteria result in a pristine
sample of 27 sightlines that are suitable for dust modelling.

Extinction probes the ISM in the foreground of the star, while
emission traces the entire sightline. \cite{Planck15} established
selection criteria for sightlines in order to obtain polarisation
measurements of the same dust grains at different wavelengths. For
this purpose, the visual extinction derived from the star's reddening
must be comparable to the visual extinction estimated from the
Planck maps. This criterion excludes sightlines with significant dust
emission originating from material located behind the star. The
limited $40'$ resolution of the Planck polarisation maps prevents a
direct comparison with starlight polarisation measurements of
individual stars at low Galactic latitudes. However, at Galactic
latitudes $|b| \gtrsim 15^{\circ}$, three stars HD~027778, HD~108927,
and HD~287150 exhibit significant starlight and Planck polarisation
with comparable extinction values, $A^{850}_{V} \sim A_{V}$
(Table~\ref{Tab1.tab}). In addition, these stars show the expected
reversal in polarisation angle between the polarised emission and the
starlight polarisation, consistent with a difference of
$90^{\circ} \pm 10^{\circ}$.

\subsection{Fiducial test cases \label{Fiducal.sec}}

%%%%%%%%%%%%%%%%
\begin{figure*} [!htb]
\begin{center}
  \includegraphics[width=6cm, clip=true,trim=4.8cm 2.0cm 2.5cm 0.cm]{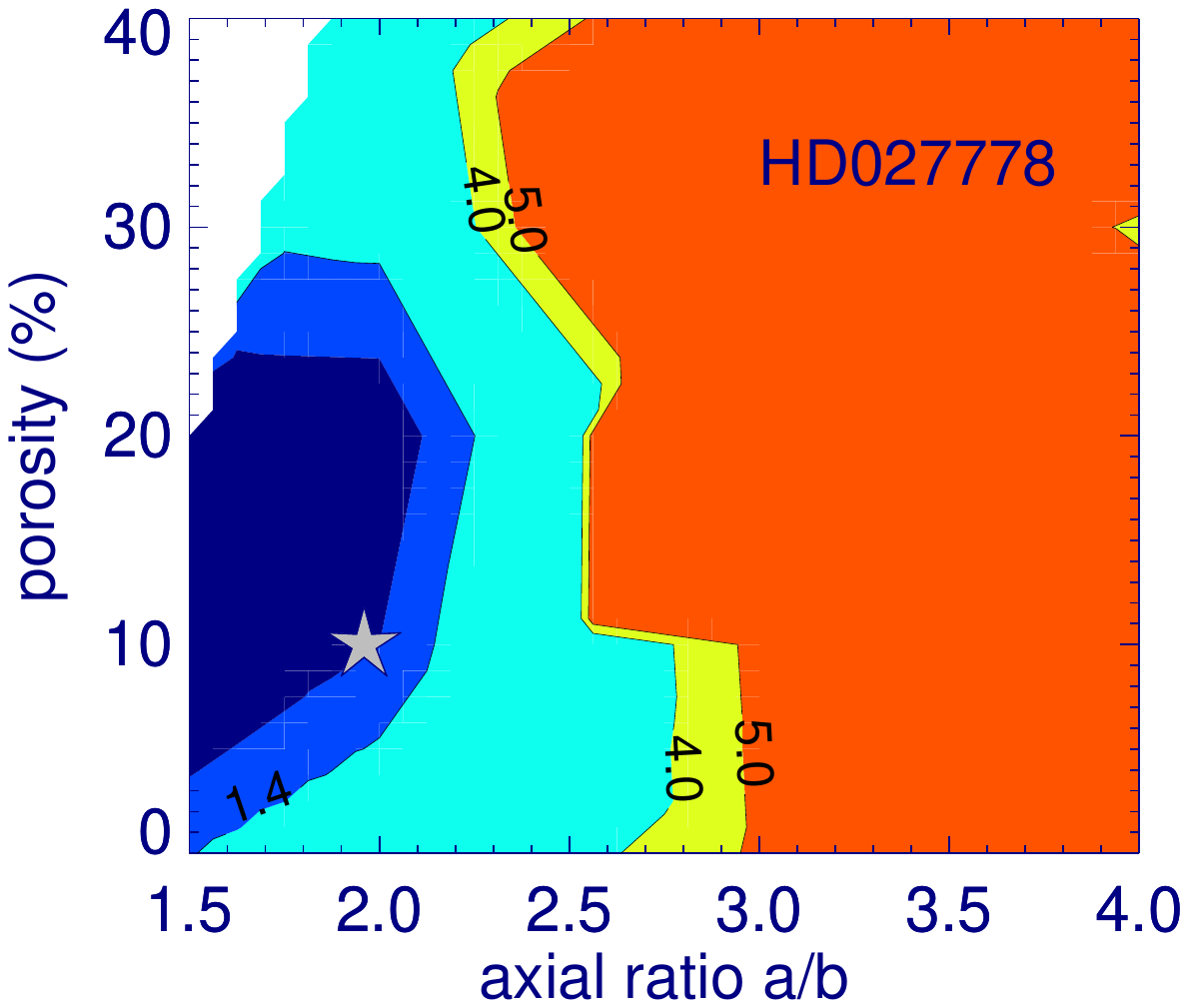}
  \includegraphics[width=6cm, clip=true,trim=4.8cm 2.0cm 2.5cm 0.cm]{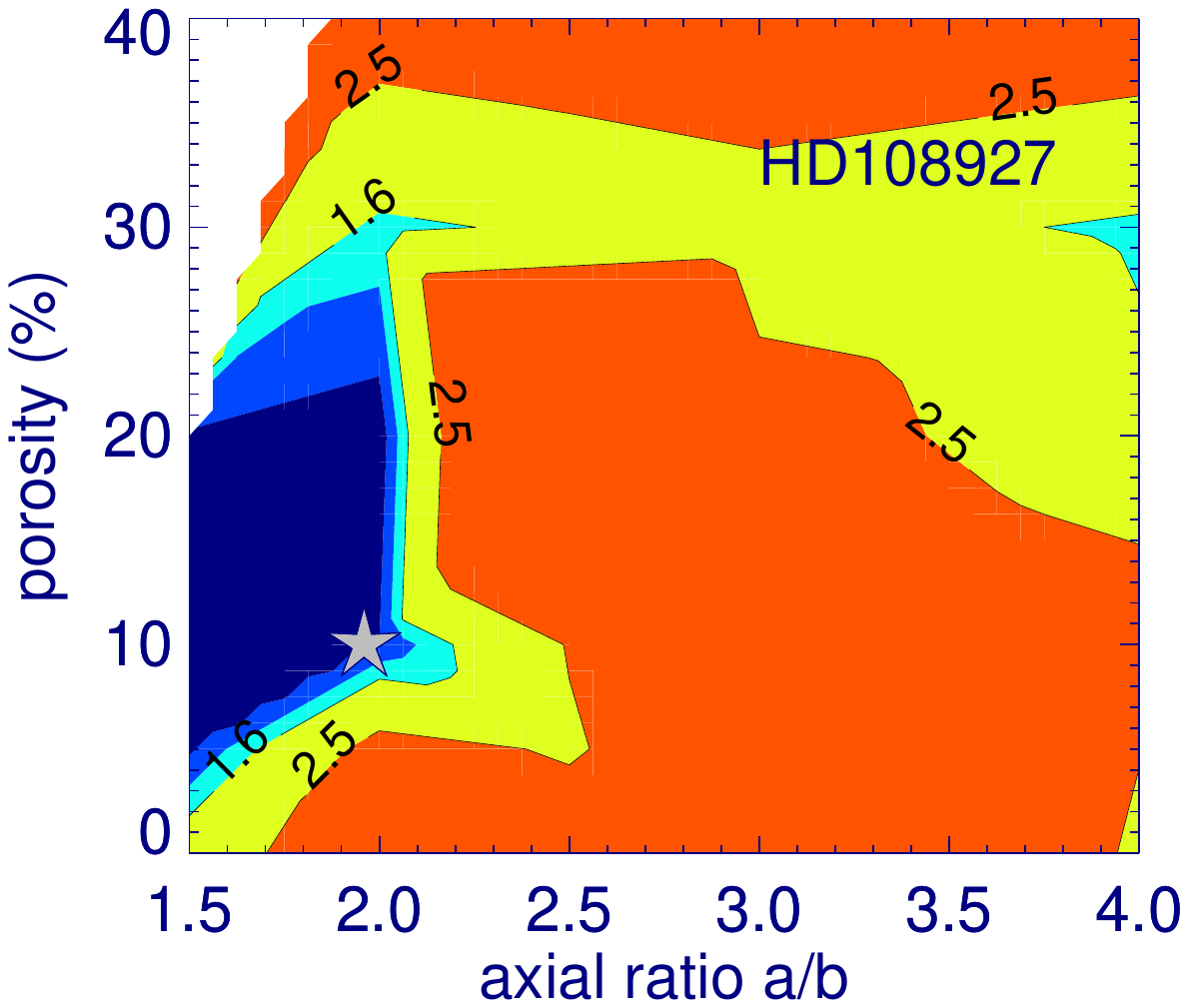}
  \includegraphics[width=6cm, clip=true,trim=4.8cm 2.0cm 2.5cm 0.cm]{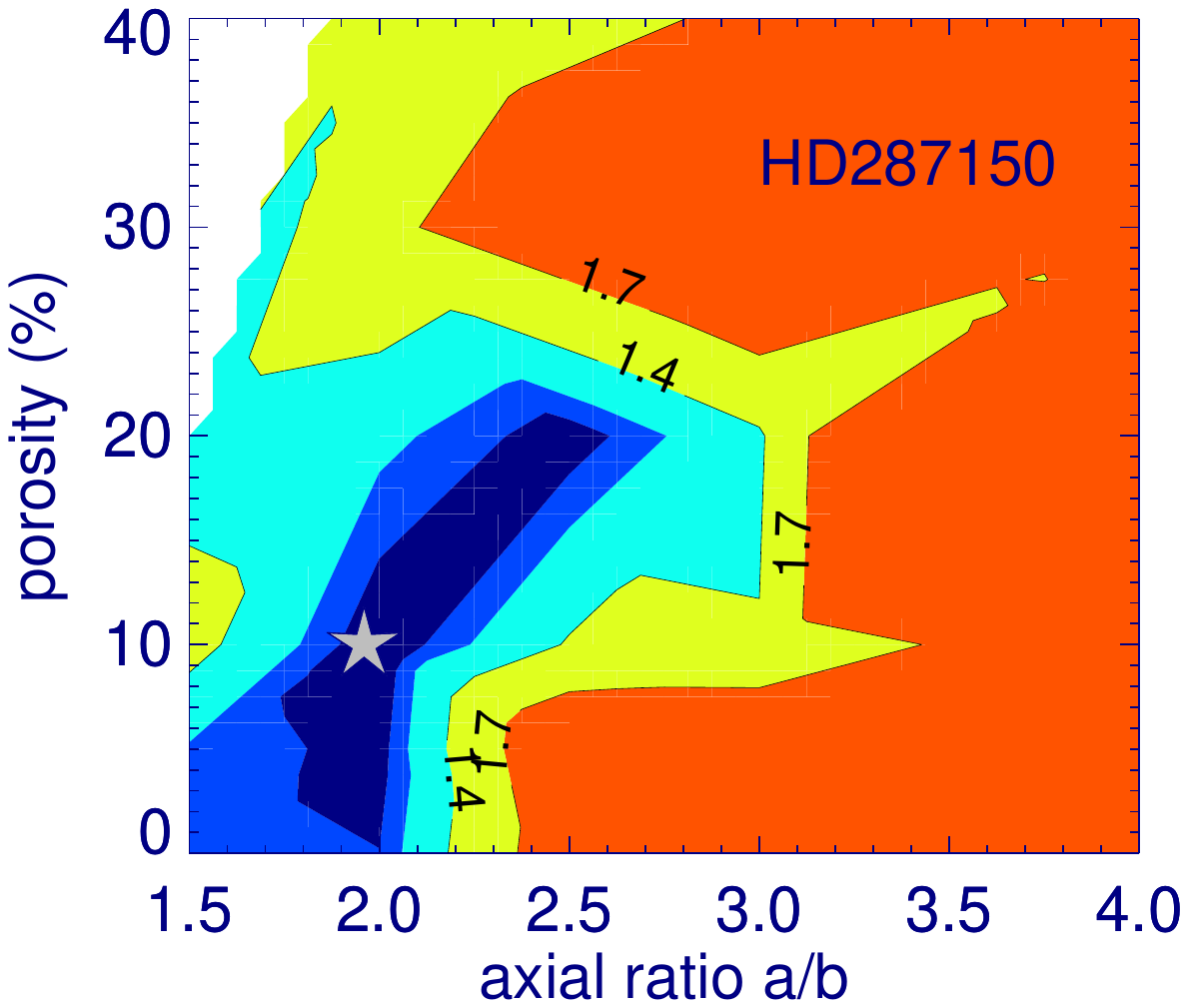}
\end{center}
\caption{The total goodness-of-fit $\chi_{\rm {{tot}}}^2$, normalised to the
  nominal model (grey symbol), as a function of axial ratio $a/b$ and
  porosity for HD\,027778 (left), HD\,108927 (middle), and HD\,287150
  (right). Regions with successful fits at $\chi_{\rm {{tot}}}^2 \simless
  1$ are shown in blue. \label{cont.pdf}}
\end{figure*}

%%%%%%%%%%%%%%%%%%%%%

The three stars HD~027778, HD~108927, and HD~287150, selected in
Sect.~\ref{sightlines.sec}, have optical polarisation spectra and
Planck-detected polarised emission measurements of grains from the
same environment. They therefore serve as fiducial test cases to
investigate the grain structure and alignment. We vary the axial ratio
of the prolates for $n_{\rm a/b} = 5$ cases of $a/b \in \{1.5, 2.0,
2.5, 3.0, 4.0\}$, and adjust the porosity, defined as the vacuum
volume fraction, of the amorphous carbon and silicate grains for
$n_{\rm Va} = 7$ cases of $V_{\rm a} \in \{0, 5, 10, 20, 30, 40,
50\}$\,(\%), as well as that of the micrometre-sized dust agglomerates
for $n_{\rm V_{\mu{\rm A}}} = 3$ cases of $V_{\rm \mu{\rm A}} \in \{5,
10, 20\}$\,(\%). We compute the volume ratios of silicate ($V_{\rm
  Si}$), carbon ($V_{\rm C}$), and vacuum ($V_{\rm \mu{\rm A}}$,
i.e. porosity) in the fluffy micrometre-sized composites to match an
abundance ratio of [Si]/[C] $\sim 3.6$ \citep{HD21}. For the three
porosity levels considered for the micrometre-sized dust, their
cross-sections are computed using volume ratios $V_{\rm Si}$:$V_{\rm
  C}$:$V_{\rm \mu{\rm A}}$ of 57:38:5, 53:37:10, and 48:32:20\,(\%).

We apply the simplified RAT alignment (Eq.~\ref{eta.eq}) unless stated
otherwise. The silicate and dark-dust materials include Fe such that
sufficiently large particles are perfectly aligned, ${f_{\rm {align}}}
= 100\,\%$. For carbon, the situation is less clear: carbon grains are
often assumed to be unaligned, ${f_{\rm {align}}} = 0$, while
\cite{Herranen21, Andersson24} find that grains without magnetic
inclusions can reach ${f_{\rm {align}}} = 50\,\%$ through RAT
alignment. We also experimented with perfect alignment of carbon
grains but found that such models fail to reproduce most of the
observations. In addition, we examined IDG alignment, noting that this
theory neglects internal alignment. The aim is to assess whether IDG
can be ruled out when fitting the data. The IDG efficiency parameter
$\delta_0$ typically varies from one sightline to another between
$\delta_0 = 0.2$ and $1\,\mu\mathrm{m}$ \citep{Das}. The maximum
alignment efficiency in IDG is approached asymptotically at $\delta_0
\gtrsim 10\,\mu\mathrm{m}$. This corresponds to alignment efficiencies
of ${f_{\rm {align}}} = 18\,\%$ for $\delta_0 = 0.2\,\mu\mathrm{m}$,
${f_{\rm {align}}} = 32\,\%$ for $\delta_0 = 1\,\mu\mathrm{m}$, and
$46\,\%$ for $\delta_0 = 10\,\mu\mathrm{m}$. In total, we inspect
$n_{{f_{\rm {align}}}} = 5$ levels of alignment efficiency. Within
this sparsely sampled parameter space of $n_{\rm a/b} \times n_{\rm
  Va} \times n_{\rm V_{\mu{\rm A}}} \times n_{{f_{\rm {align}}}} =
525$ models, we compute the dust cross-sections
(Eqs.~\ref{Cext.eq}--\ref{Cpol.eq}) of prolate particles over 300
frequencies, 30 bins of $0^{\circ} \lesssim \Omega < 90^{\circ}$, and
130 radii ranging from $6\,\rm nm$ to $3\,\mu$m. In these models, a
specific set of particle-structure parameters ($a/b$, $V_{\rm a}$,
$V_{\rm \mu{\rm A}}$) and alignment efficiency ${f_{\rm {align}}}$ is
applied uniformly across all grain types, and the fitting procedure
described in Sect.~\ref{pro.sec} is used.
    
For each star, fits to the reddening curve that respect the elemental
depletion constraints (Eq.~\ref{abu.eq}) and incorporate the Gaia
distance estimates (Eq.~\ref{Av.eq}), together with the FORS
starlight polarisation spectrum and the colour-corrected Planck total
and polarised dust-emission spectrum, are presented in
Fig.~\ref{best.pdf}. The parameter set of the cross-sections with
$a/b = 2$ and $V_{\rm a} = V_{\rm \mu{\rm A}} = 10\%$ was adopted as
the {\it nominal} model of the grain structure. In addition, for each
star, two alternative models are shown that achieve a fit of
comparable quality to the nominal model.

For HD~027778 and HD~108927, the best fits are obtained when
neglecting the alignment of carbon grains, although for HD~027778 a
comparably good fit is also achieved when assuming
${f_{\rm {align}}}(\mathrm{aC}) = 50\,\%$. In contrast, for
HD~287150, aligned carbon grains make an important contribution to
the optical polarisation and dominate the polarised emission. Models
that ignore the alignment of aC grains for HD~287150 show a too-steep
decline in the polarisation at
$\lambda > 0.7\,\mu\mathrm{m}$ to reproduce the optical polarisation
spectrum, and they underpredict $P_{850}$ by a factor of two. This
underprediction of the polarisation in the optical and
submillimetre can be compensated by a substantial increase in the
mass of the micrometre-sized grains, which, however, leads to an
overprediction of the near-IR reddening towards HD~287150.

In the optical, the polarisation cross-section $K_{\rm {pol}}$ can be
dominated by aC grains if aligned, and otherwise by aSi grains. In
this wavelength range, $\mu$A grains, if present, contribute only
weakly, whereas they dominate $K_{\rm {pol}}$ at
$\lambda > 1\,\mu$m. Because aSi and $\mu$A grains attain similar
temperatures, the polarised emission in Fig.~\ref{best.pdf} is
dominated by the micrometre-sized agglomerates when aC grains are not
aligned. Notably, the optical and submillimetre polarisation arise
from distinct grain populations. Following \cite{Fanciullo15}, we
estimate the intensity scaling parameter $G_0$ of the ISRF from
Planck \citep{Planck14} and IRAS observations, and the visual
extinction $A_{\rm V}$ (Table~\ref{Tab1.tab}) to the stars. We find
$G_0 = 0.56$, 0.37, and 0.43 for HD~027778, HD~108927, and
HD~287150. No trend is found between aC alignment and $G_0$.
 
There are degeneracies in the model grid of particle structures,
where multiple models match all datasets with similarly good quality.
In Fig.~\ref{cont.pdf}, the total goodness of fit,
$\chi_{\rm {{tot}}}^2$, normalised to the nominal model, is shown as a
function of axial ratio and porosity. The contour levels at
$\chi_{\rm {{tot}}}^2 \simless 1$ highlight regions of comparable fit
quality; these are marked in blue and correspond to models
performing similarly to the nominal model
(Fig.~\ref{cont.pdf}). Such models typically exhibit grain porosities
$V_{\rm a} \simless 20\,\%$ and axial ratios $a/b \simless 2.5$. The
nominal model yields the lowest $\chi_{\rm {{tot}}}^2$ across the
three stars for which both FORS spectra of the dichroic polarisation
and Planck polarised-emission data are available. For these models,
the upper radius of the grains, as derived from the Planck
polarisation (Eqs.~\ref{Eq.RSv}, \ref{Eq.Rpp}), is
$r^{+}_{\rm \mu{\rm A}} = 1\,\mu{\rm m}$; the other model parameters
are listed in Table~\ref{Para.tab}.

%%%%%%%%%%%%%%%%%%%%%%%%%%%%%%%%%%%%%%%%%

\begin{table*}[!htb]
  \scriptsize
  \begin{center}  
		\caption {Dust parameters of the 27 sightlines that
                  have been modelled.  \label{Para.tab}}
		\begin{tabular}{c r r c r  r r c  c  c r  r  c  | r r c}
		  \hline\hline
 1 & 2 & 3 & 4 & 5 & 6 & 7 & 8 & 9 & 10 & 11 & 12 & 13 & 14 &15 & 16 \\
\hline
 {Star} & {$a/b$} & {$V_{\rm {vac}}$}  & {$m_{\mu{\rm A}}$} & { $m_{\rm {Si}}$} & { $m_{\rm{vSi}}$} & {
   $m_{\rm{aC}}$} & { $m_{\rm{vgr}}$} & { $m_{\rm{PAH}}$ } &
 {$q$ } & {$r^-_{\text{{{pol}}, Si}}$ } & {$r^-_{\text {{pol}}, aC}$}  & {$\Omega$}  &
  {{$\frac{[\rm{Si}]}{[\rm{H}]}$ }}& {{$\frac{[\rm{C}]}{[\rm{H}]}$ }}& $\left(\frac{\tau_{\rm {s}\mu}}{\tau_{\rm {t}}}\right)_{\rm V}$\\
 & & \multicolumn{7}{c}{(\%)} &   & \multicolumn{2}{c}{(nm)} & ($^{\circ}$) & \multicolumn{2}{c}{(ppm)} & (\%) \\
			\hline
  HD~027778  &   2.0 &  10 &    25 &    37 &    28 &     5 &     3 &     1 &   2.3 &   112 &   $-$ &   38 &    42 &    78  &  20 \\
  HD~037903  &   2.0 &  10 &    33 &    31 &    18 &    13 &     4 &     1 &   2.1 &   143 &    69 &   50 &    32 &   108  &  37 \\
  HD~038023  &   1.5 &   5 &    32 &    37 &    19 &     8 &     2 &     2 &   2.6 &    25 &   $-$ &   53 &    38 &    88  &  34 \\
  HD~046223  &   1.5 &   5 &    30 &    33 &    24 &     8 &     4 &     1 &   2.6 &   136 &   $-$ &   63 &    37 &    91  &  29 \\
  HD~054439  &   1.5 &   5 &   $-$ &    58 &    21 &    13 &     5 &     3 &   2.8 &   112 &    69 &   27 &    38 &    93  &   0 \\
  HD~062542  &   1.5 &   5 &    25 &    28 &    35 &     8 &     3 &     1 &   2.8 &   143 &   $-$ &   76 &    49 &   110  &  22 \\
  HD~070614  &   2.0 &  10 &   $-$ &    52 &    22 &    18 &     6 &     2 &   2.5 &   143 &    42 &   47 &    34 &   109  &   0 \\
  HD~091824  &   2.0 &  10 &    41 &    33 &    14 &     9 &     2 &     1 &   2.3 &   102 &   $-$ &   51 &    36 &    96  &  52 \\
  HD~092044  &   2.0 &  10 &    34 &    36 &    19 &     8 &     2 &     1 &   2.6 &   112 &    88 &   39 &    38 &    89  &  48 \\
  HD~093222  &   1.5 &   5 &    35 &    34 &    13 &    15 &     2 &     1 &   2.5 &    30 &   $-$ &   41 &    32 &   110  &  38 \\
  HD~108927  &   2.0 &  10 &    30 &    32 &    20 &    13 &     4 &     1 &   2.0 &    97 &   $-$ &   46 &    33 &   106  &  23 \\
  HD~110946  &   2.0 &  10 &    19 &    39 &    27 &    10 &     4 &     1 &   2.8 &   143 &    54 &   49 &    37 &    92  &   9 \\
  HD~112607  &   1.5 &   5 &   $-$ &    39 &    39 &    13 &     6 &     3 &   2.7 &   143 &   $-$ &   37 &    35 &    96  &   0 \\
  HD~112954  &   2.0 &  10 &     8 &    32 &    44 &    10 &     5 &     1 &   2.5 &   183 &    54 &   60 &    43 &    94  &   1 \\
  HD~129557  &   1.5 &   5 &    35 &    44 &    12 &     3 &     4 &     2 &   2.0 &   102 &   $-$ &   47 &    41 &    81  &  42 \\
  HD~146285  &   1.5 &   5 &    39 &    31 &    10 &    15 &     4 &     1 &   2.0 &   143 &    69 &   61 &    28 &   114  &  33 \\
  HD~152245  &   1.5 &   5 &    35 &    38 &    14 &     8 &     3 &     2 &   2.6 &   112 &    88 &   40 &    37 &    91  &  44 \\
  HD~152249  &   1.5 &   5 &    63 &    21 &     9 &     5 &     1 &     1 &   2.3 &   201 &    92 &   40 &    38 &    94  &  44 \\
  HD~170740  &   1.5 &   5 &   $-$ &    46 &    29 &    16 &     6 &     3 &   2.8 &   183 &    26 &   72 &    35 &   103  &   0 \\
  HD~185418  &   1.5 &   5 &    27 &    43 &    13 &    10 &     5 &     2 &   2.5 &   112 &    88 &   32 &    35 &    98  &  19 \\
  HD~287150  &   2.0 &  10 &    17 &    32 &    34 &    13 &     3 &     1 &   2.8 &   143 &     8 &   36 &    36 &    97  &   6 \\
  HD~294304  &   1.5 &   5 &    27 &    36 &    23 &    10 &     3 &     1 &   2.7 &   102 &   $-$ &   34 &    37 &    95  &  23 \\
  HD~303308  &   2.0 &  10 &    26 &    38 &    23 &     9 &     3 &     1 &   2.4 &   112 &    88 &   52 &    39 &    88  &  20 \\
  HD~315021  &   1.5 &   5 &    33 &    40 &    10 &    13 &     3 &     1 &   2.4 &   118 &   $-$ &   30 &    35 &   103  &  35 \\
  HD~315023  &   1.5 &   5 &    32 &    38 &    11 &    15 &     2 &     2 &   2.5 &   143 &    88 &   53 &    33 &   108  &  32 \\
  HD~315024  &   1.5 &   5 &    67 &    15 &     7 &     8 &     2 &     1 &   2.3 &    47 &   $-$ &   53 &    45 &   170  &  46 \\
  HD~315032  &   2.0 &   5 &    34 &    33 &    17 &    11 &     4 &     1 &   2.3 &   112 &   $-$ &   16 &    34 &   104  &  43 \\
  \hline
    median &       &       &    31 &    37 &    19 &    10 &     3 &     1 &   2.5 &   118 &    69 &      &    37 &    96 &   32 \\
    sigma  &       &       &    16 &     8 &     9 &     4 &     1 &     1 &   0.3 &    23 &    25 &      &     4 &    17 &   17 \\
  \hline 
			\end {tabular}
  \end{center}
\scriptsize{\bf{Notes.}} {For HD~091824 and HD~146285 the maximum
    grain size is $r_+ = 3\,\mu$m, otherwise $r_+ = 1\,\mu$m. The
    derived C and Si abundances in the grains and the contribution of
    micrometre-sized dust agglomerates to the total optical depth are
    specified in columns 14--16. The median values and 1$\sigma$
    variations of the parameters are provided, excluding
    $r^-_{\rm {pol,Si}}$ for the two stars HD~038023 and HD~093222, as
    they exhibit peculiar polarisation spectra peaking in the
    $B$ band.}
\end {table*}

%%%%%%%%%%%%%%%%%%%%%%% 

%%%%%%%%%%%%%%%%%%%%%%%%%%%%   all models 

%  \includegraphics[width=3.5cm,clip=true,trim=3.5cm 6.cm 2.1cm 4.5cm]{./HD027778Redd.pdf}
%  \includegraphics[width=3.5cm,clip=true,trim=3.5cm 6.cm 2.1cm 4.5cm]{./HD108927Redd.pdf}
%  \includegraphics[width=3.5cm,clip=true,trim=3.5cm 6.cm 2.1cm 4.5cm]{./HD287150Redd.pdf}
%%%%%%   0 -8  %%%%%%%%%%%%%%%%%%%%%%%%%%%%%%%%%%%%%%%
\begin{figure*} [!htb]
  \begin{center}
  \includegraphics[width=3.5cm,clip=true,trim=3.25cm 6.cm 2.1cm 0cm]{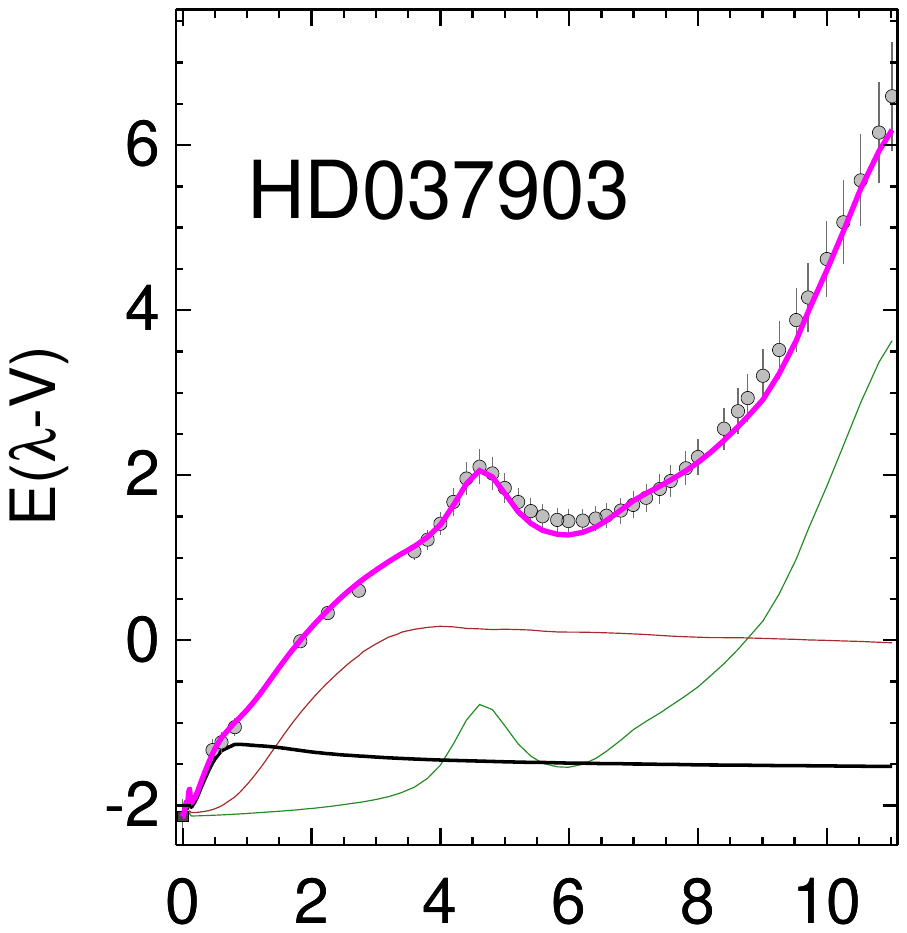}
  \includegraphics[width=3.5cm,clip=true,trim=3.25cm 6.cm 2.1cm 0cm]{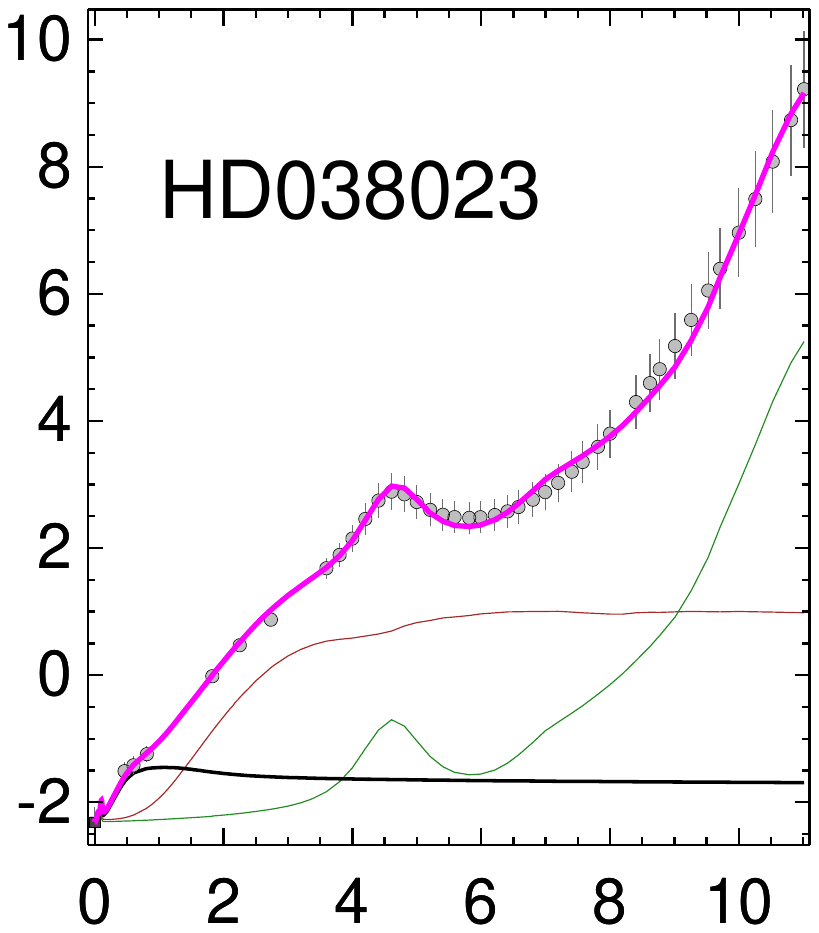}
  \includegraphics[width=3.5cm,clip=true,trim=3.25cm 6.cm 2.1cm 0cm]{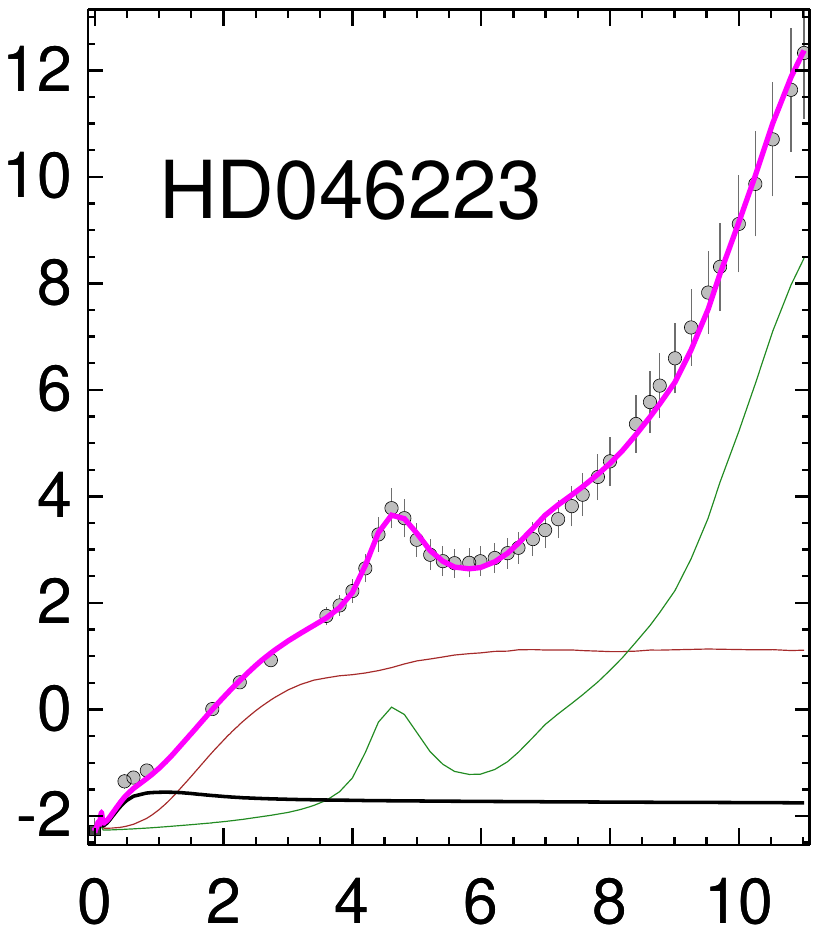}
  \includegraphics[width=3.5cm,clip=true,trim=3.25cm 6.cm 2.1cm 0cm]{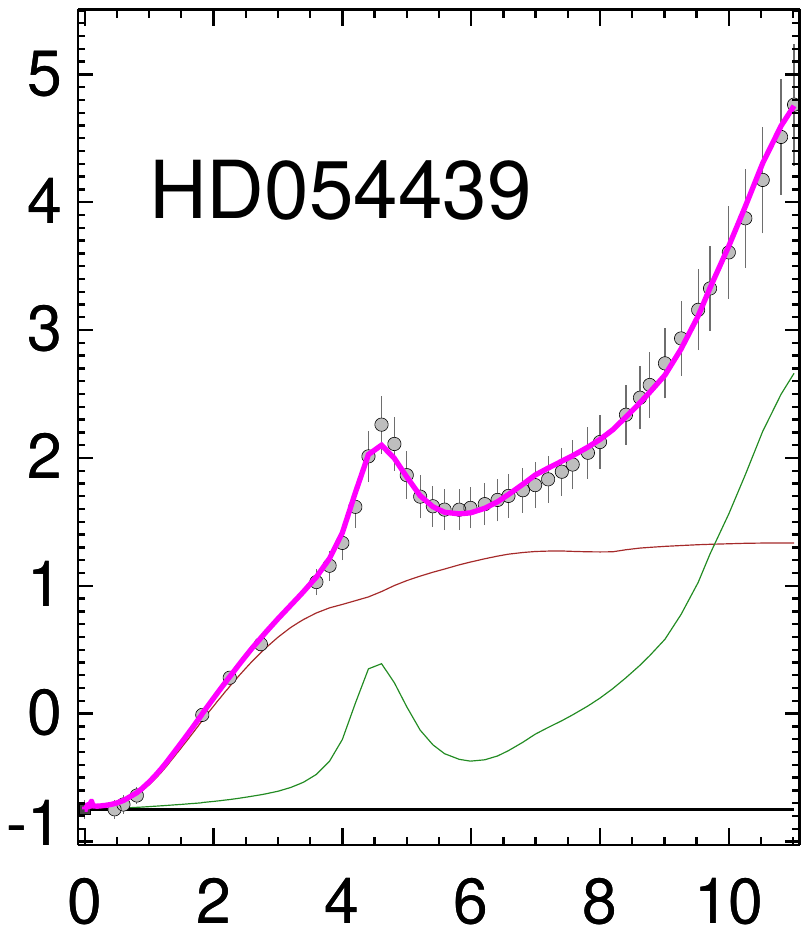}
  \includegraphics[width=3.5cm,clip=true,trim=3.25cm 6.cm 2.1cm 0cm]{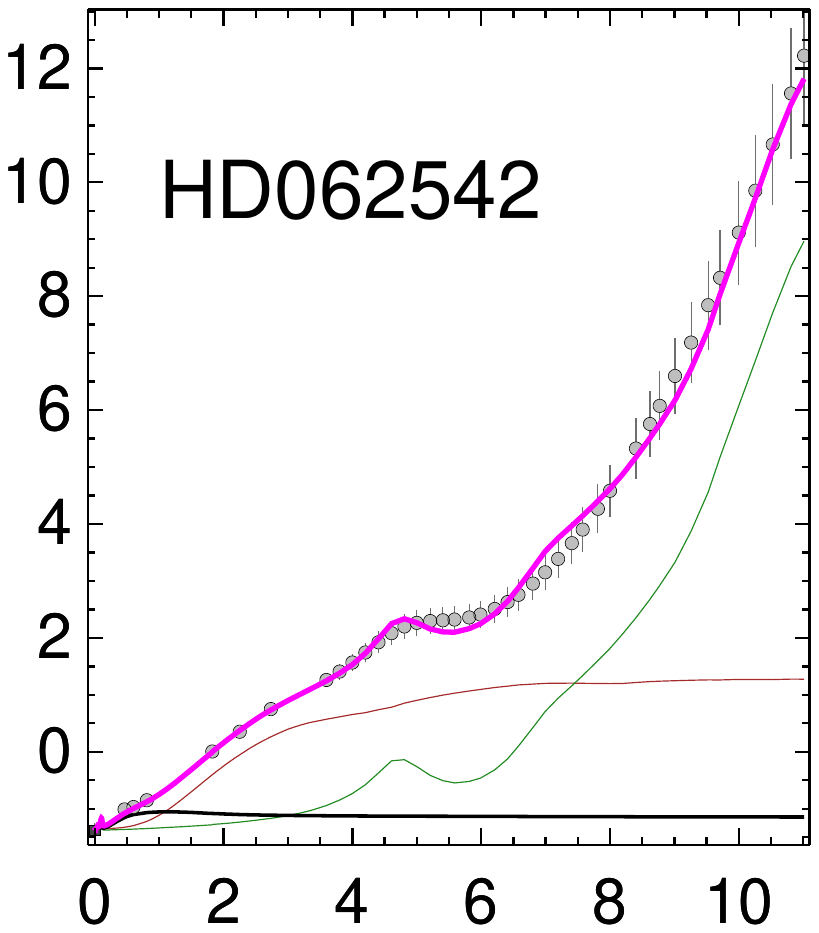}

  \includegraphics[width=3.5cm,clip=true,trim=3.25cm 6.cm 2.1cm 4.5cm]{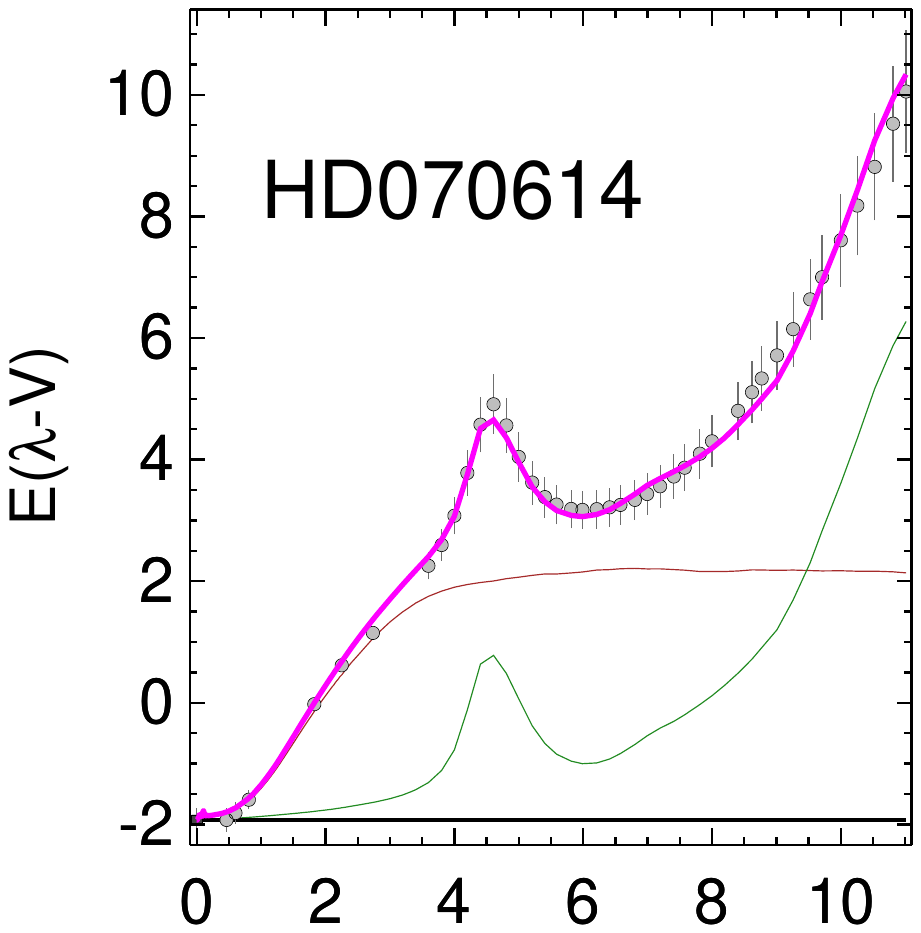}
  \includegraphics[width=3.5cm,clip=true,trim=3.25cm 6.cm 2.1cm 4.5cm]{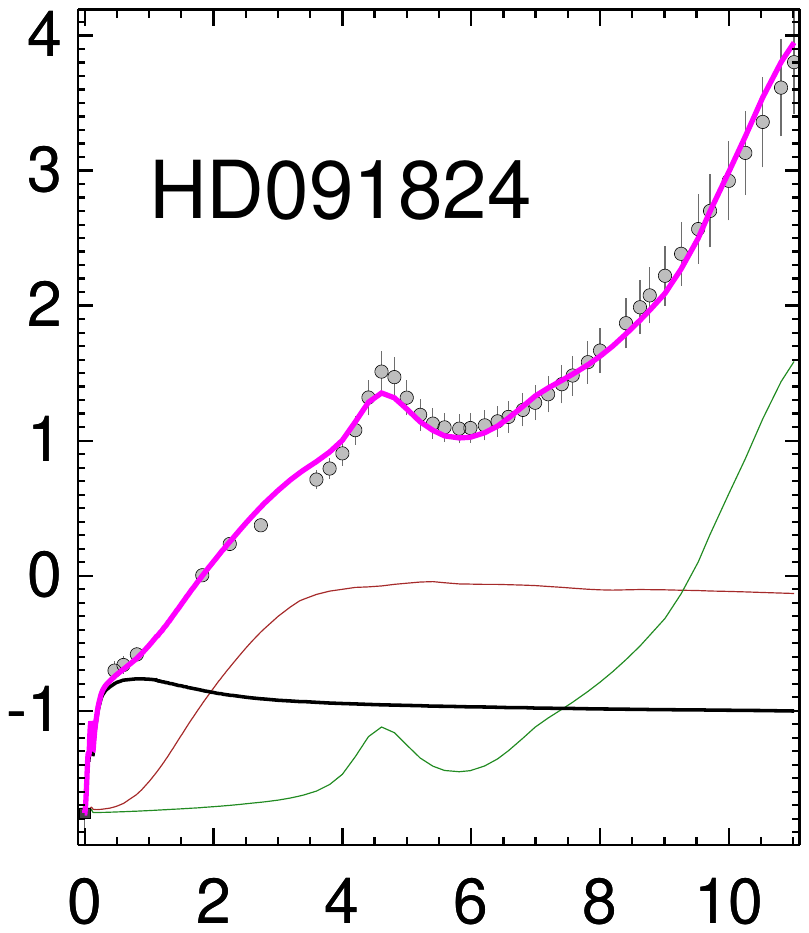}
  \includegraphics[width=3.5cm,clip=true,trim=3.25cm 6.cm 2.1cm 4.5cm]{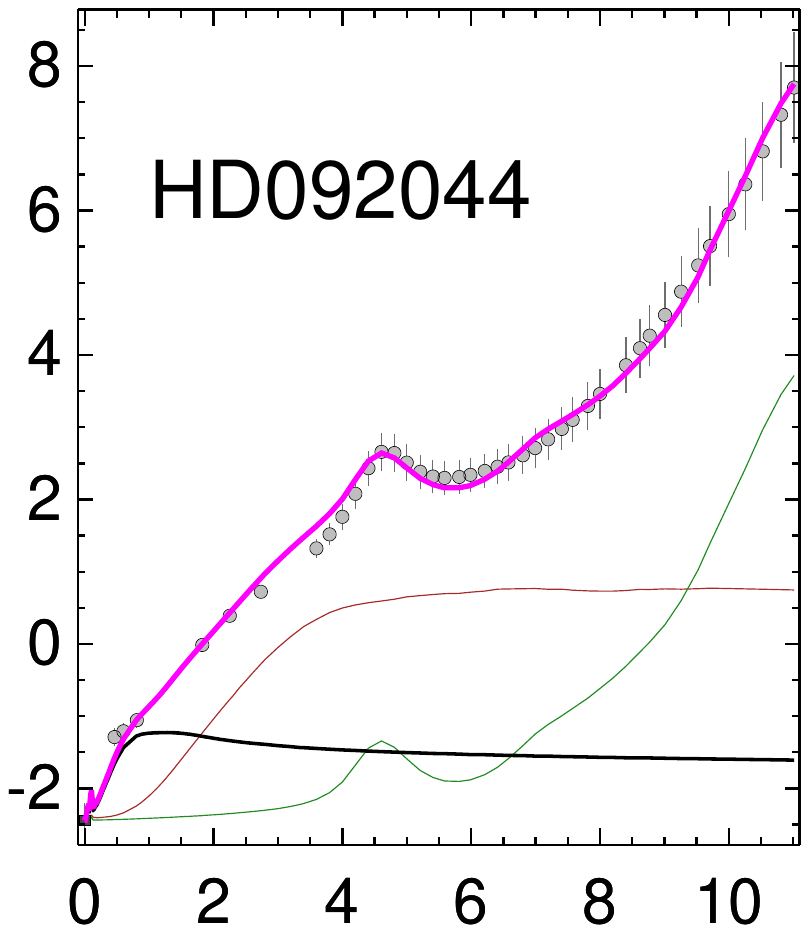}
  \includegraphics[width=3.5cm,clip=true,trim=3.25cm 6.cm 2.1cm 4.5cm]{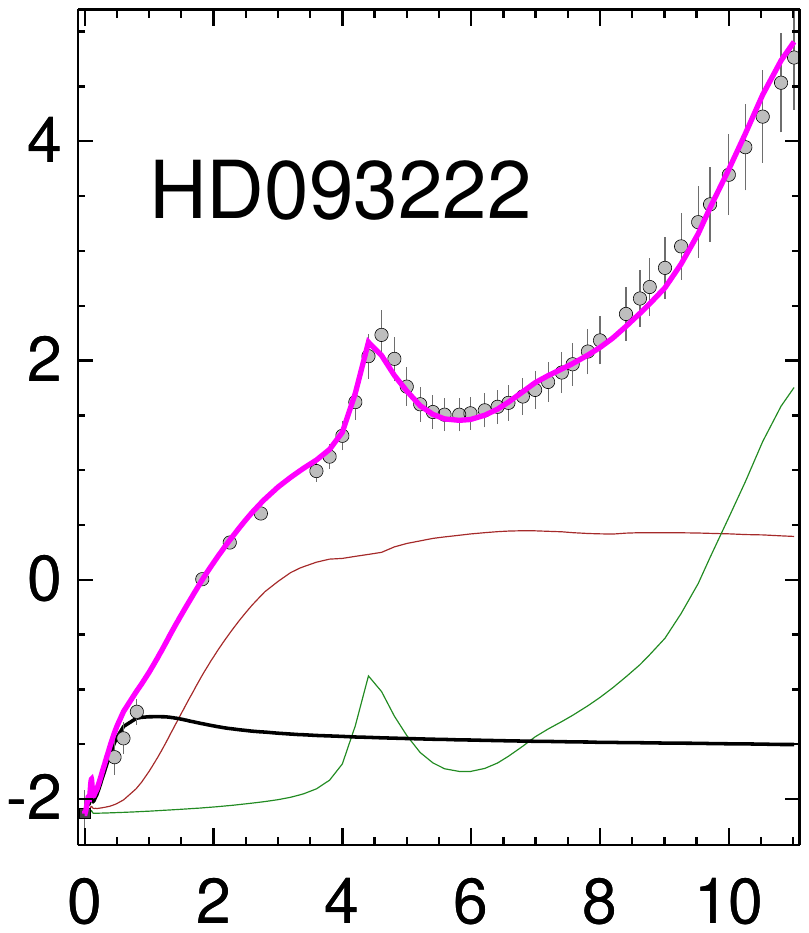}
  \includegraphics[width=3.5cm,clip=true,trim=3.25cm 6.cm 2.1cm 4.5cm]{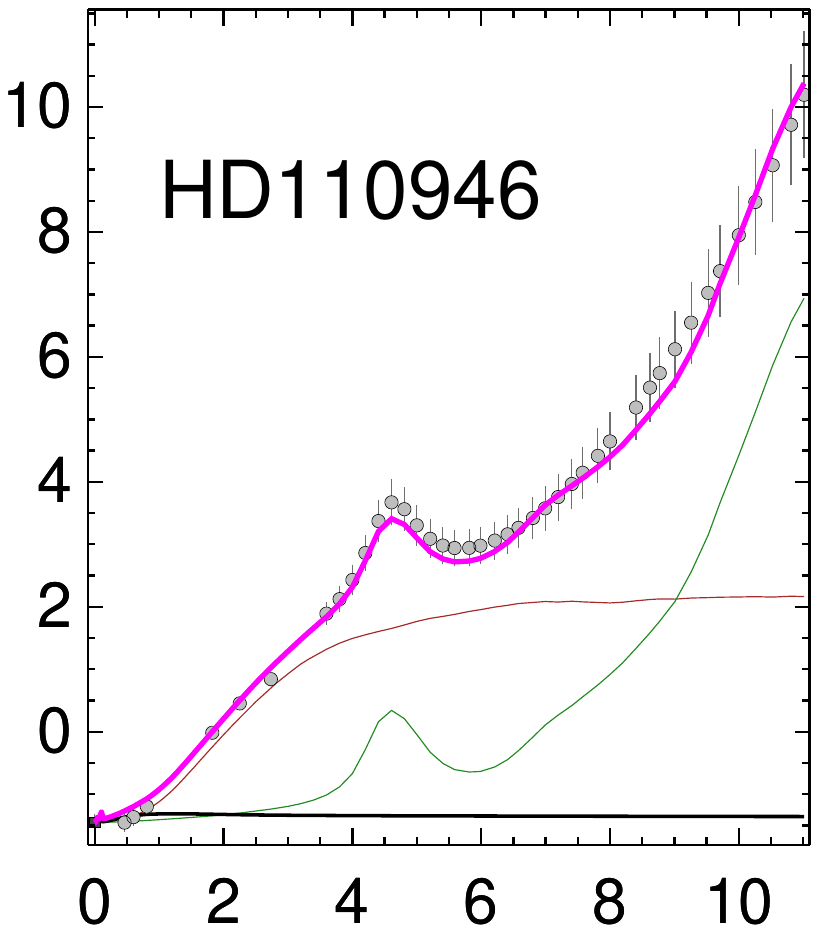}
  
  \includegraphics[width=3.5cm,clip=true,trim=3.25cm 5.cm 2.1cm 4.5cm]{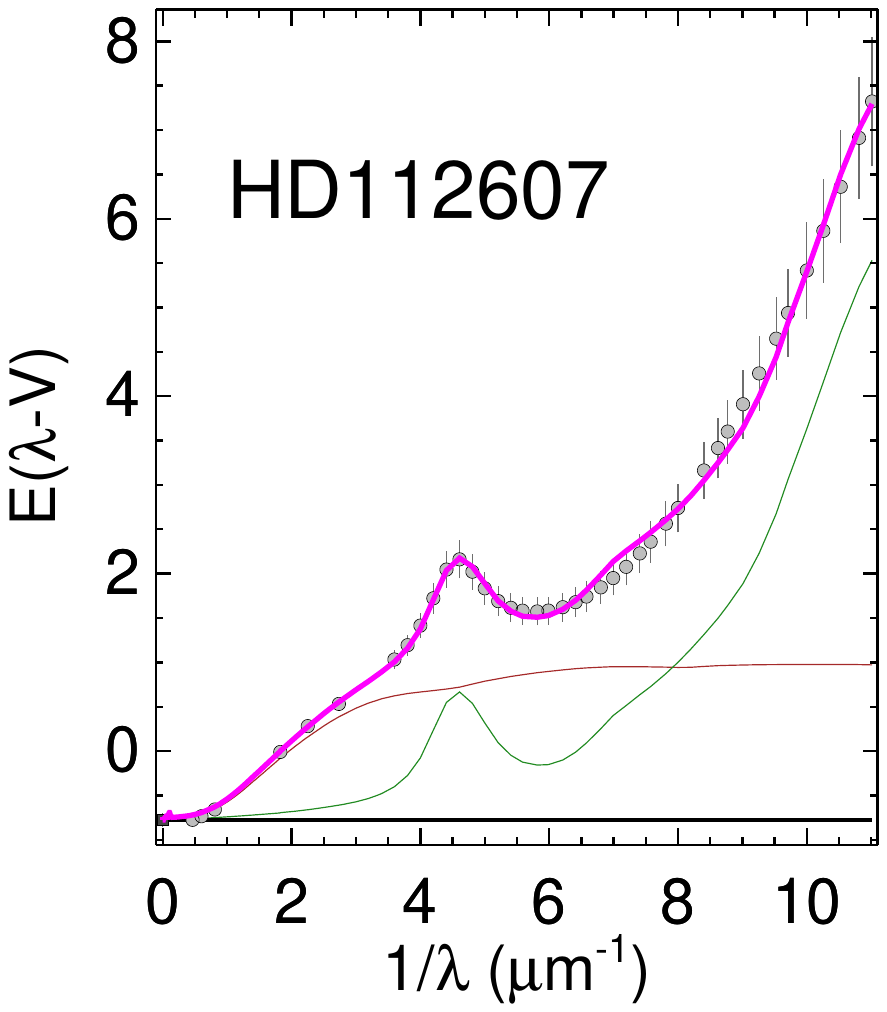}
  \includegraphics[width=3.5cm,clip=true,trim=3.25cm 5.cm 2.1cm 4.5cm]{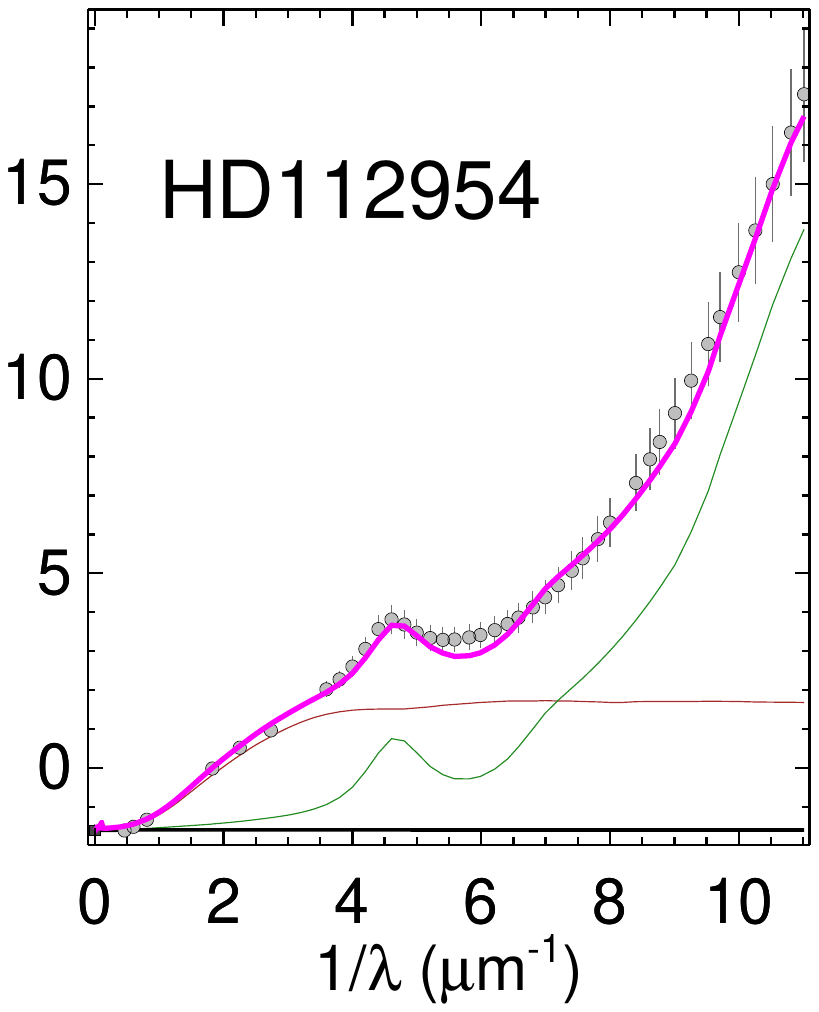}
  \includegraphics[width=3.5cm,clip=true,trim=3.25cm 5.cm 2.1cm 4.5cm]{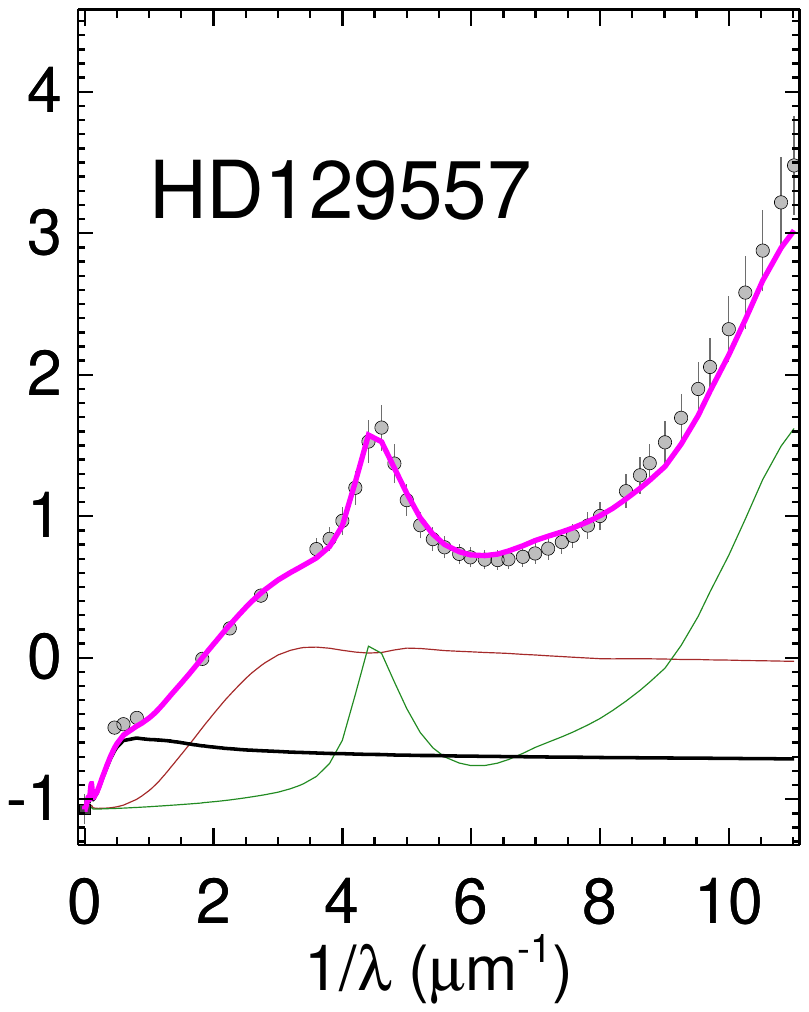}
  \includegraphics[width=3.5cm,clip=true,trim=3.25cm 5.cm 2.1cm 4.5cm]{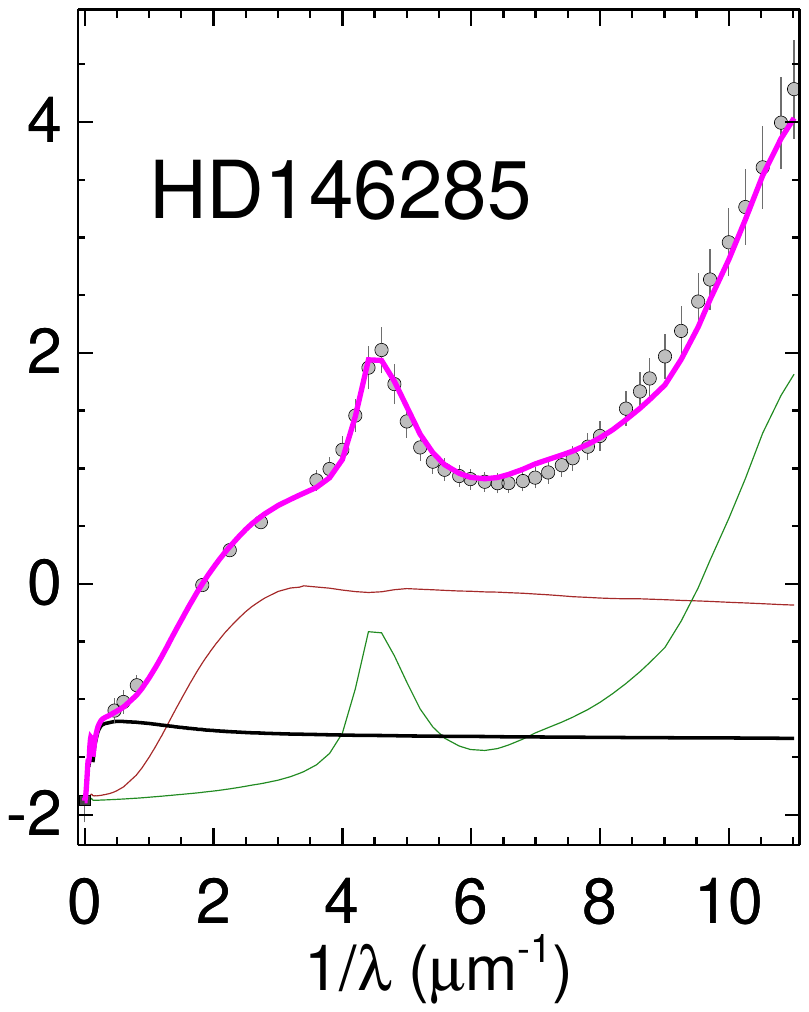}
  \includegraphics[width=3.5cm,clip=true,trim=3.25cm 5.cm 2.1cm 4.5cm]{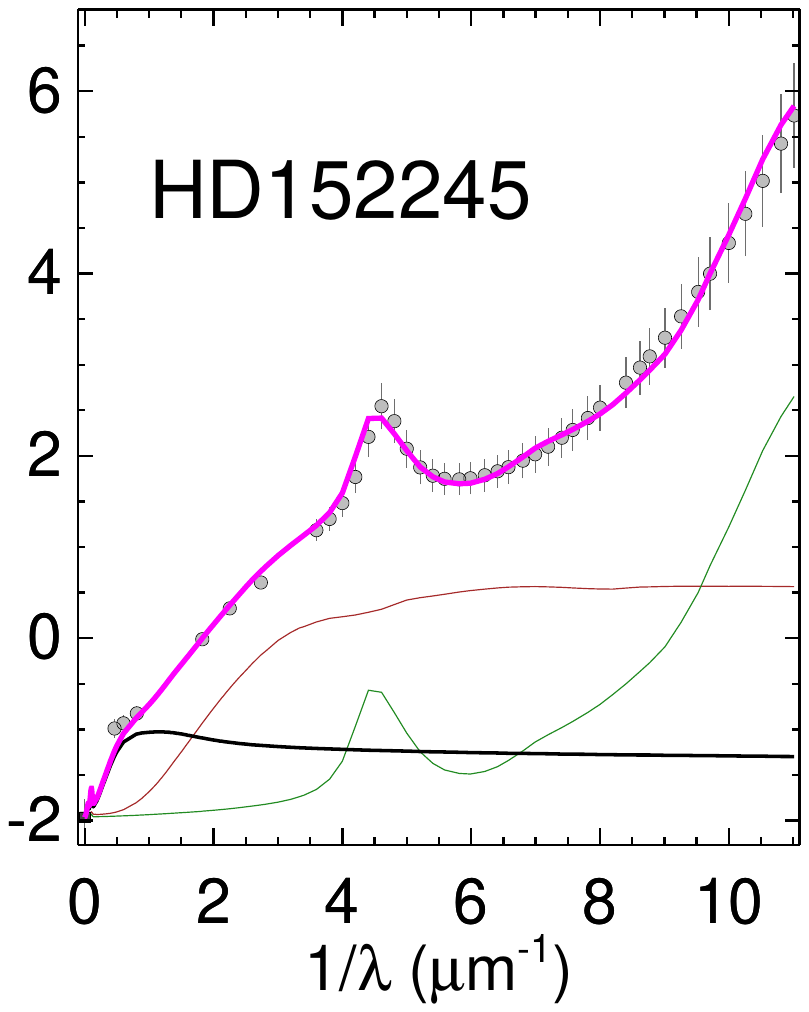}
  \end{center}
\caption{Dust model fits to the absolute reddening curves $E(\lambda
    - V)$ of 24 sightlines. Data (circles) cover $0.09 - 2.2\,\mu$m
    and are complemented at infinite wavelengths by $-A_V$
    (Table~\ref{Tab1.tab}). The best fit, with contributions from
    nanoparticles (green), amorphous silicates and carbon grains
    (brown), and micrometre-sized dust agglomerates (dark), is shown,
    with model parameters listed in Table~\ref{Para.tab}. Notably, the
    micrometre-sized grains dominate in the IR and provide a
    wavelength-independent contribution in the optical and far-UV.
    Continued in Fig.~\ref{FigRedd.cont1}.} \label{FigRedd.pdf}
\end{figure*}

  %  \includegraphics[width=3.5cm,clip=true,trim=3.5cm 6.cm 2.1cm 6.5cm]{./HD027778Pol.pdf}
  %  \includegraphics[width=3.5cm,clip=true,trim=3.5cm 6.cm 2.1cm 4.5cm]{./HD108927Pol.pdf}
  %  \includegraphics[width=3.5cm,clip=true,trim=3.5cm 6.cm 2.1cm 4.5cm]{./HD287150Pol.pdf}
%%%%%%   0 -8  %%%%%%%%%%%%%%%%%%%%%%%%%%%%%%%%%%%%%%%
\begin{figure*} [!htb]
  \begin{center}

  \includegraphics[width=3.5cm,clip=true,trim=2.7cm 6.cm 2.1cm 0.cm]{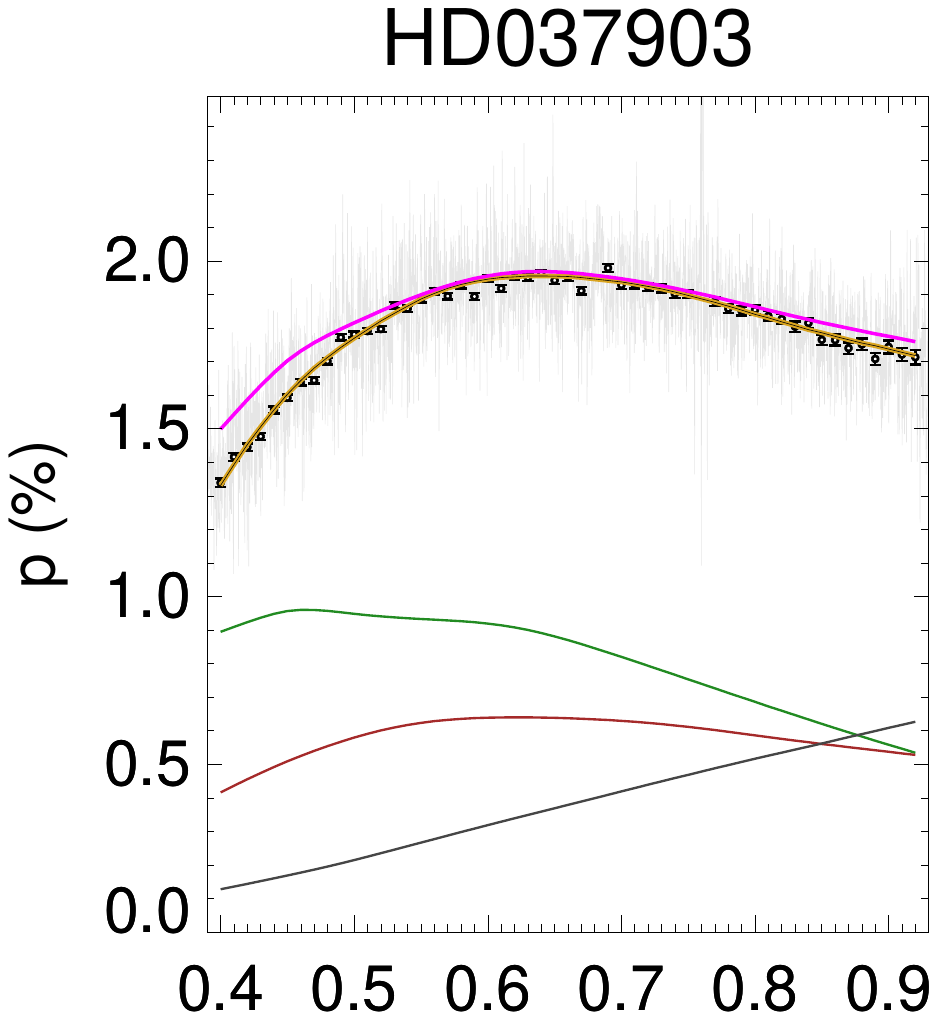}
  \includegraphics[width=3.5cm,clip=true,trim=2.7cm 6.cm 2.1cm 0.cm]{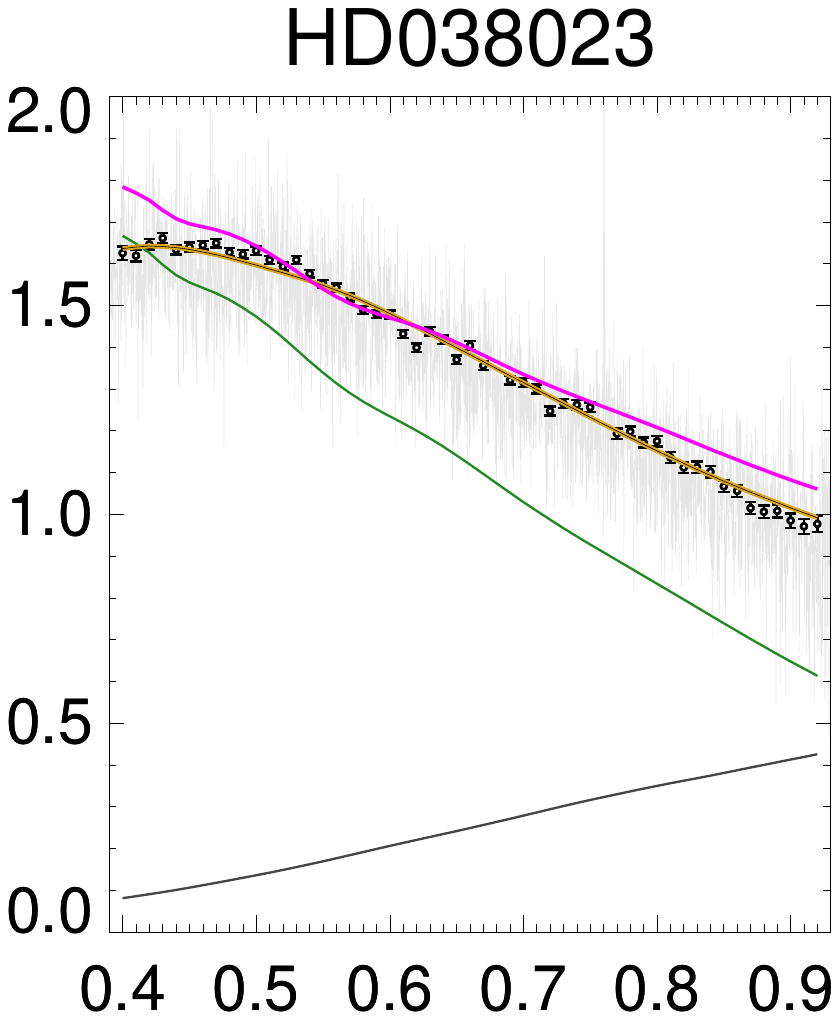}
  \includegraphics[width=3.5cm,clip=true,trim=2.7cm 6.cm 2.1cm 0.cm]{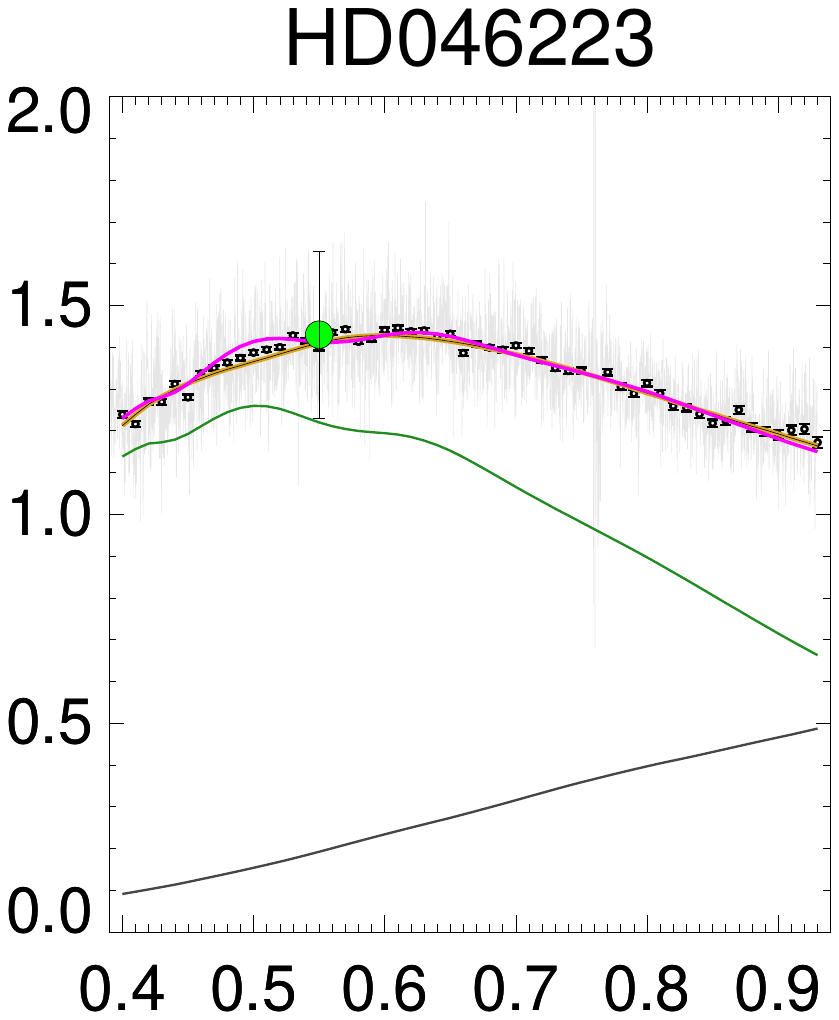}
  \includegraphics[width=3.5cm,clip=true,trim=2.7cm 6.cm 2.1cm 0.cm]{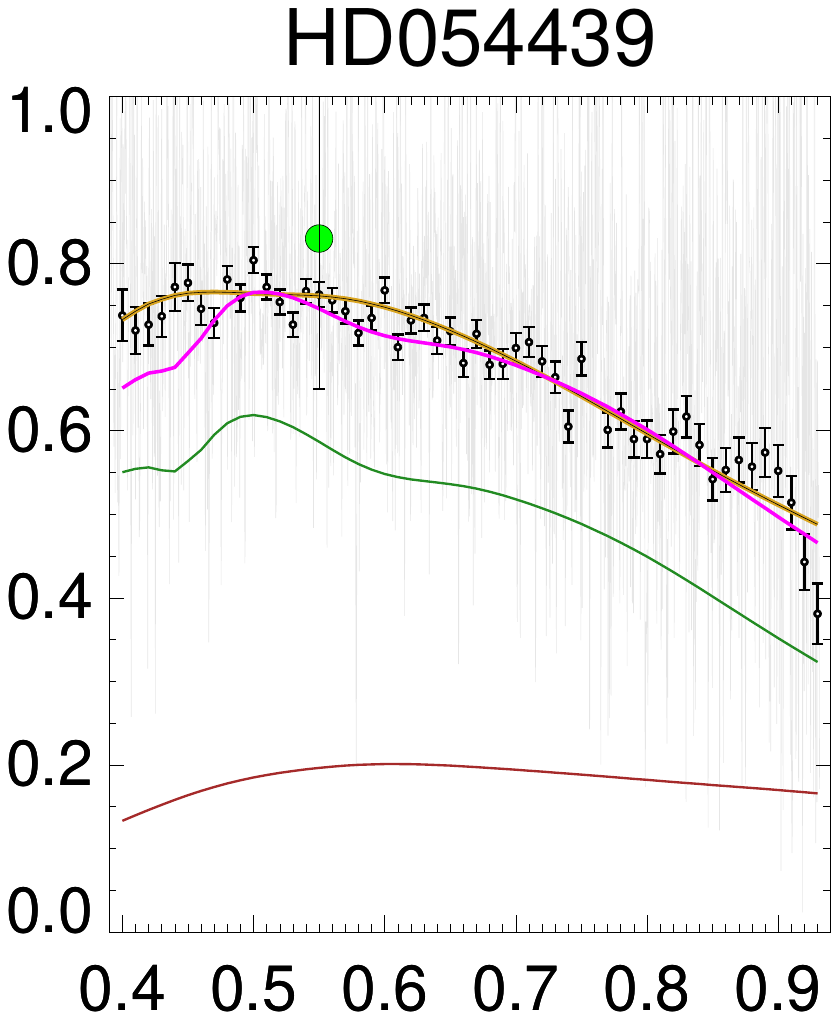}
  \includegraphics[width=3.5cm,clip=true,trim=2.7cm 6.cm 2.1cm 0.cm]{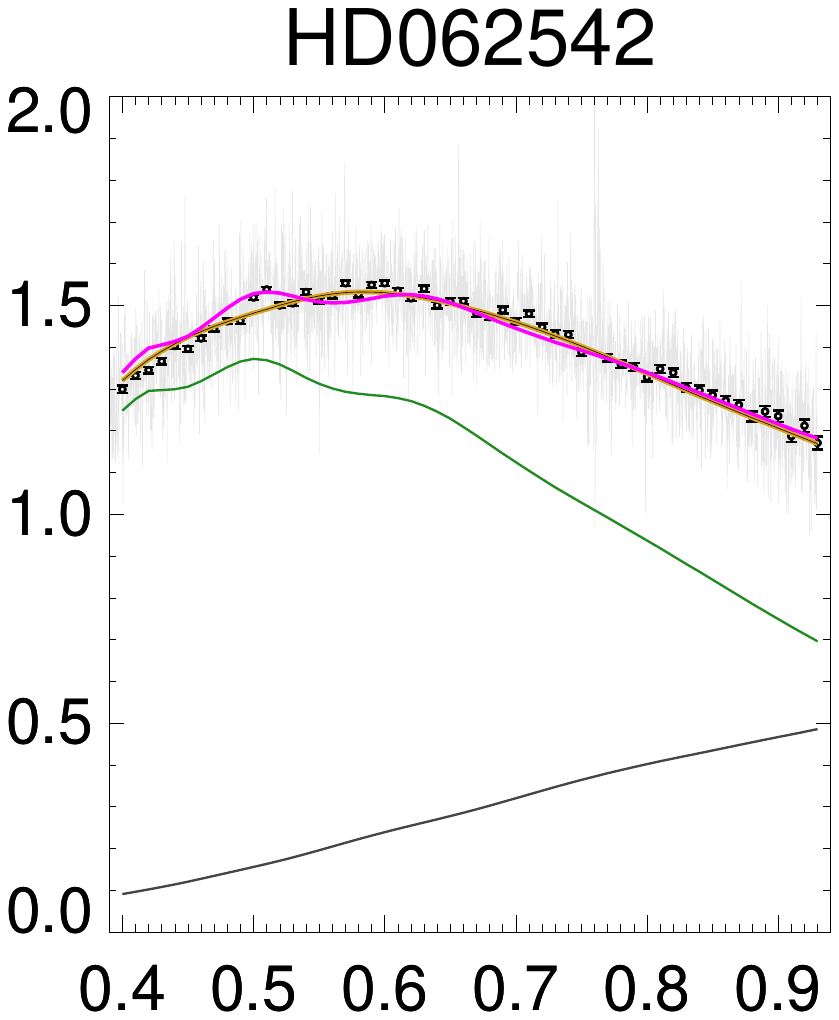}

  \includegraphics[width=3.5cm,clip=true,trim=2.7cm 6.cm 2.1cm 3.3cm]{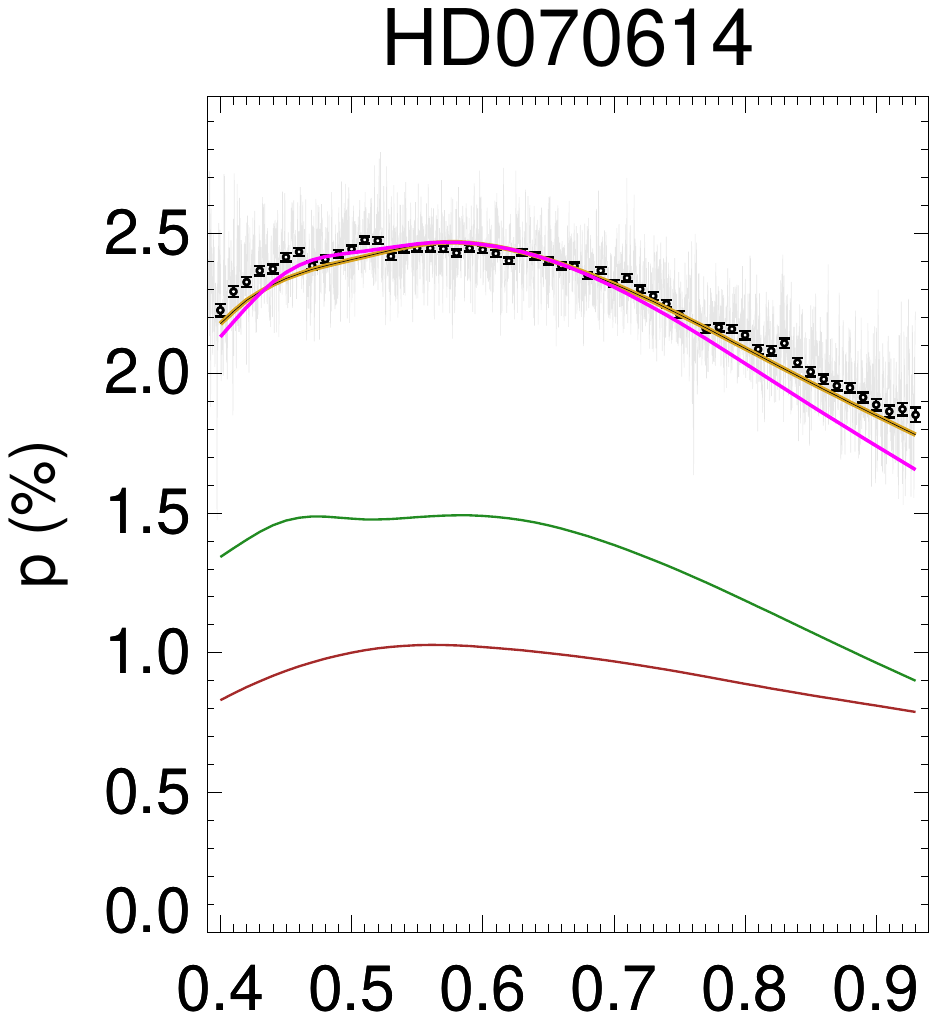}
  \includegraphics[width=3.5cm,clip=true,trim=2.7cm 6.cm 2.1cm 3.3cm]{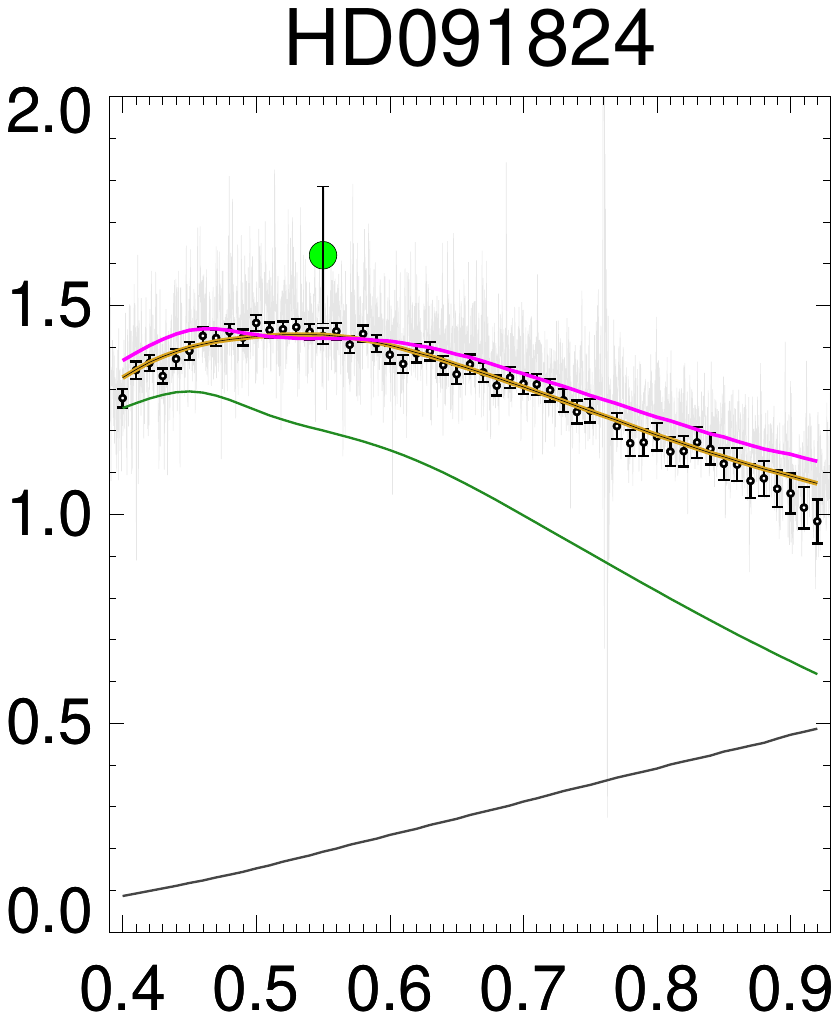}
  \includegraphics[width=3.5cm,clip=true,trim=2.7cm 6.cm 2.1cm 3.3cm]{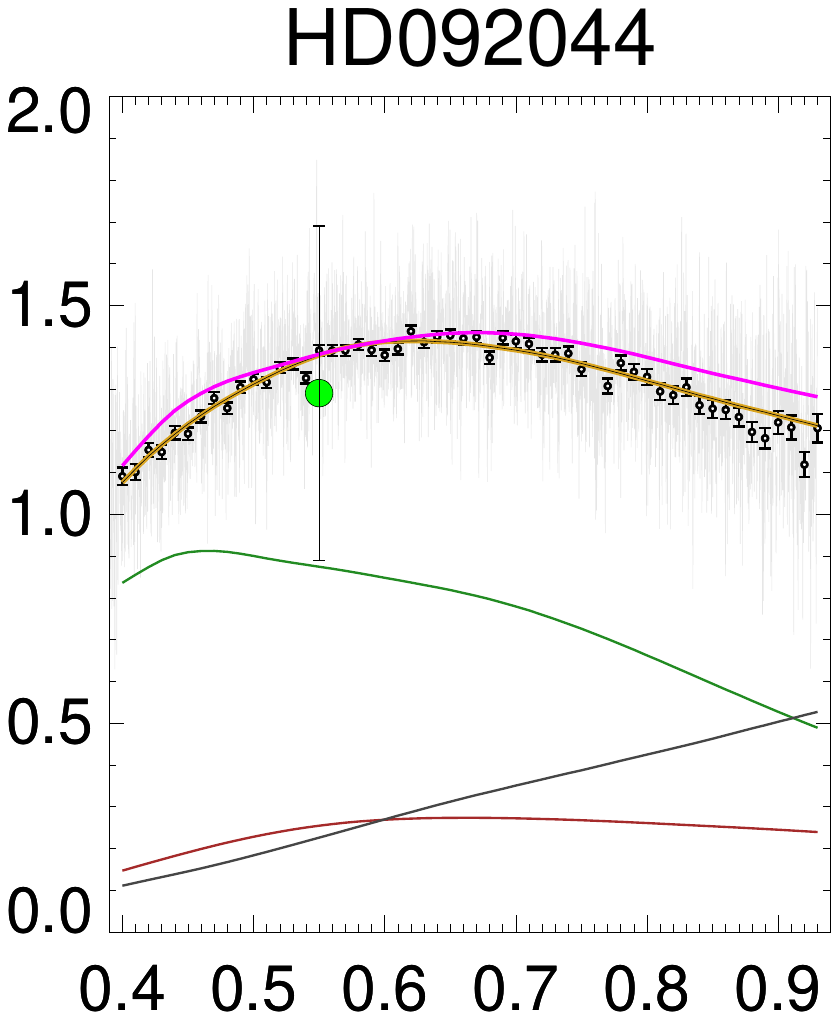}
  \includegraphics[width=3.5cm,clip=true,trim=2.7cm 6.cm 2.1cm 3.3cm]{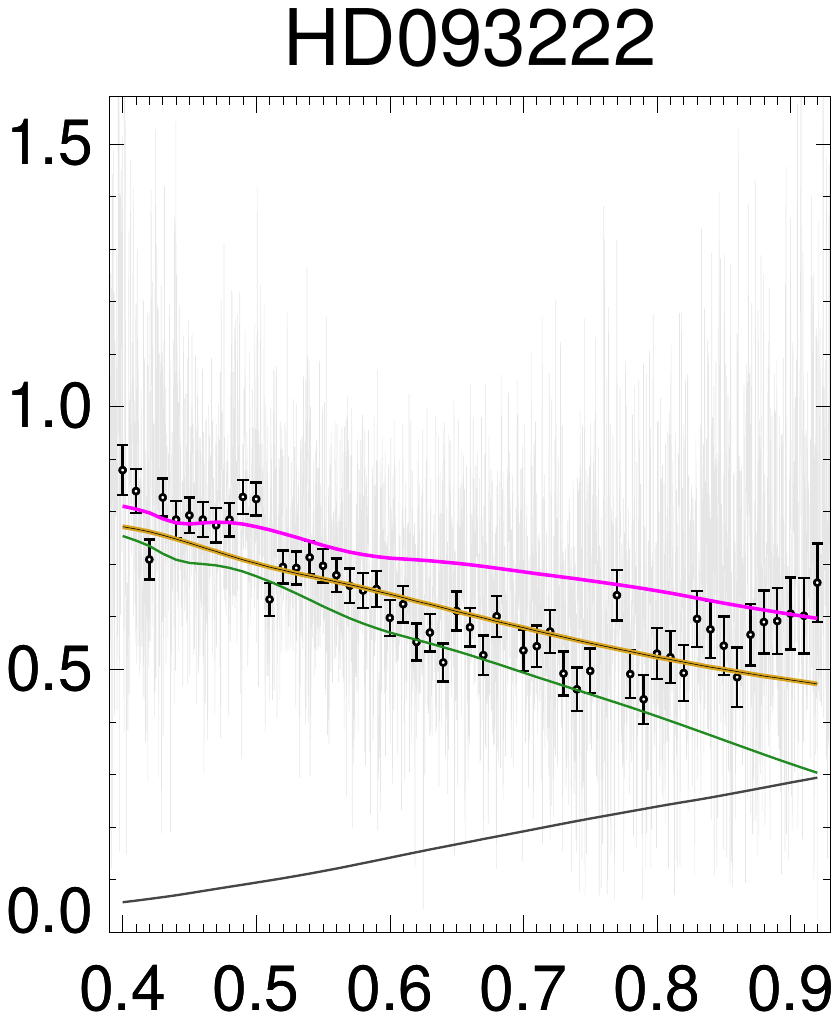}
  \includegraphics[width=3.5cm,clip=true,trim=2.7cm 6.cm 2.1cm 3.3cm]{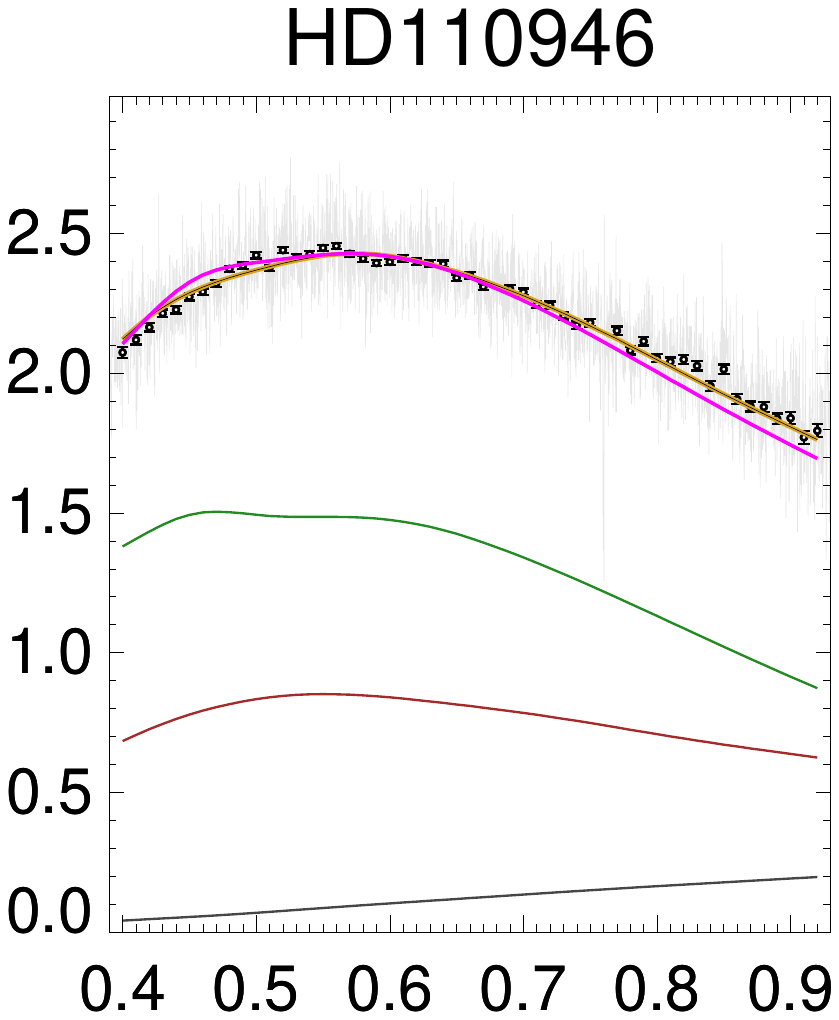}
  
  \includegraphics[width=3.5cm,clip=true,trim=2.7cm 5.cm 2.1cm 2.3cm]{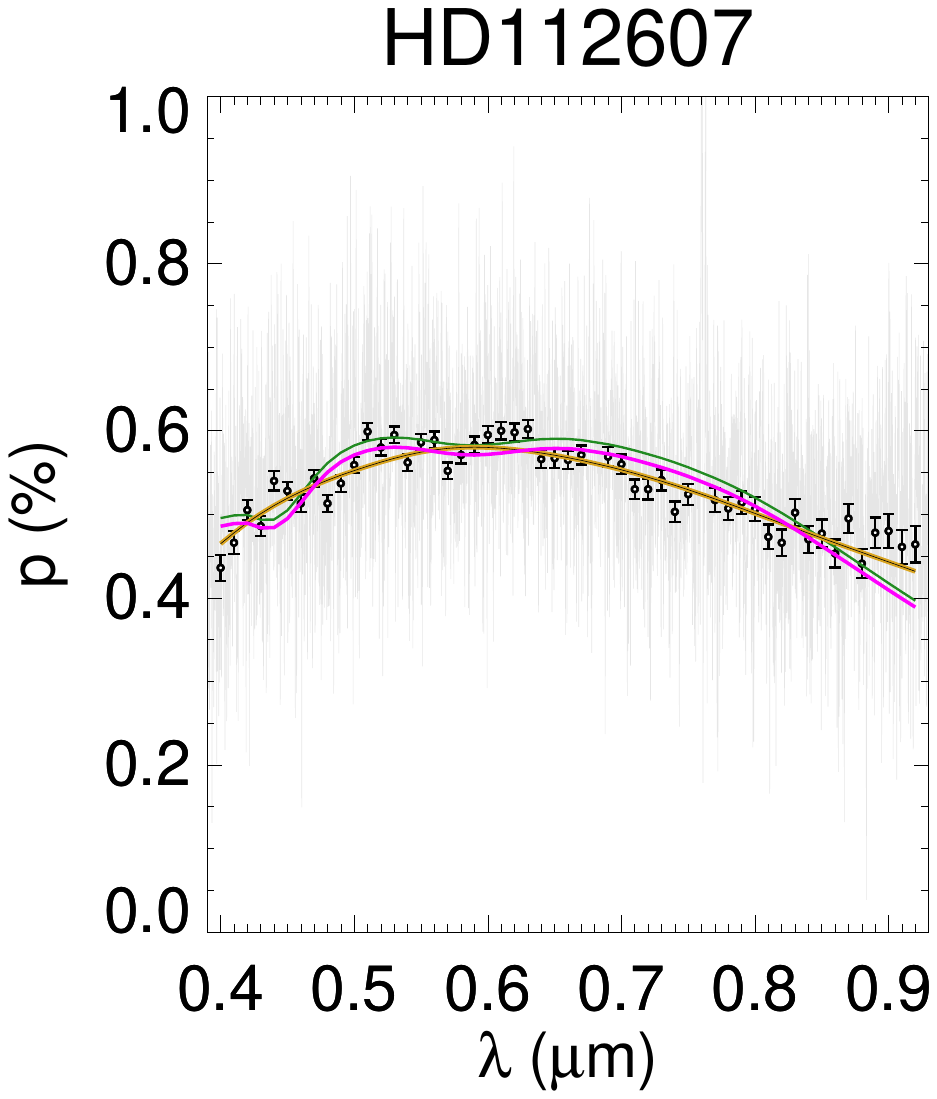}
  \includegraphics[width=3.5cm,clip=true,trim=2.7cm 5.cm 2.1cm 2.3cm]{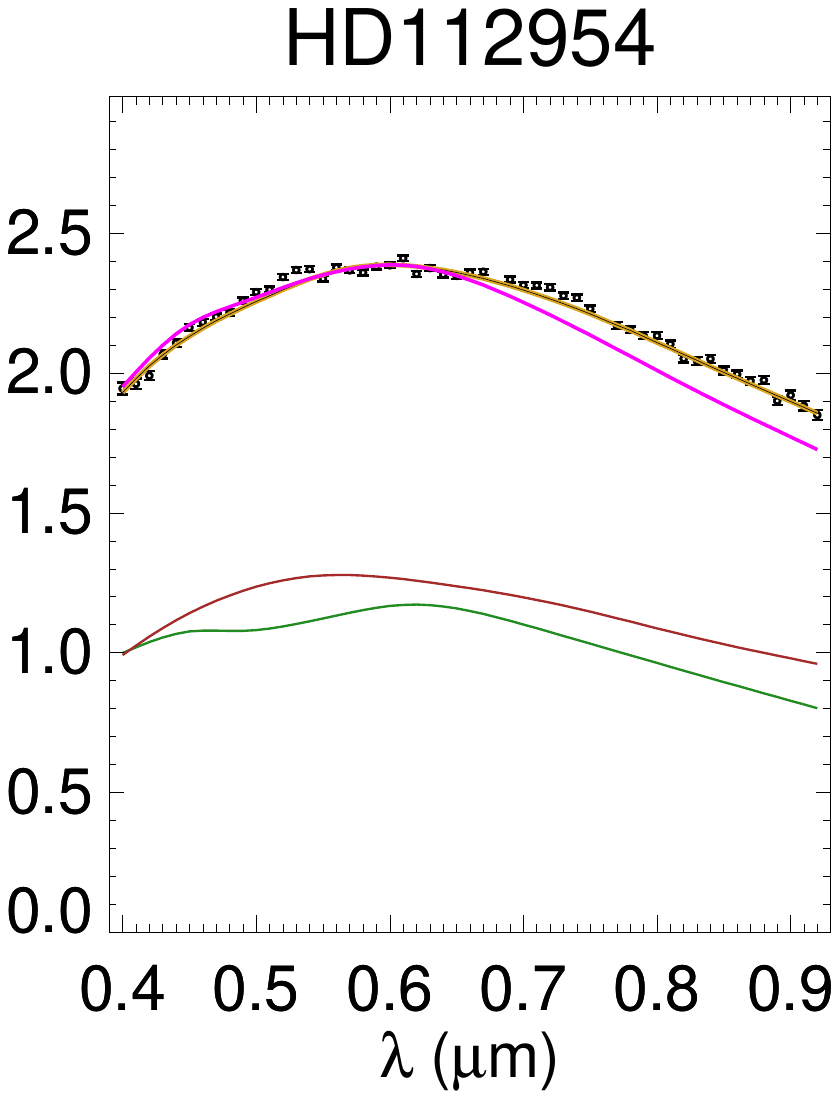}
  \includegraphics[width=3.5cm,clip=true,trim=2.7cm 5.cm 2.1cm 2.3cm]{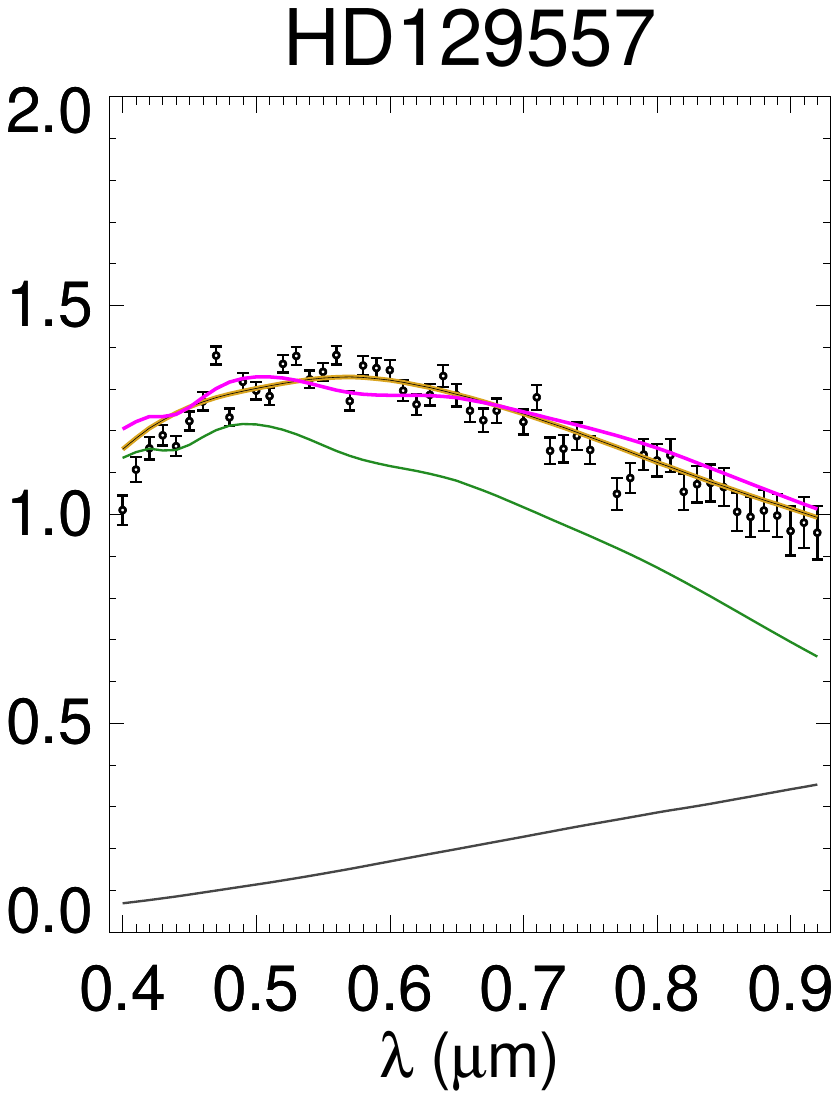}
  \includegraphics[width=3.5cm,clip=true,trim=2.7cm 5.cm 2.1cm 2.3cm]{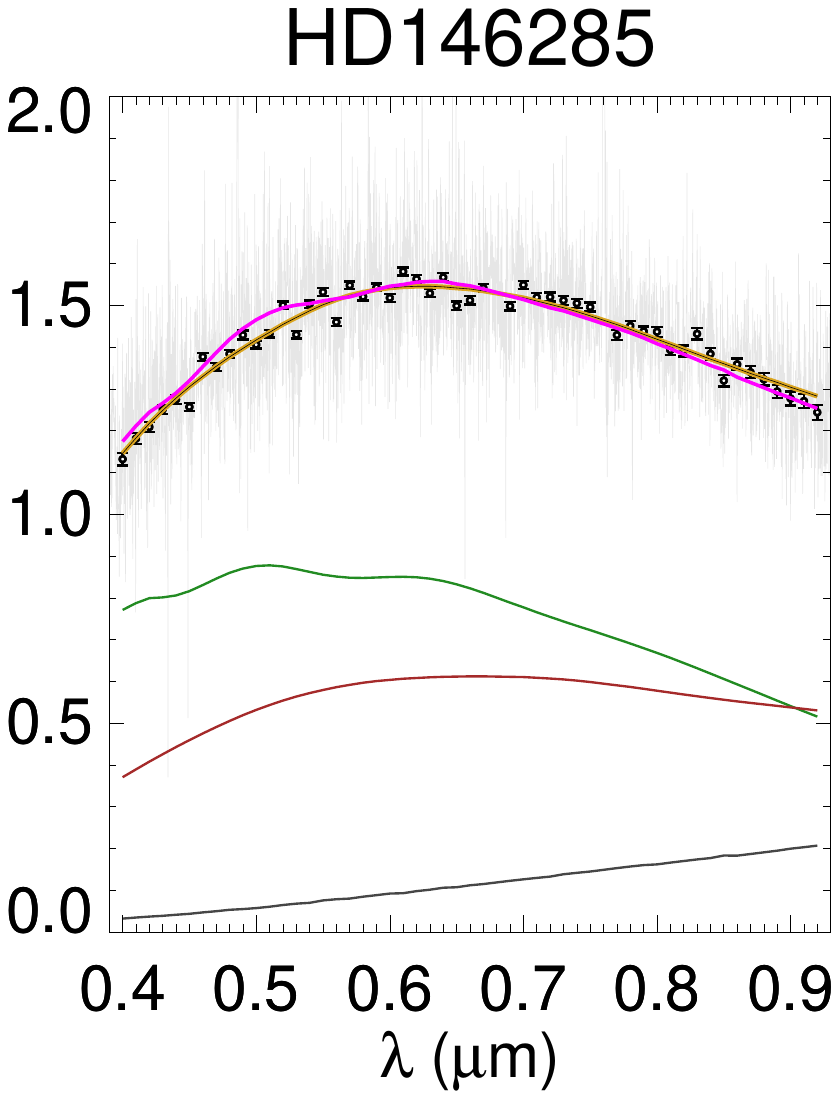}
  \includegraphics[width=3.5cm,clip=true,trim=2.7cm 5.cm 2.1cm 2.3cm]{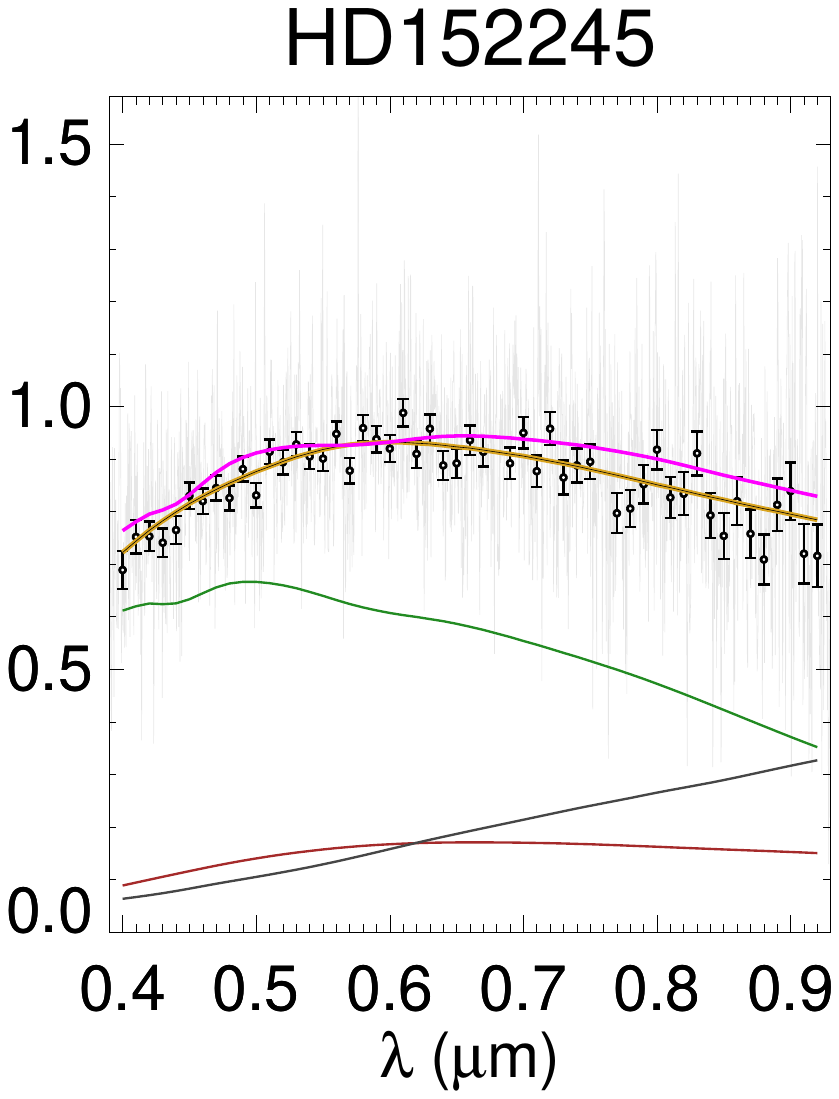}
  \end{center}
\caption{Dust model fits to optical polarisation spectra obtained
      with FORS for 24 sightlines. Observations are shown as the
      original unbinned spectrum (grey lines) and rebinned to a
      spectral resolution of $\lambda / \Delta\lambda \sim 50$
      (black open circles). The error bars associated with the
      rebinned spectra correspond to $1\,\sigma$. In each panel,
      best-fit dust models adopting maximum IDG alignment efficiency
      are shown in ochre, while models based on RAT alignment are
      shown by the magenta line, with individual contributions from
      amorphous silicates (green line), carbon grains (brown line),
      and micrometre-sized dust agglomerates (black line). Data
      available in the stellar polarisation catalogue by
      \cite{Heiles} are shown with a green circle. Continued in
      Fig.~\ref{FigPol.cont1}. \label{FigPol.pdf}}
\end{figure*}

%%%%%%%%%%%%%%%%%%%%%%%%%%%%%%%%%%%%%%%%%%%%%%%
\subsection{Dust modelling of individual sightlines \label{indi.sec}}

The dust model is applied to the remaining 24 sightlines selected in
Sect.~\ref{sightlines.sec}, for which polarised emission data from the
same grains responsible for the optical polarisation are not
available. These stars are located at $|b| < 15^{\circ}$, where the
Planck observations suffer from low ($40'$) spatial resolution. The
expected $90^{\circ}$ flip in the polarisation angle between dichroic
polarisation and polarised emission is also not observed
(Table~\ref{Tab1.tab}).

The reddening curves and starlight polarisation spectra are fitted
using Steps~1 and~2 of the fitting procedure
(Sect.~\ref{pro.sec}). The grain structures of the nominal model ($a/b
= 2$, $V = 10\,\%$), together with the three other combinations of
$a/b \in \{1.5, 2\}$ and $V \in \{5, 10\}$\,(\%), are used. We adopt as
default RAT alignment with perfectly aligned silicates,
${f_{\rm {align}}} = 100\,\%$, and include or exclude alignment of
carbon grains with ${f_{\rm {align}}}({\rm aC}) \in \{0, 50\}$\,\%, as
well as alignment of micrometre-sized dust with
${f_{\rm {align}}} = 100\,\%$. For comparison, IDG alignment is also
applied using a size dependence similar to Eq.~\ref{eta.eq}
\citep{S14}. The model fits and the contributions of the different
grain components to the absolute reddening and the optical
polarisation spectra are shown in Fig.~\ref{FigRedd.pdf} and
Fig.~\ref{FigPol.pdf}, respectively.

The dust parameters and three derived quantities are listed for all 27
sightlines in Table~\ref{Para.tab}. The axial ratios and porosities are
given in cols.~2--3. The model with $a/b = 2$ and a porosity of
$10\,\%$ is preferred for 11 sightlines (63\%), while the model with
$a/b = 1.5$ and a porosity of $5\,\%$ provides a slightly lower
$\chi^2_{\rm tot}$ for 16 sightlines. Except for two cases, the
maximum grain radius is $r^{+}_{\rm \mu{\rm A}} = 1\,\mu{\rm m}$.

The relative masses $m_{\text{vgr}}$, $m_{\text{vSi}}$,
$m_{\text{PAH}}$, $m_{\text{aC}}$, $m_{\text{aSi}}$, and
$m_{\mu{\rm A}}$ in 1\,g of dust (\%) for the individual dust
components are listed in Table~\ref{Para.tab}, cols.~4--9. The
percentage contribution of the micrometre-sized grains to the total
extinction in the optical is given in col.~16. Typically, about
one-third of the dust mass resides in micrometre-sized dust
agglomerates, which also contribute about one-third of the total
extinction $A_{\rm V}$ (Table~\ref{Para.tab}).

The exponent of the dust size distribution $q$, the minimum alignment
radii of amorphous silicate and carbon grains,
$r^-_{\rm {pol,Si}}$ and $r^-_{\rm {pol,aC}}$, and the derived magnetic
field orientation $\Omega$ are given in Table~\ref{Para.tab},
cols.~10--13. Additional derived quantities include the total Si and C
abundances in dust relative to H (in ppm; cols.~14--15). The typical
abundances, [Si]/[H] $\sim 37$\,ppm and [C]/[H] $\sim 96$\,ppm, agree
with estimates for the diffuse ISM by \citet{HD21}. The median values
and the $1\sigma$ scatter of the parameters are provided at the bottom
of the table. The large scatter indicates substantial variation across
individual sightlines.

As shown in Fig.~\ref{FigRedd.pdf}, sightlines containing a
significant amount of dark dust are dominated in the near-IR
reddening by these micrometre-sized grains, which produce grey
(constant) extinction in the optical. The amorphous grains produce a
linear rise in extinction in the optical and add grey extinction in
the far-UV. Nanoparticles are responsible for the 2175\,\AA{}
extinction bump and the steep far-UV rise.

The polarisation spectra (Fig.~\ref{FigPol.pdf}) generally follow the
Serkowski law with $\lambda_{\rm max}$ in the visible. Seven stars
peak in the $B$ band, and twelve stars peak at
$\lambda_{\rm max} \simgreat 0.65\,\mu{\rm m}$
(Table~\ref{Tab1.tab}). The Serkowski fit is usually interpreted as
tracing the characteristic size of aligned grains in a single
absorbing cloud, although this is a simplification
\citep{Andersson11}. The polarisation results from the sum of
contributions from silicate grains, dark dust, and (when aligned)
amorphous carbon particles. These components peak at different
wavelengths and may arise from different clouds along the sightline
\citep{Mandarakas25}.

A noticeable deviation from the Serkowski curve is seen towards
HD~093222, where the polarisation peaks at
$\lambda_{\rm max} = 0.43\,\mu$m and, after an initial decline, rises
again at $0.7\,\mu$m towards the IR (Fig.~\ref{FigPol.pdf}). A strong
wavelength gradient in the polarisation angle,
${\rm d}\theta/{\rm d}\lambda = 97^{\circ}/\mu$m, is also observed
(Table~\ref{Tab1.tab}).

In a single-cloud scenario with a common magnetic field, the
polarisation angle should remain constant and the polarisation should
peak for typical ISM grain sizes at
$\lambda_{\rm max} = 0.55\,\mu$m. Indeed, \citet{Mandarakas25}
reproduce the observed wavelength dependence of the polarisation
angle towards HD~093222 using a two-cloud model. Two additional
multiple-cloud sightlines in our sample, HD~037903 and HD~152245,
show variations in polarisation angle, with
${\rm d}\theta/{\rm d}\lambda = 5.3$ and $6.8^{\circ}/\mu$m and maxima
at $\lambda_{\rm max} = 0.66$ and $0.62\,\mu$m, respectively. Five
single-cloud sightlines\footnote{HD~054439, HD~046223, HD~092044,
HD~038023, and HD~294304} show variations of
$6.6 \lesssim {\rm d}\theta/{\rm d}\lambda \lesssim 18.5$
($^{\circ}/\mu{\rm m}$), while the remaining sightlines exhibit a weak
wavelength dependence with
${\rm d}\theta/{\rm d}\lambda < 5^{\circ}/\mu$m
(Table~\ref{Tab1.tab}, Fig.~\ref{FigTheta.pdf}).

The model fits the optical polarisation spectra typically within
$1\sigma$ of the FORS data rebinned to
$\lambda/\Delta\lambda \sim 50$ (Fig.~\ref{FigPol.pdf}). The
contributions of the individual grain components to the optical
polarisation spectra are shown in Fig.~\ref{FigPol.pdf}. Most of the
polarisation, especially in the $V$ band and at shorter wavelengths,
is produced by silicate grains with a minimum alignment radius of
$r^-_{\rm {pol,Si}} = 118 \pm 23$\,nm. For half of the sightlines, the
polarisation spectra are better reproduced when aligned carbon
particles are included, with
$r^-_{\rm {pol,aC}} = 69 \pm 25$\,nm. The contribution from
micrometre-sized grains rises steadily from $p \sim 0.1$\,\% in the UV
to $p \simless 0.5$\,\% at $0.9\,\mu$m. These grains dominate the
polarised emission in the (sub)millimetre regime
(Fig.~\ref{best.pdf}).

The IDG model predicts lower alignment efficiency than RAT theory.
Fits to the optical polarisation spectra using IDG systematically
require the maximal alignment efficiency allowed by this model, with
both silicate and carbon grains being aligned. Note that the lower
alignment efficiencies in IDG are compensated in the fits (orange
lines in Fig.~\ref{FigPol.pdf}) by increasing the axial ratio to
typically $a/b = 2.5$, the porosity to $10\,\%$, and by overestimating
the magnetic-field orientation $\Omega$, typically by a factor of
two.

%%%%%%%%%%%%%%%%%%%%%%%%%%%%%%%%%%%%%%%%
\section{Conclusion}

We completed the Large Interstellar Polarisation Survey (LIPS), which
obtained FORS spectropolarimetry in the $0.38$–$0.92\,\mu$m wavelength
range for 161 sightlines through the diffuse ISM. Sixty polarisation
spectra are presented in this work. The LIPS sample was selected based
on the availability of reddening curves: in the far-UV from the IUE
and FUSE satellite missions, in the optical from ground-based
photometry, and in the near-IR from 2MASS. High-resolution spectra
were obtained with UVES/VLT to verify the spectral types and
luminosity classes of the stars used for deriving the reddening
curves and to probe the number of clouds along individual
sightlines. Gaia parallaxes were used to estimate the visual
extinction $A_V$ necessary for reconciling the derived luminosity
distances with the trigonometric distance estimates for the same
stars. The starlight polarisation spectra were complemented by
Planck 850\,$\mu$m polarimetry.

This dataset is used to constrain the properties of grains in the
diffuse ISM using a three-component model by \cite{S23} that includes
nanoparticles, amorphous grains, and micrometre-sized grains. The
nanoparticles are responsible for the far-UV rise in the reddening
curve, the 2175\,\AA\ bump, and the mid-IR emission bands. The
amorphous grains produce grey (constant) extinction in the far-UV,
an almost linear decline towards longer wavelengths, and the far-IR
emission. The micrometre-sized particles contribute grey extinction
in the optical, a linear decline in the near-IR, and the
submillimetre emission. The optical polarisation is dominated by
amorphous grains, while the 850\,$\mu$m polarisation is dominated by
the micrometre-sized dust agglomerates unless amorphous carbon
particles are also aligned.

Within the sample, three sightlines exhibit both significant
starlight and Planck polarisation, with comparable extinction
values, $A^{850}_V \sim A_V$, and the expected $90^{\circ}$ reversal
in the polarisation angle between polarised emission and starlight
polarisation. The polarisation data for these sightlines provide an
almost orthogonal perspective on the aligned grains, enabling
constraints on their particle shape, porosity, and alignment
efficiency. Although degeneracies exist among the dust model
parameters, a good fit to all data for these three sightlines is
obtained using an axial ratio $a/b = 2$ and porosities of $10\,\%$
for both the amorphous and micrometre-sized grains.

We applied RAT theory \citep{HL_16} in the simplified formulation of
\citet{Reissl20}, assuming perfect alignment for silicate and
micrometre-sized dust. For carbon grains, we considered two cases,
either including alignment with
${f_{\rm {align}}}({\rm aC}) = 50\,\%$ or assuming no alignment. This
nominal model is applied to 24 additional sightlines that possess
high-quality reddening curves and optical polarisation spectra, but
for which the Planck polarisation data are ambiguous. For half of
these sightlines, the best fits include aligned carbon grains. A
single-cloud model generally provides fits consistent within the
$1\sigma$ uncertainties. The IDG models, which ignore internal
alignment, reproduce the optical polarisation only when forced to
their asymptotic maximum alignment for both silicate and carbon
grains, at the cost of overestimating parameters such as the magnetic
field orientation $\Omega$.

The contribution of micrometre-sized dust is most evident in the
near-IR extinction and, with few exceptions, is marginal in the
optical spectropolarimetry. Overall, we find that micrometre-sized
dust is responsible for approximately one-third of the total
extinction and comprises one-third of the total dust mass.
Significant variations in dust abundances persist from cloud to
cloud.

The present analysis is limited by the low spatial resolution of the
Planck maps, which leads to source confusion. A follow-up
submillimetre survey with a high spatial-resolution polarimeter is
required, once it becomes available, to combine polarised emission
and starlight polarisation measurements through the diffuse ISM.

\section{Code and data availability}

The Fortran code of the dust model described in \citet{SH26} and the
corresponding library of dust cross-sections for spheroidal grains
\citep{SH26b} are publicly available. The processed FORS polarisation
spectra presented in this work are accessible at the CDS
(\url{https://cdsarc.cds.unistra.fr}).

\begin{acknowledgements}
We thank the referee for the valuable and constructive comments, in
particular regarding the consideration of radiative torque alignment
(RAT) theory. This research has made use of the services of the ESO
Science Archive Facility and of the SIMBAD database, operated at the
CDS, Strasbourg, France. This work is partially based on observations
collected at the European Southern Observatory under ESO programme
102.C-0040.
\end{acknowledgements}

\bibliographystyle{aa}
\bibliography{References}

\begin{appendix}
\onecolumn
\section{Tables and figures}

%%%%%%   0 -8  %%%%%%%%%%%%%%%%%%%%%%%%%%%%%%%%%%%%%%%
\begin{figure*} [!htb]
  \begin{center}
 \includegraphics[width=3.6cm,clip=true,trim=2.7cm 5.0cm 1.5cm 3.0cm]{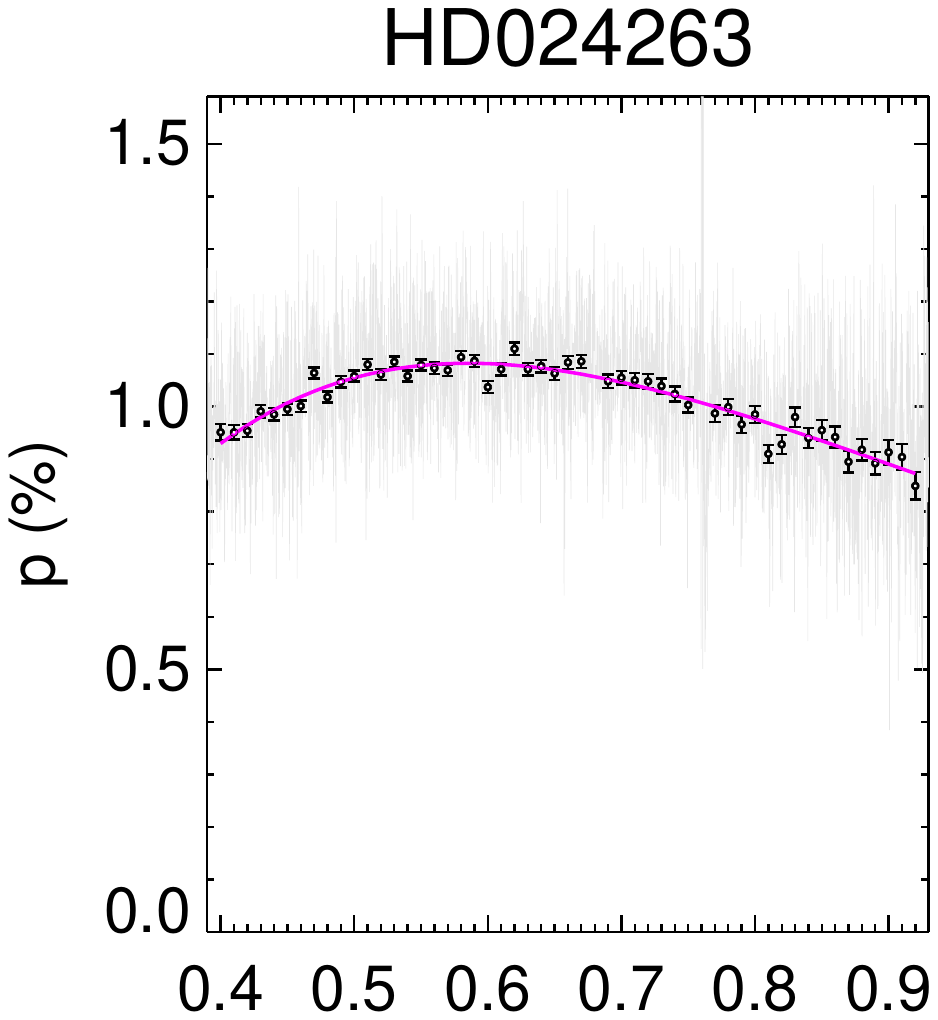}
 \includegraphics[width=3.6cm,clip=true,trim=2.7cm 5.0cm 1.5cm 3.0cm]{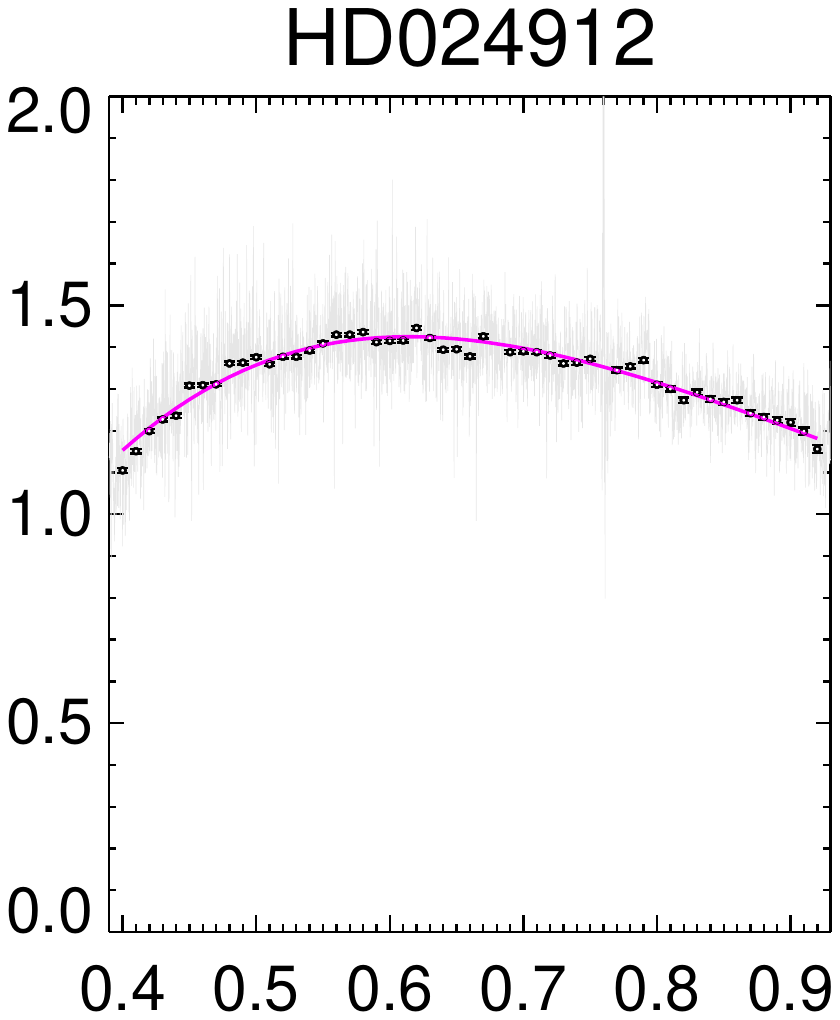}
 \includegraphics[width=3.6cm,clip=true,trim=2.7cm 5.0cm 1.5cm 3.0cm]{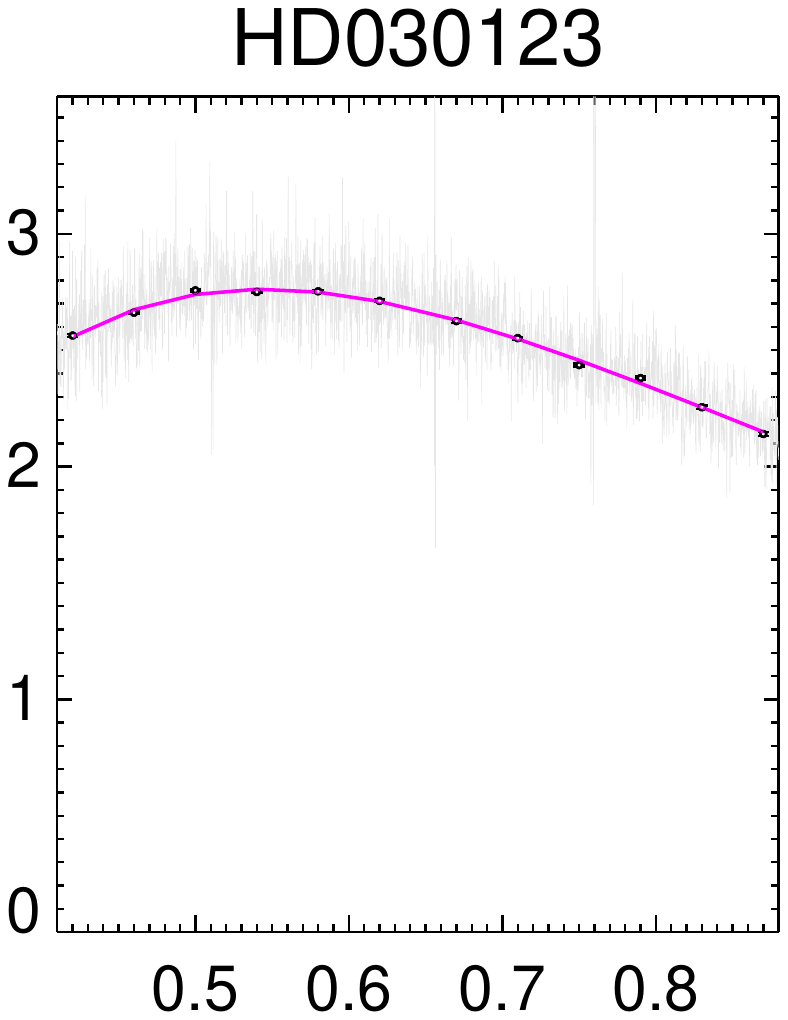}
 \includegraphics[width=3.6cm,clip=true,trim=2.7cm 5.0cm 1.5cm 3.0cm]{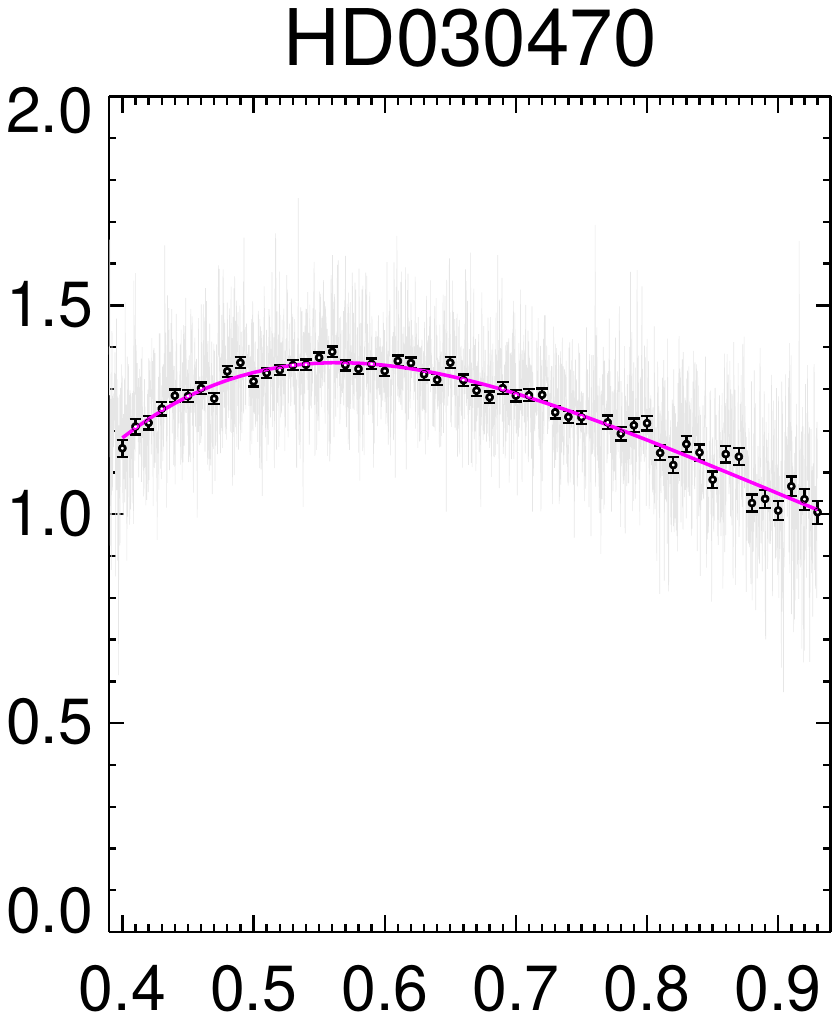}
 \includegraphics[width=3.6cm,clip=true,trim=2.7cm 5.0cm 1.5cm 3.0cm]{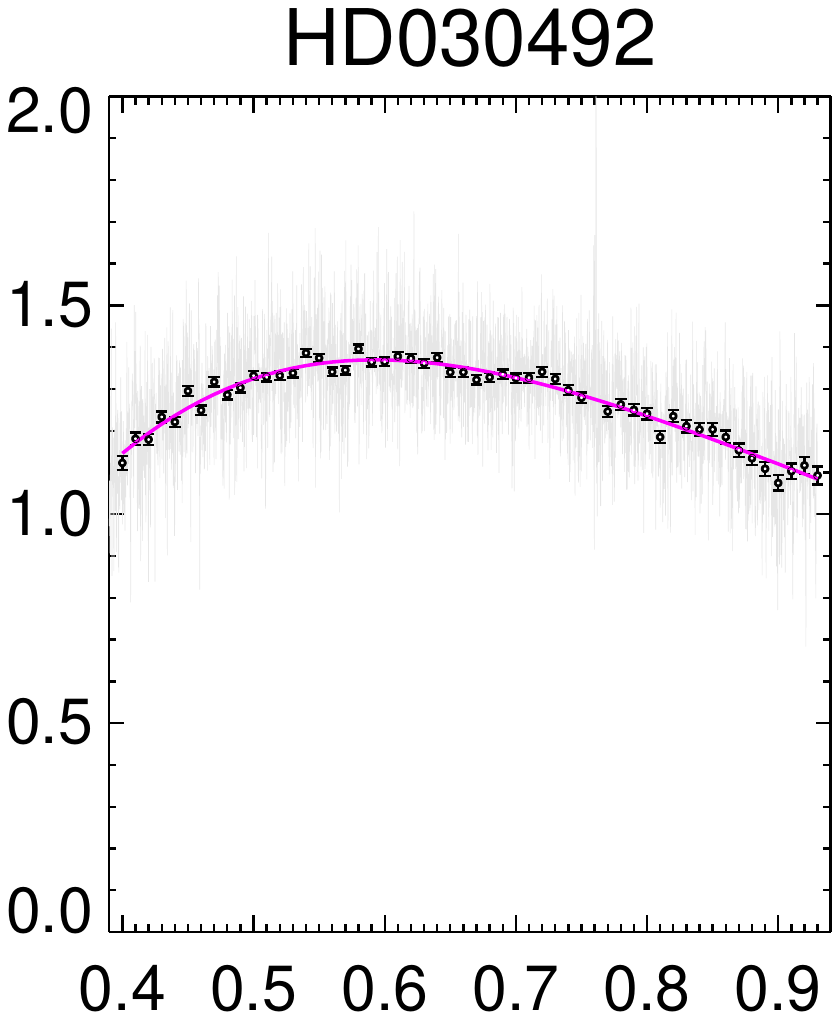}
 \includegraphics[width=3.6cm,clip=true,trim=2.7cm 5.0cm 1.5cm 3.0cm]{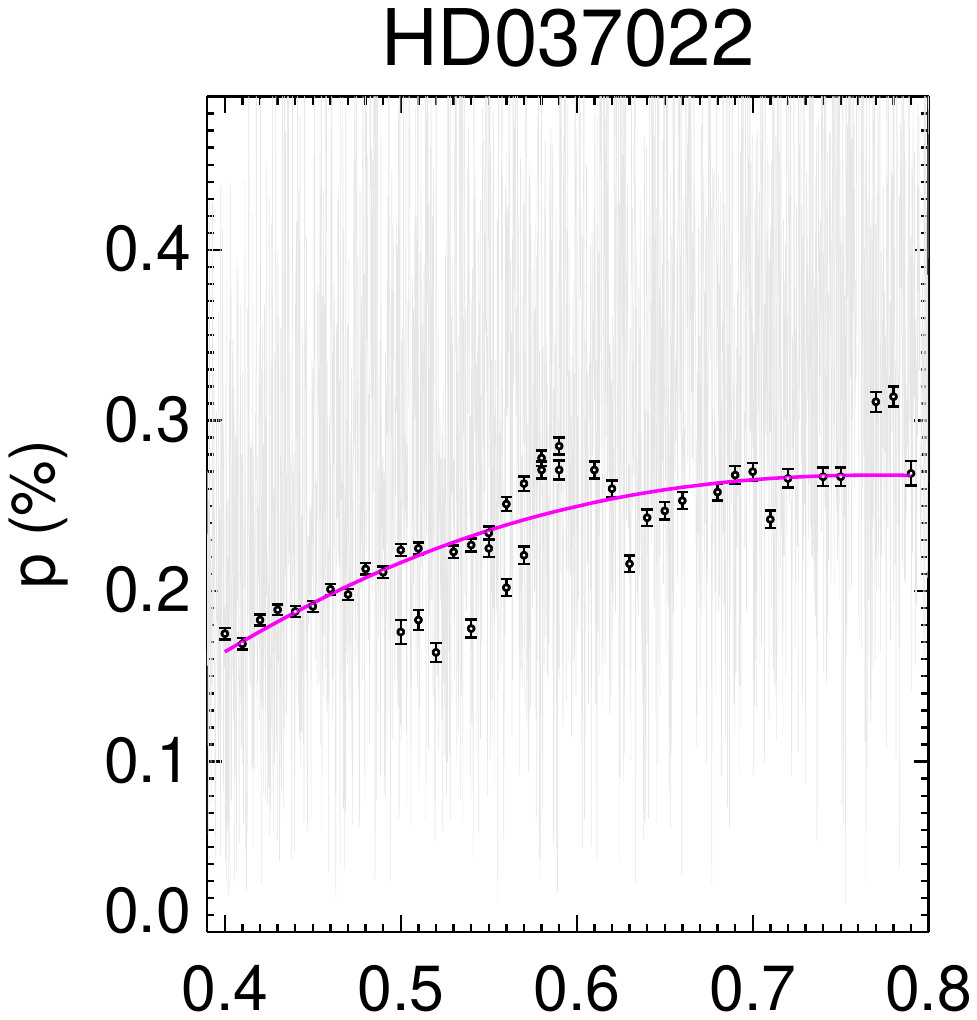}
 \includegraphics[width=3.6cm,clip=true,trim=2.7cm 5.0cm 1.5cm 3.0cm]{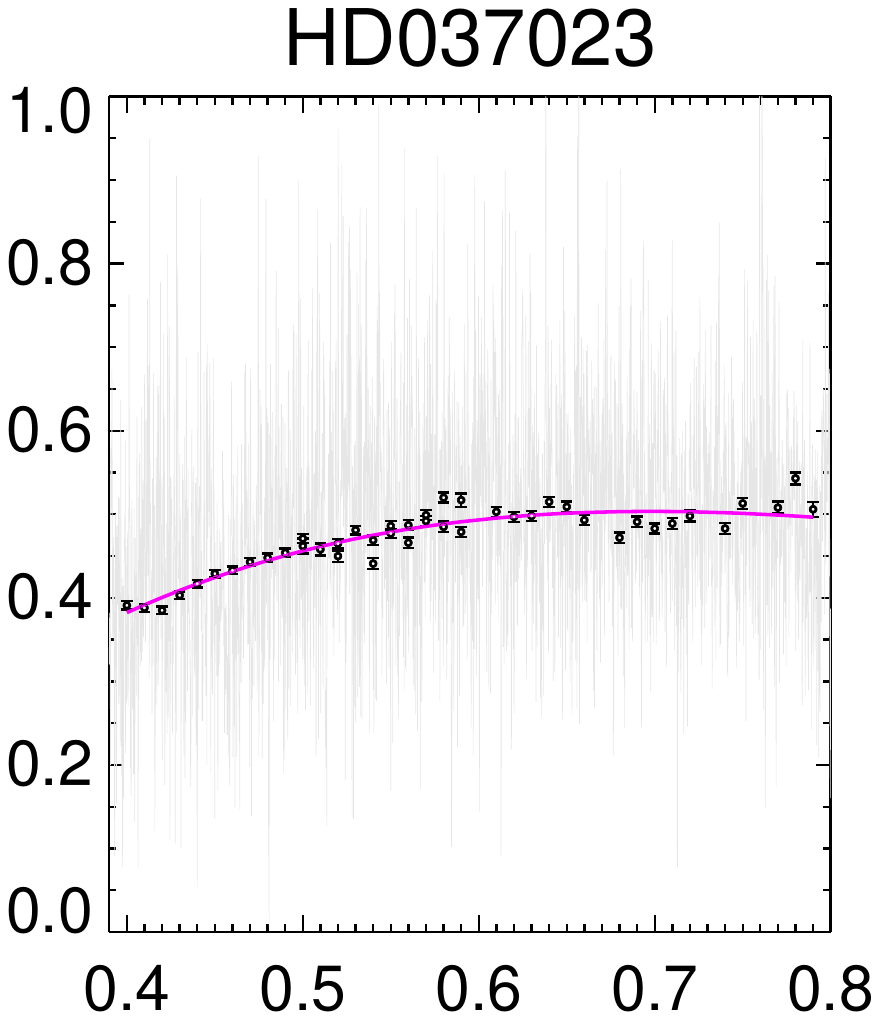}
 \includegraphics[width=3.6cm,clip=true,trim=2.7cm 5.0cm 1.5cm 3.0cm]{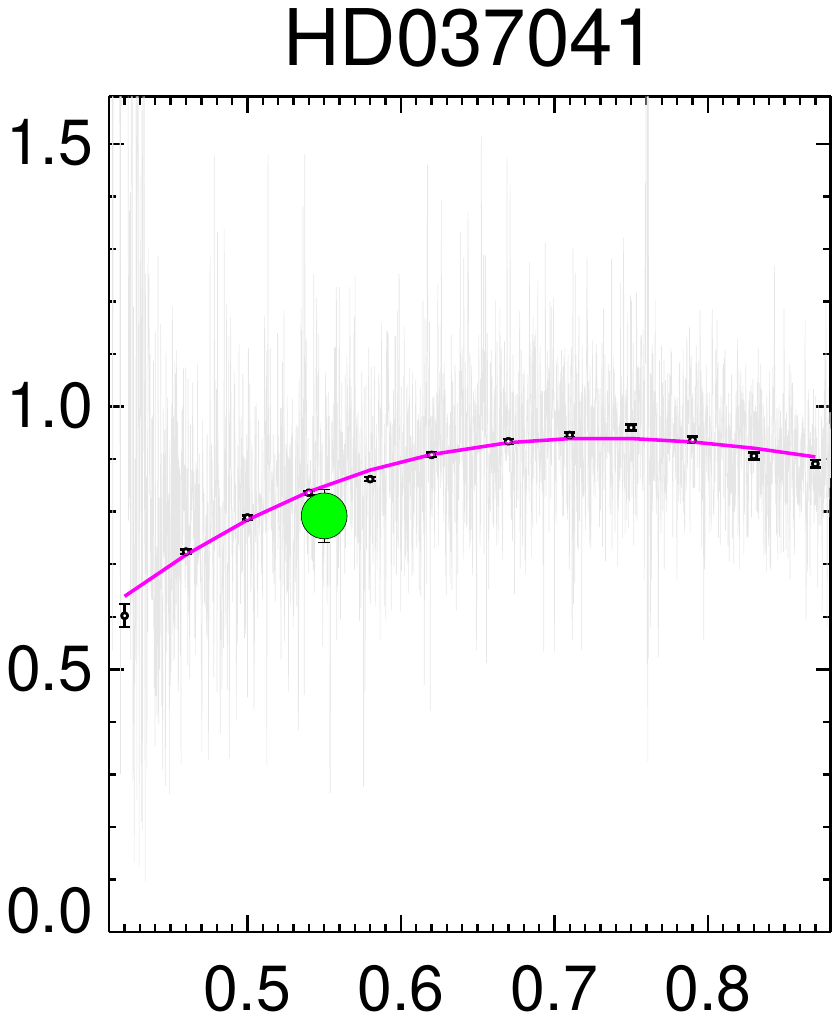}
 \includegraphics[width=3.6cm,clip=true,trim=2.7cm 5.0cm 1.5cm 3.0cm]{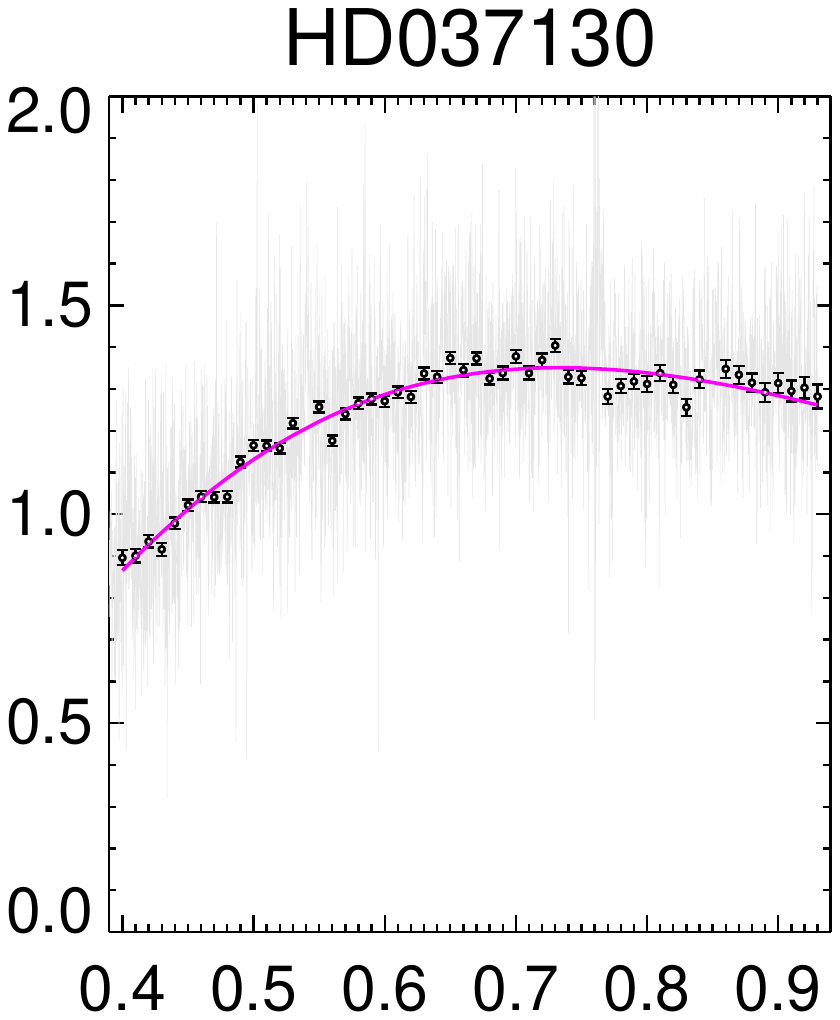}
 \includegraphics[width=3.6cm,clip=true,trim=2.7cm 5.0cm 1.5cm 3.0cm]{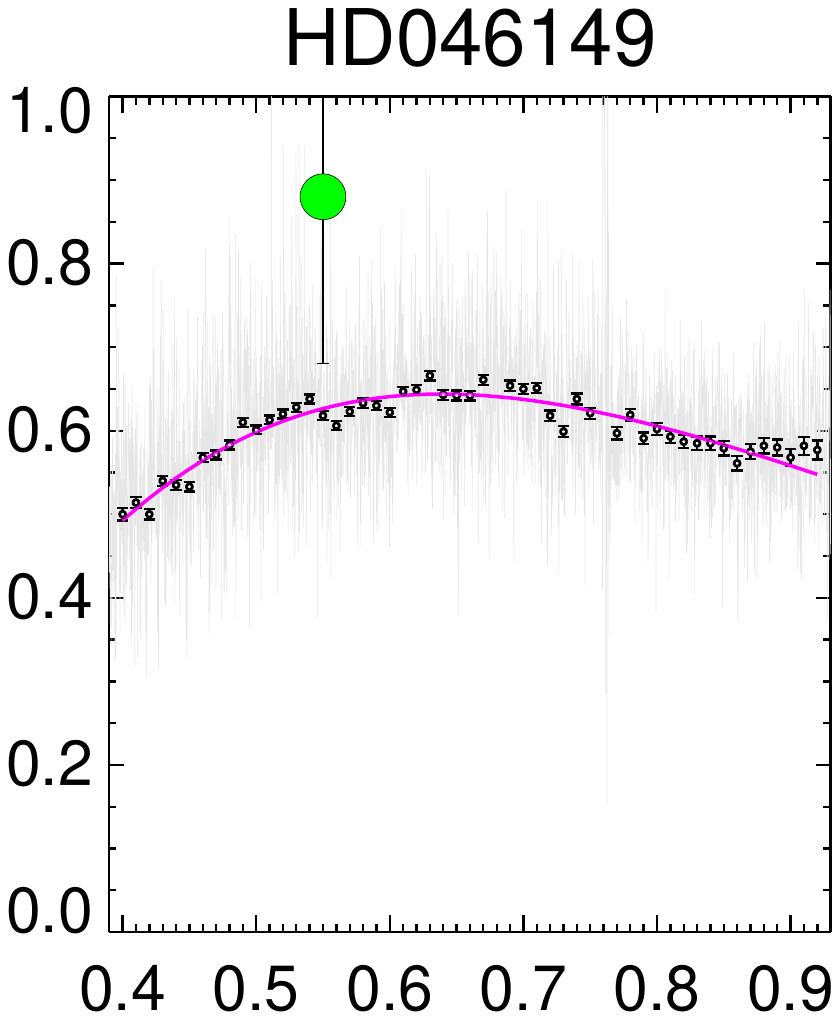}
 \includegraphics[width=3.6cm,clip=true,trim=2.7cm 5.0cm 1.5cm 3.0cm]{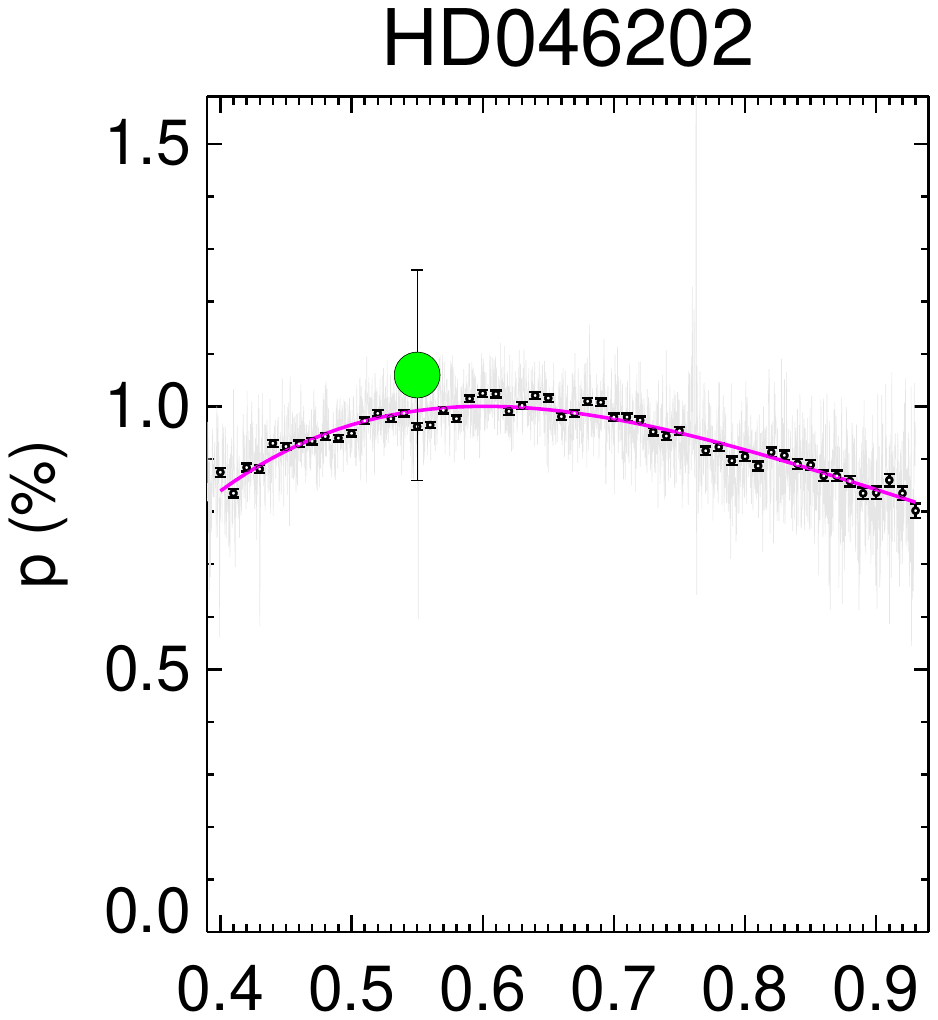}
 \includegraphics[width=3.6cm,clip=true,trim=2.7cm 5.0cm 1.5cm 3.0cm]{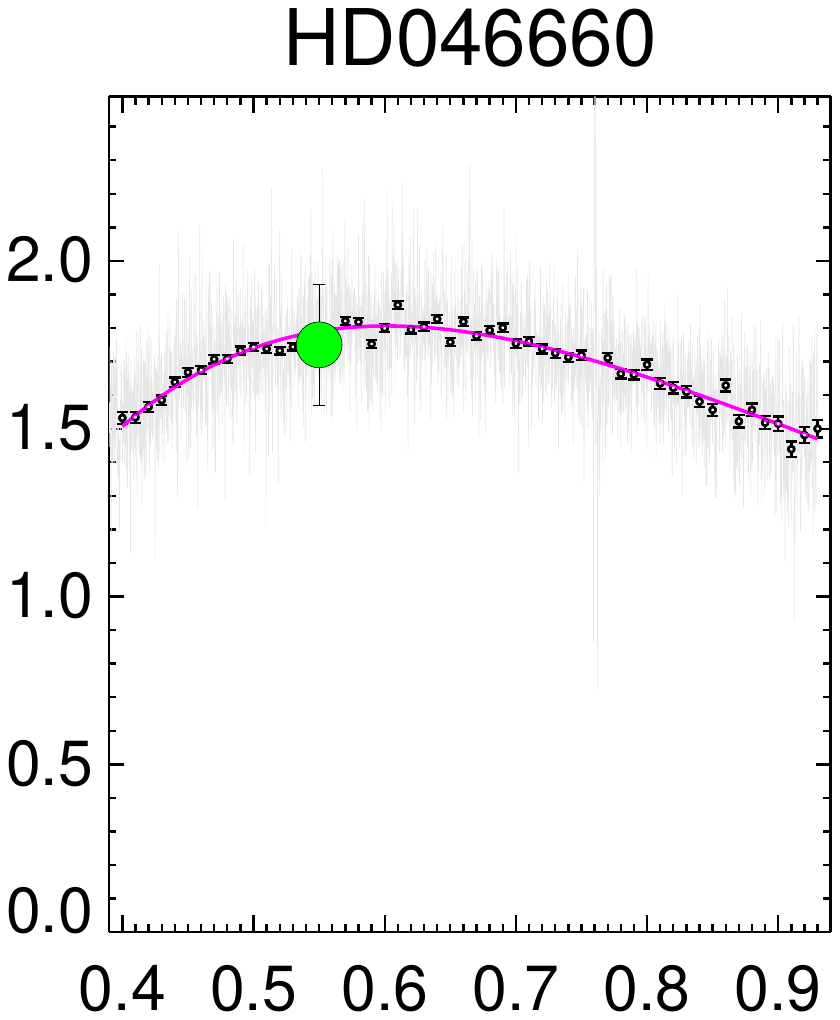}
 \includegraphics[width=3.6cm,clip=true,trim=2.7cm 5.0cm 1.5cm 3.0cm]{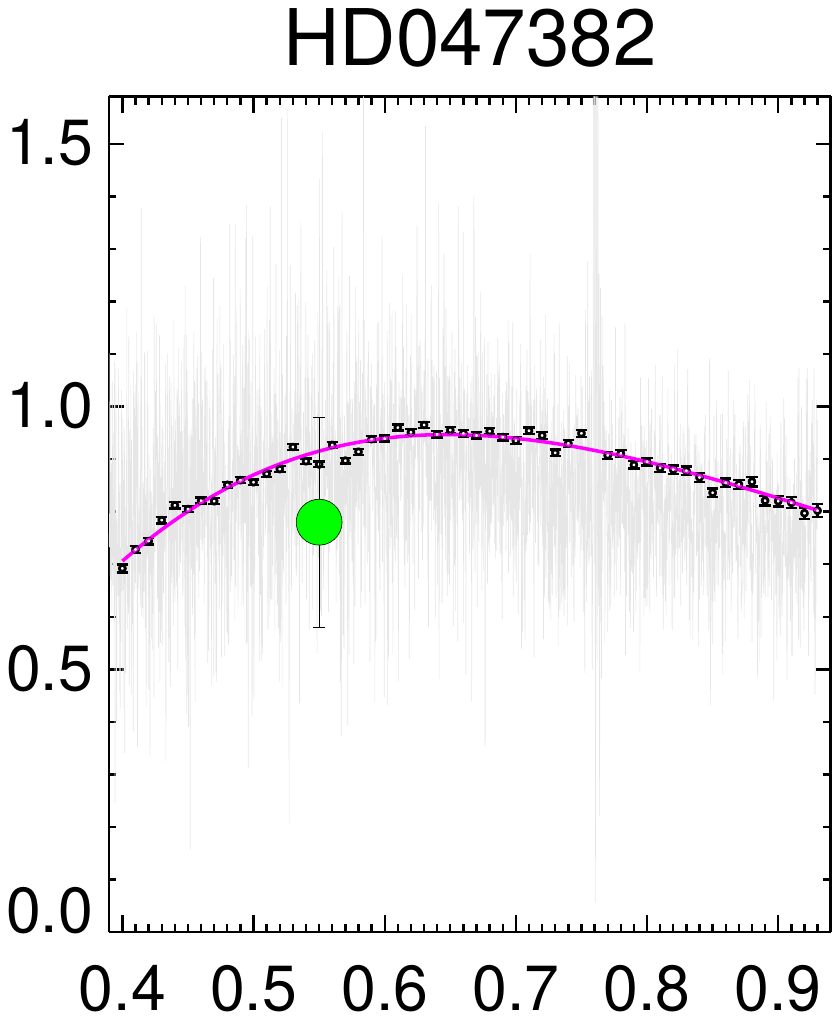}
 \includegraphics[width=3.6cm,clip=true,trim=2.7cm 5.0cm 1.5cm 3.0cm]{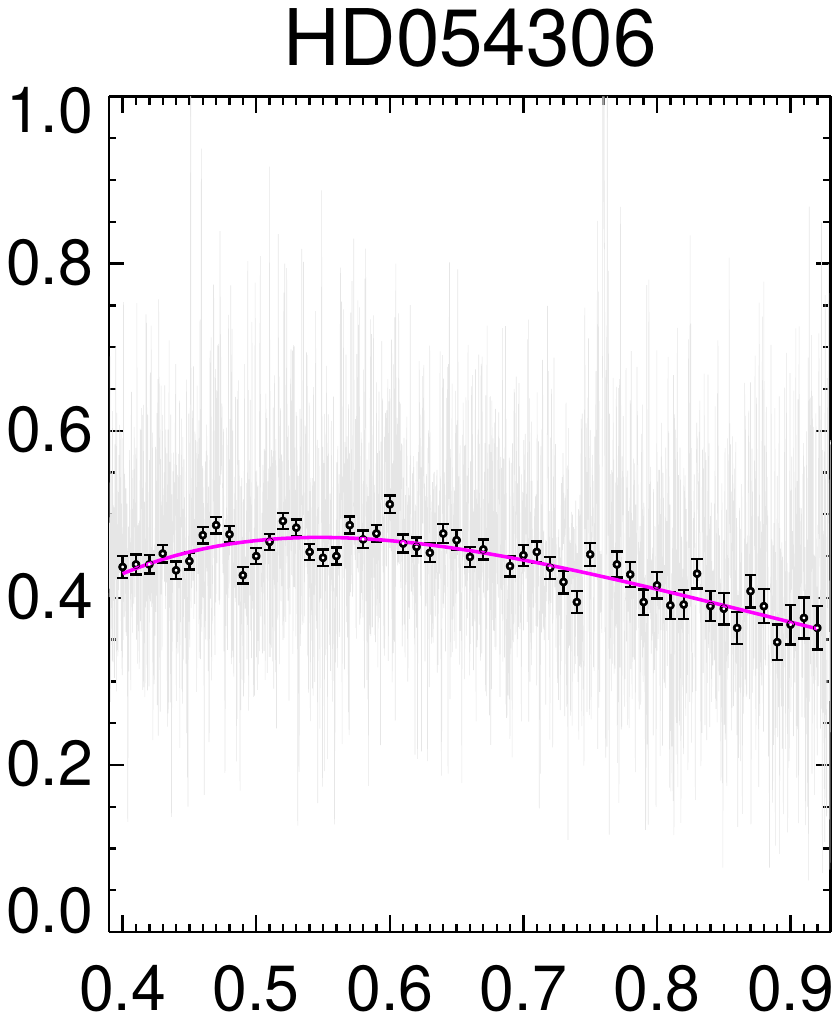}
 \includegraphics[width=3.6cm,clip=true,trim=2.7cm 5.0cm 1.5cm 3.0cm]{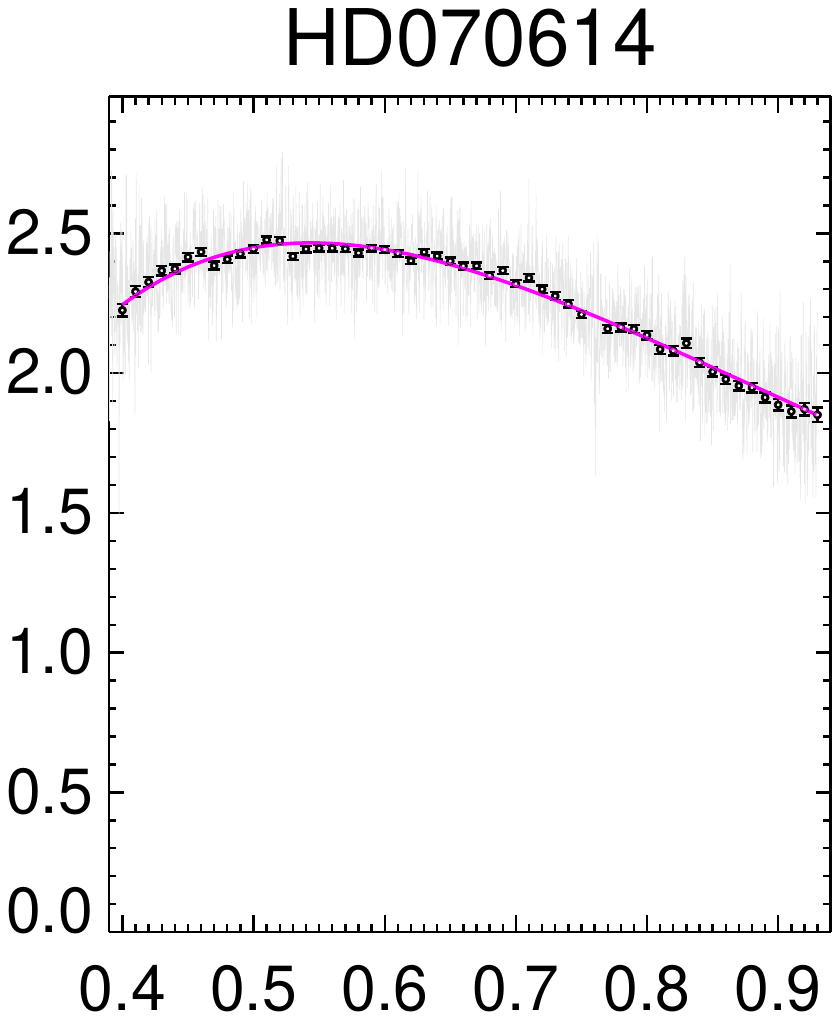}
 \includegraphics[width=3.6cm,clip=true,trim=2.7cm 5.0cm 1.5cm 3.0cm]{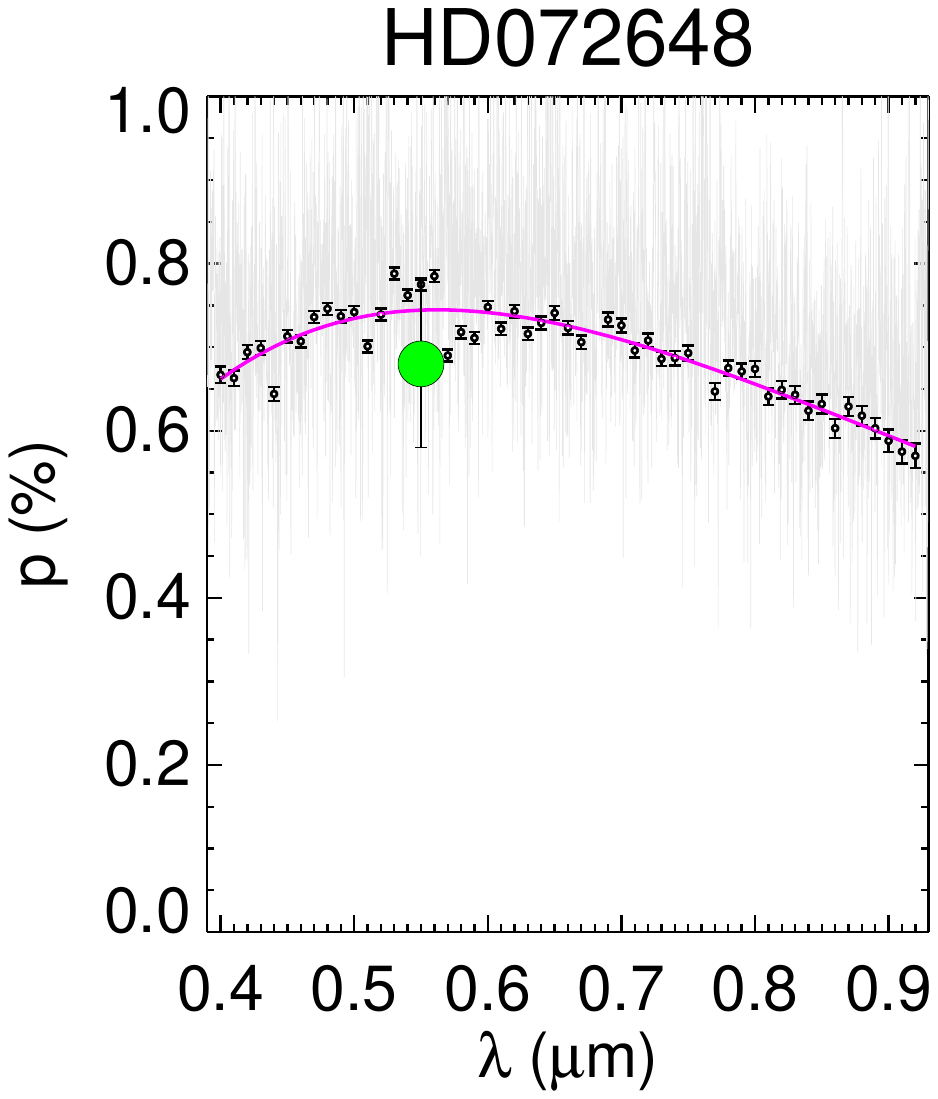}
 \includegraphics[width=3.6cm,clip=true,trim=2.7cm 5.0cm 1.5cm 3.0cm]{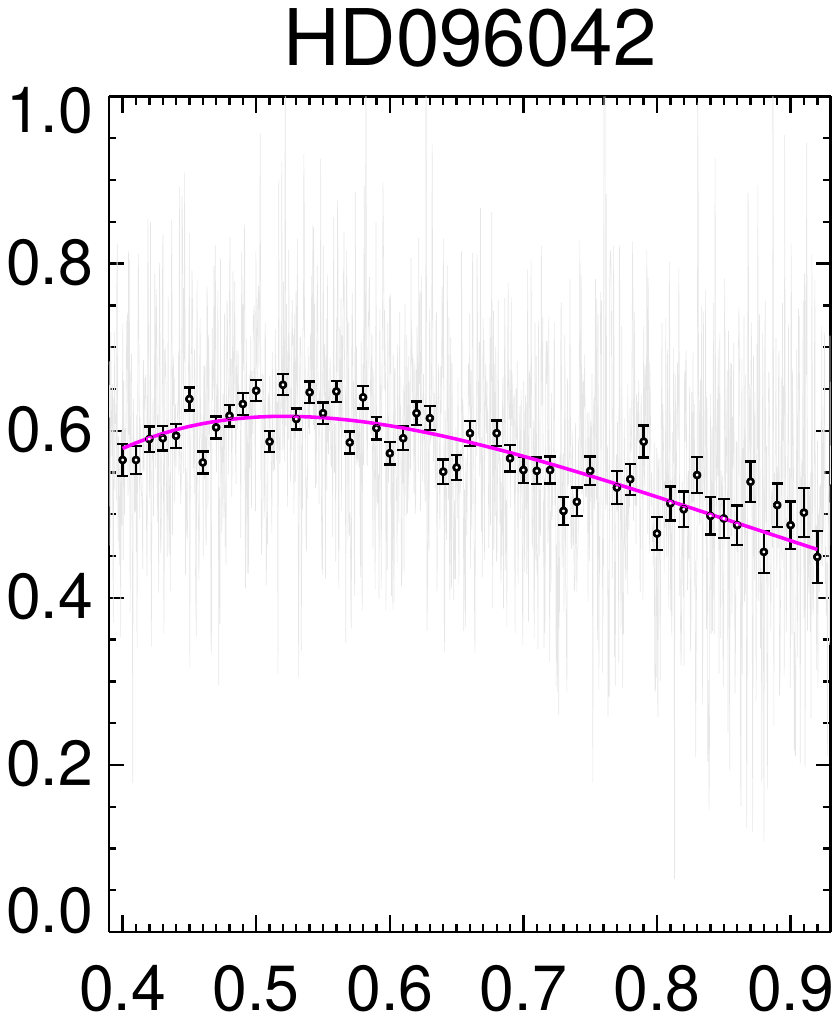}
 \includegraphics[width=3.6cm,clip=true,trim=2.7cm 5.0cm 1.5cm 3.0cm]{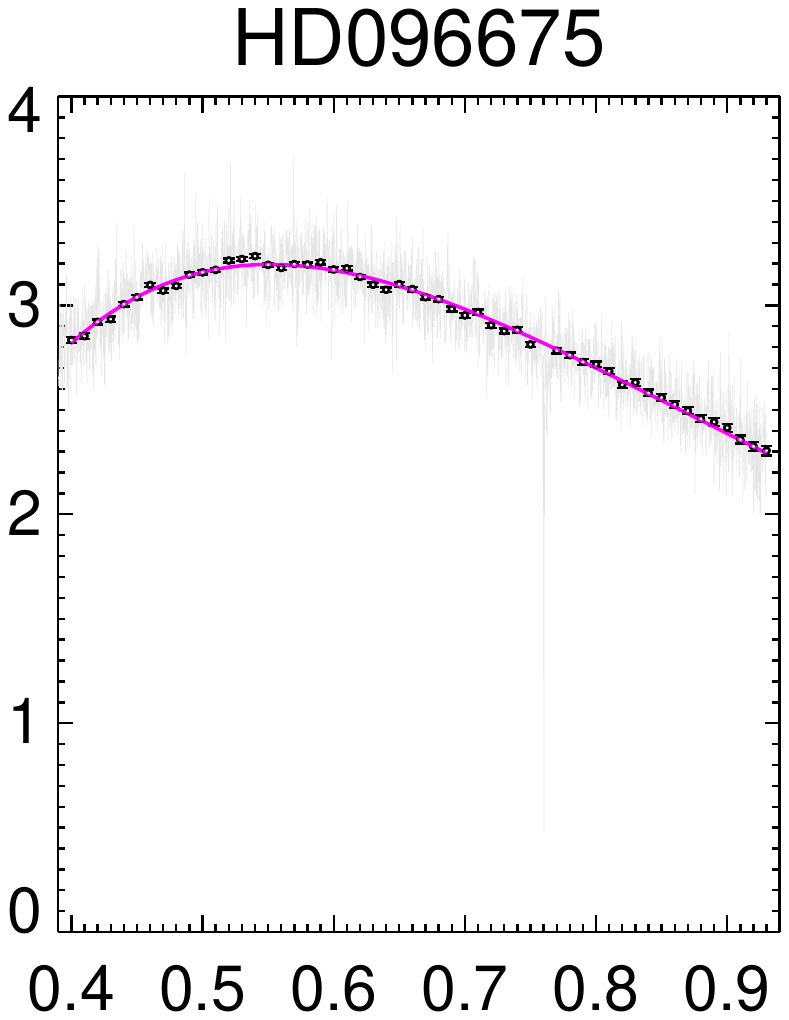}
 \includegraphics[width=3.6cm,clip=true,trim=2.7cm 5.0cm 1.5cm 3.0cm]{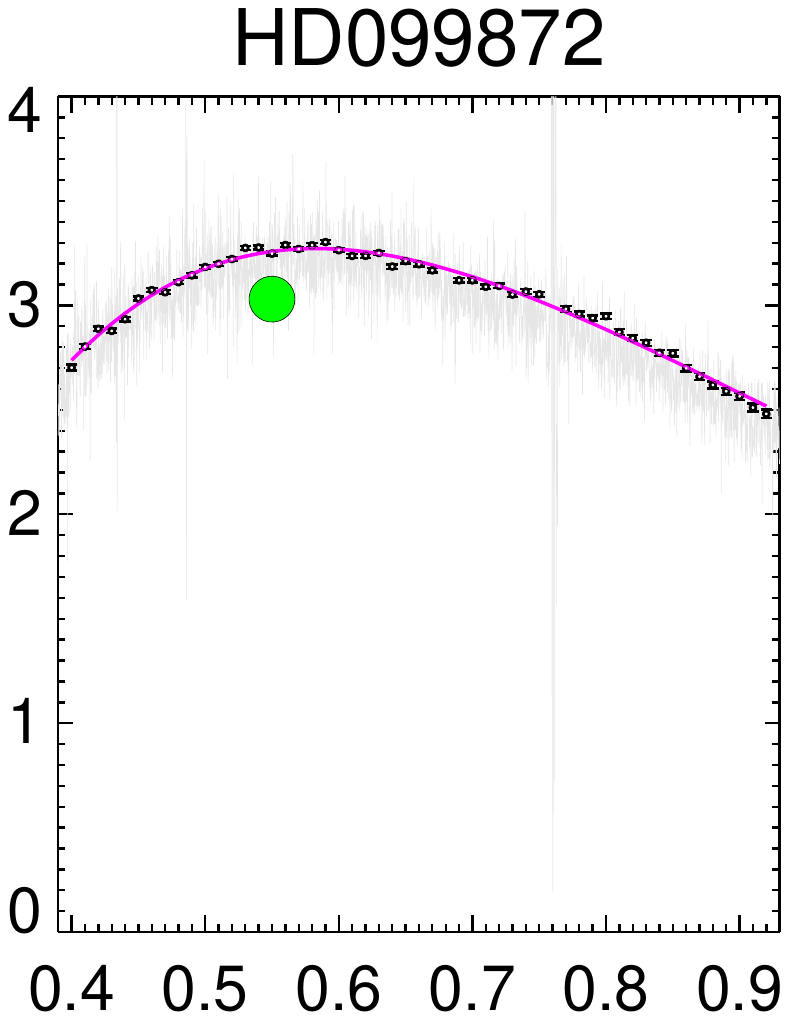}
 \includegraphics[width=3.6cm,clip=true,trim=2.7cm 5.0cm 1.5cm 3.0cm]{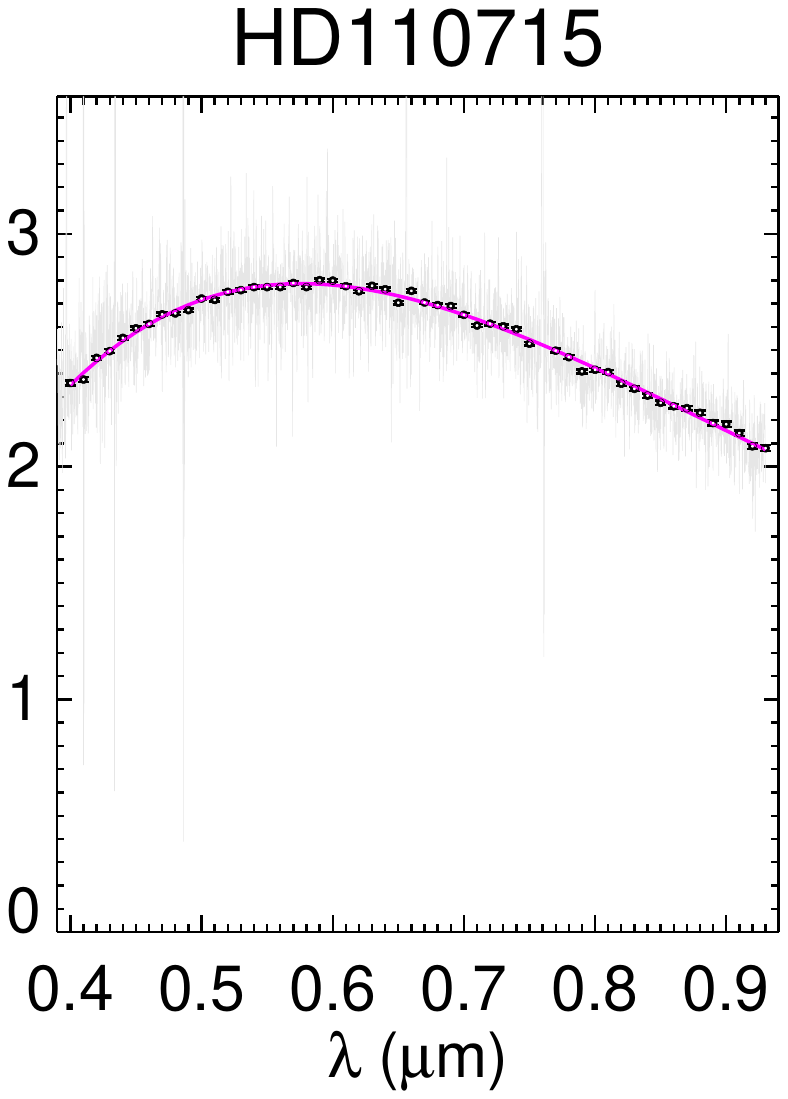}
 \includegraphics[width=3.6cm,clip=true,trim=2.7cm 5.0cm 1.5cm 3.0cm]{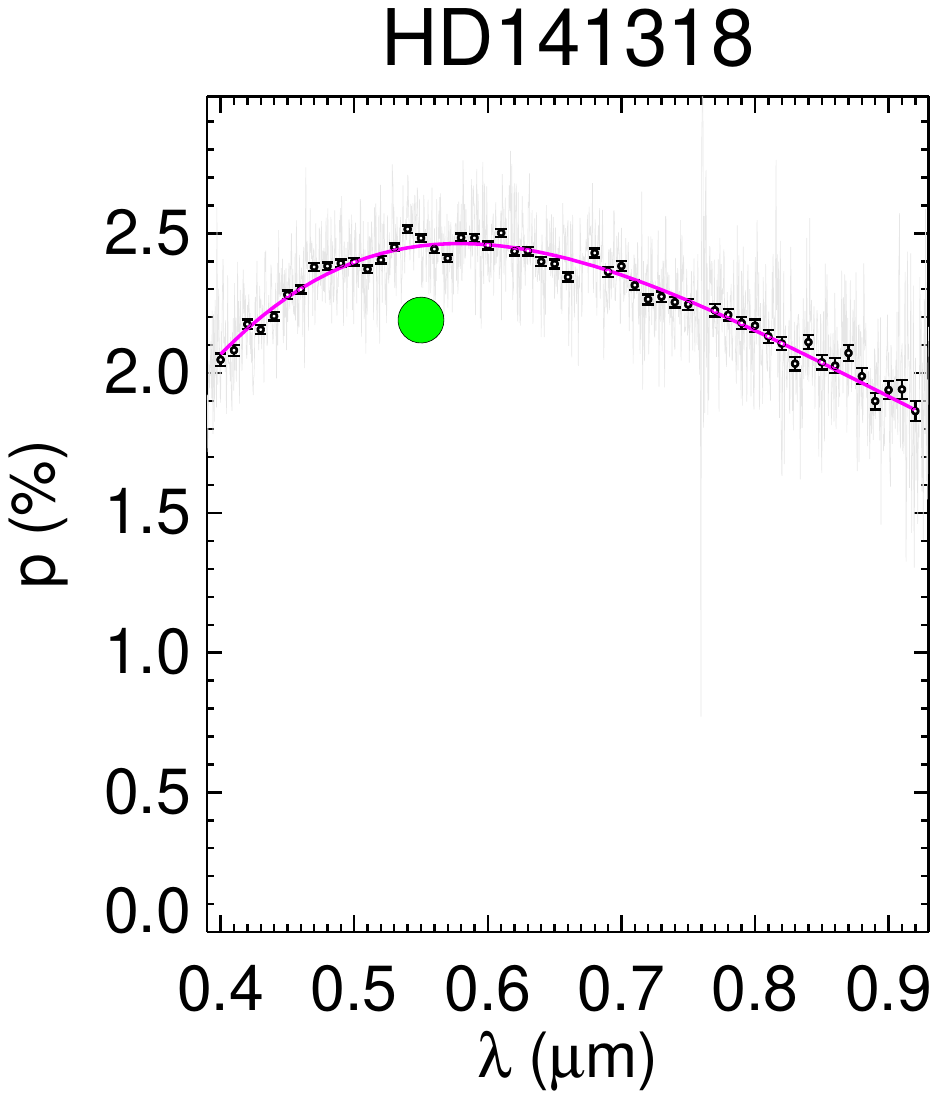}
 \includegraphics[width=3.6cm,clip=true,trim=2.7cm 5.0cm 1.5cm 3.0cm]{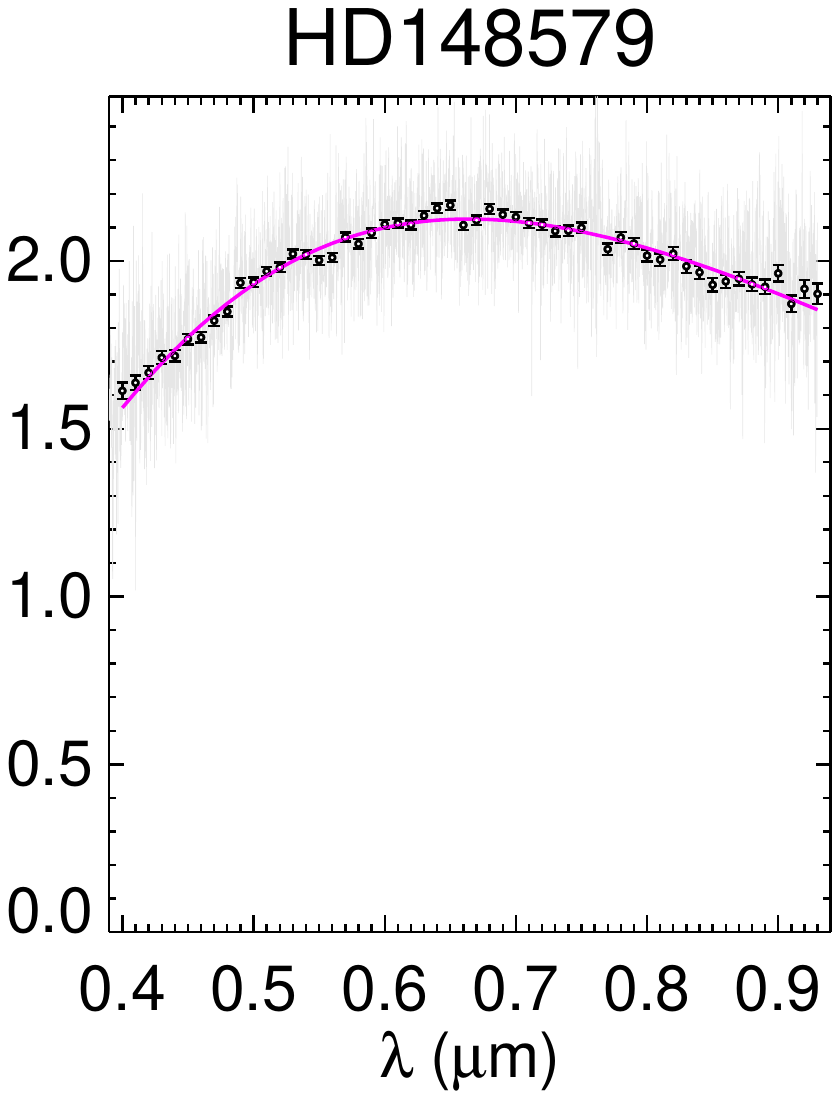}
 \includegraphics[width=3.6cm,clip=true,trim=2.7cm 5.0cm 1.5cm 3.0cm]{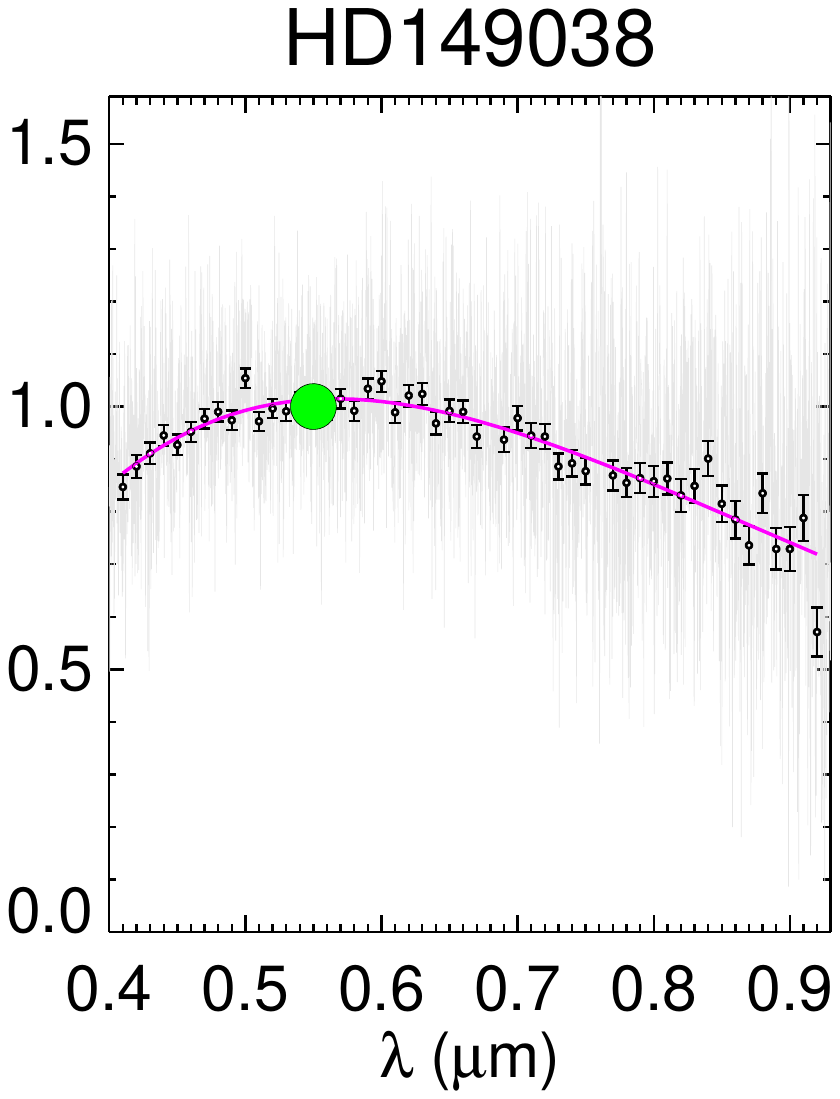}
 \includegraphics[width=3.6cm,clip=true,trim=2.7cm 5.0cm 1.5cm 3.0cm]{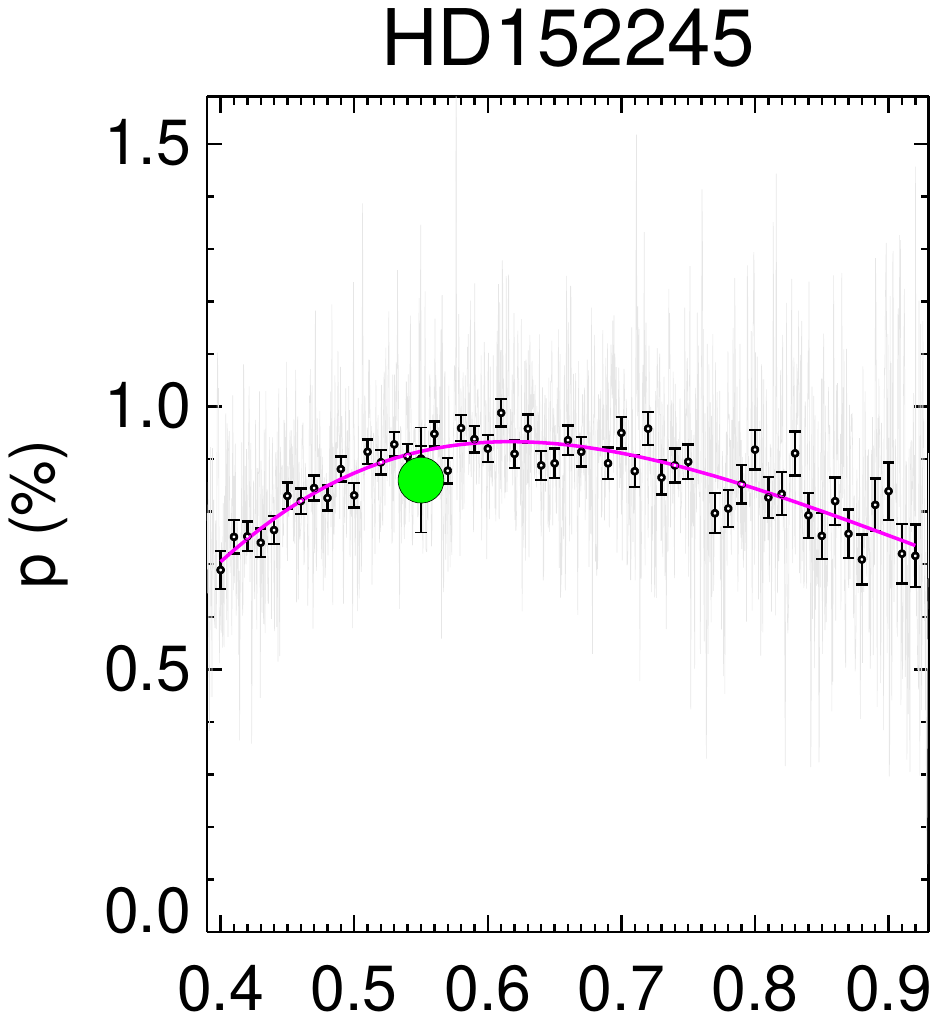}
 \includegraphics[width=3.6cm,clip=true,trim=2.7cm 5.0cm 1.5cm 3.0cm]{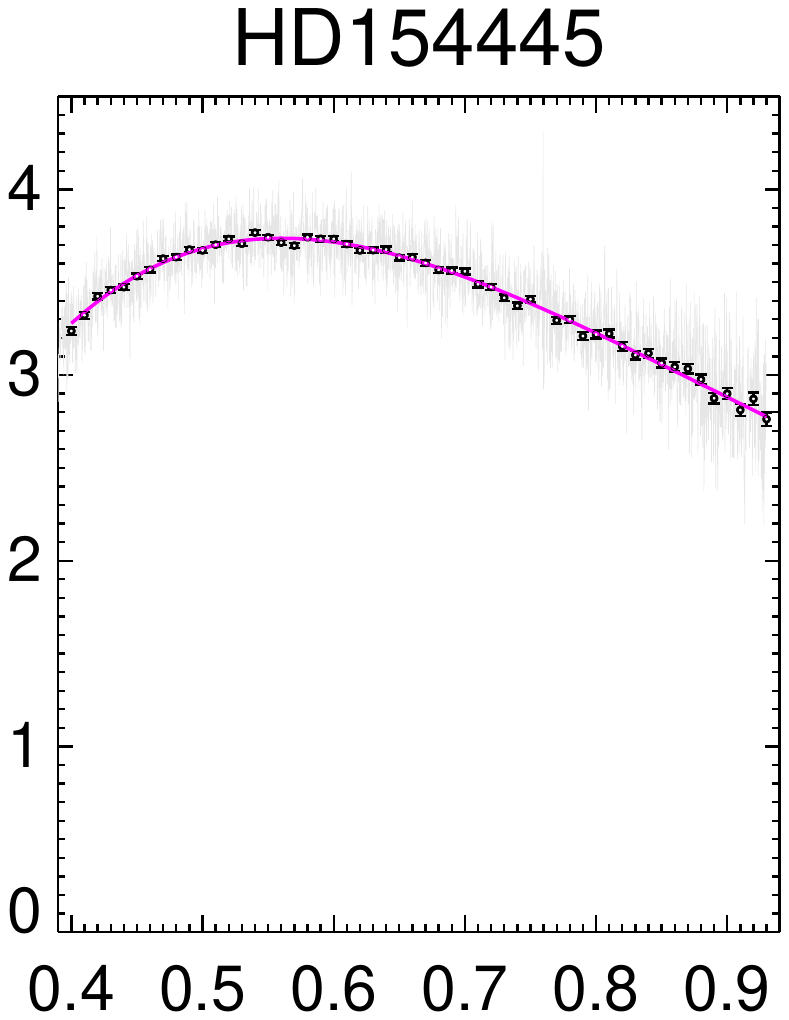}
   \end{center}
  \caption{FORS polarisation spectra of 43 stars. The grey lines show the original
data, while the black open circles (with $1\,\sigma$ error bars)
represent the data rebinned to a spectral resolution of
$\lambda / \Delta\lambda \sim 50$. The green circles represent the
measurements available in the catalogue by \citet{Heiles}. The
magenta lines show the best fits obtained with the Serkowski
formula. Continued in Fig.~\ref{Fig1.cont2}.}  \label{Fig1.pdf}
\end{figure*}

%%%%%%%%%%%%%%%%%%%%%%%%%%%%%%%%%%%%%%%%%

 \begin{figure*} 
   \begin{center}
 \includegraphics[width=3.6cm,clip=true,trim=2.7cm 5.0cm 1.5cm 3.0cm]{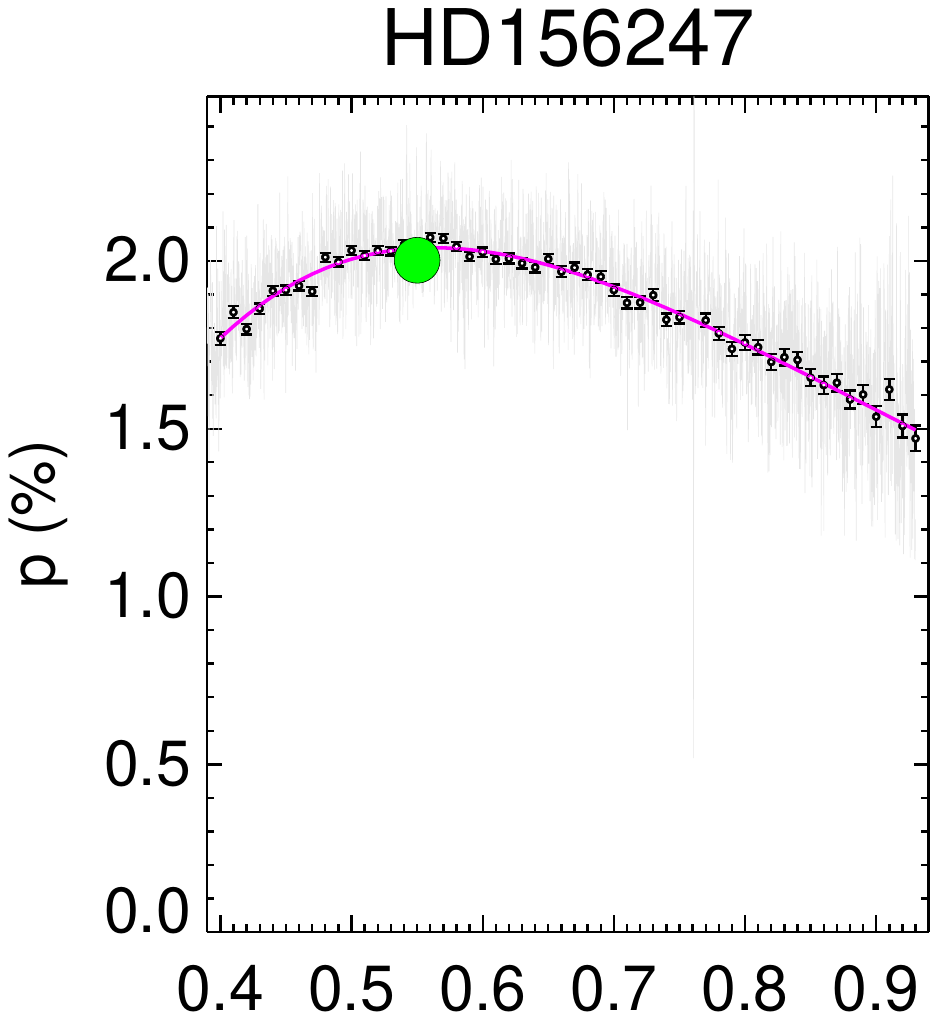}
 \includegraphics[width=3.6cm,clip=true,trim=2.7cm 5.0cm 1.5cm 3.0cm]{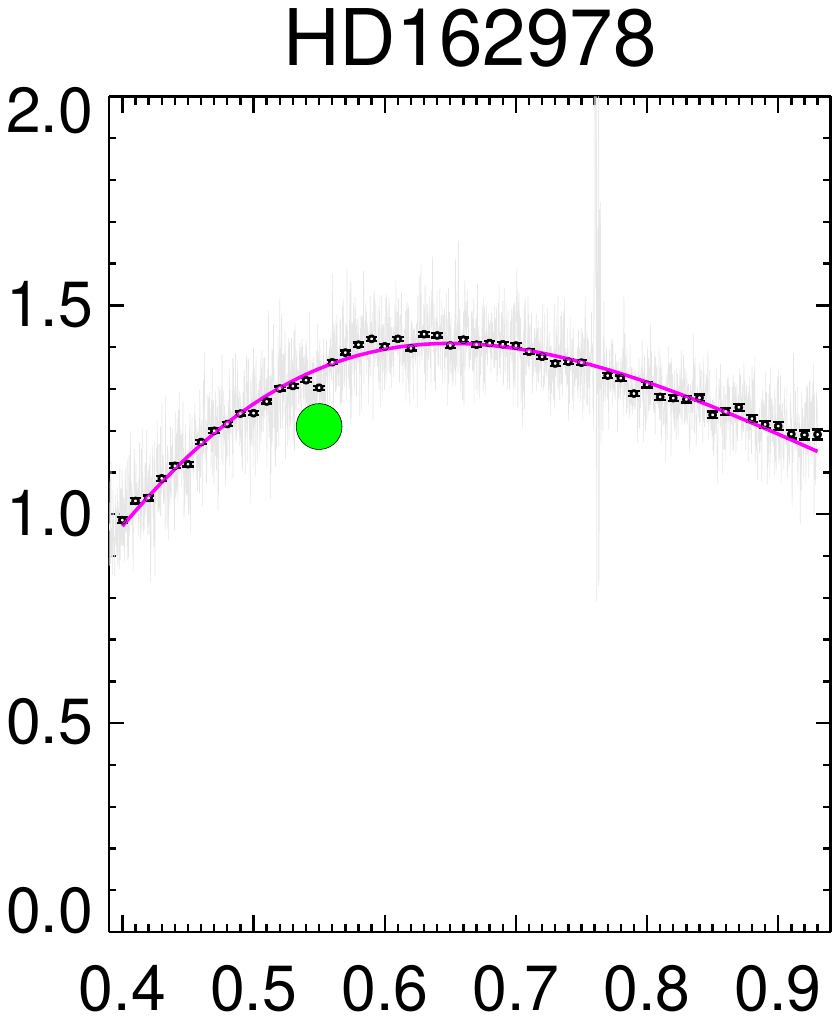}
 \includegraphics[width=3.6cm,clip=true,trim=2.7cm 5.0cm 1.5cm 3.0cm]{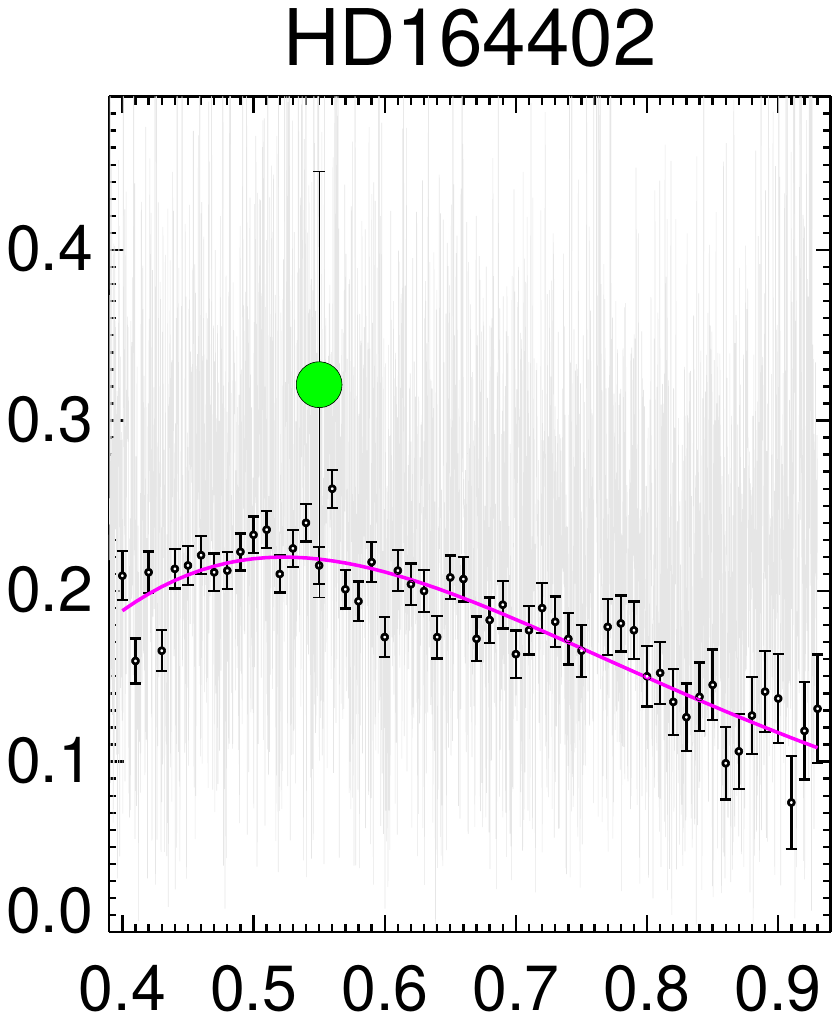}
 \includegraphics[width=3.6cm,clip=true,trim=2.7cm 5.0cm 1.5cm 3.0cm]{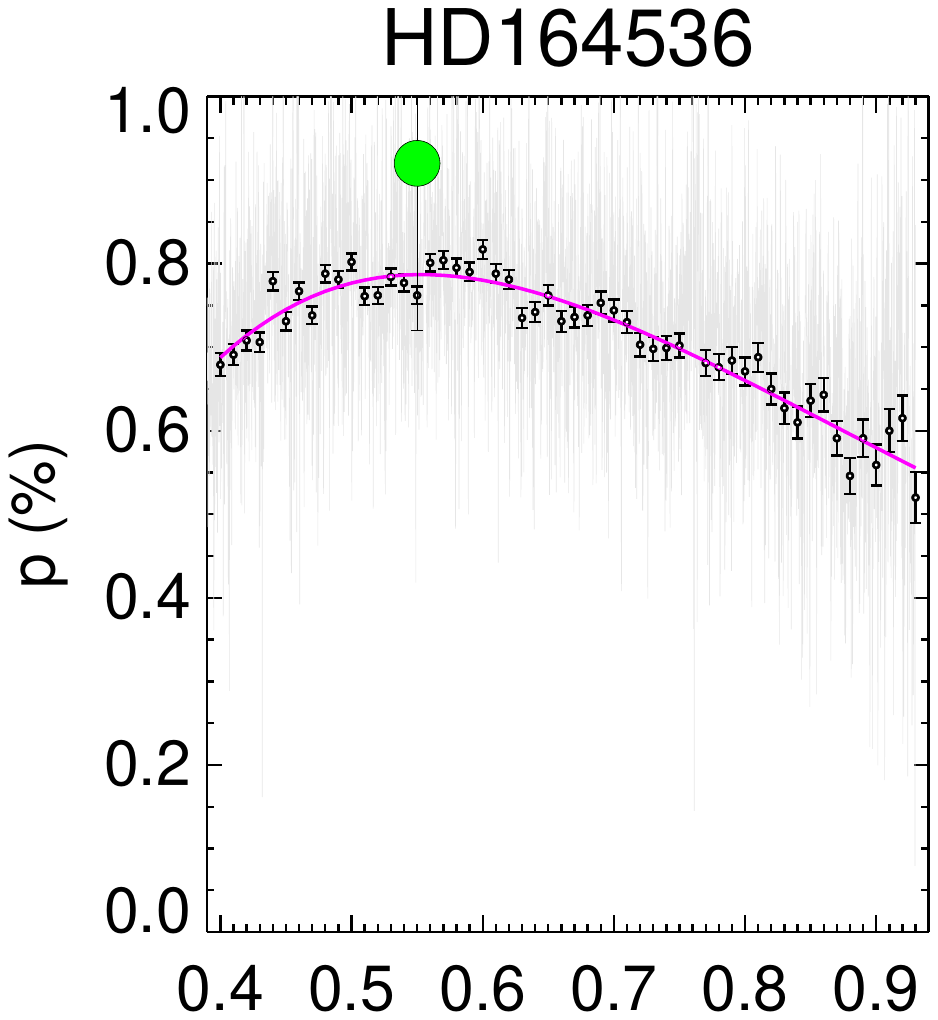}
 \includegraphics[width=3.6cm,clip=true,trim=2.7cm 5.0cm 1.5cm 3.0cm]{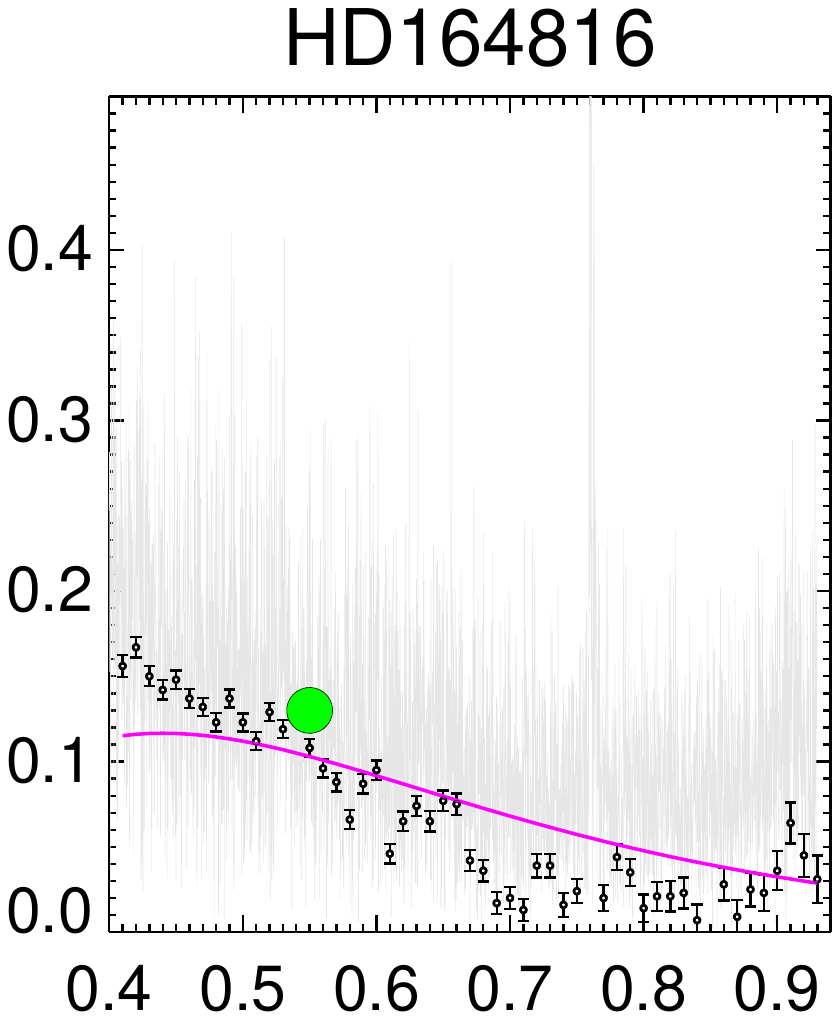}     
 \includegraphics[width=3.6cm,clip=true,trim=2.7cm 5.0cm 1.5cm 3.0cm]{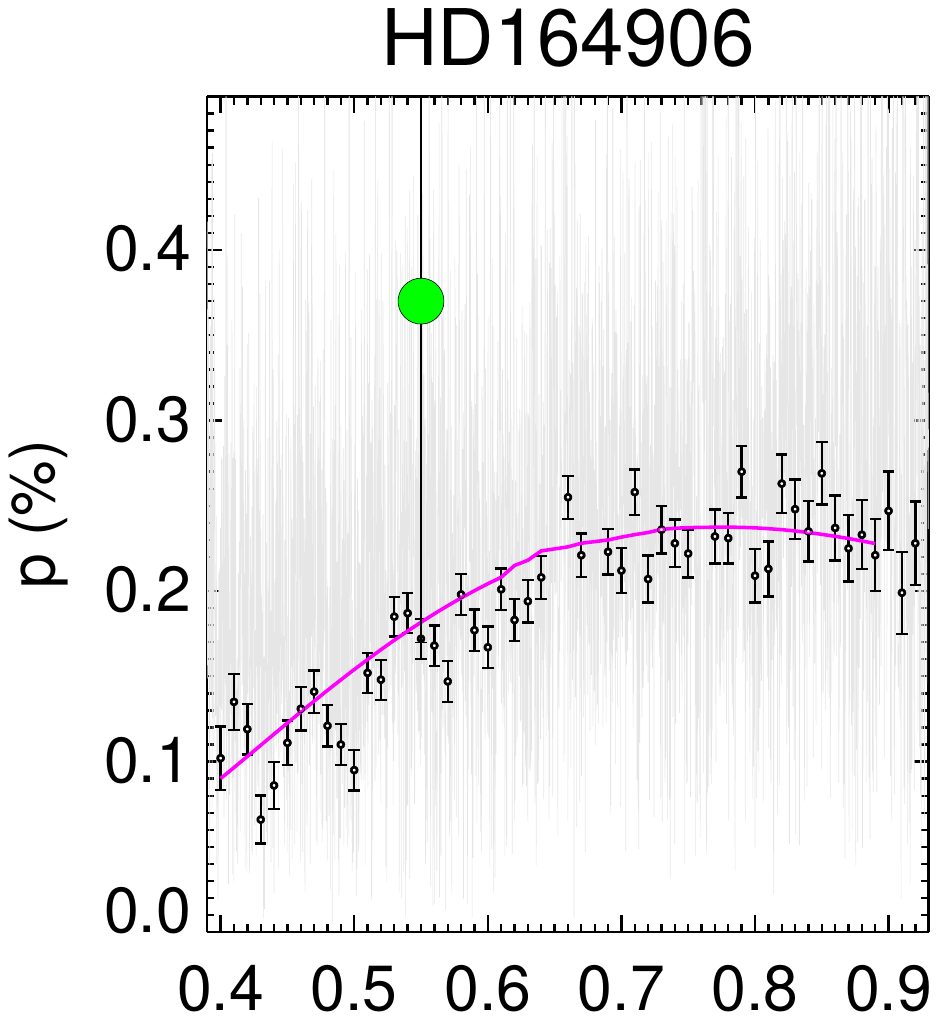}
 \includegraphics[width=3.6cm,clip=true,trim=2.7cm 5.0cm 1.5cm 3.0cm]{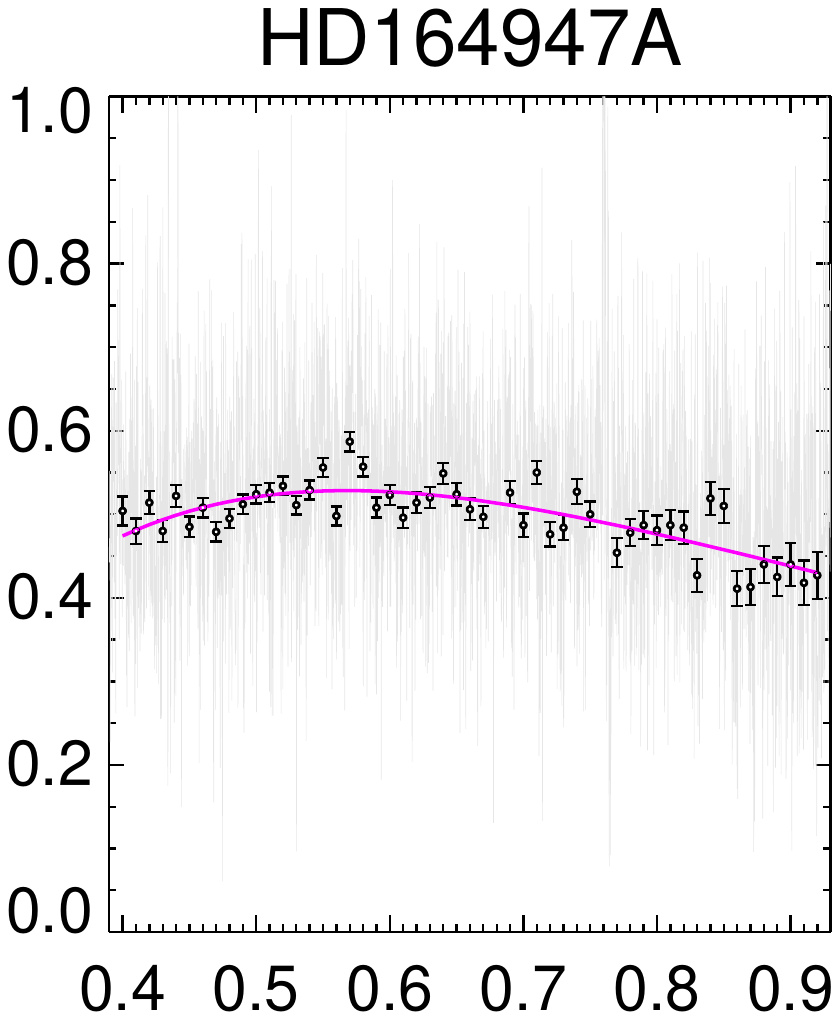}
 \includegraphics[width=3.6cm,clip=true,trim=2.7cm 5.0cm 1.5cm 3.0cm]{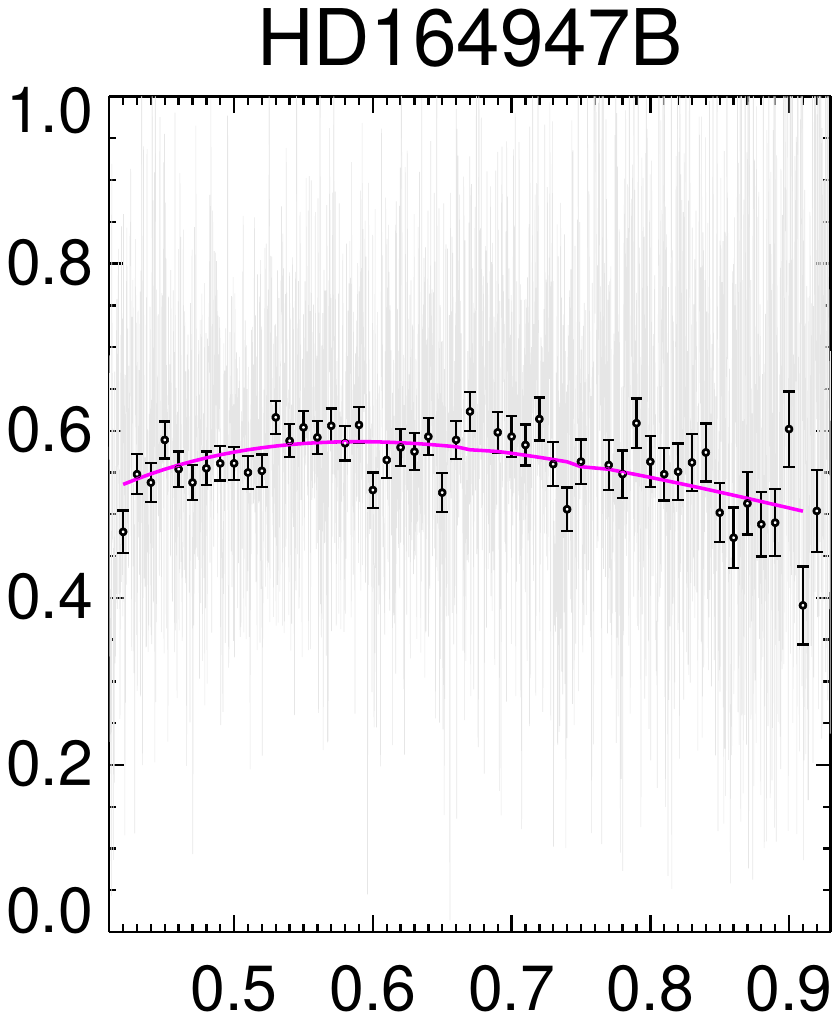}
 \includegraphics[width=3.6cm,clip=true,trim=2.7cm 5.0cm 1.5cm 3.0cm]{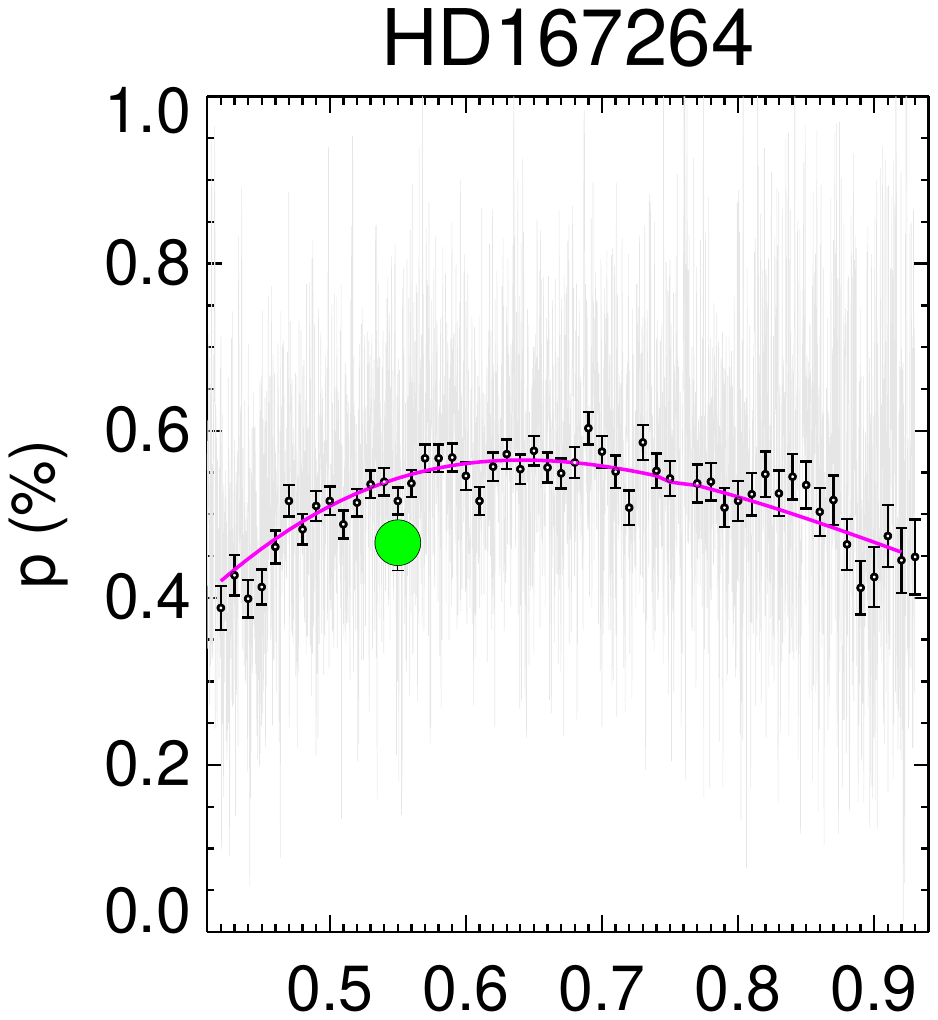}
 \includegraphics[width=3.6cm,clip=true,trim=2.7cm 5.0cm 1.5cm 3.0cm]{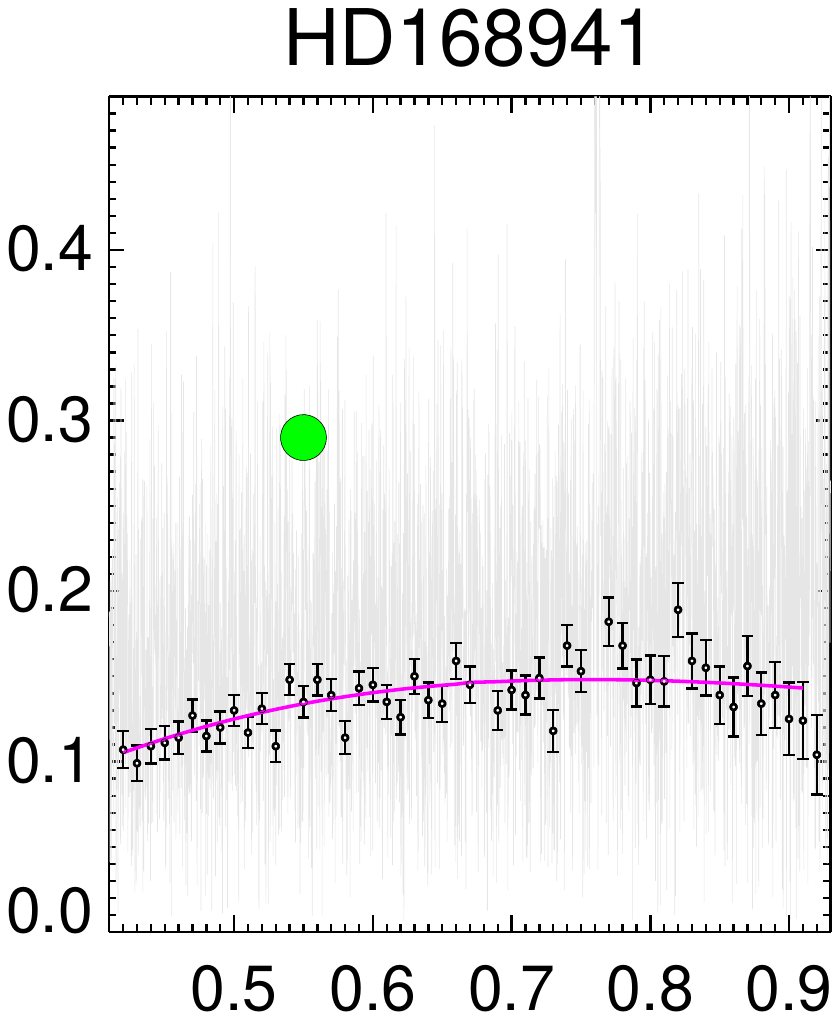}
 \includegraphics[width=3.6cm,clip=true,trim=2.7cm 5.0cm 1.5cm 3.0cm]{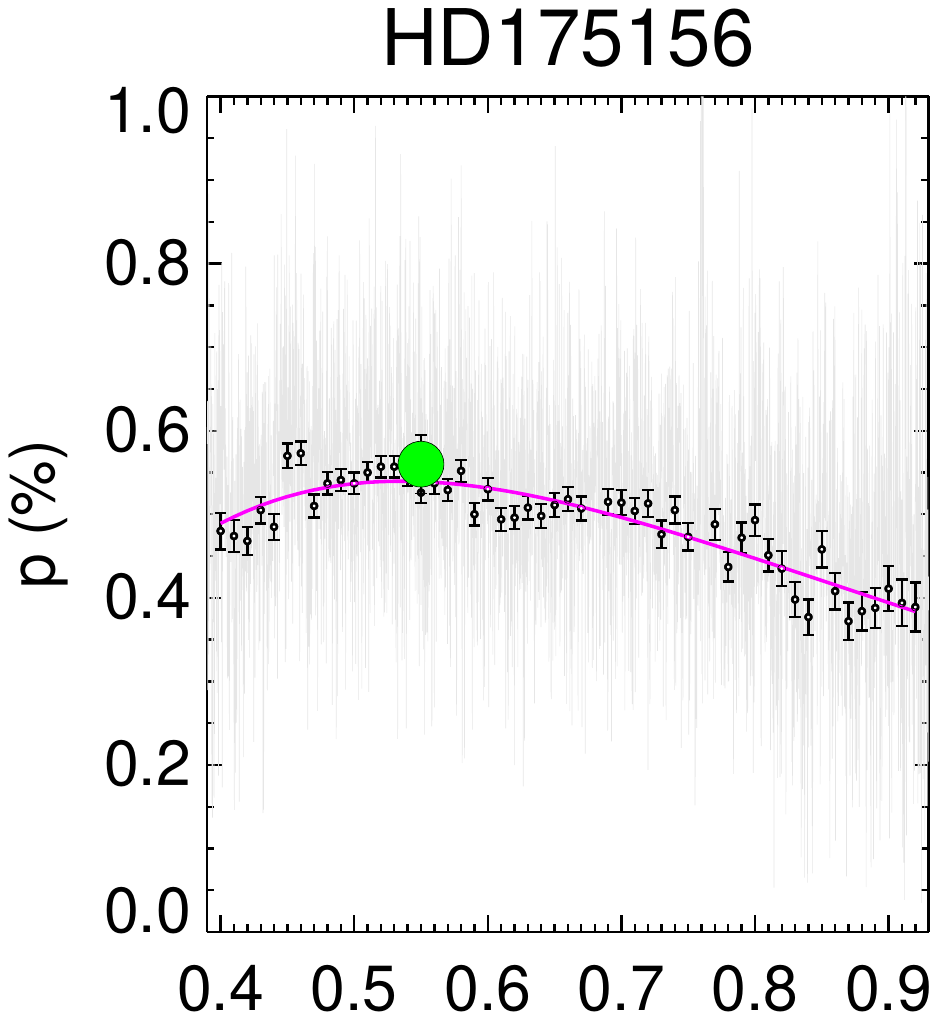}
 \includegraphics[width=3.6cm,clip=true,trim=2.7cm 5.0cm 1.5cm 3.0cm]{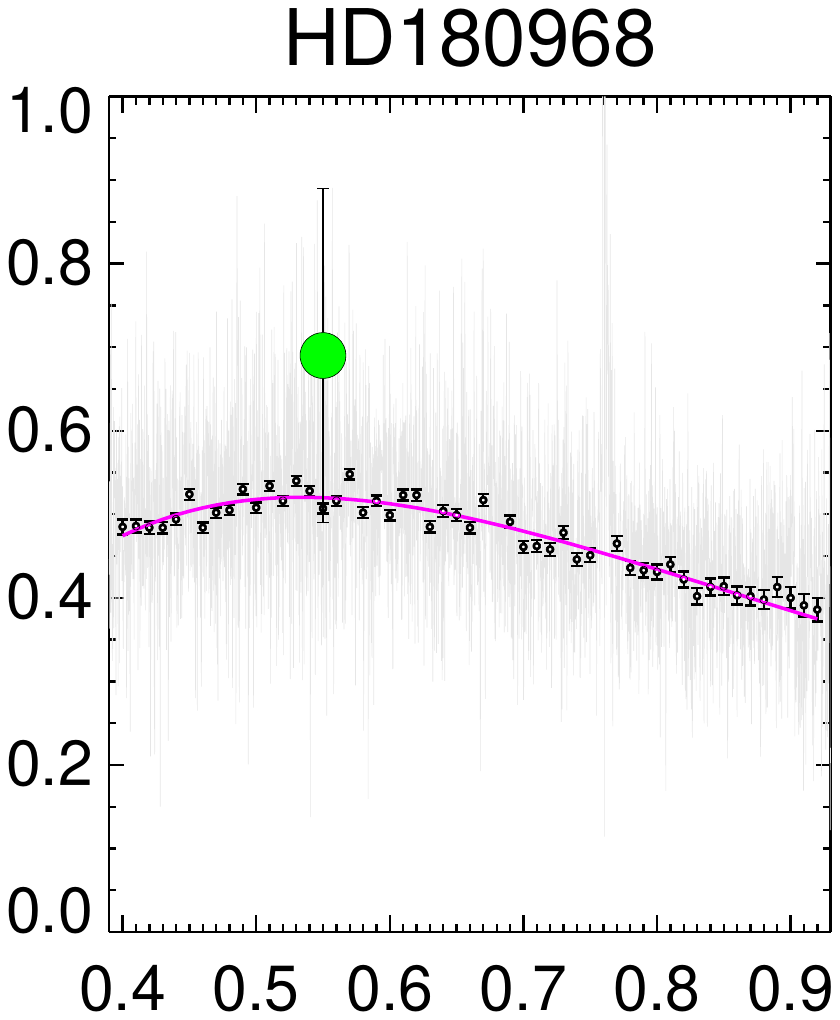}
 \includegraphics[width=3.6cm,clip=true,trim=2.7cm 5.0cm 1.5cm 3.0cm]{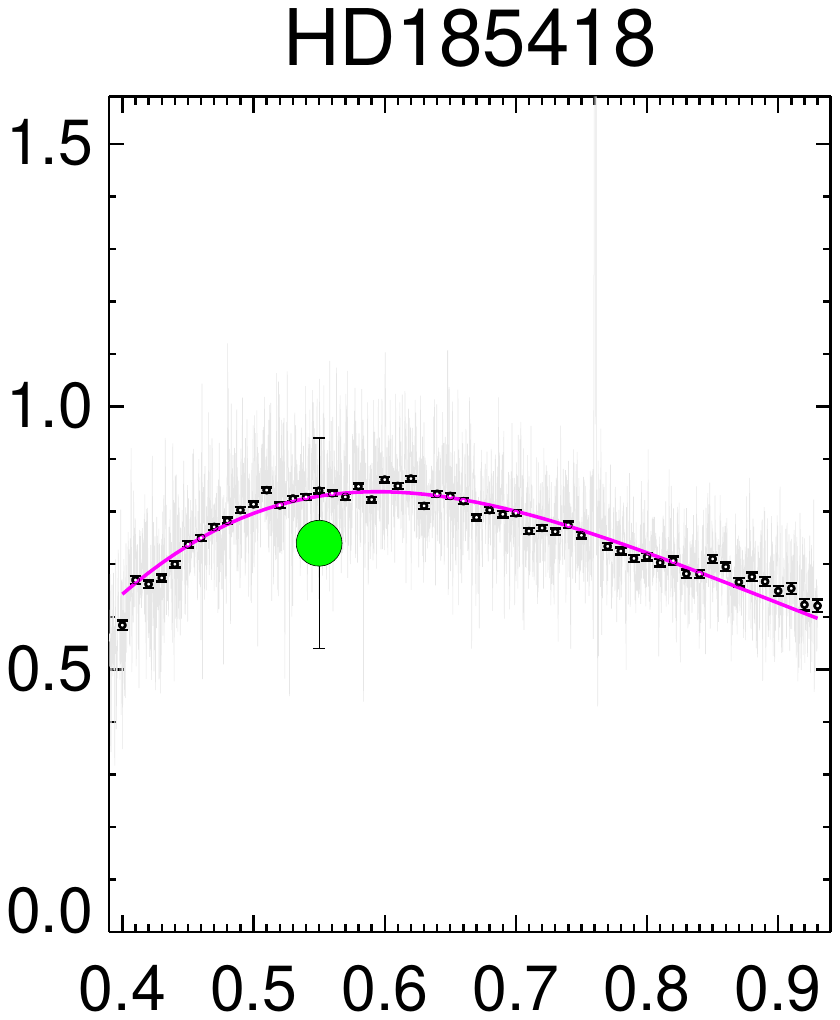}
 \includegraphics[width=3.6cm,clip=true,trim=2.7cm 5.0cm 1.5cm 3.0cm]{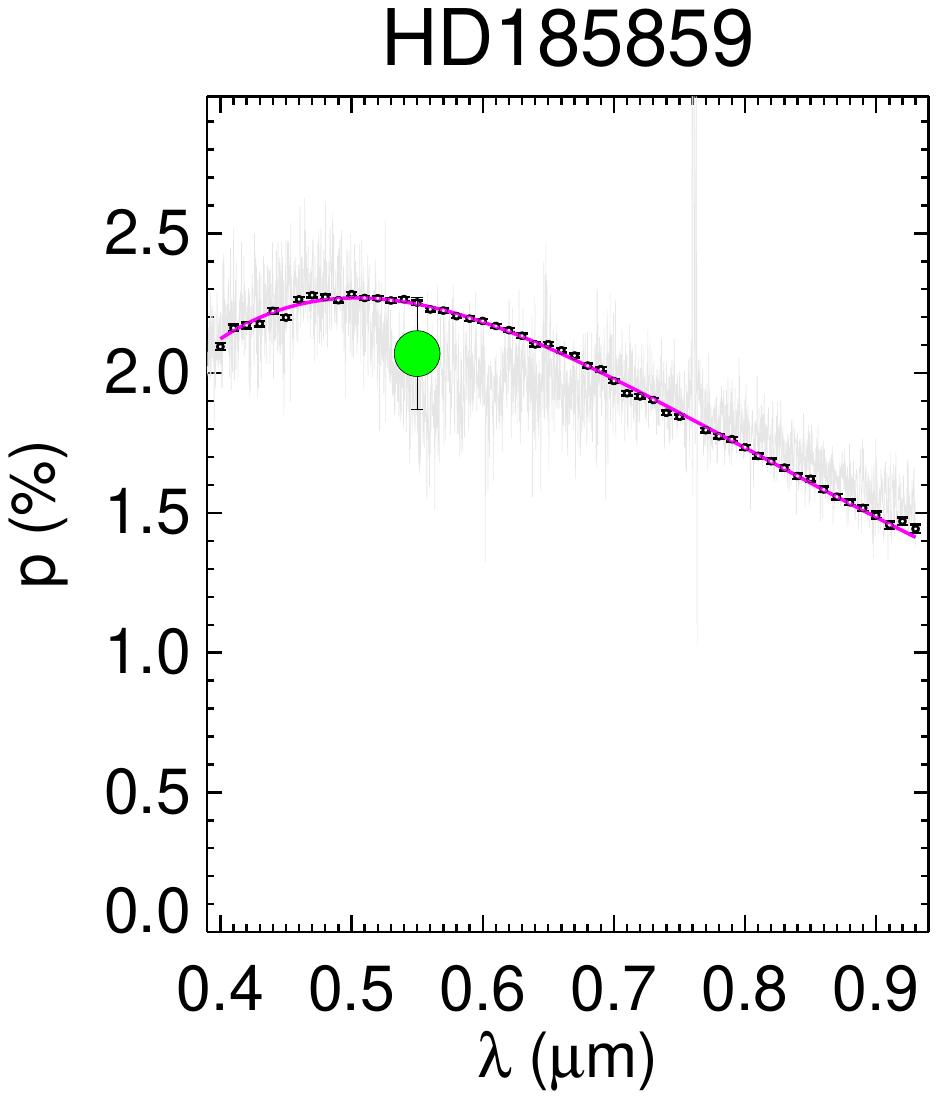}
 \includegraphics[width=3.6cm,clip=true,trim=2.7cm 5.0cm 1.5cm 3.0cm]{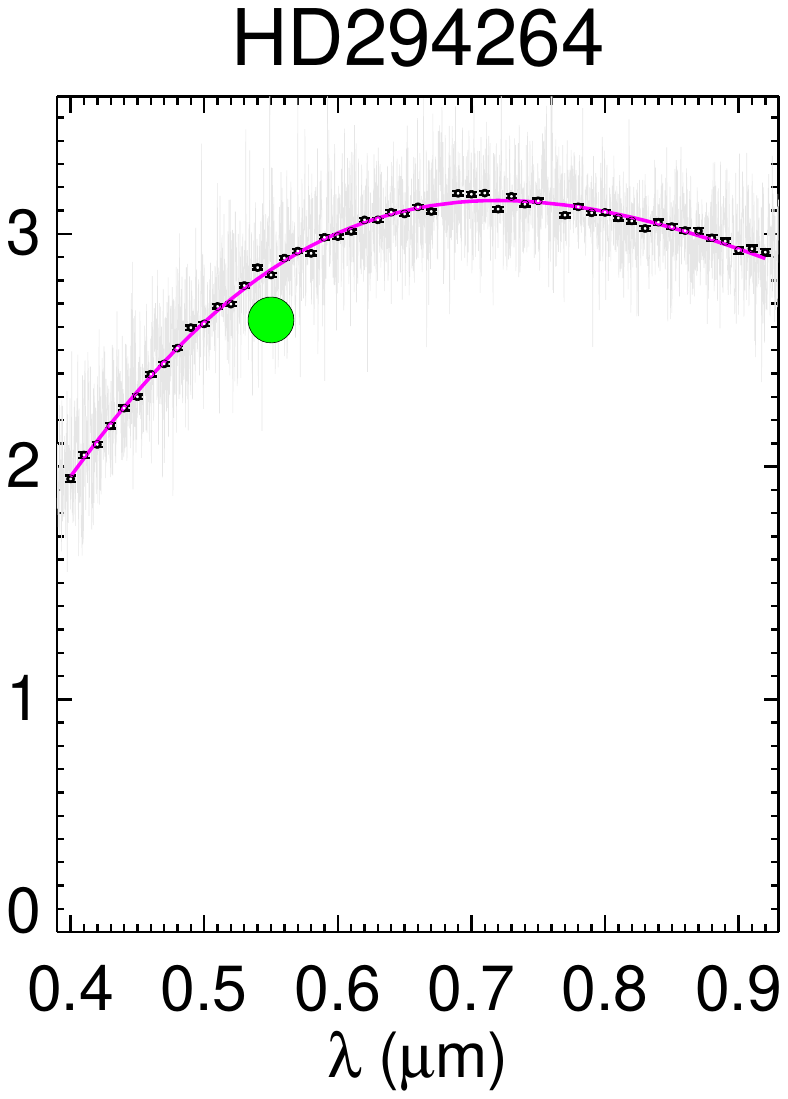}
 \includegraphics[width=3.6cm,clip=true,trim=2.7cm 5.0cm 1.5cm 3.0cm]{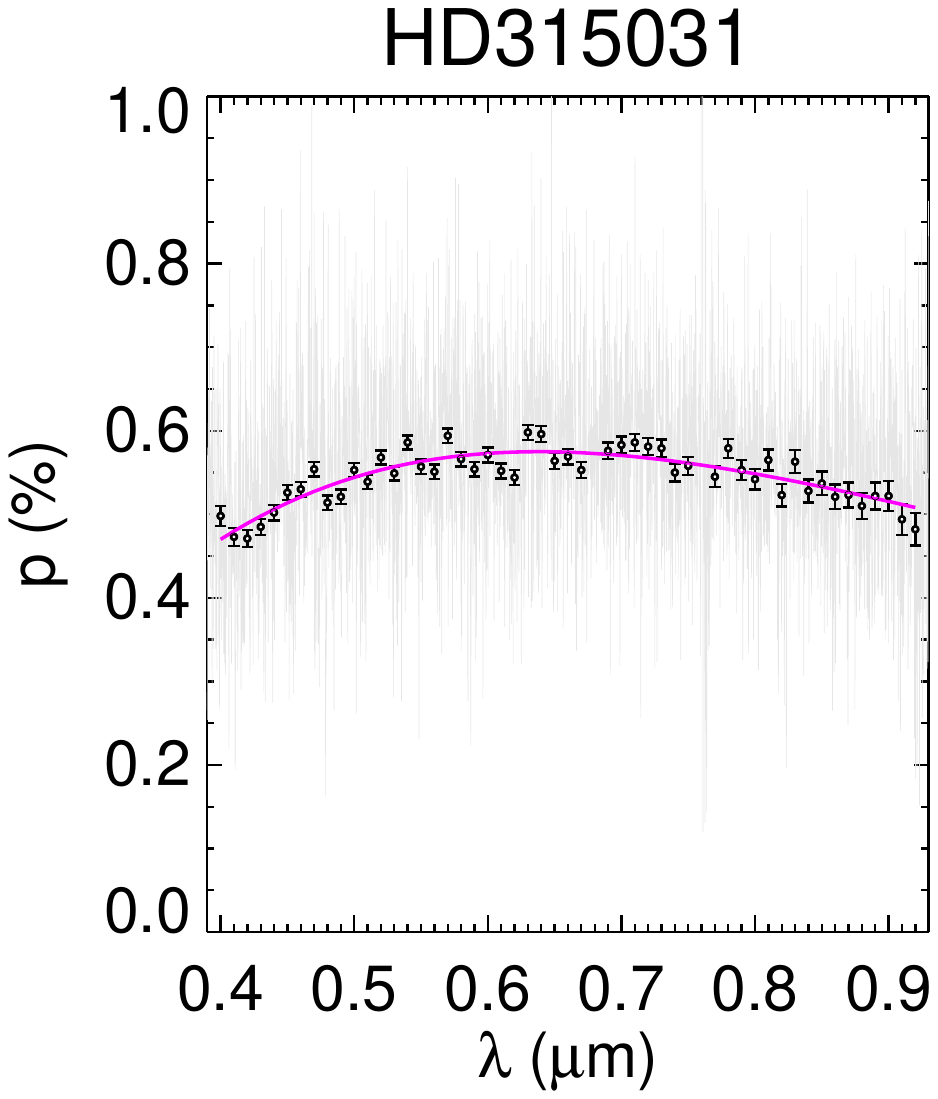}
 \includegraphics[width=3.6cm,clip=true,trim=2.7cm 5.0cm 1.5cm 3.0cm]{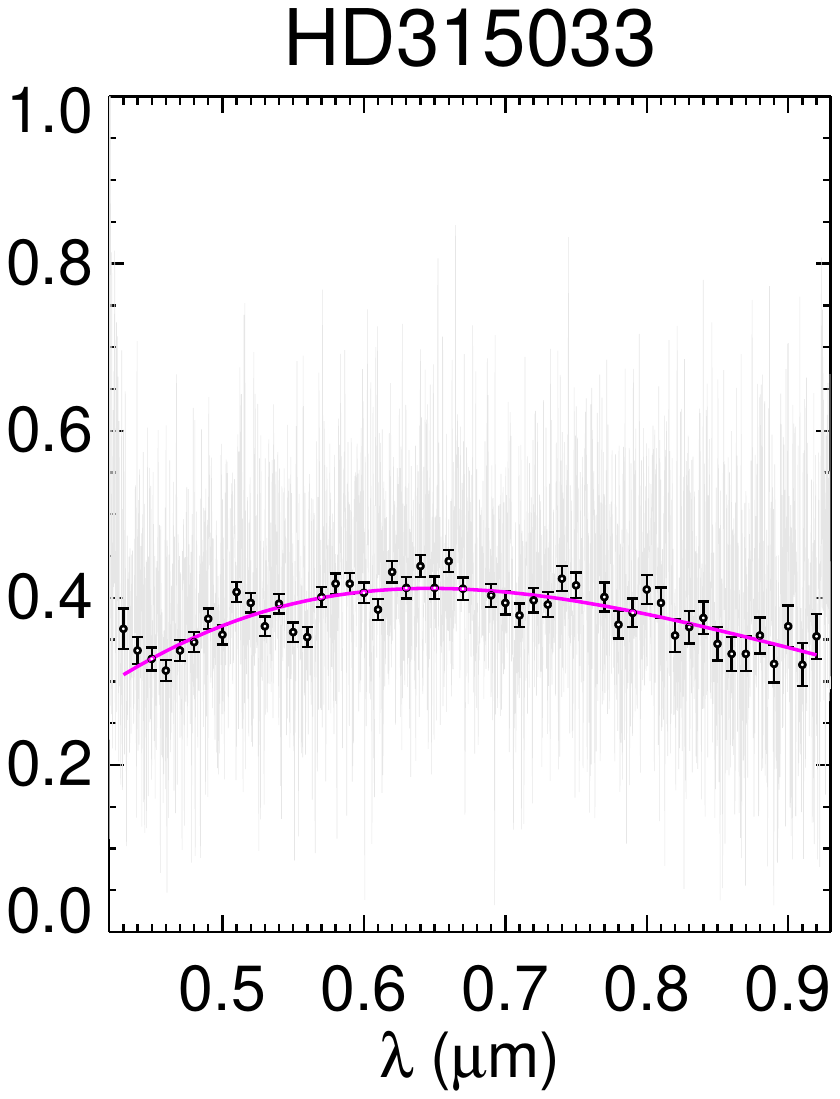}
 \includegraphics[width=3.6cm,clip=true,trim=2.7cm 5.0cm 1.5cm 3.0cm]{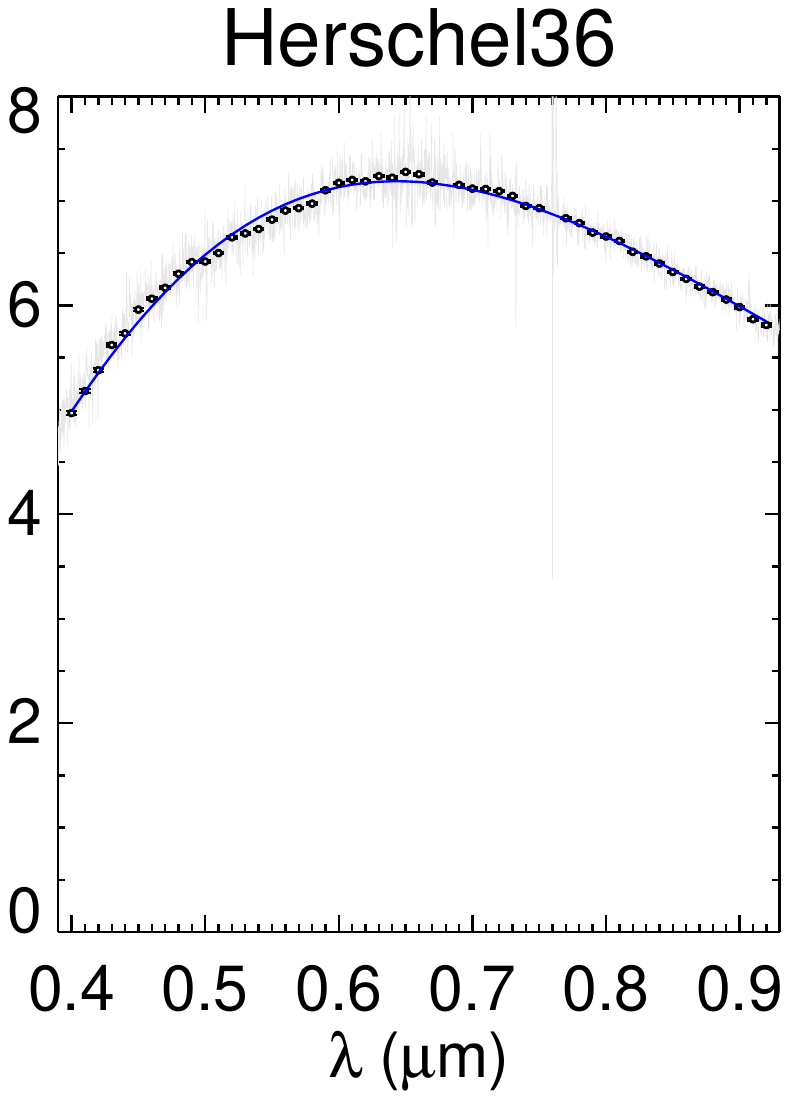}
\end{center}
\caption{FORS polarisation spectra, continued from Fig.~\ref{Fig1.pdf}.}
\label{Fig1.cont2}
\end{figure*}

 %

%%%%%%%%%%%%%%%%%%%%%%%%%%%%%%%%%%%%%%%%
\begin{figure*} [!htb]
\begin{center}
\includegraphics[width=18cm, clip=true,trim=7.6cm 0.1cm 4.2cm 0.cm]{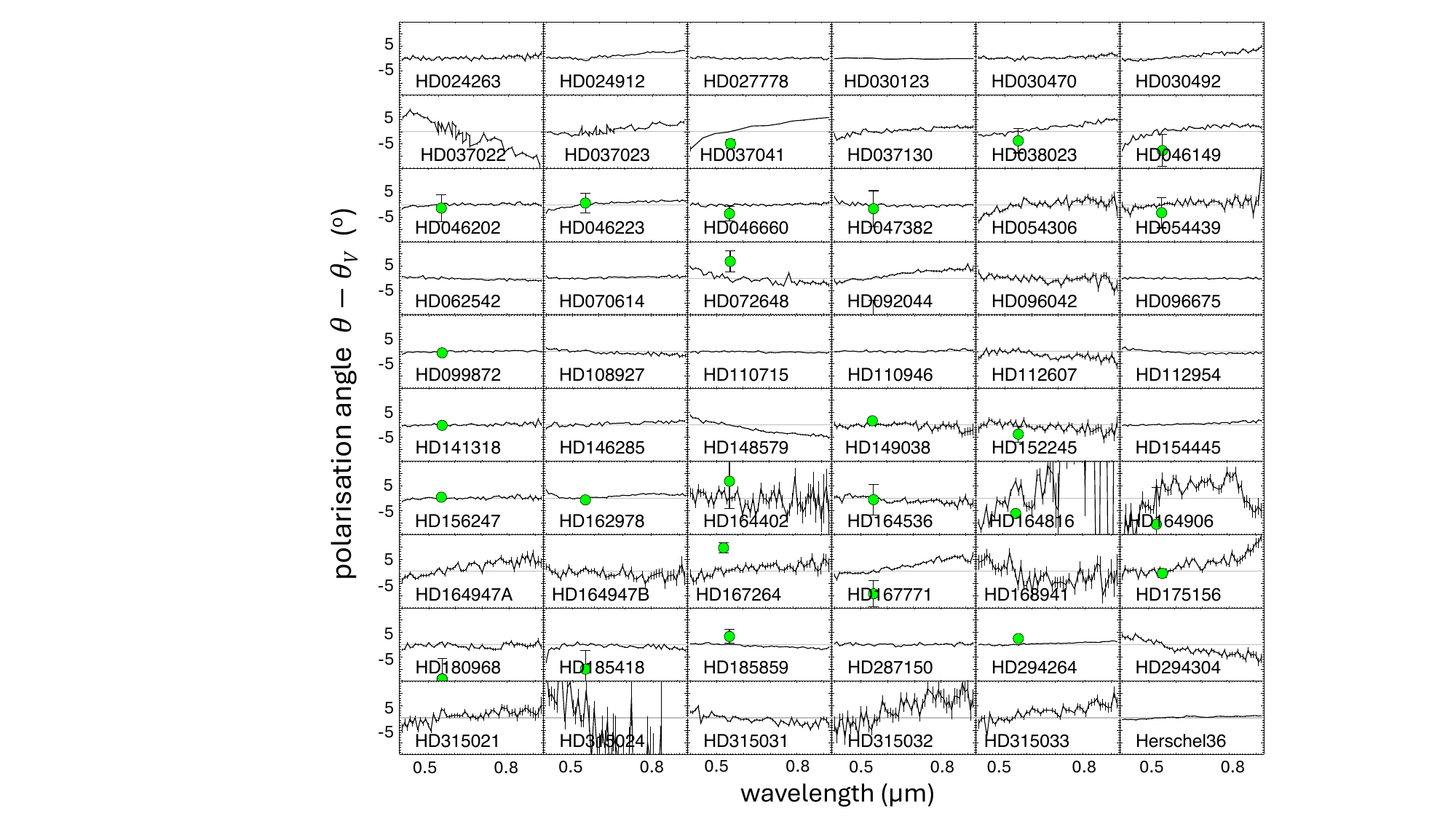}
\end{center}
\caption{\label{FigTheta.pdf} The position angle of the polarisation for the 60 stars in the LIPS
sample analysed in this work, offset with respect to its value in the
$V$ optical filter, $\theta - \theta_V$ (see
Table~\ref{Tab1.tab}). Data available in the catalogue by
\citet{Heiles} are shown with green circles.}
\end{figure*}

%%%%%%%%%%%%%%%%%%%%%%%%%%%%%%%%%%%%%%%%%

\begin{figure*} [!htb]
  \begin{center} 
  \includegraphics[width=3.5cm,clip=true,trim=3.25cm 5.cm 2.1cm 4.5cm]{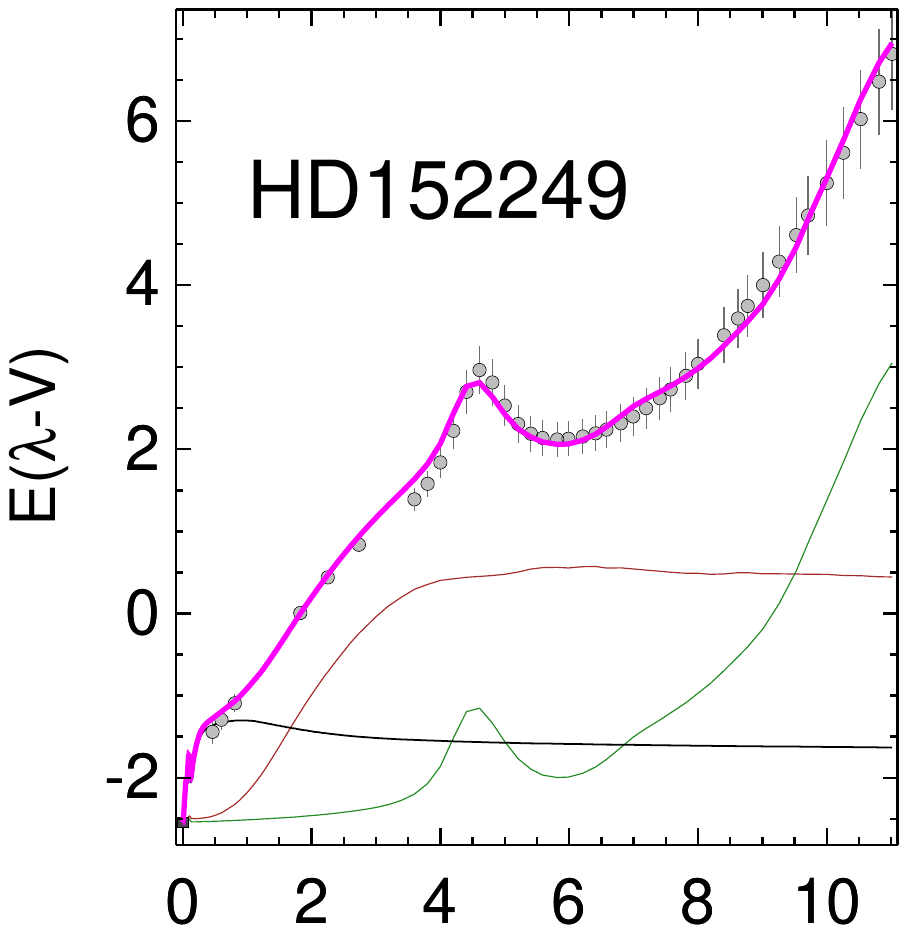}
  \includegraphics[width=3.5cm,clip=true,trim=3.25cm 5.cm 2.1cm 4.5cm]{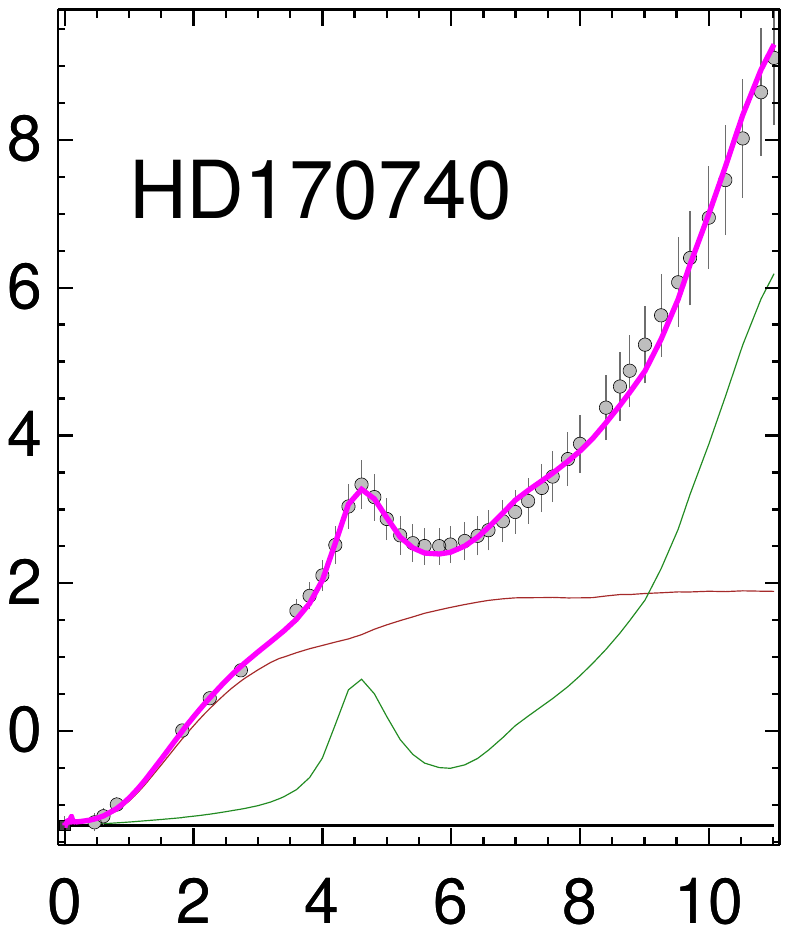}
  \includegraphics[width=3.5cm,clip=true,trim=3.25cm 5.cm 2.1cm 4.5cm]{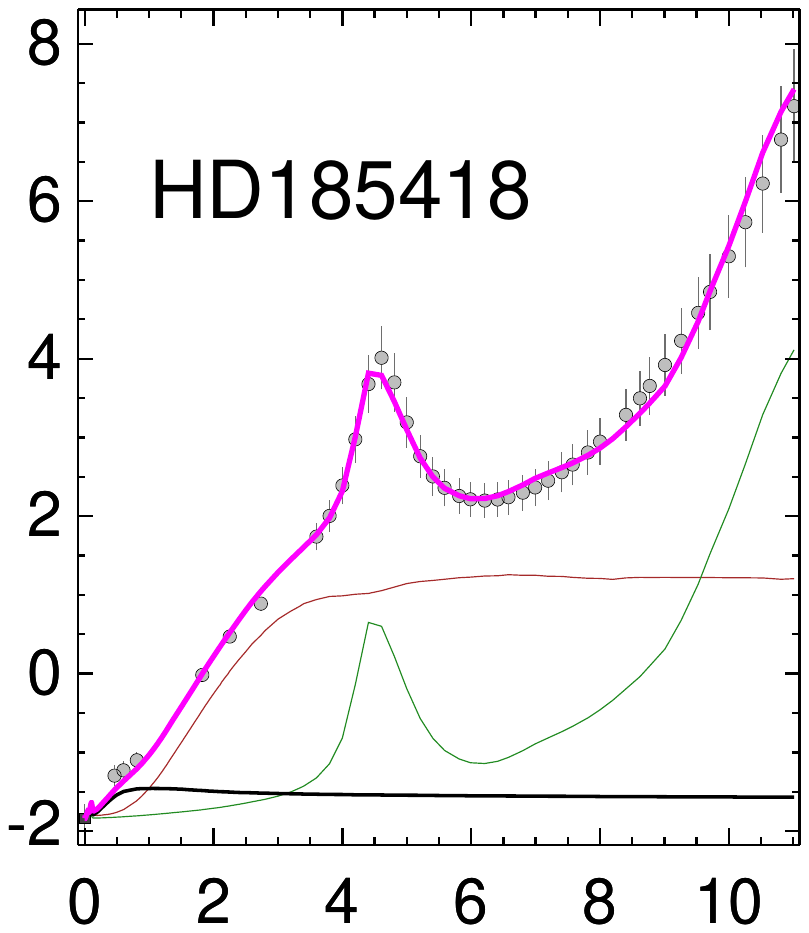}
  \includegraphics[width=3.5cm,clip=true,trim=3.25cm 5.cm 2.1cm 4.5cm]{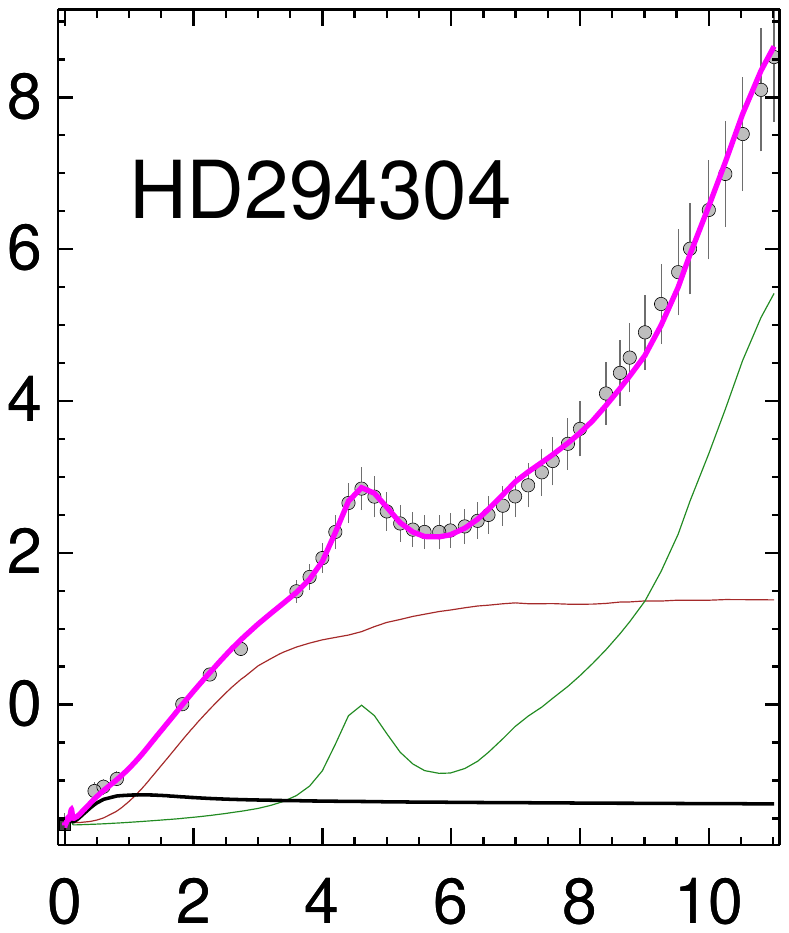}
  \includegraphics[width=3.5cm,clip=true,trim=3.25cm 5.cm 2.1cm 4.5cm]{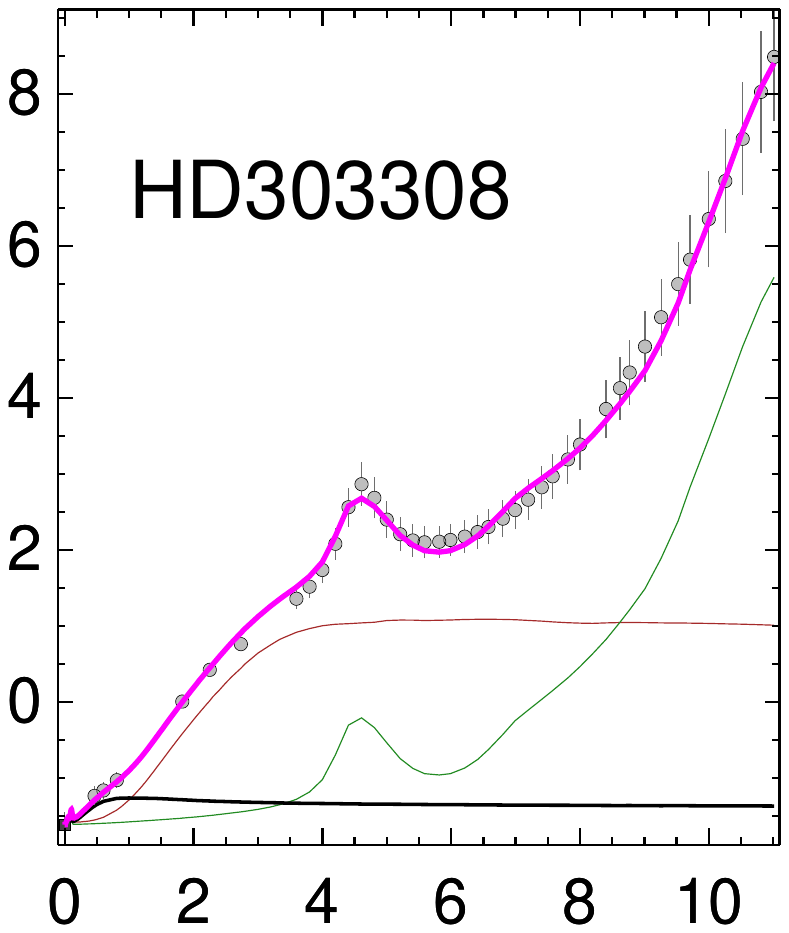}
  
  \includegraphics[width=3.5cm,clip=true,trim=3.25cm 5.cm 2.1cm 4.5cm]{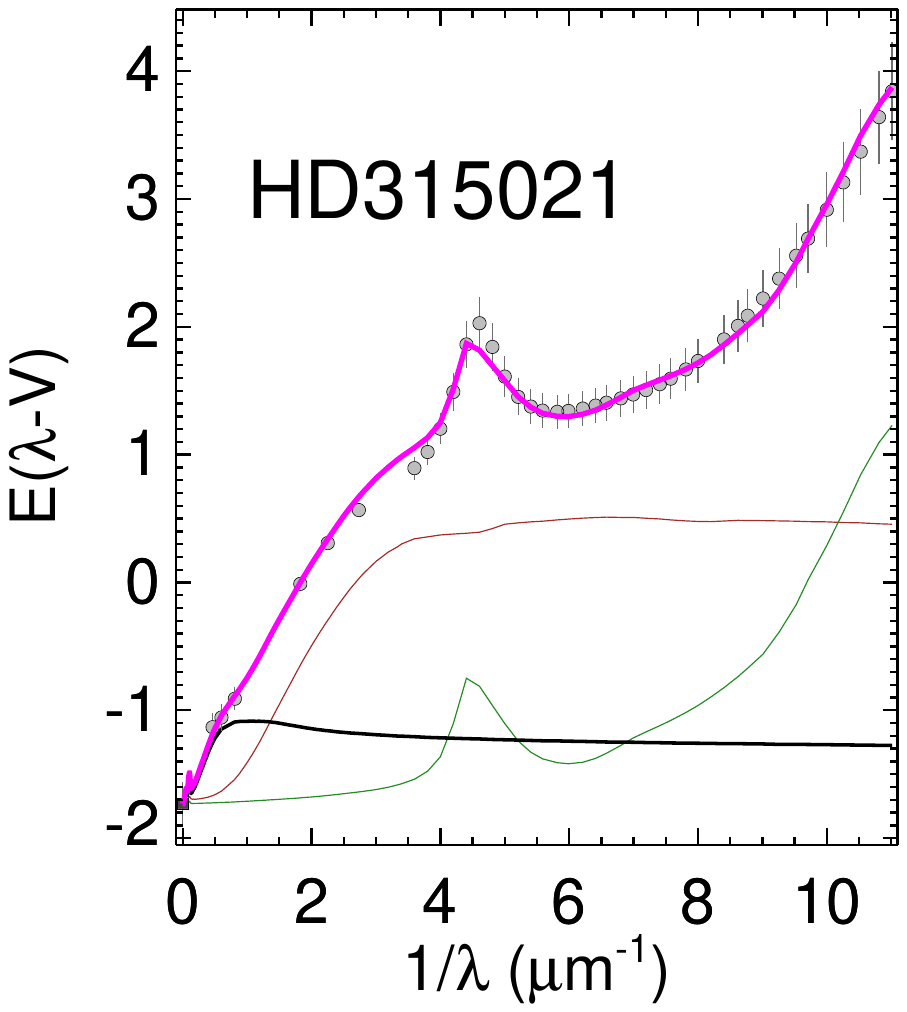}
  \includegraphics[width=3.5cm,clip=true,trim=3.25cm 5.cm 2.1cm 4.5cm]{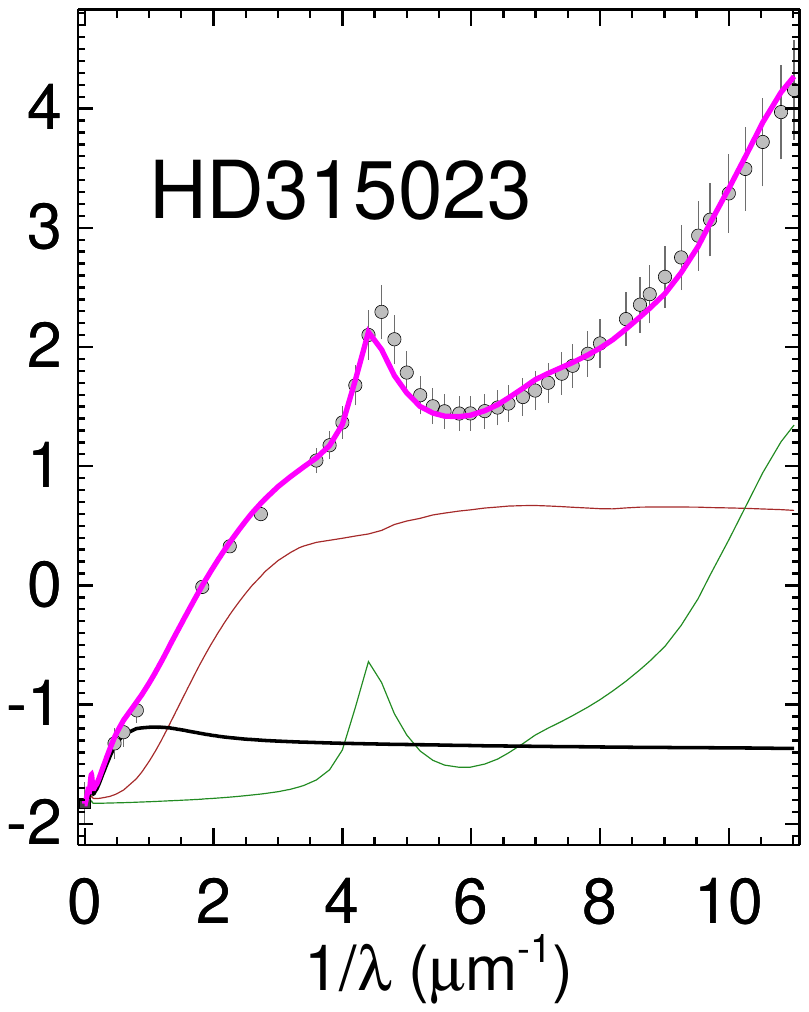}
  \includegraphics[width=3.5cm,clip=true,trim=3.25cm 5.cm 2.1cm 4.5cm]{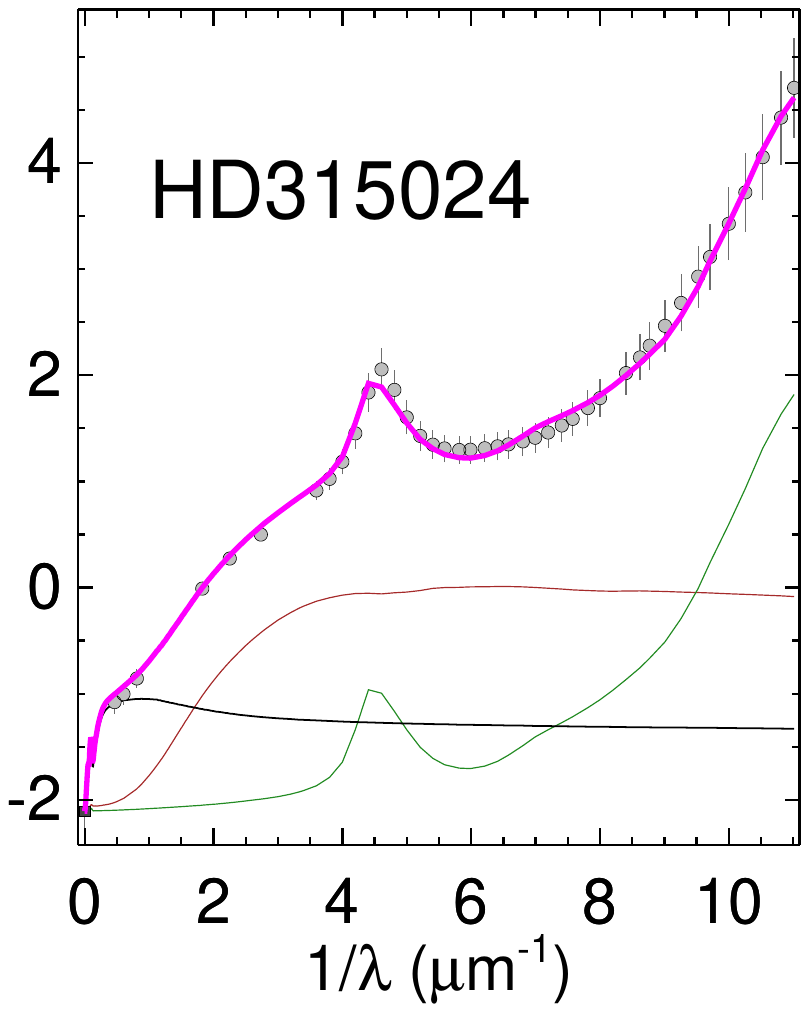}
  \includegraphics[width=3.5cm,clip=true,trim=3.25cm 5.cm 2.1cm 4.5cm]{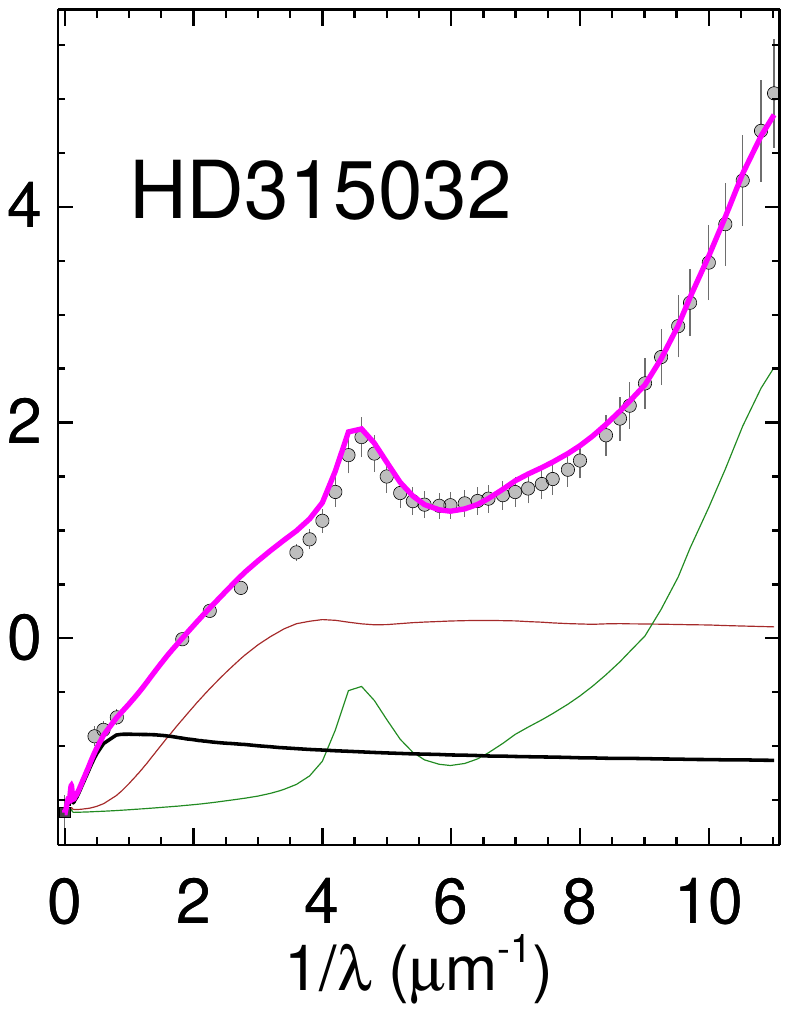}
  \end{center}
\caption{Dust model fits to the absolute reddening curves, continued
from Fig.~\ref{FigRedd.pdf}.}
\label{FigRedd.cont1}
  \end{figure*}  
%%%%%%%%%%%%%%%%%%%%%%%%%%%%%%%%%%%%%%%%%

\begin{figure*} [!htb]
  \begin{center}
  \includegraphics[width=3.5cm,clip=true,trim=2.7cm 5.cm 2.1cm 3.3cm]{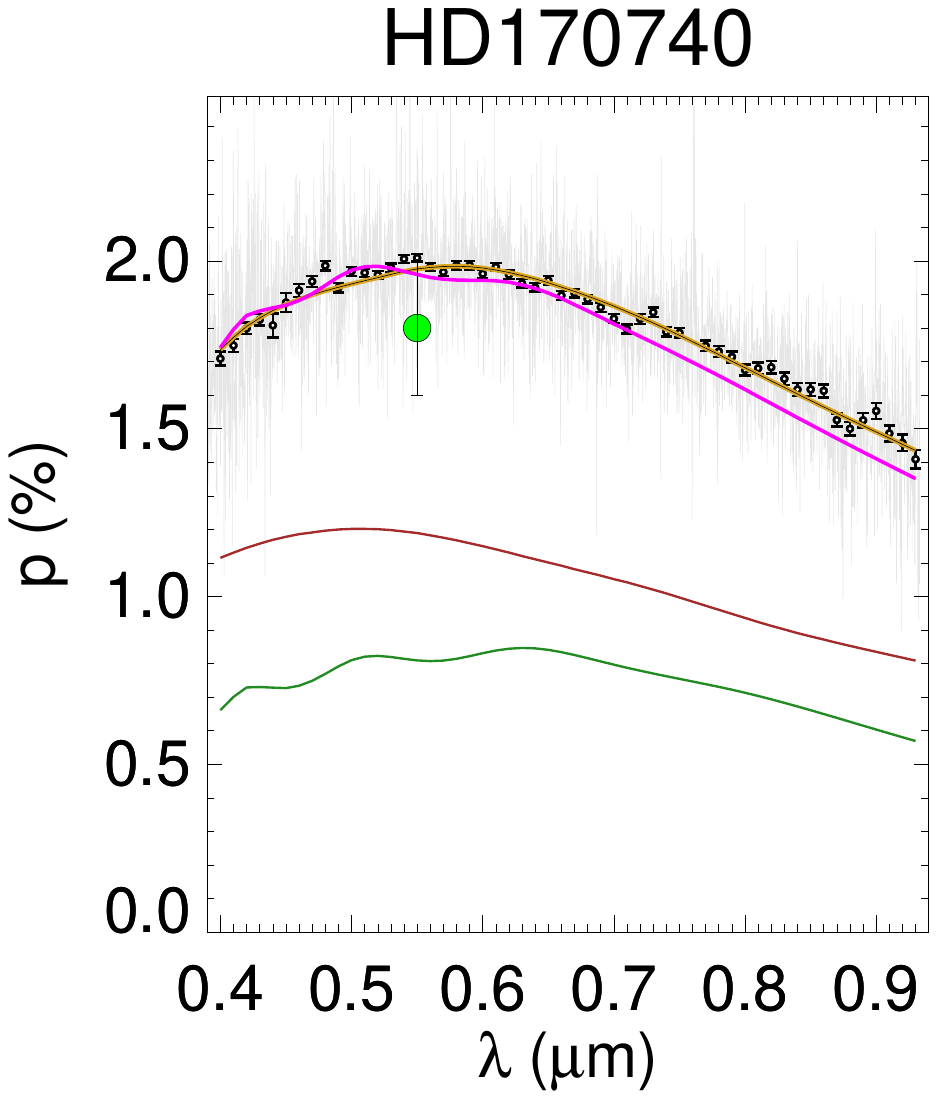}
  \includegraphics[width=3.5cm,clip=true,trim=2.7cm 5.cm 2.1cm 3.3cm]{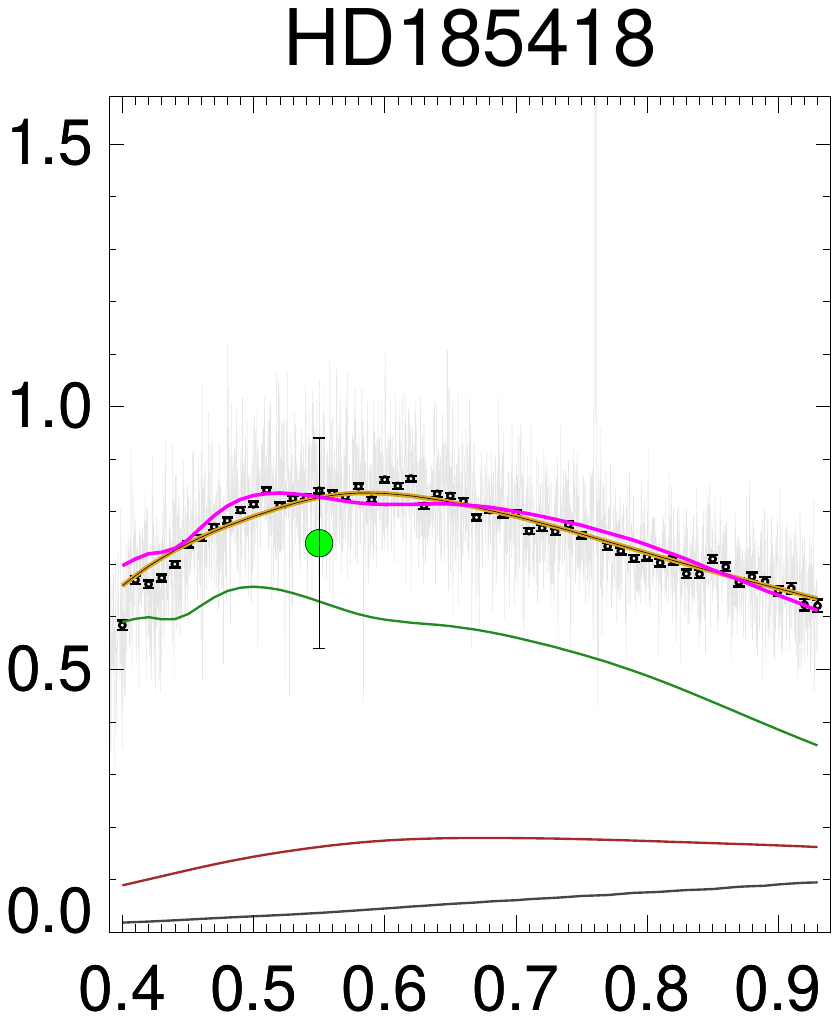}
  \includegraphics[width=3.5cm,clip=true,trim=2.7cm 5.cm 2.1cm 3.3cm]{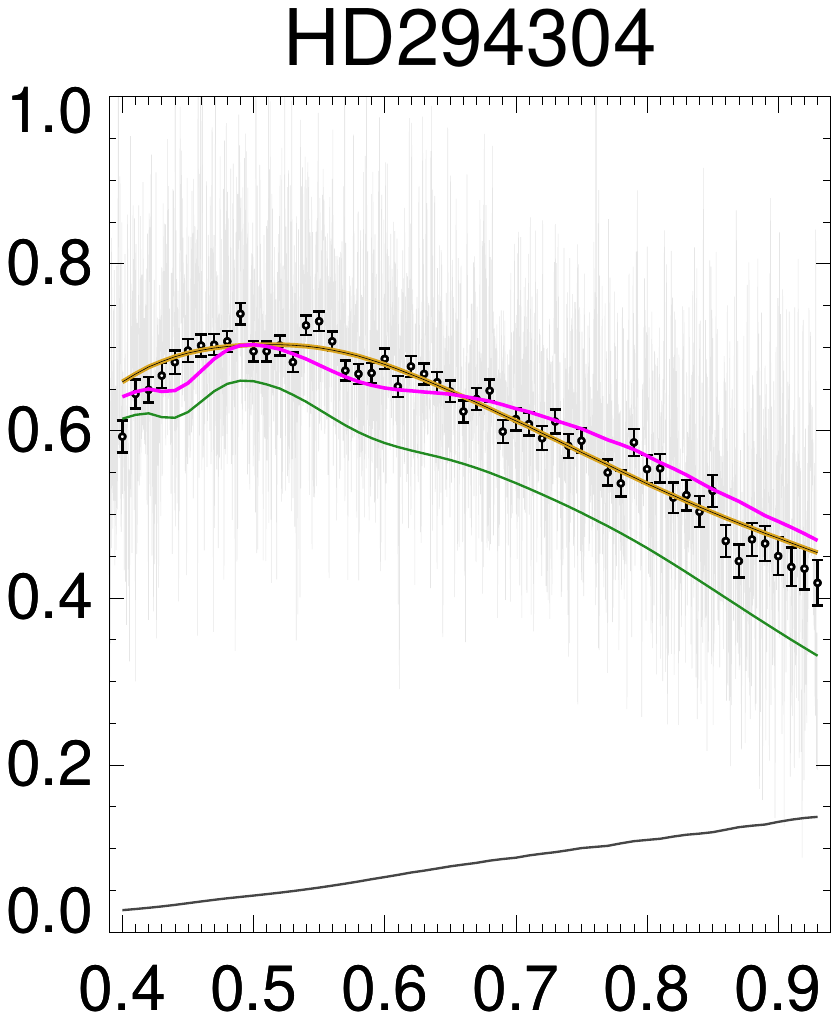}
  \includegraphics[width=3.5cm,clip=true,trim=2.7cm 5.cm 2.1cm 3.3cm]{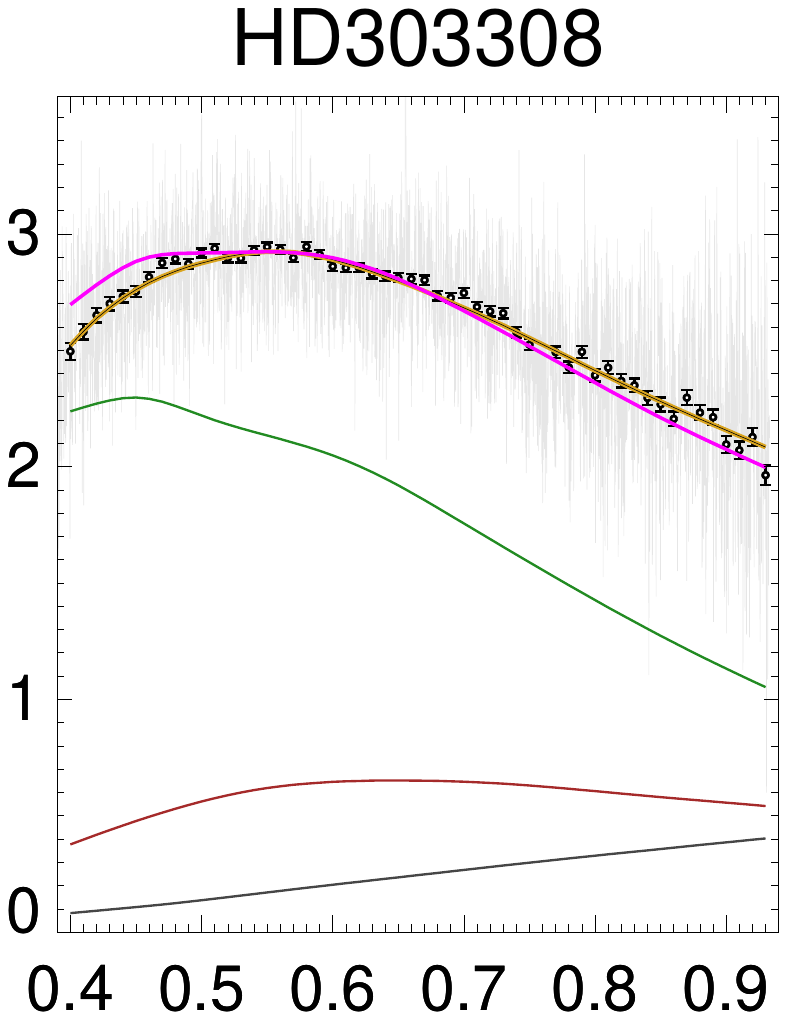}
  \includegraphics[width=3.5cm,clip=true,trim=2.7cm 5.cm 2.1cm 3.3cm]{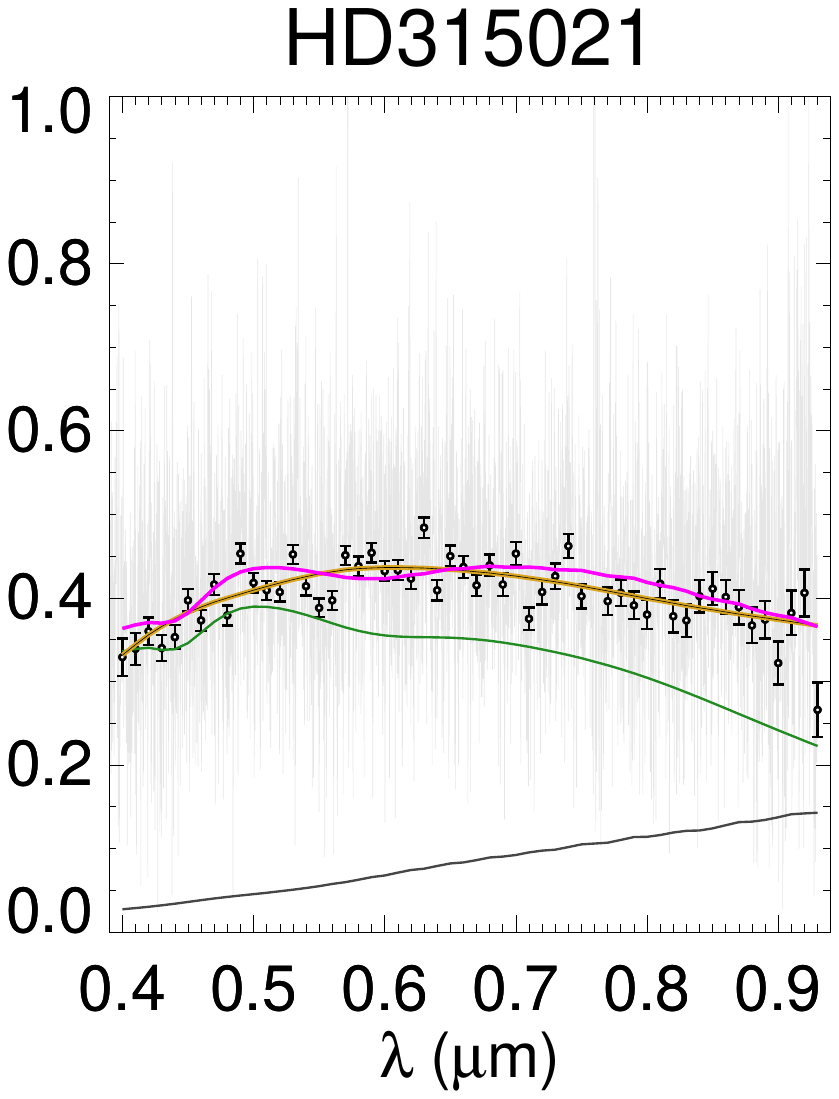}
  \includegraphics[width=3.5cm,clip=true,trim=2.7cm 5.cm 2.1cm 3.3cm]{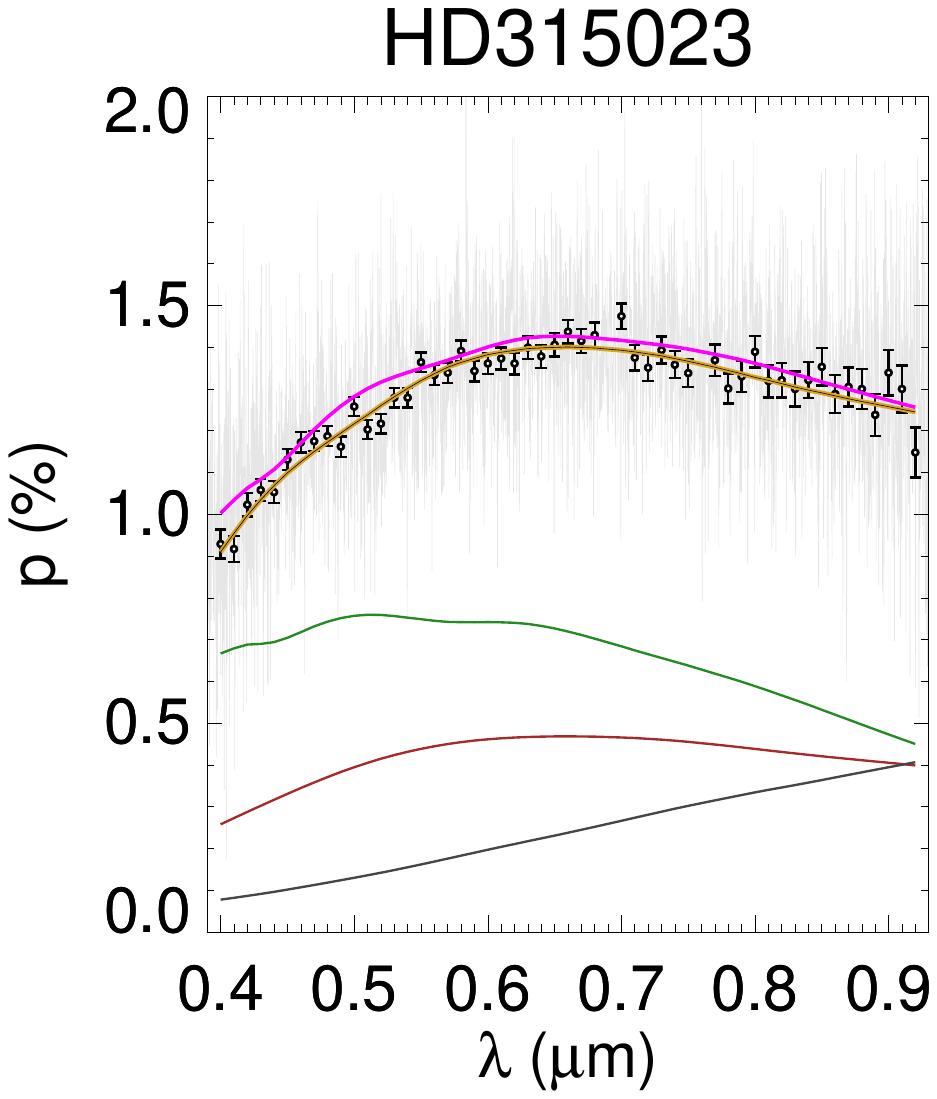}
  \includegraphics[width=3.5cm,clip=true,trim=2.7cm 5.cm 2.1cm 3.3cm]{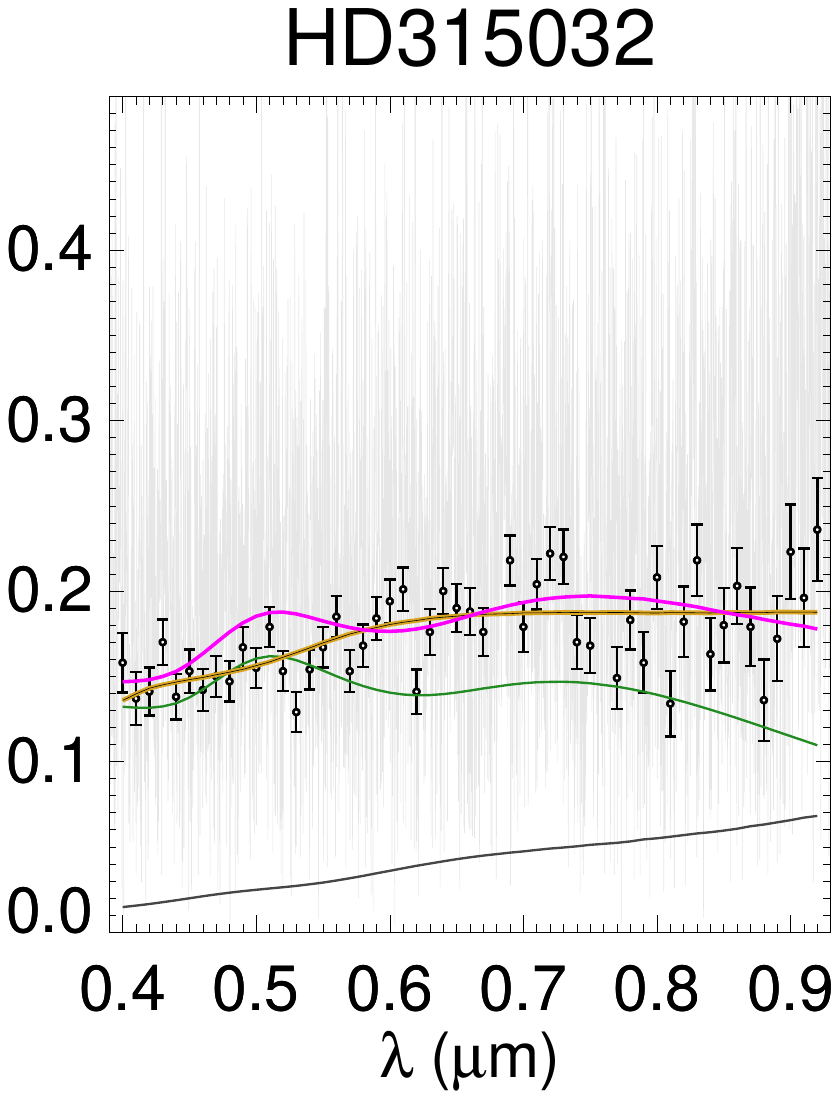}
  \end{center}
  \caption{FORS polarisation spectra and models continued from Fig.~\ref{FigPol.pdf}.} 
\label{FigPol.cont1} 
\end{figure*}

%%%%%%%%%%%%%%%%%%%%%%%%%%%%%%%%%%%%%%%%

\begin{table*}[!htb]
\scriptsize   
\begin{center} 
  \caption {Stars with derived Planck, reddening, FORS, and Serkowski
    fit parameters. \label {Tab1.tab}}
  \begin{tabular}{l  r | c c c c |l c r  r  | c  c  c  c | c  c  c }
%                 1  2   3 4 5 6  7 8 9 10    11 12 13 14  15 16 17 \\
    \hline\hline
    1 & 2& 3& 4& 5& 6& 7& 8& 9& 10& 11& 12& 13& 14& 15& 16 & 17 \\
    \hline
    \multicolumn{2}{c|}{Star}  &   \multicolumn{4}{c|}{PLANCK}  & \multicolumn{4}{c|}{Reddening}  &   \multicolumn{4}{c|}{FORS}   & \multicolumn{3}{c}{Serkowski} \\
\hline    
Name     & $||b||$  & $I_{850}$ & $p_{850}$  & $\theta_{850}$ & $A^{850}_{V}$ & $A_{V}$ & $A_{V}^{\rm ref}$& Ref & SM & Date & $p_{V}$ & $\theta_{V}$ & ${\rm{d}\theta}/{\rm{d}\lambda}$ &
   $p_{\rm {max}}$ & $\lambda_{\rm {max}}$ &  $k_{\rm {pol}}$ \\
         &  &MJy/sr  &\% & $^{\circ}$ &mag&mag&mag & & & &\% &$^{\circ}$ & $^{\circ}/\mu$m&\%&$\mu$m& \\
 \hline 
HD~024263  & 35 &   1.00 &    6.0$\pm$   2.4 &    77$\pm$   18 &    0.8 &   $-$ &    0.7 &   V & S &  2019-02-24 &    1.1  $\pm$  0.1 &   149  $\pm$  0.5 &      2 $\pm$  0.7 &    1.08 &    0.58 &    1.04 \\
HD~024912  & 13 &   1.28 &    6.9$\pm$   2.4 &    27$\pm$    6 &    1.1 &   $-$ &    1.0 &   V & M &  2015-12-25 &    1.4  $\pm$  0.1 &   111  $\pm$  0.6 &      7 $\pm$  1.1 &    1.43 &    0.62 &    1.16 \\
 \quad \quad $"$ & & & & & & & & & & 2018-11-17 & & & & & & \\
 \quad \quad $"$ & & & & & & & & & & 2018-11-17 & & & & & & \\
 HD~027778  & 17 &   1.33 &    6.5$\pm$   2.4 &   152$\pm$    7 &  1.2 & $1.2$ &    1.1 &   G & M &  2015-12-23 &    1.6  $\pm$  0.1 &    69  $\pm$  0.3 &     -1 $\pm$  0.3 &    1.65 &    0.52 &    1.08 \\
 \quad \quad $"$ & & & & & & & & & & 2018-11-14 & & & & & & \\
HD~030123  & 17 &   1.96 &    6.8$\pm$   2.4 &   179$\pm$    4 &    1.6 &   $-$ &    1.6 &   F & M &  2020-10-02 &    2.8  $\pm$  0.1 &    85  $\pm$  0.2 &     -1 $\pm$  0.2 &    2.76 &    0.54 &    1.14 \\
HD~030470  & 21 &   1.26 &    6.5$\pm$   2.7 &   178$\pm$   12 &    1.1 &   $-$ &    1.1 &   F & S &  2019-03-17 &    1.4  $\pm$  0.1 &    76  $\pm$  0.5 &      2 $\pm$  0.6 &    1.36 &    0.56 &    1.20 \\
HD~030492  & 21 &   1.26 &    6.7$\pm$   2.8 &   171$\pm$    4 &    1.1 &   $-$ &    1.2 &   F & S &  2019-03-18 &    1.4  $\pm$  0.1 &    77  $\pm$  0.4 &      9 $\pm$  1.5 &    1.37 &    0.59 &    1.16 \\
HD~037022  & 19 &   353 &    2.9$\pm$   2.7 &    40$\pm$   16 &   294 &   $-$ &    1.9 &   F & S &  2015-12-23 &    0.2  $\pm$  0.1 &   146  $\pm$  2.9 &    -52 $\pm$  5.7 &    0.27 &    0.77 &    1.25 \\
 \quad \quad $"$ & & & & & & & & & & 2020-10-02 & & & & & & \\
HD~037023  & 19 &   353 &    2.9$\pm$   2.7 &    40$\pm$   68 &   294 &   $-$ &    1.7 &   V & S &  2015-12-23 &    0.5  $\pm$  0.1 &    61  $\pm$  1.3 &     13 $\pm$  1.7 &    0.51 &    0.71 &    0.82 \\
 \quad \quad $"$ & & & & & & & & & & 2020-10-01 & & & & & & \\
HD~037041  & 19 &   255 &    2.7$\pm$   2.6 &    40$\pm$   29 &   212 &   $-$ &    1.1 &   V & S &  2020-10-25 &    0.8  $\pm$  0.1 &   101  $\pm$  0.7 &     20 $\pm$  3.6 &    0.94 &    0.72 &    1.45 \\
HD~037130  & 19 &   5.34 &    7.4$\pm$   2.3 &    30$\pm$   17 &    4.5 &   $-$ &    1.3 &   F & S &  2019-03-18 &    1.2  $\pm$  0.1 &   138  $\pm$  0.7 &      7 $\pm$  1.4 &    1.35 &    0.74 &    1.17 \\
HD~037367  &  1 &   4.46 &    4.7$\pm$   4.7 &    70$\pm$   35 &    3.7 &   $-$ &    1.5 &   V & M &         B17 &    1.0  $\pm$  0.1 &    15  $\pm$  0.9 &      0 $\pm$  1.2 &    1.02 &    0.63 &    1.13 \\
HD~037903  & 17 &   35.4 &    2.2$\pm$   2.0 &    65$\pm$   34 &    30 &   $-$ &    1.5 &   G & M &  2015-12-25 &    1.9  $\pm$  0.1 &   121  $\pm$  0.3 &      5 $\pm$  0.9 &    1.96 &    0.66 &    1.42 \\
 \quad \quad $"$ & & & & & & & & & & 2018-11-12 & & & & & & \\
HD~038023  & 19 &   11.7 &    3.8$\pm$   2.0 &    20$\pm$   23 &    9.7 &    2.3 &    1.6 &   F & S &  2019-02-23 &    1.6  $\pm$  0.1 &    87  $\pm$  0.7 &     13 $\pm$  2.1 &    1.64 &    0.43 &    0.93 \\
 \quad \quad $"$ & & & & & & & & & & 2019-03-18 & & & & & & \\
HD~046149  &  2 &   4.89 &    4.6$\pm$   2.1 &    65$\pm$   30 &    4.1 &   $-$ &    1.3 &   F & M &  2018-12-07 &    0.6  $\pm$  0.1 &     5  $\pm$  1.0 &     13 $\pm$  2.4 &    0.64 &    0.64 &    1.17 \\
HD~046202  &  2 &   4.85 &    4.2$\pm$   2.1 &    62$\pm$   26 &    4.0 &   $-$ &    1.5 &   G & M &  2018-12-08 &    1.0  $\pm$  0.1 &   178  $\pm$  0.5 &      2 $\pm$  0.5 &    1.00 &    0.60 &    1.05 \\
HD~046223  &  2 &   6.26 &    4.2$\pm$   2.1 &    64$\pm$   14 &    5.2 &    2.3 &    1.5 &   V & S &  2018-12-07 &    1.4  $\pm$  0.1 &   168  $\pm$  0.6 &      8 $\pm$  1.2 &    1.43 &    0.59 &    1.05 \\
HD~046660  &  1 &   3.75 &    5.6$\pm$   2.3 &    87$\pm$   19 &    3.1 &   $-$ &    1.7 &   F & M &  2020-10-02 &    1.8  $\pm$  0.1 &    17  $\pm$  0.3 &      1 $\pm$  0.4 &    1.81 &    0.60 &    1.09 \\
HD~047382  &  1 &   3.76 &    5.8$\pm$   2.0 &    69$\pm$    5 &    3.1 &   $-$ &    1.4 &   F & M &  2018-12-11 &    0.9  $\pm$  0.1 &   155  $\pm$  0.6 &     -2 $\pm$  0.8 &    0.95 &    0.65 &    1.28 \\
 \quad \quad $"$ & & & & & & & & & & 2018-12-14 & & & & & & \\
HD~054306  &  2 &   3.39 &    3.4$\pm$   1.9 &    56$\pm$    2 &    2.8 &   $-$ &    0.6 &   F & M &  2019-02-23 &    0.5  $\pm$  0.1 &   147  $\pm$  1.5 &     12 $\pm$  2.3 &    0.47 &    0.55 &    0.98 \\
HD~054439  &  2 &   3.07 &    3.3$\pm$   1.9 &    56$\pm$    7 &    2.6 &    0.7 &    0.8 &   F & S &  2015-12-23 &    0.8  $\pm$  0.1 &   139  $\pm$  0.6 &      7 $\pm$  2.3 &    0.77 &    0.51 &    1.17 \\
HD~062542  &  9 &   3.17 &    0.3$\pm$   1.4 &   147$\pm$   31 &    2.6 &    1.4 &    1.2 &   G & S &  2018-12-07 &    1.5  $\pm$  0.1 &    26  $\pm$  0.3 &     -3 $\pm$  0.5 &    1.53 &    0.58 &    1.21 \\
HD~070614  &  3 &   5.87 &    2.2$\pm$   1.5 &    32$\pm$   65 &    4.9 &   $-$ &    2.1 &   F & M &  2019-02-06 &    2.4  $\pm$  0.1 &    58  $\pm$  0.2 &      2 $\pm$  0.4 &    2.47 &    0.54 &    1.01 \\
HD~072648  &  2 &   7.38 &    0.1$\pm$   1.6 &    57$\pm$   36 &    6.1 &   $-$ &    1.2 &   F & M &  2019-02-24 &    0.7  $\pm$  0.1 &     3  $\pm$  0.9 &    -11 $\pm$  2.0 &    0.74 &    0.56 &    1.03 \\
 \quad \quad $"$ & & & & & & & & & & 2019-03-17 & & & & & & \\
HD~073882  &  1 &   10.8 &    1.7$\pm$   0.2 &    72$\pm$    2 &    9.0 &   $-$ &    2.5 &   G & M &         B17 &    1.9  $\pm$  0.1 &   164  $\pm$  0.5 &     -1 $\pm$  0.8 &    2.08 &    0.69 &    1.30 \\
HD~075309  &  2 &   5.05 &    2.3$\pm$   1.6 &   114$\pm$   31 &    4.2 &   $-$ &    0.9 &   F & M &         B17 &    0.6  $\pm$  0.1 &    54  $\pm$  1.9 &     -7 $\pm$  3.6 &    0.62 &    0.51 &    1.33 \\
HD~079186  &  2 &   3.89 &    1.0$\pm$   1.6 &   102$\pm$   36 &    3.2 &   $-$ &    1.3 &   V & S &         B17 &    2.6  $\pm$  0.1 &    47  $\pm$  0.3 &     -2 $\pm$  0.5 &    2.61 &    0.52 &    1.19 \\
HD~089137  &  4 &   1.14 &    3.1$\pm$   1.7 &   101$\pm$   28 &    0.9 &   $-$ &    0.7 &   V & S &  2019-02-06 &    0.4  $\pm$  0.1 &    39  $\pm$  1.3 &     -4 $\pm$  1.6 &    0.41 &    0.64 &    1.09 \\
HD~091824  &  0 &   16.0 &    2.3$\pm$   1.6 &    20$\pm$   13 &    13 &   $-$ &    0.8 &   F & M &  2018-12-14 &    1.4  $\pm$  0.1 &    97  $\pm$  0.4 &      4 $\pm$  0.8 &    1.43 &    0.53 &    1.08 \\
HD~091983  &  0 &   17.6 &    1.0$\pm$   1.5 &    74$\pm$   33 &    15 &   $-$ &    0.9 &   F & S &         B17 &    1.1  $\pm$  0.1 &   131  $\pm$  1.0 &     18 $\pm$  3.2 &    1.11 &    0.56 &    0.95 \\
HD~092044  &  0 &   18.8 &    2.0$\pm$   1.5 &    89$\pm$   19 &    16 &    2.4 &    1.4 &   F & S &  2020-10-04 &    1.4  $\pm$  0.1 &   160  $\pm$  0.5 &     12 $\pm$  1.9 &    1.42 &    0.63 &    1.33 \\
HD~093205  &  1 &   33.5 &    0.3$\pm$   1.5 &    17$\pm$    8 &    28 &   $-$ &    1.2 &   V & M &         B17 &    2.1  $\pm$  0.1 &   100  $\pm$  0.4 &     -5 $\pm$  1.0 &    2.10 &    0.55 &    1.16 \\
HD~093222  &  1 &   22.0 &    0.2$\pm$   1.4 &    55$\pm$   11 &    18 &   $-$ &    1.8 &   G & M &         B17 &    0.7  $\pm$  0.1 &   134  $\pm$  3.9 &     97 $\pm$ 12.2 &    0.77 &    0.43 &    1.45 \\
HD~093632  &  1 &   28.5 &    1.0$\pm$   1.4 &   151$\pm$    8 &    24 &   $-$ &    2.3 &   V & M &         B17 &    1.1  $\pm$  0.1 &    53  $\pm$  1.1 &    -20 $\pm$  4.9 &    1.47 &    0.84 &    1.29 \\
HD~094493  &  1 &   11.6 &    0.3$\pm$   1.4 &   146$\pm$   51 &    9.7 &   $-$ &    0.8 &   V & M &         B17 &    0.6  $\pm$  0.1 &   107  $\pm$  2.2 &     12 $\pm$  4.5 &    0.67 &    0.43 &    1.04 \\
HD~096042  &  1 &   7.68 &    1.8$\pm$   1.5 &    97$\pm$   73 &    6.4 &   $-$ &    0.9 &   V & M &  2015-01-02 &    0.6  $\pm$  0.1 &   115  $\pm$  0.9 &     -4 $\pm$  1.5 &    0.61 &    0.52 &    0.85 \\
HD~096675  & 15 &   1.76 &   10.6$\pm$   1.4 &    19$\pm$   21 &    1.5 &   $-$ &    1.0 &   G & S &  2019-03-14 &    3.2  $\pm$  0.1 &   130  $\pm$  0.2 &      0 $\pm$  0.2 &    3.19 &    0.55 &    1.20 \\
HD~097484  &  1 &   11.0 &    1.2$\pm$   1.4 &    44$\pm$   74 &    9.2 &   $-$ &    1.5 &   V & M &         B17 &    0.9  $\pm$  0.1 &    60  $\pm$  1.4 &     10 $\pm$  2.2 &    0.96 &    0.52 &    1.18 \\
HD~099872  & 11 &   1.28 &   13.9$\pm$   1.5 &    31$\pm$    3 &    1.1 &   $-$ &    1.1 &   G & M &         S14 &    3.2  $\pm$  0.1 &   118  $\pm$  0.3 &      1 $\pm$  0.3 &    3.27 &    0.58 &    1.27 \\
 \quad \quad $"$ & & & & & & & & & & 2019-03-09 & & & & & & \\
HD~103779  &  1 &   10.8 &    1.8$\pm$   1.6 &   170$\pm$    5 &    9.0 &   $-$ &    0.7 &   G & M &         B17 &    0.6  $\pm$  0.1 &    75  $\pm$  1.7 &    -26 $\pm$  5.2 &    0.62 &    0.52 &    1.76 \\
HD~104705  &  0 &   17.0 &    1.2$\pm$   2.0 &    66$\pm$   70 &    14 &   $-$ &    1.2 &   F & S &         B17 &    0.8  $\pm$  0.2 &    86  $\pm$  3.5 &     -7 $\pm$  7.2 &    0.77 &    0.62 &    0.89 \\
HD~108927  & 15 &   0.92 &    7.4$\pm$   1.8 &    29$\pm$    3 &    0.8 &    1.1 &    0.7 &   F & S &  2019-03-09 &    1.5  $\pm$  0.1 &   122  $\pm$  0.4 &     -5 $\pm$  0.8 &    1.53 &    0.52 &    1.15 \\
HD~110715  &  2 &   5.12 &    5.9$\pm$   2.3 &   168$\pm$    3 &    4.3 &   $-$ &    1.3 &   F & S &  2019-03-08 &    2.8  $\pm$  0.1 &    75  $\pm$  0.2 &     -1 $\pm$  0.3 &    2.79 &    0.58 &    1.28 \\
 \quad \quad $"$ & & & & & & & & & & 2019-03-14 & & & & & & \\
HD~110946  &  2 &   5.10 &    5.4$\pm$   2.1 &   166$\pm$    3 &    4.3 &    1.5 &    1.6 &   F & S &  2019-02-11 &    2.4  $\pm$  0.1 &    79  $\pm$  0.3 &      1 $\pm$  0.4 &    2.43 &    0.56 &    1.30 \\
HD~112607  &  1 &   16.7 &    1.5$\pm$   2.5 &   146$\pm$    6 &    14 &    0.7 &    0.8 &   F & S &  2019-03-04 &    0.6  $\pm$  0.1 &    63  $\pm$  0.6 &     -7 $\pm$  1.5 &    0.58 &    0.59 &    1.47 \\
HD~112954  &  0 &   29.2 &    0.4$\pm$   2.2 &    41$\pm$   87 &    24 &    1.5 &    1.7 &   F & S &  2019-02-24 &    2.3  $\pm$  0.1 &    45  $\pm$  0.3 &     -4 $\pm$  0.7 &    2.39 &    0.60 &    1.35 \\
HD~122879  &  2 &   8.07 &    2.0$\pm$   1.9 &   173$\pm$   13 &    6.7 &   $-$ &    1.1 &   V & M &         B17 &    1.8  $\pm$  0.1 &    70  $\pm$  0.5 &     -5 $\pm$  0.9 &    1.81 &    0.55 &    1.52 \\
HD~129557  &  4 &   2.32 &    6.2$\pm$   2.0 &   171$\pm$    0 &    1.9 &    1.1 &    0.5 &   V & S &         B17 &    1.3  $\pm$  0.1 &    80  $\pm$  0.7 &      1 $\pm$  1.0 &    1.33 &    0.57 &    1.54 \\
HD~134591  & 20 &   0.76 &    4.2$\pm$   2.3 &   131$\pm$   73 &    0.6 &   $-$ &    0.6 &   V & M &         B17 &    0.3  $\pm$  0.1 &   114  $\pm$  5.4 &     24 $\pm$ 11.6 &    0.31 &    0.43 &    0.54 \\
HD~141318  &  1 &   33.8 &    2.5$\pm$   2.3 &   149$\pm$    8 &    28 &   $-$ &    0.8 &   V & M &  2014-10-10 &    2.4  $\pm$  0.1 &    51  $\pm$  0.2 &      1 $\pm$  0.5 &    2.46 &    0.58 &    1.27 \\
HD~146285  & 18 &   2.12 &    4.9$\pm$   2.6 &   123$\pm$   15 &    1.8 &    1.9 &    1.2 &   F & S &  2019-03-19 &    1.5  $\pm$  0.1 &    18  $\pm$  0.5 &      4 $\pm$  0.8 &    1.55 &    0.63 &    1.47 \\
HD~147888  & 18 &   4.59 &    7.2$\pm$   2.5 &   132$\pm$   12 &    3.8 &   $-$ &    2.0 &   G & S &         B17 &    3.3  $\pm$  0.1 &    54  $\pm$  0.2 &     -3 $\pm$  0.6 &    3.49 &    0.66 &    1.49 \\
HD~147889  & 17 &   17.6 &    1.1$\pm$   2.6 &   123$\pm$   37 &    15 &   $-$ &    4.3 &   V & S &         B17 &    3.4  $\pm$  0.2 &   177  $\pm$  0.4 &     -7 $\pm$  1.4 &    4.20 &    0.81 &    1.30 \\
HD~148379  &  2 &   8.01 &    1.8$\pm$   2.3 &   132$\pm$   12 &    6.7 &   $-$ &    2.4 &   V & M &         B17 &    1.9  $\pm$  0.1 &    30  $\pm$  0.6 &     -1 $\pm$  0.7 &    1.95 &    0.58 &    0.94 \\
HD~148579  & 16 &   3.91 &    2.5$\pm$   2.5 &   139$\pm$   30 &    3.3 &   $-$ &    1.4 &   F & S &  2021-01-23 &    2.0  $\pm$  0.1 &    79  $\pm$  0.7 &    -15 $\pm$  2.4 &    2.12 &    0.66 &    1.16 \\
HD~149038  &  3 &   5.42 &    1.4$\pm$   2.4 &   161$\pm$   43 &    4.5 &   $-$ &    1.1 &   V & M &  2019-04-02 &    1.0  $\pm$  0.1 &    29  $\pm$  0.6 &     -1 $\pm$  1.2 &    1.02 &    0.57 &    1.47 \\
HD~151804  &  2 &   5.95 &    3.0$\pm$   2.3 &   148$\pm$   15 &    5.0 &   $-$ &    1.3 &   V & M &         B17 &    1.1  $\pm$  0.1 &    43  $\pm$  0.8 &     -8 $\pm$  1.6 &    1.12 &    0.57 &    1.21 \\
HD~152235  &  1 &   17.9 &    2.1$\pm$   2.4 &   138$\pm$   67 &    15 &   $-$ &    2.2 &   V & M &         B17 &    0.8  $\pm$  0.1 &   115  $\pm$  1.5 &     -6 $\pm$  2.7 &    0.79 &    0.47 &    1.54 \\
HD~152245  &  2 &   6.15 &    0.7$\pm$   2.4 &   142$\pm$    3 &    5.1 &   $-$ &    1.1 &   V & M &  2015-02-06 &    0.9  $\pm$  0.1 &    49  $\pm$  1.1 &     -7 $\pm$  1.7 &    0.93 &    0.62 &    1.48 \\
HD~152249  &  1 &   15.9 &    2.2$\pm$   2.4 &   144$\pm$   10 &    13 &    2.5 &    1.6 &   G & S &         B17 &    0.3  $\pm$  0.1 &    64  $\pm$  3.9 &    -27 $\pm$  6.2 &    0.30 &    0.65 &    2.37 \\
HD~153919  &  2 &   5.28 &    1.4$\pm$   2.3 &    80$\pm$   20 &    4.4 &   $-$ &    2.0 &   V & M &         B17 &    2.7  $\pm$  0.1 &    10  $\pm$  0.3 &      7 $\pm$  1.2 &    2.67 &    0.57 &    1.09 \\
HD~154445  & 23 &   1.27 &   16.6$\pm$   2.1 &   180$\pm$    1 &    1.1 &   $-$ &    1.2 &   F & S &  2020-10-03 &    3.7  $\pm$  0.1 &    91  $\pm$  0.2 &      3 $\pm$  0.6 &    3.74 &    0.56 &    1.15 \\
HD~156247  & 22 &   0.78 &   15.1$\pm$   2.1 &   178$\pm$    1 &    0.7 &   $-$ &    0.7 &   F & S &  2020-10-03 &    2.0  $\pm$  0.1 &    87  $\pm$  0.2 &      2 $\pm$  0.5 &    2.04 &    0.56 &    1.20 \\
 HD~162978  &  0 &   40.7 &    2.1$\pm$  2.4 &   126$\pm$   36 &    34 &   $-$ &    1.2 &   V & M &  2018-11-06 &    1.3  $\pm$  0.1 &    -1  $\pm$  0.3 &      3 $\pm$  0.8 &    1.41 &    0.65 &    1.53 \\
 HD~163181  &  4 &   3.71 &    2.6$\pm$  2.4 &    75$\pm$   10 &    3.1 &   $-$ &    2.4 &   V & M &         B17 &    1.4  $\pm$  0.1 &   175  $\pm$  1.2 &     19 $\pm$  3.0 &    1.44 &    0.47 &    0.38 \\
HD~164073  & 13 &   0.80 &    8.5$\pm$   2.1 &    97$\pm$    5 &    0.7 &   $-$ &    1.1 &   F & S &         B17 &    1.1  $\pm$  0.1 &     1  $\pm$  0.7 &     -3 $\pm$  1.2 &    1.10 &    0.63 &    1.11 \\
HD~164402  &  0 &   57.5 &    2.4$\pm$   2.2 &   113$\pm$   19 &    48 &   $-$ &    0.7 &   V & S &  2020-10-03 &    0.2  $\pm$  0.1 &     4  $\pm$  2.8 &     -8 $\pm$  3.8 &    0.22 &    0.52 &    2.10 \\
HD~164536  &  1 &   31.8 &    2.1$\pm$   2.2 &   106$\pm$   38 &    26 &   $-$ &    0.9 &   F & S &  2020-10-03 &    0.8  $\pm$  0.1 &   159  $\pm$  0.7 &     -7 $\pm$  1.2 &    0.79 &    0.55 &    1.28 \\
 HD~164816  &  1 &   36.7 &    1.3$\pm$ 2.1 &    98$\pm$   43 &    31 &   $-$ &    1.0 &   G & S &  2019-02-27 &    0.1  $\pm$  0.1 &    51  $\pm$  3.7 &     61 $\pm$ 42.3 &    0.11 &    0.44 &    2.50 \\
 \hline
\end{tabular}
\end{center}
\scriptsize{\bf{Notes.}} {The columns are explained in
  Sect.~\ref{sample.sec}. B17 refers to \citet{B17}, and S14 to
  \citet{S14}.}
\end{table*}

\setcounter{table}{0}
\begin{table*}[!htb]
\scriptsize   
\begin{center} 
  \caption { - continued - }
  \begin{tabular}{l  r | c c c c |l c r  r  | c  c  c  c | c  c  c }
%                 1  2   3 4 5 6  7 8 9 10    11 12 13 14  15 16 17 \\
    \hline\hline
    1 & 2& 3& 4& 5& 6& 7& 8& 9& 10& 11& 12& 13& 14& 15& 16 & 17 \\
    \hline
    \multicolumn{2}{c|}{Star}  &   \multicolumn{4}{c|}{PLANCK}  & \multicolumn{4}{c|}{Reddening}  &   \multicolumn{4}{c|}{FORS}   & \multicolumn{3}{c}{Serkowski} \\
\hline    
Name     & $||b||$  & $I_{850}$ & $p_{850}$  & $\theta_{850}$ & $A^{850}_{V}$ & $A_{V}$ & $A_{V}^{\rm ref}$& Ref & SM & Date & $p_{V}$ & $\theta_{V}$ & ${\rm{d}\theta}/{\rm{d}\lambda}$ &
   $p_{\rm {max}}$ & $\lambda_{\rm {max}}$ &  $k_{\rm {pol}}$ \\
         &  &MJy/sr  &\% & $^{\circ}$ &mag&mag&mag & & & &\% &$^{\circ}$ & $^{\circ}/\mu$m&\%&$\mu$m& \\
 \hline 
 HD~164906  &  1 &   36.2 &    1.3$\pm$ 2.1 &   109$\pm$   14 &    30 &   $-$ &    2.2 &   G & S &  2020-10-03 &    0.2  $\pm$  0.1 &     5  $\pm$  4.5 &     19 $\pm$  8.2 &    0.24 &    0.80 &    2.50 \\
 HD~164947A &  1 &   23.9 &    1.4$\pm$   2.2 &   111$\pm$   51 &    20 &   $-$ &    1.1 &   F & S &  2020-10-03 &    0.5  $\pm$  0.1 &    72  $\pm$  1.2 &     17 $\pm$  2.8 &    0.53 &    0.56 &    0.88 \\
HD~164947B &  1 &   23.8 &    1.4$\pm$   2.2 &   111$\pm$   28 &    20 &   $-$ &    1.1 &   F & S &  2020-10-03 &    0.6  $\pm$  0.1 &    49  $\pm$  1.0 &     -5 $\pm$  1.8 &    0.59 &    0.60 &    0.90 \\
HD~167264  &  2 &   9.02 &    1.7$\pm$   2.2 &   151$\pm$   41 &    7.5 &   $-$ &    1.0 &   V & S &  2020-10-03 &    0.5  $\pm$  0.1 &   102  $\pm$  1.3 &     11 $\pm$  2.1 &    0.57 &    0.65 &    1.80 \\
HD~167771  &  1 &   13.2 &    1.2$\pm$   2.3 &    59$\pm$   78 &    11 &    2.2 &    1.5 &   G & S &  2019-02-20 &    0.5  $\pm$  0.1 &    47  $\pm$  0.9 &     16 $\pm$  2.6 &    0.57 &    0.68 &    1.24 \\
\quad \quad $"$ & & & & & & & & & & 2019-03-19 & & & & & & \\
HD~167838  &  0 &   24.7 &    2.3$\pm$   2.1 &   122$\pm$   66 &    21 &   $-$ &    2.1 &   V & M &         B17 &    0.3  $\pm$  0.1 &    97  $\pm$  2.7 &     48 $\pm$  8.7 &    0.32 &    0.57 &    1.56 \\
HD~168076  &  1 &   40.5 &    2.4$\pm$   2.2 &   146$\pm$   11 &    34 &   $-$ &    2.6 &   V & M &         B17 &    3.4  $\pm$  0.1 &    66  $\pm$  0.3 &     -5 $\pm$  0.9 &    3.41 &    0.58 &    1.33 \\
HD~168941  &  6 &   1.47 &    0.8$\pm$   2.3 &    80$\pm$   60 &    1.2 &   $-$ &    1.2 &   G & M &  2019-03-03 &    0.1  $\pm$  0.1 &    50  $\pm$  3.9 &    -18 $\pm$  4.2 &    0.15 &    0.73 &    1.35 \\
HD~169454  &  1 &   36.6 &    2.4$\pm$   2.2 &   119$\pm$   14 &    30 &   $-$ &    3.6 &   V & S &         B17 &    2.1  $\pm$  0.1 &    15  $\pm$  0.8 &     -9 $\pm$  1.6 &    2.12 &    0.58 &    1.26 \\
HD~170740  &  1 &   25.9 &    2.2$\pm$   2.1 &   129$\pm$   38 &    22 &   $-$ &    1.4 &   F & M &         B17 &    2.0  $\pm$  0.1 &    77  $\pm$  0.5 &     -5 $\pm$  0.9 &    1.98 &    0.56 &    1.23 \\
HD~175156  &  8 &   1.33 &    1.8$\pm$   2.4 &    80$\pm$   52 &    1.1 &   $-$ &    1.1 &   V & S &  2020-10-03 &    0.5  $\pm$  0.1 &    42  $\pm$  1.1 &     18 $\pm$  3.7 &    0.54 &    0.54 &    1.23 \\
HD~180968  &  5 &   2.98 &    2.3$\pm$   1.7 &    30$\pm$   84 &    2.5 &   $-$ &    0.8 &   F & S &  2019-04-08 &    0.5  $\pm$  0.1 &    36  $\pm$  0.7 &     -1 $\pm$  0.9 &    0.52 &    0.53 &    1.05 \\
HD~185418  &  2 &   4.03 &    2.7$\pm$   1.9 &   148$\pm$   34 &    3.4 &   $-$ &    1.4 &   G & M &  2019-03-25 &    0.8  $\pm$  0.1 &    24  $\pm$  0.4 &     -1 $\pm$  1.2 &    0.84 &    0.60 &    1.65 \\
HD~185859  &  1 &   7.09 &    4.3$\pm$   1.8 &   111$\pm$   16 &    5.9 &   $-$ &    1.6 &   V & S &  2020-10-04 &    2.2  $\pm$  0.1 &     5  $\pm$  0.2 &     -4 $\pm$  0.7 &    2.27 &    0.50 &    1.25 \\
 \quad \quad $"$ & & & & & & & & & & 2020-10-06 & & & & & & \\
HD~203532  & 32 &   1.14 &    7.4$\pm$   1.7 &    48$\pm$   12 &    1.0 &   $-$ &    0.9 &   F & S &  2018-11-05 &    1.4  $\pm$  0.1 &   126  $\pm$  0.4 &     -1 $\pm$  0.6 &    1.41 &    0.58 &    1.21 \\
HD~210121  & 44 &   0.85 &    8.8$\pm$   2.0 &    46$\pm$   20 &    0.7 &   $-$ &    0.8 &   F & S &         B17 &    1.3  $\pm$  0.1 &   156  $\pm$  1.1 &      9 $\pm$  1.7 &    1.35 &    0.44 &    0.48 \\
HD~287150  & 21 &   1.23 &    8.5$\pm$   2.7 &   172$\pm$   11 &    1.1 &    1.1 &    1.2 &   F & S &  2019-03-19 &    1.5  $\pm$  0.1 &    72  $\pm$  0.3 &      0 $\pm$  0.4 &    1.51 &    0.56 &    0.97 \\
HD~294264  & 19 &   19.9 &    3.0$\pm$   2.0 &    13$\pm$   26 &    17 &   $-$ &    2.8 &   F & M &  2019-03-18 &    2.8  $\pm$  0.1 &    78  $\pm$  0.3 &      3 $\pm$  0.5 &    3.14 &    0.72 &    1.37 \\
 \quad \quad $"$ & & & & & & & & & & 2019-03-19 & & & & & & \\
HD~294304  & 17 &   2.16 &    1.8$\pm$   2.0 &    68$\pm$   18 &    1.8 &    1.6 &    1.2 &   F & S &  2020-10-01 &    0.7  $\pm$  0.1 &   140  $\pm$  1.5 &    -18 $\pm$  2.8 &    0.70 &    0.52 &    1.47 \\
HD~303308  &  1 &   44.0 &    0.6$\pm$   1.4 &   148$\pm$   41 &    37 &   $-$ &    1.4 &   V & M &         B17 &    2.9  $\pm$  0.1 &    99  $\pm$  0.2 &      0 $\pm$  0.4 &    2.92 &    0.55 &    1.32 \\
HD~315021  &  1 &   25.6 &    1.4$\pm$   2.1 &   110$\pm$   36 &    21 &    1.7 &    1.2 &   F & S &  2019-02-10 &    0.4  $\pm$  0.1 &    56  $\pm$  2.3 &     11 $\pm$  2.4 &    0.44 &    0.62 &    1.38 \\
HD~315023  &  1 &   25.0 &    0.9$\pm$   2.0 &   122$\pm$   61 &    21 &    1.8 &    1.5 &   F & S &         B17 &    1.3  $\pm$  0.1 &   151  $\pm$  0.7 &      1 $\pm$  0.9 &    1.40 &    0.68 &    1.41 \\
HD~315024  &  1 &   28.3 &    1.4$\pm$   2.1 &   109$\pm$   16 &    24 &    2.1 &    1.2 &   F & S &  2020-10-04 &    0.1  $\pm$  0.1 &   183  $\pm$  9.2 &   -130 $\pm$ 29.3 &    0.10 &    0.43 &    2.50 \\
HD~315031  &  1 &   32.3 &    1.4$\pm$   2.1 &   109$\pm$   26 &    27 &   $-$ &    1.2 &   F & S &  2020-10-04 &    0.6  $\pm$  0.1 &    45  $\pm$  0.8 &     -9 $\pm$  1.8 &    0.58 &    0.64 &    0.95 \\
HD~315032  &  1 &   41.7 &    1.4$\pm$   2.1 &   109$\pm$   48 &    35 &    1.6 &    1.0 &   F & S &  2020-10-04 &    0.2  $\pm$  0.1 &    66  $\pm$  3.8 &     29 $\pm$  5.5 &    0.19 &    0.87 &    0.50 \\
HD~315033  &  1 &   41.0 &    1.0$\pm$   2.1 &   102$\pm$   83 &    34 &   $-$ &    1.4 &   V & S &  2020-10-04 &    0.4  $\pm$  0.1 &    95  $\pm$  1.5 &     17 $\pm$  3.1 &    0.41 &    0.64 &    1.65 \\
Herschel~36 &  1 &   58.9 &    1.1$\pm$   2.3 &    91$\pm$   87 &    49 &   $-$ &    0.7 &   F & S &  2020-10-05 &    6.8  $\pm$  0.2 &    95  $\pm$  0.2 &      3 $\pm$  0.5 &    7.19 &    0.64 &    1.61 \\
Walker~67  &  1 &   58.9 &    1.1$\pm$   2.3 &    91$\pm$   16 &    49 &   $-$ &    0.7 &   F & S &         B17 &    4.1  $\pm$  0.3 &    17  $\pm$  0.6 &     -6 $\pm$  1.2 &    5.17 &    0.81 &    1.47 \\
 \hline
\end{tabular}
\end{center}
\end{table*}

\end{appendix}
\end{document}